\documentclass[reqno,12pt]{amsart}
 \usepackage{amsmath}
 \usepackage{amsthm}
 \usepackage{amssymb}
 \usepackage{latexsym,longtable}
 \usepackage{graphicx}
 \usepackage{multicol}
 \usepackage{mathrsfs}
 \usepackage{young}
 \usepackage[vcentermath,enableskew,stdtext]{youngtab}
 \usepackage[left=1.1in,right=1.1in,top=1.1in,bottom=1.14in]{geometry}
 \usepackage[all]{xy}
    \SelectTips{cm}{10}     
    \everyxy={<2.5em,0em>:} 
 \usepackage{fancyhdr}      
 \usepackage{multicol}
 \usepackage{multirow}
 \usepackage{enumitem}
 \usepackage{stmaryrd}
 \usepackage{etex}
 \usepackage{tikz}
 \usepackage{float}
 \usepackage{array}
 \restylefloat{figure}
 \usetikzlibrary{shapes}
 \usetikzlibrary{positioning}
 \usetikzlibrary{snakes}
 \usetikzlibrary{decorations.pathmorphing}
 \usetikzlibrary{decorations.markings}

\usepackage{amsfonts,epsfig,color}
\newcounter{ctr}
\theoremstyle{plain}
\newtheorem{theorem}{Theorem}[section]
\newtheorem{lemma}[theorem]{Lemma}
\newtheorem{corollary}[theorem]{Corollary}
\newtheorem{proposition}[theorem]{Proposition}
\newtheorem{conjecture}[theorem]{Conjecture}
\newtheorem{problem}[theorem]{Problem}
\newtheorem{propdef}[theorem]{Proposition-Definition}

\theoremstyle{definition}
\newtheorem{definition}[theorem]{Definition}

\newtheorem{remark}[theorem]{Remark}
\newtheorem{example}[theorem]{Example}

\newcommand{\ignore}[1]{}
\newcommand{\mb}[1]{{\ensuremath{\mathbf{#1}}}}
\newcommand{\bracket}[1]{{|#1\rangle}}

\newcommand{\xsym}{{\nsbr{X}_{\mbox {sym}}}}
\newcommand{\xwedge}{{\nsbr{X}_\wedge}}

\newcommand{\aut}{\text{\rm Aut}\,}
\newcommand{\A}{{\ensuremath{\mathscr{A}}}}

\newcommand{\stack}[2]{{\genfrac{}{}{0pt}{2}{#1}{#2}}}
\newcommand{\B}{\ensuremath{\mathscr{B}}}
\renewcommand{\b}{\ensuremath{\flat}}
\newcommand{\CC}{\ensuremath{\mathbb{C}}}

\newcommand{\E}{\ensuremath{\mathscr{E}}}

\newcommand{\End}{\text{\rm End}}

\newcommand{\F}{\ensuremath{\mathscr{F}}}

\newcommand{\g}{\ensuremath{\mathfrak{g}}}
\newcommand{\gl}{\ensuremath{\mathfrak{gl}}}
\newcommand{\G}{\ensuremath{\mathscr{G}}}
\newcommand{\gr}{\text{\rm gr}}
\newcommand{\grin}{\text{\rm in}}
\renewcommand{\H}{\ensuremath{\mathscr{H}}}
\newcommand{\cH}{\ensuremath{\mathcal{H}}}

\renewcommand{\hom}{\text{\rm Hom}}

\newcommand{\id}{\text{\rm Id}}
\newcommand{\idelm}{\ensuremath{id}}

\newcommand{\im}{\text{\rm im\,}}

\renewcommand{\L}{\ensuremath{\mathscr{L}}}

\newcommand{\mat}[1]{\ensuremath{\left( %
        \begin{array}{cccccccccccccccccccccccccccc} #1 \end{array}\right)}}
\newcommand{\matb}[2]{\ensuremath{\left( %
       \begin{array}{#1} #2 \end{array}\right)}}
\newcommand{\Mod}{\ensuremath{\mathbf{Mod}}}

\renewcommand{\O}{\ensuremath{\mathscr{O}}}

\newcommand{\QQ}{\ensuremath{\mathbb{Q}}}
\newcommand{\R}{\ensuremath{\mathscr{R}}}

\renewcommand{\sl}{\ensuremath{\mathfrak{sl}}}

\newcommand{\U}{\ensuremath{\mathcal{U}}}

\newcommand{\X}{{\ensuremath{\mathcal{X}}}}
\newcommand{\ZZ}{\ensuremath{\mathbb{Z}}}

\newcommand{\be}{\begin{equation}}
\newcommand{\ee}{\end{equation}}
\newcommand{\Res}{\text{\rm Res}}
\renewcommand{\S}{\ensuremath{\mathcal{S}}}
\newcommand{\tsr}{\ensuremath{\otimes}}
\newcommand{\tsrvw}{\ensuremath{\star}}
\newcommand{\tsrdual}{\ensuremath{*}}
\newcommand{\C}{\ensuremath{C^{\prime}}} 

\newcommand{\liftc}{\ensuremath{\tilde{c}}}
\newcommand{\liftb}{\ensuremath{\tilde{b}}}

\newcommand{\br}[1]{\ensuremath{\overline{#1}}}
\newcommand{\nsbr}[1]{{\ensuremath{\check{#1}}}}

\renewcommand{\u}{\ensuremath{q}}  
\newcommand{\ui}{\ensuremath{q^{-1}}} 

\newcommand{\leftexp}[2]{{\vphantom{#2}}^{#1}{#2}}

\newcommand{\Tab}{\ensuremath{\mathcal{T}}}

 \newcommand{\dual}[1]{\ensuremath{{#1}^\vee}}
\newcommand{\dualcnv}{} 

\newcommand{\y}{\ensuremath{Y}} 

\DeclareMathOperator{\rank}{rank}

\newcommand{\gd}{\ensuremath{\trianglerighteq}} 
\newcommand{\gdneq}{\ensuremath{\triangleright}}
\newcommand{\ld}{\ensuremath{\trianglelefteq}}
\newcommand{\ldneq}{\ensuremath{\triangleleft}}
\newcommand{\tto}{\ensuremath{\rightsquigarrow}}

\newcommand{\sh}{\text{\rm sh}}

\newcommand{\transpose}[1]{{#1}^T}

\renewcommand{\P}{\ensuremath{\mathbf{P}}}
\newcommand{\NP}{\ensuremath{\mathbf{NP}}}

\newcommand{\nsH}{\nsbr{\H}}
\newcommand{\nsP}{\nsbr{\mathscr{P}}}

\newcommand{\bv}{v}
\newcommand{\bw}{w}
\newcommand{\bx}{x}
\newcommand{\by}{y}
\DeclareMathOperator{\trace}{tr}
\newcommand{\two}{[2]}
\newcommand{\sP}{\mathcal{P}}
\newcommand{\sQ}{\mathcal{Q}}
\newcommand{\field}{\ensuremath{K}}
\newcommand{\klo}[1]{\ensuremath{\leq_{#1}}}
\newcommand{\klocov}[1]{\ensuremath{\preceq_{#1}}}
\newcommand{\kloneq}[1]{\ensuremath{<_{#1}}}

\newcommand{\bT}{\ensuremath{\mathbf{T}}}

\DeclareMathOperator{\sort}{sort}
\newcommand{\Wedge}{\ensuremath{\Lambda}}
\newcommand{\tgraph}{\ensuremath{\mathcal{TG}}}
\newcommand{\cvw}[2]{\ensuremath{c_{\substack{{#1} \\ {#2}}}}}
\newcommand{\vw}[2]{\ensuremath{{\hspace{-.8mm}\tiny \ \substack{{#1} \\[-.45mm] {#2}}}}}
\newcommand{\zz}[2]{\ensuremath{z^{#2}_{#1}}}
\newcommand{\tzz}[2]{\ensuremath{\tilde {z}^{#2}_{#1}}}
\newcommand{\liftcvw}[2]{\ensuremath{\tilde{c}_{\substack{{#1} \\ {#2}}}}}

\newcommand{\heart}{\heartsuit}
\newcommand{\heartproj}{{\ensuremath{\tilde{\heart}}}}
\newcommand{\heartvw}{{\ensuremath{\heart}}}
\newcommand{\heartp}{{\ensuremath{\heart_{\nsbr{p}_-}}}}
\newcommand{\heartl}{{\ensuremath{\nsbr{\heart}}}}
\newcommand{\heartns}{{\ensuremath{\nsbr{\heart}}}}
\newcommand{\heartg}{{\ensuremath{\nsbr{\heart}_\gdneq}}}

\newcommand{\dv}{{\ensuremath{d_V}}}
\newcommand{\dw}{{\ensuremath{d_W}}}
\newcommand{\dx}{{\ensuremath{d_X}}}
\newcommand{\wl}{{\ensuremath{\mathbf{X}}}} 
\newcommand{\myvcenter}[1]{\ensuremath{\vcenter{\hbox{#1}}}}
\newcommand{\myunderbrace}[1]{\ensuremath{\underbrace{\hbox{\ensuremath{#1}}}}}

\newcommand{\arc}[1]{\ensuremath{\cup^{\text{#1}}}}
\newcommand{\exfreev}{\ensuremath{\text{ext-free}_V}}
\newcommand{\exfreew}{\ensuremath{\text{ext-free}_W}}

\newcommand{\unitary}{{\ensuremath{{\tt U}}}}
\newcommand{\Uq}{{U_q(\g_V)}}
\newcommand{\Uqt}{{\ensuremath{U_{q}^\tau}}}
\newcommand{\Uqvw}{{\ensuremath{U_q(\g_V \oplus \g_W)}}}
\newcommand{\Oqt}{{\ensuremath{\O_{q}^\tau}}}
\newcommand{\QQA}{{\ensuremath{\QQ[\u,\ui]}}}
\newcommand{\Oint}[1]{\ensuremath{\O_{\text{int}}^{\geq 0}(#1)}}
\newcommand{\Ointtau}{\ensuremath{\Oint{\Uqt}}}
\newcommand{\pNSTC}{\ensuremath{ \text{$+$HNSTC} }}
\newcommand{\pSNST}{\ensuremath{ \text{$+$HSNST} }}
\newcommand{\overbtwo}{\ensuremath{{\textstyle \frac{1}{[2]}}}}
\def\Tiny{\fontsize{6pt}{6pt}\selectfont}
\newcommand{\TinyV}{{\hspace{0mm}\Tiny \text{V}}}
\newcommand{\TinyW}{{\hspace{0mm}\Tiny \text{W}}}
\newcommand{\nssym}[2]{\ensuremath{\nsbr{S}^{#1}\nsbr{#2}}}
\newcommand{\nswedge}[2]{\ensuremath{\nsbr{\Wedge}^{#1}\nsbr{#2}}}
\newcommand{\nssymalgebra}[1]{\ensuremath{\nsbr{S}(\nsbr{X}#1)}}
\newcommand{\nswedgealgebra}[1]{\ensuremath{\nsbr{\Wedge}(\nsbr{X}#1)}}
\newcommand{\nsSchur}[1]{\ensuremath{\nsbr{\mathscr{S}}(\nsbr{X},#1)}}
\newcommand{\ssym}[2]{\ensuremath{S_q^{#1}#2}}
\newcommand{\swedge}[2]{\ensuremath{\Wedge_q^{#1}#2}}
\newcommand{\ssymalgebra}[1]{\ensuremath{S_q({#1})}}
\newcommand{\swedgealgebra}[1]{\ensuremath{\Wedge_q({#1})}}
\newcommand{\Ap}{{\ensuremath{\mathbf{A}'}}}
\newcommand{\Cbasis}{\ensuremath{\mathcal{C}}}

\newcommand{\crystal}[1]{\ensuremath{G\tilde{#1}^{\hspace{.1pt}\text{up}}}}
\newcommand{\crystalusual}[1]{\ensuremath{\tilde{#1}^{\hspace{-.8pt}\text{up}}}}
\newcommand{\rootset}{\ensuremath{[\dv-1]}}

        \newlength{\cellsizeCol}
        \newcommand\column[1]{
        \vtop{
        \let\\=\cr
        \baselineskip=-16000pt
        \lineskiplimit=16000pt
        \lineskip=0pt
        \halign{& \columncell{##} \cr \noalign{\hrule height \arrayrulewidth width \cellsizeCol} #1 \cr \noalign{\hrule height \arrayrulewidth width \cellsizeCol} \crcr} }
        \hspace{-.73ex}}

\newcommand\columnL[1]{
        \vtop{
        \let\\=\cr
        \baselineskip=-16000pt
        \lineskiplimit=16000pt
        \lineskip=0pt
        \halign{& \columncellFull{##} \cr \noalign{\hrule height \arrayrulewidth width \cellsizeCol} #1 \cr \noalign{\hrule height \arrayrulewidth width \cellsizeCol} \crcr} }
        \hspace{-.73ex}}

\newcommand\columnBB[1]{
        \vtop{
        \let\\=\cr
        \baselineskip=-16000pt
        \lineskiplimit=16000pt
        \lineskip=0pt
        \halign{& \columncellFullB{##} \cr \noalign{\hrule height \arrayrulewidth width \cellsizeCol} #1 \cr \noalign{\hrule height \arrayrulewidth width \cellsizeCol} \crcr} }
        \hspace{-.73ex}}
\newcommand\columnB[1]{
        \vtop{
        \let\\=\cr
        \baselineskip=-16000pt
        \lineskiplimit=16000pt
        \lineskip=0pt
        \halign{& \columncellFullBB{##} \cr \noalign{\hrule height .3mm width \cellsizeCol} #1 \cr \noalign{\hrule height .3mm width \cellsizeCol} \crcr} }
        \hspace{-.73ex}}

\newcommand\ctableaua[1]{
\myvcenter{\ensuremath{\column{#1}}}
}
\newcommand\ctableau[2]{
\myvcenter{\ensuremath{\column{#1}  \column{#2}}}
}
\newcommand\ctableauc[3]{
\myvcenter{\ensuremath{\column{#1}  \column{#2}  \column{#3}}}
}
\newcommand\ctableaud[4]{
\myvcenter{\ensuremath{\column{#1}  \column{#2}  \column{#3} \column{#4}}}
}

\newcommand\ctableauf[6]{
\myvcenter{\ensuremath{\column{#1}  \column{#2}  \column{#3}  \column{#4}  \column{#5}  \column{#6}}}
}

\newcommand\ctableauh[8]{
\myvcenter{\ensuremath{\column{#1}  \column{#2}  \column{#3}  \column{#4}  \column{#5}  \column{#6} \column{#7} \column{#8}}}
}

\newlength{\oldSize}
\newcommand{\ctableausmalla}[1]{
\setlength{\oldSize}{\cellsizeCol}
\cellsizeCol=1.5ex
\myvcenter{\tiny\ensuremath{\columnL{#1}}}
\setlength{\cellsizeCol}{\oldSize}
}
\newcommand{\ctableausmall}[2]{
\setlength{\oldSize}{\cellsizeCol}
\cellsizeCol=1.5ex
\myvcenter{\tiny\ensuremath{\columnL{#1}  \columnL{#2}}}
\setlength{\cellsizeCol}{\oldSize}
}
\newcommand{\ctableausmallc}[3]{
\setlength{\oldSize}{\cellsizeCol}
\cellsizeCol=1.5ex
\myvcenter{\tiny\ensuremath{\columnL{#1} \columnL{#2} \columnL{#3}}}
\setlength{\cellsizeCol}{\oldSize}
}
\newcommand{\ctableausmalld}[4]{
\setlength{\oldSize}{\cellsizeCol}
\cellsizeCol=1.5ex
\myvcenter{\tiny\ensuremath{\columnL{#1} \columnL{#2}\columnL{#3} \columnL{#4}}}
\setlength{\cellsizeCol}{\oldSize}
}

\newcommand{\ctableausmallf}[6]{
\setlength{\oldSize}{\cellsizeCol}
\cellsizeCol=1.5ex
\myvcenter{\tiny\ensuremath{\columnL{#1} \columnL{#2}\columnL{#3}\columnL{#4}\columnL{#5} \columnL{#6}}}
\setlength{\cellsizeCol}{\oldSize}
}

\newcommand{\ctableausmallh}[8]{
\setlength{\oldSize}{\cellsizeCol}
\cellsizeCol=1.5ex
\myvcenter{\tiny\ensuremath{\columnL{#1} \columnL{#2}\columnL{#3}\columnL{#4}\columnL{#5}\columnL{#6}\columnL{#7} \columnL{#8}}}
\setlength{\cellsizeCol}{\oldSize}
}

\newcommand\mybox[1]{
\vcenter{
\let\\=\cr
\baselineskip=-16000pt \lineskiplimit=16000pt \lineskip=0pt
\halign{&\boxcell{##}\cr\vline#1\vline\crcr}}}

\newcommand{\boxcell}[1]{{%
\unitlength=\cellsizeCol
\begin{picture}(1,1)
\put(0,0){\makebox(1,1){$#1$}}
\put(0,0){\line(1,0){1}}
\put(0,1){\line(1,0){1}}
\end{picture}%
}}

\newcommand\pad[1]{
\vtop{
\let\\=\cr
\baselineskip=-16000pt
\lineskiplimit=16000pt
\lineskip=0pt
\halign{& \inviscell{##} \cr #1 \crcr} }
\hspace{-.73ex}}

\newcommand{\inviscell}[1]{{%
\unitlength=\cellsizeCol
\begin{picture}(1,1)
\put(0,0){\makebox(1,1){$#1$}}
\end{picture}%
}}

\newcommand{\columncell}[1]{{%
\unitlength=\cellsizeCol
\begin{picture}(1,1)
\put(0,0){\makebox(1,1){\footnotesize \centering $#1$}}
\put(0,0){\line(0,1){1}}
\put(1,0){\line(0,1){1}}
\end{picture}%
}}

\newcommand{\columncellFull}[1]{{%
\unitlength=\cellsizeCol
\begin{picture}(1,1)
\put(0,0){\makebox(1,1){\centering $#1$}}
\put(0,0){\line(0,1){1}}
\put(1,0){\line(0,1){1}}
\end{picture}%
}}

\newcommand{\columncellFullB}[1]{{%
\unitlength=\cellsizeCol
\begin{picture}(1,1)
\put(0,0){\makebox(1,1){$\mathbf{#1}$}}
\put(0,0){\line(0,1){1}}
\put(1,0){\line(0,1){1}}
\end{picture}%
}}

\newcommand{\columncellFullBB}[1]{{%
\unitlength=\cellsizeCol
\begin{picture}(1,1)
\linethickness{0.3mm}
\put(0,0){\makebox(1,1){$\mathbf{#1}$}}
\put(0,0){\line(0,1){1}}
\put(1,0){\line(0,1){1}}
\end{picture}%
}}

\newlength{\cellsize}
\newcommand\tableau[1]{
\vcenter{
\let\\=\cr
\baselineskip=-16000pt \lineskiplimit=16000pt \lineskip=0pt
\halign{&\tableaucell{##}\cr#1\crcr}}}

\newcommand{\tableaucell}[1]{{%
\def \arg{#1}\def \void{}%
\ifx \void \arg
\vbox to \cellsize{\vfil \hrule width \cellsize height 0pt}%
\else \unitlength=\cellsize
\begin{picture}(1,1)
\put(0,0){\makebox(1,1){$#1$}}
\put(0,0){\line(1,0){1}}
\put(0,1){\line(1,0){1}}
\put(0,0){\line(0,1){1}}
\put(1,0){\line(0,1){1}}
\end{picture}%
\fi}}

\newlength{\colskip}
\newlength{\dwidth}
\newcommand{\newpad}[1]{
\settowidth{\dwidth}{$#1$}
\setlength{\colskip}{\cellsizeCol}
\addtolength{\colskip}{-\dwidth}
\hspace{.5\colskip} {#1} \hspace{.5\colskip}
}
\newcommand{\downtsr}{\newpad{\raisebox{-10pt}{\tsr}}}
\newcommand{\downheart}{\newpad{\heart}}
\newcommand{\downdots}{\newpad{{ \atop \dots}}}

\newcommand{\ttaut}{\A}
\newcommand{\dualtheta}{\ensuremath{\#}}
\newcommand{\dualone}{\ensuremath{\diamond}}

\title[GCT IV: nonstandard quantum group for the Kronecker problem]{Geometric Complexity Theory IV: nonstandard quantum group for the Kronecker problem}

\keywords{Kronecker problem, complexity theory, canonical basis, quantum group, Hecke algebra, graphical calculus}

\begin{document}

\author{Jonah Blasiak}
\address{Department of Mathematics, Drexel University, Philadelphia, PA 19104}
\email{jblasiak@gmail.com}
\thanks{2010 \emph{Mathematics Subject Classification}. Primary 33D80, 20C30, 05E10; Secondary 16S80, 11Y16.}
\thanks{J. Blasiak was supported by an NSF postdoctoral fellowship.}

\author{Ketan D. Mulmuley}
\address{The University of Chicago}
\email{mulmuley@cs.uchicago.edu}

\author{Milind Sohoni}
\address{I.I.T., Mumbai}
\email{sohoni@cse.iitb.ac.in}

\dedicatory{Dedicated to Sri Ramakrishna}

\begin{abstract}
The Kronecker coefficient $g_{\lambda \mu \nu}$ is the multiplicity of the
$GL(V)\times GL(W)$-irreducible $V_\lambda \otimes W_\mu$ in the restriction of the $GL(X)$-irreducible $X_\nu$ via the natural map $GL(V)\times GL(W) \to GL(V \otimes W)$, where  $V, W$ are  $\mathbb{C}$-vector spaces and $X = V \otimes W$.
A fundamental open problem in algebraic combinatorics is to
find a positive combinatorial formula for these coefficients.


We construct two quantum objects for this problem, which we call the nonstandard quantum group and nonstandard Hecke algebra.
We show that the nonstandard quantum group has a compact real form and its representations are completely reducible, that the nonstandard Hecke algebra is semisimple,  and that they satisfy an analog of quantum Schur-Weyl duality.

Using these nonstandard objects as a guide, we follow the approach of Adsul, Sohoni, and Subrahmanyam \cite{QuantumDeformations} to construct, in the case  $\dim(V) = \dim(W) =2$, a
representation $\nsbr{X}_\nu$ of the nonstandard quantum group that specializes to $\Res_{GL(V) \times GL(W)} X_\nu$ at  $q=1$.
We then define a global crystal basis $\pNSTC(\nu)$ of $\nsbr{X}_\nu$ that solves the two-row Kronecker problem: the number of highest weight elements of $\pNSTC(\nu)$ of weight  $(\lambda,\mu)$ is the Kronecker coefficient $g_{\lambda \mu \nu}$.
We go on to develop the beginnings of a graphical calculus for this basis, along the lines of the $U_q(\sl_2)$ graphical calculus from \cite{FK}, and use this to organize
the crystal components of  $\pNSTC(\nu)$ into eight families.
This yields a fairly simple, positive formula for two-row Kronecker coefficients, generalizing a formula in \cite{BWZ}.  As a byproduct of the approach, we also obtain
a rule for the decomposition of $\Res_{GL_2 \times GL_2 \rtimes \S_2} X_\nu$ into irreducibles.
\end{abstract}

\maketitle
\tableofcontents

\section{Introduction}
\subsection{The Kronecker problem}
This is a continuation of the series of articles \cite{GCT1,GCT2,GCT3}
on geometric complexity theory (GCT),
an approach to $\P$ vs. $\NP$ and related problems using geometry and representation theory.
A basic philosophy of this approach
is called the {\em flip}; see \cite{GCToverview,GCTflip,GCT6} for its
detailed exposition.
The flip suggests that separating the classes $\P$ and $\NP$ will require solving
difficult positivity problems in algebraic geometry and representation theory.
A central positivity problem arising here is the following fundamental problem
in the representation theory of the symmetric group.

Let  $\S_r$ denote the symmetric group on $r$ letters and let $M_\nu$ denote the $\S_r$-irreducible corresponding to the partition $\nu$.
Given three partitions $\lambda,\mu,\nu$ of  $r$,
the Kronecker coefficient $g_{\lambda \mu \nu}$ is defined to be the multiplicity of
$M_\nu$ in the tensor product $M_\lambda \otimes M_\mu$. As explained in \textsection\ref{ss basis theoretic version of Kronecker}, this is also equal to the multiplicity of the
$GL(V)\times GL(W)$-irreducible $V_\lambda \otimes W_\mu$ in the restriction of the $GL(X)$-irreducible $X_\nu$ via the natural map $GL(V)\times GL(W) \to GL(X)$, where $X = V \otimes W$.

\begin{problem}[Kronecker problem] \label{pintrokronecker}
Find a positive combinatorial formula for the Kronecker coefficients
$g_{\lambda \mu \nu}$.
\end{problem}


There are two precise related problems in complexity theory that arise in the flip:
(1) find a (positive) $\#\P$ formula for Kronecker coefficients, and harder, (2) find a polynomial time algorithm to determine whether a Kronecker coefficient is zero.

Although the Kronecker problem has been studied since the early twentieth century,
its  general case still seems out of reach.
A combinatorial interpretation for Kronecker coefficients in the case that two of the partitions are hooks was first given by Lascoux \cite{Lascoux},  and other formulae were later given by Remmel \cite{Remmel} and Rosas \cite{Rosas}.
An explicit combinatorial formula for Kronecker coefficients in the case that $\lambda$ and  $\mu$ have at most two rows, which we refer to as the two-row case,
was given by Remmel and Whitehead in \cite{RW}.  Later, a formula for this case, not obviously equivalent to Remmel and Whitehead's, was given by Rosas \cite{Rosas}.  Using Rosas's work, Briand, Orellana, and Rosas give a piecewise quadratic quasipolynomial formula for the
two-row case \cite{BOR}. Though these formulae for the two-row case are quite explicit, none of them is positive and hence do not  solve the Kronecker problem in this case.
Briand-Orellana-Rosas \cite{BOR, BOR2} and Ballantine-Orellana \cite{BO} have also made progress on the Kronecker problem for
the special case of reduced Kronecker coefficients, sometimes called the stable limit, in which the first part of the partitions $\lambda,\mu,\nu$ is large.


In addition to the connections to complexity theory discussed in \cite{GCT1,GCT2,GCT3}, the Kronecker problem also has connections to quantum information theory \cite{Bravyi, BWZ} and the geometry of the  $GL_a \times GL_b \times GL_c$-variety  $\CC^a \tsr \CC^b \tsr \CC^c$ \cite{LandsbergBook, LandsbergManivel, BI, AS}.
See \cite{macdonald,stanley,St} for more on its history and significance.

In this paper we focus on a stronger, basis-theoretic version of the Kronecker problem which, to our knowledge,
has not been studied  in the literature (even in the two-row case).
As will be described in more precision and detail in \textsection\ref{ss basis theoretic version of Kronecker}--\ref{ss towards an upper canonical basis for nsX}, this version asks for a canonical basis for
$\Res_{GL(V) \times GL(W)} X_\nu$ (actually, a quantization of this module)
such that the labels of the highest weight basis elements of weight $(\lambda, \mu)$ give a combinatorial formula for  $g_{\lambda \mu \nu}$.
We believe this basis-theoretic strengthening to be important because (1) it is what is ultimately needed in GCT (see \cite{GCT6}), (2) it may be useful for better understanding the $GL_a \times GL_b \times GL_c$-variety  $\CC^a \tsr \CC^b \tsr \CC^c$, (3) the structure coefficients for the action of the Chevalley generators on the basis may have certain positivity properties and an interesting geometric interpretation, and (4) making more demands on combinatorial objects that count Kronecker coefficients may make them easier to find.

In this paper we give an approach to the   basis-theoretic version of the
Kronecker problem and implement it successfully in the two-row case. The approach uses two new quantum objects,
the nonstandard quantum group and nonstandard Hecke algebra. In the two-row case, we construct a representation $\nsbr{X}_\nu$ of the nonstandard quantum group that specializes to $\Res_{GL(V) \times GL(W)} X_\nu$ at  $q=1$.  We then define a canonical basis for $\nsbr{X}_\nu$ and use this to obtain an explicit formula for two-row Kronecker coefficients.
Much of the machinery developed here extends to more general cases than the two-row case.
Additionally, the sequels \cite{GCT7,canonical} describe a
nonstandard quantum group, nonstandard Hecke algebra, and a conjectural scheme
for constructing positive canonical bases of their representations for the more general
plethysm problem \cite{macdonald,stanley} of which the Kronecker problem considered in
this article is a special case.
We have not yet been able to use the machinery developed in this paper or the sequels to solve the Kronecker problem outside the two-row case or the plethysm problem because explicit computation of the canonical bases
is much harder than in the two-row case. Nonetheless, we hope that the concrete implementation in the two-row case
illustrates and supports the approach in general.
The remainder of the introduction
summarizes the approach and its implementation in the two-row case.


\subsection{The basis-theoretic version of the Kronecker problem}
\label{ss basis theoretic version of Kronecker}

Let $V, W$ be  $\QQ$-vector spaces of dimensions $\dv,\dw$, respectively, considered as the natural representations of $U(\g_V), U(\g_W)$, respectively, where $\g_V$ denotes the Lie algebra  $\gl(V)$. Set $X = V \tsrvw W$, where
$\tsrvw$ is the symbol we use for tensor product between objects associated to $V$ and objects associated to $W$, to distinguish these from other tensor products.  There is a natural algebra homomorphism
\be\label{e gV gW to gX}
U(\g_V \oplus \g_W) = U(\g_V) \tsrvw U(\g_W) \to U(\g_X)
\ee
corresponding to the group homomorphism  $GL(V) \times GL(W) \to GL(X),\ (g,g') \mapsto g \tsrvw g'$.

The vector space  $X^{\tsr r}$ becomes a left $U(\g_X)$-module via the coproduct of  $U(\g_X)$ and this left action commutes with the right action of $\S_r$ given by permuting tensor factors.  Schur-Weyl duality says that, as an  $(U(\g_X),\QQ \S_r)$-bimodule,
\be \label{e Schur-Weyl intro X}
X^{\tsr r} \cong \bigoplus_{\nu \vdash_\dx r} X_{\nu} \tsr M_{\nu},
\ee
where $X_{\nu}$ is the irreducible $U(\g_X)$-module of highest weight $\nu$ and  $\nu \vdash_\dx r$ means that $\nu$ is a partition of  $r$ with  at most  $\dx := \dv \dw$ parts.
We can also apply Schur-Weyl duality for $V^{\tsr r}$ and  $W^{\tsr r}$ to obtain
\be \label{e Schur-Weyl intro VW}
V^{\tsr r} \tsrvw W^{\tsr r} \cong
\bigoplus_\stack{\lambda \vdash_\dv r}{\mu \vdash_\dw r} (V_{\lambda}\tsr M_{\lambda}) \tsrvw (W_{\mu} \tsr M_{\mu})
\cong \bigoplus_\stack{\lambda \vdash_\dv r,\, \mu \vdash_\dw r}{\nu \vdash_\dx r} (V_\lambda \tsrvw W_\mu \tsr M_\nu)^{\oplus g_{\lambda \mu \nu}}.
\ee
Putting \eqref{e Schur-Weyl intro X} and \eqref{e Schur-Weyl intro VW} together, we obtain the $(U(\g_V \oplus \g_W),\QQ \S_r)$-bimodule isomorphism
\be
\label{e Schur-Weyl intro}
\bigoplus_{\nu \vdash_\dx r} X_{\nu} \tsr M_{\nu} \cong \bigoplus_\stack{\lambda \vdash_\dv r,\, \mu \vdash_\dw r}{\nu \vdash_\dx r} (V_\lambda \tsrvw W_\mu \tsr M_\nu)^{\oplus g_{\lambda \mu \nu}}.
\ee
Thus it is easily seen here
that the Kronecker coefficient  $g_{\lambda \mu \nu}$ is also the multiplicity of  $V_\lambda \tsrvw W_\mu$ in $\Res_{U(\g_V \oplus \g_W)} X_\nu$, where the restriction is via the map \eqref{e gV gW to gX}.

We have decided that the isomorphism \eqref{e Schur-Weyl intro} coming from Schur-Weyl duality is a good setting to study the Kronecker problem because it allows both descriptions of Kronecker coefficients to be seen simultaneously.
It also suggests a way to make more demands on a combinatorial formula for Kronecker coefficients---\emph{in the hopes that demanding more structure on the combinatorial objects will make them easier to find}.
We would like to obtain, not only a set of objects that count Kronecker coefficients, but stronger, a bijection between objects indexing both sides of \eqref{e Schur-Weyl intro}, which amounts to a bijection
\be \label{e bijection goal}
\bigsqcup_{\nu} SSYT_{\dx}(\nu) \times SYT(\nu) \cong \bigsqcup_{\lambda, \mu, \nu} SSYT_{\dv}(\lambda) \times SSYT_{\dw}(\mu) \times SYT(\nu)\times [g_{\lambda \mu \nu}],
\ee
where  $[k]$ denotes the set  $\{1,\ldots,k\},$  SSYT$_l(\nu)$ denotes the set of semistandard Young tableaux of shape $\nu$ and with entries in  $[l]$, and SYT$(\nu)$ denotes the set of standard Young tableaux of shape $\nu$.
Stronger still, we would like to find a basis for $X^{\tsr r}$ whose cells (cells are defined as a general notion for any module with basis in  \textsection\ref{ss cells}) correspond to the decompositions in \eqref{e Schur-Weyl intro} and whose labels are indexed by either side of \eqref{e bijection goal}; this is explained in more detail in the next subsections.

However, nothing easy seems to work. One difficulty is that there does not seem to be a way to obtain a bijection between the weight basis $x_1, \dots, x_{\dx}$ of $X$ and the weight basis $\{v_i\tsrvw w_j\}_{i \in [\dv], j \in [\dw]}$ of $V\tsrvw W$ that is compatible with the Kronecker problem.
The approach seems to be lost without some additional structure. So to aid it, we add structure from quantum groups and Hecke algebras, and try to apply the theory of canonical bases.

\subsection{Canonical bases connect quantum Schur-Weyl duality with RSK}
To get an idea of the basis-theoretic solution to the Kronecker problem we are after, let us see how the canonical basis of  $V^{\tsr r}$ nicely connects quantum Schur-Weyl duality with the RSK correspondence.  From this picture, we can also see two different ways that canonical bases yield a combinatorial formula for Littlewood-Richardson coefficients, which is another reason we have turned to canonical bases for a solution to the Kronecker problem.

Let $\Uq$ be the quantized enveloping algebra over $\field= \QQ(q)$ and $\H_r$ the type $A_{r-1}$ Hecke algebra over $\mathbf{A} = \ZZ[\u, \ui]$ (see  \textsection\ref{ss quantized enveloping algebra} and  \textsection\ref{s Canonical bases of the type A Hecke algebra} for precise definitions and conventions).
From now on, we write $V_\lambda$ (resp.  $M_\lambda$) for the irreducible $\Uq$-module (resp.  $\field\H_r$-module) corresponding to $\lambda$ and let  $V_\lambda|_{q=1}$ (resp.  $M_\lambda|_{q=1}$) denote the corresponding $U(\g_V)$-module (resp.  $\QQ\S_r$-module).
Schur-Weyl duality generalizes nicely to the quantum setting:
\begin{theorem}[Jimbo \cite{Jimbo}]\label{c intro Schur-Weyl duality basic}
As a $(\Uq, \field\H_r)$-bimodule, $V^{\tsr r}$ decomposes into irreducibles as
\be \label{e intro quantum Schur-Weyl}
V^{\tsr r} \cong \bigoplus_{\lambda \vdash_\dv r}  V_\lambda \tsr M_\lambda.
\ee
\end{theorem}
This algebraic decomposition has a combinatorial underpinning, which is the bijection
\be \label{e intro RSK correspondence}
 [\dv]^r \cong \bigsqcup_{\lambda \vdash_\dv r}\text{SSYT}_{\dv}(\lambda) \times \text{SYT}(\lambda),\ \mathbf{k} \mapsto (P(\mathbf{k}), Q(\mathbf{k})),
 \ee
given by the RSK correspondence, where $P(\mathbf{k})$ (resp.   $Q(\mathbf{k})$) denotes the insertion (resp. recording) tableau of the word $\mathbf{k}$.

Now the upper canonical basis $B^r_V := \{c_\mathbf{k}: \mathbf{k} \in [\dv]^r\}$ of  $V^{\tsr r}$ can be defined by $c_\mathbf{k} := v_{k_1} \heart \dots \heart v_{k_r}, \ \mathbf{k} = k_1 ,\ldots, k_r \in  [\dv]^r$, where  $\heart$ is like the $\diamond$ of  \cite{LBook} for tensoring based modules, adapted to upper canonical bases, as explained in \cite{BProjected, Brundan} and reviewed in \textsection\ref{ss upper canonical basis of bT}.

The basis $B^r_V$ has cells corresponding to the decomposition \eqref{e intro quantum Schur-Weyl} and labels to \eqref{e intro RSK correspondence}, as the following theorem makes precise.
\begin{theorem}[\cite{GL} (see {\cite[Corollary 5.7]{BProjected}} and Theorem \ref{t standard Schur-Weyl duality2})]
\label{t intro standard Schur-Weyl duality}
\
\begin{list}{\emph{(\roman{ctr})}} {\usecounter{ctr} \setlength{\itemsep}{1pt} \setlength{\topsep}{2pt}}
\item The $ \H_r$-module with basis $(V^{\tsr r}, B^r_V)$ decomposes into $ \H_r$-cells as
\[B^r_V = \bigsqcup_{\lambda \vdash_\dv r, \ T \in \text{SSYT}_{\dv}(\lambda)} \Gamma_T, \quad \text{where }\ \Gamma_T := \{c_\mathbf{k} : P(\mathbf{k}) = T \}.\]
\item The $ \H_r$-cellular subquotient spanned by $\Gamma_T$ is isomorphic to $M_{\sh(T)}$,
where $\sh(T)$ denotes the shape of $T$.
\item The  $\Uq$-module with basis $(V^{\tsr r}, B^r_V)$ decomposes into $\Uq$-cells as
\[B^r_V = \bigsqcup_{\lambda \vdash_\dv r, \ T \in \text{SYT}(\lambda)} \Lambda_T, \quad \text{where }\ \Lambda_T  = \{c_\mathbf{k} : Q(\mathbf{k}) = T \}.\]
\item The $\Uq$-cellular subquotient spanned by $ \Lambda_T$ is isomorphic to $V_{\sh(T)}$.
\end{list}
\end{theorem}

\begin{figure}[H]
\begin{tikzpicture}[xscale = 4.7,yscale = 2.8, xshift = -.7]
\tikzstyle{vertex}=[inner sep=0pt, outer sep=3pt, fill = white]
\tikzstyle{edge} = [draw, thick, ->,black]
\tikzstyle{LabelStyleH} = [text=black, anchor=south]
\tikzstyle{LabelStyleV} = [text=black, anchor=east]

    \node[vertex] (111) at (3,0) {$c_{111} = \bv_{111} $};

    \node[vertex] (112) at (2,-1) {$c_{112} = \bv_{112}$};
    \node[vertex] (121) at (2, 0) {$c_{121} = \bv_{121} - \ui \bv_{112} $};
    \node[vertex] (211) at (2,1) {$c_{211}= \bv_{211}- \ui \bv_{121}$};

    \node[vertex] (122) at (1, -1) {$c_{122}= \bv_{122} $};
    \node[vertex] (212) at (1 , 0) {$c_{212}  = \bv_{212}- \ui \bv_{122} $};
    \node[vertex] (221) at (1, 1) {$c_{221}= \bv_{221}- \ui \bv_{212}$};

    \node[vertex] (222) at (0, 0) {$c_{222}= \bv_{222} $};

\draw[edge] (111) to node[LabelStyleH]{\footnotesize $[3]$} (112);
\draw[edge] (111) to node[LabelStyleH]{\footnotesize $[2]$} (121);
\draw[edge] (111) to node[LabelStyleH]{\footnotesize $ 1$} (211);
\draw[edge] (112) to node[LabelStyleH]{\footnotesize $ [2]$} (122);
\draw[edge] (112) to node[LabelStyleH]{\footnotesize $ 1$} (212);
\draw[edge] (121) to node[LabelStyleH, near start]{\footnotesize $ 1$} (221);
\draw[edge] (211) to node[LabelStyleH, near start]{\footnotesize $ 1$} (212);
\draw[edge] (122) to node[LabelStyleH]{\footnotesize $ 1$} (222);

\foreach \y in {-.2}{
  \node[vertex] (t111) at (3,0+\y) { $\left(\tiny \tableau{ 1 & 1 & 1}, \tableau{1 & 2 & 3}\right)$};

    \node[vertex] (t112) at (2,-1+\y) { $\left(\tiny \tableau{ 1 & 1 & 2}, \tableau{1 & 2 & 3}\right)$};
    \node[vertex] (t121) at (2, 0+1.65*\y) { $\left(\tiny \tableau{ 1 & 1 \cr 2}, \tableau{1 & 2 \cr 3}\right)$};
    \node[vertex] (t211) at (2, 1+1.65*\y) { $\left(\tiny \tableau{ 1 & 1 \cr 2}, \tableau{1 & 3\cr 2}\right)$};

    \node[vertex] (t122) at (1, -1+\y) { $\left(\tiny \tableau{ 1 & 2 & 2}, \tableau{1 & 2 & 3}\right)$};
    \node[vertex] (t212) at (1 , 0 +1.65*\y) { $\left(\tiny \tableau{ 1 & 2\cr 2}, \tableau{1 & 3\cr 2}\right)$};
    \node[vertex] (t221) at (1, 1 +1.65*\y) {$\left(\tiny \tableau{ 1 & 2\cr 2}, \tableau{1 & 2 \cr 3}\right)$};

    \node[vertex] (t222) at (0, 0+\y) {$\left(\tiny \tableau{2 & 2 & 2}, \tableau{1 & 2 & 3}\right)$};
}
\end{tikzpicture}
\caption{An illustration of Theorem \ref{t intro standard Schur-Weyl duality} for  $r=3,\ \dv=2$.  The notation $\bv_\mathbf{k}$,  $\mathbf{k} \in [\dv]^r$, denotes the tensor monomial
 $v_{k_1} \tsr \cdots \tsr v_{k_r}$.
The pairs of tableaux are of the form $(P(\mathbf{k}), Q(\mathbf{k}))$.  The arrows and their coefficients give the action of $F \in \Uq$ on the upper canonical basis  $B^3_V$, where
 $[k] := \frac{\u^k - \u^{-k}}{\u - \ui}$.}
\label{f c_111 example upper}
\end{figure}
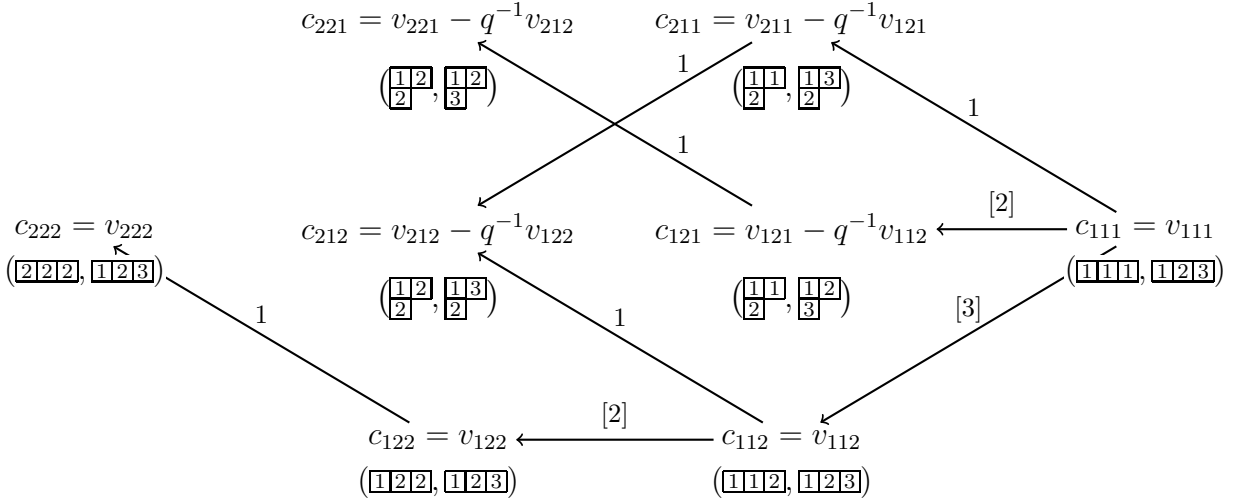

We can also use Theorem \ref{t intro standard Schur-Weyl duality} to obtain two formulae for the Littlewood-Richardson coefficients $c_{\lambda \, \mu}^\nu$: one comes from reading off the $\Uq$-cells of shape $\nu$ in a tensor product $\field \Lambda_{Q^1} \tsr \field \Lambda_{Q^2}$,  $\sh(Q^1) = \lambda, \ \sh(Q^2) = \mu$, and the other from reading off the  $\H_r$-cells of shape  $\nu$ in an induced module $\H_r \tsr_{\H_{k} \tsr \H_{r-k}} (\mathbf{A} \Gamma_{P^1} \tsr \mathbf{A} \Gamma_{P^2})$, where  $\lambda \vdash k$,  $\mu \vdash r-k$,
$\sh(P^1) = \lambda,$ and $\sh(P^2) = \mu$
 (see \cite[\textsection4.2]{B0}).

Our refined goal is now to find a canonical basis of $X^{\tsr r}$ with cells corresponding to
\eqref{e Schur-Weyl intro} and labels to \eqref{e bijection goal}, but now $X$ will be a $\field$-vector space
so that the basis can perhaps be defined by a globalization procedure like that used for quantum groups in \cite{Kas1,Kas2}.
To do this, we first need to give $X^{\tsr r}$ the structure of a bimodule for quantum objects that are suited to this problem.  This is addressed in the next subsection.
Then in \textsection\ref{ss towards an upper canonical basis for nsX} and  \textsection\ref{ss intro global crystal basis for Xnu}, we return to the construction of the basis.


\subsection{The nonstandard quantum group and Hecke algebra}
\label{ss intro nonstandard quantum group and Hecke algebra}
We seek quantum objects that play an analogous role for $X^{\tsr r}$ that $\H_r$ and $\Uq$ do for $V^{\tsr r}$.
We also require that these objects be compatible with the commuting actions of $\H_r \tsr \H_r$ and $\Uqvw$ on $X^{\tsr r}$.
The resulting quantized objects we have arrived at are the nonstandard Hecke algebra $\nsH_r$ and the nonstandard quantum group $GL_q(\nsbr{X})$. These objects are, in a certain sense, the best possible quantizations satisfying these requirements.

We point out that new quantum objects are necessary for this problem. The commuting actions of $\H_r$ and $U_q(\g_X)$ on $X^{\tsr r}$ are not satisfactory quantizations of the commuting $\S_r$ and $U(\g_X)$ actions, given the compatibility requirements just mentioned. On the Hecke algebra side, this is because the Hecke algebra is not a Hopf algebra in any natural way. Similarly, on the quantum group side, it can be shown \cite{hayashi} that the homomorphism
\eqref{e gV gW to gX} cannot be quantized in the category of Drinfel$'$d-Jimbo quantum groups.

The \emph{nonstandard Hecke algebra} $\nsH_r$ is the subalgebra of $\H_r \tsr \H_r$ generated by the elements
\be
\sP_i := \C_{s_i} \tsr \C_{s_i} + C_{s_i} \tsr C_{s_i},\ i \in \rootset,
\ee
where $\C_{s_i}$ and $C_{s_i}$ are the simplest lower and upper Kazhdan-Lusztig basis elements, which are proportional to the trivial and sign idempotents of the parabolic sub-Hecke algebra $(\H_r)_{\{s_i\}}$.
We think of the inclusion $\nsbr{\Delta}:\nsH_r \hookrightarrow   \H_r \tsr \H_r$ as a deformation of the coproduct
$\Delta_{\ZZ \S_r} :\ZZ \S_r \to \ZZ \S_r \tsr \ZZ \S_r$,  $w \mapsto w \tsr w$.
As is explained more precisely in Remark \ref{r nsH as small as possible},  the nonstandard Hecke algebra is the subalgebra of $\H_r \tsr \H_r$ making
$\nsbr{\Delta}$ as close as possible to $\Delta_{\ZZ \S_r}$ at  $\u =1$.

To define the nonstandard quantum group $GL_q(\nsbr{X})$, we follow the approach of \cite{RTF,KS} to quantum groups.
We now recall a few of the relevant concepts, leaving a thorough review to \textsection\ref{ss FRT algebras}--\ref{ss compact}.
The quantum group $GL_q(V)$ is not an actual group, but just a virtual object associated to the quantized enveloping algebra $\Uq$ and the quantum coordinate algebra  $\O(GL_q(V))$.  These are dually paired Hopf algebras, which implies that any  $\O(GL_q(V))$-comodule is a $\Uq$-module.  In this paper, we are only interested in the $V_\lambda$ for partitions $\lambda$, which are both $\Uq$-modules and $ \O(GL_q(V))$-comodules, so (at least for our purposes) $\Uq$ and  $\O(GL_q(V))$ provide dual approaches to the same objects.
The quantum coordinate algebra $\O(M_q(V))$ is defined to be the FRT-algebra $A(\R_{V,V})$ \cite{RTF} associated to the $\R$-matrix $\R_{V,V}\in\End(V^{\tsr 2})$, which is a quotient of the tensor bialgebra $T(U)=\bigoplus_{r\geq 0} U^{\tsr r}$, $U:= V\tsr V^*$, that specializes to $\O(M(V))$ at $q=1$.
The Hopf algebra $\O(GL_q(V))$ is then defined from $\O(M_q(V))$ by inverting the quantum determinant.

The nonstandard quantum group $GL_q(\nsbr{X})$ is a virtual object associated to the \emph{nonstandard coordinate algebra}  $\O(GL_q(\nsbr{X}))$ (we have yet
to construct a nonstandard enveloping algebra dual to  $ \O(GL_q(\nsbr{X}))$, but we think this is possible). To define  $\O(GL_q(\nsbr{X}))$, we first define the \emph{nonstandard coordinate algebra} $\O(M_q(\nsbr{X}))$.  This is most quickly defined as the FRT-algebra $A(P_+^{\nsbr{X}})$, where $P_+^{\nsbr{X}} \in \End(\nsbr{X}^{\tsr 2})$ is equal to the action of $\frac{1}{[2]^2} \sP_1$ on  $\nsbr{X}^{\tsr 2}$ ($\sP_1$ acts on  $\nsbr{X}^{\tsr 2}$ by quantum Schur-Weyl duality for  $V^{\tsr 2}$ and  $W^{\tsr 2}$); see  \textsection\ref{ss definition of OnsX} for details.  Here and throughout the paper, $\nsbr{X}$ is the same as $X$, the decoration indicating that it is associated to a nonstandard object.


Much of the abstract theory of the standard quantum group $GL_q(V)$ can be replicated in the nonstandard case, but explicit computations become significantly harder.
For instance, we can define nonstandard symmetric and exterior algebras $\nssymalgebra{}$ and $\nswedgealgebra{}$ (\textsection\ref{ss definitions ns symmetric exterior}), which are $\O(M_q(\nsbr{X}))$-comodule algebras and specialize to the symmetric and exterior algebras of $X$ at $q=1$.
However, $\nssymalgebra{}$ is already isomorphic to  $\O(M_q(V))$ when  $W = V^*$, and thus explicitly determining the multiplication in this algebra, in terms of the Gelfand-Tsetlin basis, say, has been intensively studied and is still not completely understood (see \cite{KS,vilenkin} and  \textsection\ref{sexpproduct}). 
Understanding  $\O(M_q(\nsbr{X}))$ explicitly is yet another level of difficulty beyond this. We show (Appendix \ref{s reduction system}) that a natural reduction system for this coordinate algebra does not satisfy the diamond property.

Let $\nswedge{r}{X}$ denote the degree $r$ part of  $\nswedgealgebra{}$.
The \emph{nonstandard determinant} $\nsbr{D}$ is defined to be the matrix coefficient of the comodule $\nswedge{\dx}{X}$. This object is somewhat mysterious in that we do not understand it explicitly (in the monomial basis, say). Nonetheless, we show (\textsection\ref{s nonstandard quantum groups GLq}) that $\O(M_q(\nsbr{X}))[\frac{1}{\nsbr{D}}]$ can be given a Hopf algebra structure. The result is the \emph{nonstandard coordinate algebra} $\O(GL_q(\nsbr{X}))$. We now state our main theorem about this object, which is proved in \textsection\ref{squantumdet} and \textsection\ref{s nonstandard quantum groups GLq}.

\begin{theorem}[Theorem \ref{tcmqg}] \label{t intro cmqg}
Assume that all objects are over $\CC$ and $q$ is real and transcendental. Then

\noindent (a)
The Hopf algebra $\O(GL_q(\nsbr{X}))$ can be made into a Hopf $*$-algebra.
This is considered to be the coordinate ring of the compact real form of the nonstandard quantum group $GL_q(\nsbr{X})$. This virtual compact real form is
denoted $\unitary_q(\nsbr{X})$, which is a compact quantum group
 in the sense of Woronowicz \cite{wor1}.

\noindent (b) There is a Hopf  $*$-algebra homomorphism
\[
\tilde{\psi}: \O(GL_q(\nsbr{X})) \rightarrow \O(GL_q(V)) \otimes \O(GL_q(W)),
\]

\noindent (c) Every finite-dimensional representation of $\unitary_q(\nsbr{X})$ is unitarizable, and
hence, is a direct sum of irreducible representations.

\noindent (d) An analog of the Peter-Weyl theorem holds:
\[
\O(GL_q(\nsbr{X})) = \bigoplus_{\alpha \in \nsP} \nsbr{\X}_\alpha^* \otimes \nsbr{\X}_\alpha,
\]
where $\nsP$ is an index set for the irreducible right comodules of $\O(GL_q(\nsbr{X}))$ and $\nsbr{\X}_\alpha$
is the comodule labeled by $\alpha$.
\end{theorem}

In a similar spirit, we show (Proposition \ref{p nsH semisimple}) that the nonstandard Hecke algebra $\field \nsH_r$ is semisimple.
As far as the representation theory of $\nsH_r$ and $\O(GL_q(\nsbr{X}))$ are concerned, we have had more luck understanding that of $\nsH_r$.
Fortunately, as the next result shows, we can transfer our knowledge of $\field\nsH_r$-irreducibles to $\O(GL_q(\nsbr{X}))$-irreducibles.

Just as in the standard case, there are commuting actions of the nonstandard Hecke algebra and nonstandard quantum group on  $\nsbr{X}^{\tsr r}$.
Since we do not yet have a nonstandard enveloping algebra dual to  $ \O(GL_q(\nsbr{X}))$, we instead work with the \emph{nonstandard Schur algebra}, denoted $\field \nsSchur{r}$, which is defined to be the algebra dual to the coalgebra $\O(M_q(\nsbr{X}))_r$. We have the following nonstandard analog of quantum Schur-Weyl duality.
\begin{theorem}[Theorem \ref{t nonstandard schur-weyl duality}] \label{t intro nonstandard schur-weyl duality}
As a $(\field \nsSchur{r}, \field \nsH_r)$-bimodule,  $ \nsbr{X}^{\tsr r}$ decomposes into irreducibles as
\be \label{e intro nonstandard Schur-Weyl duality}
\nsbr{X}^{\tsr r} \cong \bigoplus_{\alpha \in \nsP_r} \nsbr{\X}_\alpha \tsr \nsbr{M}_\alpha,
\ee
where $\nsP_r$ is an index set so that $\nsbr{\X}_\alpha$ ranges over $\field \nsSchur{r}$-irreducibles and $\nsbr{M}_\alpha$ ranges over $\nsH_r$-irreducibles.
\end{theorem}
We deduce that there are nonnegative integers $n^{\lambda,\mu}_\alpha= n_{\lambda,\mu}^\alpha$ that correspond to the multiplicities in the following two decomposition problems:
\be \label{e intro n lambda mu multiplicities}
\nsbr{\X}_\alpha \cong \bigoplus_{\lambda, \mu}(V_\lambda \tsrvw W_\mu)^{\oplus n^{\lambda,\mu}_\alpha}, \quad  \Res_{\nsH_r} M_\lambda \tsrvw M_\mu \cong \bigoplus_{\alpha} \nsbr{M}_\alpha^{\oplus n_{\lambda,\mu}^\alpha}.
\ee

The representation theory of the nonstandard Hecke algebra and quantum group thus decompose the Kronecker problem into two steps:
\begin{list}{(\roman{ctr})} {\usecounter{ctr} \setlength{\itemsep}{1pt} \setlength{\topsep}{2pt}}
\item Determine the multiplicity $n_{\lambda,\mu}^\alpha $ of the irreducible $\nsH_r$-module  $\nsbr{M}_\alpha$ in the tensor product $M_\lambda \otimes M_\mu$.
Equivalently, determine the multiplicity $n^{\lambda,\mu}_\alpha$ of the irreducible  $\O(GL_q(V))\tsrvw\O(GL_q(W))$-comodule $V_\lambda \tsrvw W_\mu$ in
the irreducible  $ \O(M_q(\nsbr{X}))$-comodule  $\nsbr{\X}_\alpha$.
\item Determine the multiplicity $m_{\alpha\nu}$ of the  $\S_r$-irreducible  $M_\nu|_{q=1}$ in $\nsbr{M}_\alpha|_{\u =1}$.
\end{list}
The resulting formula for Kronecker coefficients is
\be \label{e ns Kronecker decomposition}
g_{\lambda \mu \nu}= \sum_\alpha n_{\lambda,\mu}^\alpha m_{\alpha\nu}.
\ee
Thus a positive combinatorial formula for $n_{\lambda,\mu}^\alpha$ and $m_{\alpha\nu}$ would yield  one for Kronecker coefficients.

Unfortunately, this does not get us very far. Despite its being as small as possible, $\nsH_r$ has dimension much larger than that of $\S_r$.
Similarly, despite its being as large as possible, $\O(M_q(\nsbr{X}))_r$ has dimension much smaller than that of $\O(M(X))_r$ (see Remark \ref{r ns coordinate algebra as large as possible} and Proposition \ref{ppoincare}).
It turns out that in the two-row ($\dv = \dw =2$) case,  $\nsH_{r,2}$ is quite close to  $S^2 \H_r$ and the nonstandard coordinate algebra  $ \O(GL_q(\nsbr{X}))$ is close to
the smash coproduct $\Oqt := \O(GL_q(V))\tsrvw\O(GL_q(W)) \rtimes \F(\S_2)$,
where  $\nsH_{r,2}$ is the image of  $\nsH_r$ in  $\End(\nsbr{X}^{\tsr r})$ when  $\dv = \dw = 2$ and
$\F(\S_2)$ is the Hopf algebra of functions on $\S_2$; see \cite{B4}, \textsection\ref{ss nonstandard two-row case}, and Appendix \ref{s appendix Oqt} for details.
Thus most of the work is left to determining the multiplicities in (ii).

However, we have gained \emph{something}. In addition to the slight help that \eqref{e ns Kronecker decomposition} provides, finding a basis for $\nsbr{M}_\alpha$ whose cells are compatible with the decomposition  $\nsbr{M}_\alpha|_{q=1} \cong \bigoplus_\nu (M_\nu|_{q=1})^{\oplus m_{\alpha\nu}}$ has significantly more structure than finding a basis for $M_\lambda\tsr M_\mu|_{q=1}$ compatible with the decomposition  $M_\lambda\tsr M_\mu|_{q=1} \cong \bigoplus_\nu (M_\nu|_{q=1})^{\oplus g_{\lambda \mu \nu}}$, despite the fact that $\nsbr{M}_\alpha$ is typically equal to some $\Res_{\nsH_r} M_\lambda\tsr M_\mu$; that is, finding a basis \emph{before} specializing  $q=1$ has more structure than finding it \emph{after} specializing.

Also, it is shown in \cite{B4} that the restriction of an  $\nsH_{r,2}$-irreducible to $\nsH_{r-1,2}$ is multiplicity-free.  The seminormal basis (in the sense of \cite{RamSeminormal}) of some $\nsbr{M}_\alpha = \Res_{\nsH_{r,2}} M_\lambda\tsr M_\mu$ coming from restricting along the chain  $\nsH_{1,2} \subseteq \cdots \subseteq \nsH_{r-1,2} \subseteq \nsH_{r,2}$ is significantly different from the one coming from the chain $\H_{1,2} \tsr \H_{1,2} \subseteq \cdots \subseteq \H_{r-1,2} \tsr \H_{r-1,2} \subseteq \H_{r,2} \tsr \H_{r,2}$.  Here,  $\H_{r,2}$ denotes the Temperley-Lieb algebra (see \textsection\ref{ss nonstandard two-row case}).
The article \cite{canonical} suggests a conjectural scheme for constructing
a canonical basis of $\nsbr{M}_\alpha$ using the former chain, but we
have not been able to prove its correctness.
We therefore follow a different path to construct a canonical basis for the two-row Kronecker problem, described below.
Along similar lines, it is illustrated in  \textsection\ref{ss two-column relations} that, though the difference between $\Oqt$ and $\O(GL_q(V))\tsrvw\O(GL_q(W))$ is small, the little bit of extra structure added by considering $\Oqt$-comodules can be quite important.

\subsection{Towards an upper canonical basis for  $\nsbr{X}^{\tsr r}$}
\label{ss towards an upper canonical basis for nsX}
Our goal is now to construct a basis of $\nsbr{X}^{\tsr r}$ with  $\field \nsSchur{r}$-cells and  $\nsH_r$-cells that are compatible with the decomposition
\be
\nsbr{X}^{\tsr r} \cong \bigoplus_{\alpha \in \nsP_r} \nsbr{\X}_\alpha \tsr \nsbr{M}_\alpha,
\ee
and so that after specializing $\u=1$, the cells are compatible with the decomposition
\be \label{e intro X tsr M}
\bigoplus_{\nu \vdash_\dx r} X_\nu|_{q=1} \tsr M_\nu|_{q=1}.
\ee
See Conjecture \ref{cj canonical basis X^r} for a more precise and detailed statement.
After many failed attempts, we succeeded in constructing such a basis in the two-row case for  $r$ up to 4 (see Examples \ref{ex ns Schur-Weyl duality canonical basis r3} and \ref{ex ns Schur-Weyl duality canonical basis r4}) and for some highest weight spaces of  $\nsbr{X}^{\tsr 5}$ and $\nsbr{X}^{\tsr 6}$.  We are therefore quite hopeful that such a basis exists in general.  However, the construction of global crystal bases from a balanced triple \cite{Kas1, Kas2} and the similar theory of based modules \cite{LBook} does not seem to be enough here.

Though the construction of a basis for all of $\nsbr{X}^{\tsr r}$ remains unfinished, we have been able to construct one so-called \emph{fat cell}  $\nsbr{X}_\nu$ of $\nsbr{X}^{\tsr r}$, which is defined in Conjecture \ref{cj canonical basis X^r} to be a union of $\field \nsSchur{r}$-cells that corresponds to a copy of $X_\nu|_{q=1}$ in \eqref{e intro X tsr M}.
Let $\nu'$ be the conjugate of the partition $\nu$.
The fat cell $\nsbr{X}_\nu$ we can construct is the one corresponding to the recording SYT $\transpose{(Z_{\nu'}^*)}$ in the left-hand side of \eqref{e bijection goal}, where $\transpose{(Z_{\nu'}^*)}$ is the SYT with $1,\ldots,\nu'_1$ in its first column, $\nu'_1+1,\ldots,\nu'_1+\nu'_2$ in its second column, etc.
This is enough to give a nice basis-theoretic solution to the two-row Kronecker problem.

\subsection{The approach of Adsul, Sohoni, and Subrahmanyam}
\label{ss the approach of Adsul, Sohoni, and Subrahmanyam}
Our construction of $\nsbr{X}_\nu$ follows the construction of Adsul, Sohoni, and Subrahmanyam \cite{QuantumDeformations} of a similar quantum object for the Kronecker problem.  Their construction can be viewed as a quantum version of the robust characteristic-free definition of Schur modules from \cite{ABWeyman} (see \cite[\textsection2.1]{Weyman}). We next recall this definition.


Let $X$ be a free module over a commutative ring, $\nu' \vdash_l r$, and set
\[Y_{\nu'}:= \Wedge^{\nu'_1}X \tsr \cdots \tsr \Wedge^{\nu'_l}X.\]
The Schur module $L_{\nu'} X$ \cite[\textsection2.1]{Weyman} is first defined in the $l=2$ case to be $Y_{\nu'}/Y_{\gdneq \nu'}$ where $Y_{\gdneq \nu'}$ is defined in terms of the product and coproduct on the exterior algebra $\Wedge(X)$---we do not need to know the details for our application except that $L_{\nu'} X$ agrees with what we have been calling $X_\nu|_{q=1}$ in this $l=2$ case.
In general, $L_{\nu'} X$ is defined to be the quotient of $Y_{\nu'}$
by the (generally, not direct) sum over all  $i \in [l-1]$ of
\be\label{e Weyman Schur Functor}
Y_{\gdneq^i \nu'}:= Y_{(\nu'_1,\ldots,\nu'_{i-1})} \tsr Y_{\gdneq (\nu'_{i},\nu'_{i+1})} \tsr Y_{(\nu'_{i+2},\ldots,\nu'_l)}.
\ee
In the case that $X =\QQ^\dx$, the Schur module $L_{\nu'}X$ is equal to the  $U(\g_X)$-irreducible $X_{\nu}|_{q=1}$.

The $\O(GL_q(\nsbr{X}))$-comodule
$\nsbr{X}_\nu$ is defined in a similar way.
Although the irreducible $\O(GL_q(\nsbr{X}))$-comodules are in general much smaller than those of $\O(GL(X))$, the nonstandard exterior algebra $\nswedgealgebra{}$ specializes to $\Wedge(X)$ at $q=1$.
So, as above, let $\nu' \vdash_l r$ and define
\be
\nsbr{Y}_{\nu'} := \nswedge{{\nu'_1}}{X} \tsr \nswedge{{\nu'_2}}{X} \tsr \dots \tsr \nswedge{{\nu'_l}}{X}.
\ee
Next, we restrict to the two-row ($\dv = \dw =2$) case, and define the submodule $\nsbr{Y}_{\gdneq \nu'}$ of  $\nsbr{Y}_{\nu'}$ ``by hand'' for  $l=\ell(\nu')=2$ (the reader may now wish to take a look at Figures \ref{f straightening11}--\ref{f straightening33}, where the $\nsbr{Y}_{\gdneq \nu'}$ are defined). Then for $\nu'\vdash_l r$, $\nsbr{Y}_{\gdneq^i \nu'}$ is defined just as in \eqref{e Weyman Schur Functor} and
\[
\nsbr{X}_\nu := \nsbr{Y}_{\nu'}/\left(\textstyle \sum_{i=1}^{l-1}\nsbr{Y}_{\gdneq^i \nu'} \right).
\]
This is a $\O(GL_q(\nsbr{X}))$-comodule and therefore a $\Uqvw$-module, 
and it specializes to $\Res_{U(\g_V \oplus \g_W)} (X_\nu|_{q=1})$ at  $q=1$, 
though some care is required to define the correct integral form of  $\nsbr{X}_\nu$ to make sense of this specialization.

We point out that it is possible to define a version of $\nsbr{X}_{\nu}$ for general $\dv,\dw$ and $l=2$ (see  \textsection\ref{ss Defining X nu outside the two-row case}).
Extending this to  $l > 2$ yields  $\O(GL_q(\nsbr{X}))$-comodules $\nsbr{X}_\nu$ that in general have $\field$-dimension less than $\dim_\QQ(X_\nu|_{q=1})$.  Similar difficulties are encountered in \cite{QuantumDeformations}.  Berenstein and Zwicknagl \cite{BerensteinZ,Zwicknagl} investigate a quantum approach to the plethysms $S^r V_\lambda$,  $\Wedge^r V_\lambda$ and encounter similar difficulties.

\subsection{A global crystal basis for $\nsbr{X}_\nu$}
\label{ss intro global crystal basis for Xnu}
Now we come to our new results in crystal basis theory and combinatorics.
To define a basis of $\nsbr{X}_\nu$, we first define (\textsection\ref{s A canonical basis for Yalpha}) a global crystal basis of $\nswedge{r}{X}$,
whose elements are labeled by what we call nonstandard columns of height-$r$ $(\text{NSC}^{\, r})$.
We then define a canonical basis of $\nsbr{Y}_\alpha$ by putting the bases $\text{NSC}^{\alpha_i}$ together using Lusztig's construction for tensoring based modules \cite[Theorem 27.3.2]{LBook}.  This basis is labeled by nonstandard tabloids (NST), which are just sequences of nonstandard columns.
We show that the image of (a rescaled version of) a certain subset of NST$(\nu')$ in  $\nsbr{X}_\nu$ yields a well-defined basis $\pNSTC(\nu)$ of $\nsbr{X}_\nu$, thus obtaining
\begin{theorem}
\label{t intro main theorem advertisement}
The set $\pNSTC(\nu)$ is a global crystal basis of $\nsbr{X}_\nu$ that solves the two-row Kronecker problem: the number of highest weight elements of $\pNSTC(\nu)$ of weight  $(\lambda,\mu)$ is the Kronecker coefficient $g_{\lambda \mu \nu}$.
\end{theorem}
We comment here that for this part of the paper, we mostly work with $\Uqt := \Uqvw \rtimes \S_2$ modules instead of $\O(GL_q(\nsbr{X}))$-comodules.  We do not lose much and gain convenience by doing this because $\Oqt$ is close to $\O(GL_q(\nsbr{X}))$ in the two-row case and $\Uqt$ is Hopf dual to  $\Oqt$.
Moreover, with slight modifications of the usual theory, we have a theory of based modules for $\Uqt$.
The  $ \O(GL_q(\nsbr{X}))$-comodule $\nsbr{X}_\nu$ is a  $\Uqt$-module, so in addition to obtaining a rule for two-row Kronecker coefficients, we also obtain  a rule for what we call the symmetric and exterior two-row Kronecker coefficients---the \emph{symmetric} (resp. \emph{exterior}) \emph{Kronecker coefficient} $g_{+  \lambda\nu}$ (resp. $g_{-  \lambda\nu}$) is the multiplicity of $M_\nu|_{q=1}$ in $S^2 M_\lambda|_{q=1}$ (resp. $\Wedge^2 M_\lambda|_{q=1}$).
See Theorem \ref{t main canonical basis}, the stronger and more technical version Theorem \ref{t intro main theorem advertisement}, and \eqref{e intro Kronecker generating function}, below,  for this rule.

There are some subtleties that arise in the construction of  $\pNSTC(\nu)$ and the proof of this theorem.
In order to obtain the global crystal basis  $\pNSTC(\nu)$, the rescaling of the NST$(\nu')$ must be chosen carefully.
Each NST $T$ of size  $r$ has a $V$-column (resp.  $W$-column) reading word  $\mathbf{k} \in \{1,2\}^r$ (resp.  $\mathbf{l}\in \{1,2\}^r$).  The word  $\mathbf{k}$ (resp.  $\mathbf{l}$) is naturally associated to the canonical basis element $c_\mathbf{k} \in B^r_V \subseteq V^{\tsr r}$ (resp.  $c_\mathbf{l} \in B^r_W \subseteq W^{\tsr r}$).  These basis elements are nicely depicted as a diagram of arcs according to the $U_q(\sl_2)$ graphical calculus of \cite{FK}.
We define the degree  $\deg(T)$ of an  NST  $T$ in terms of the diagrams of its  $V$ and  $W$-column reading words (for the full definition, see Definition \ref{d invariants}).  The rescaled version of  $T$ is then $({\textstyle -\frac{1}{[2]}})^{\deg(T)} T$.
Experts on global crystal bases may find this to be the most interesting part of the paper.
Another difficulty (which is closely related to the need for rescaling) is that  $\nsbr{Y}_{\gdneq^i {\nu'}}$ is not easily expressed in terms of the basis NST$({\nu'})$.  To remedy this we define a canonical basis for $\nsbr{Y}_{\gdneq^i {\nu'}}$ and prove some general results about how tensoring based modules is compatible with projections.

In \textsection\ref{ss Kronecker graphical calculus} we give a description of the  crystal components of $(\nsbr{X}_\nu,\pNSTC(\nu))$ in terms of arcs of the reading words of NST, which is independent of a $\pNSTC$ in the component and the rescaled NST representing the $\pNSTC$.
This graphical description of the crystal components helps us organize and count them.   We show that the degree 0 crystal components (degree for NST gives rise to a well-defined notion of degree for crystal components) can be grouped into eight different one-parameter families depending on the heights of the columns that the arcs connect (see Figure \ref{f kronecker graphical calculus}), and counting crystal components easily reduces to the degree 0 case.
This description helps us obtain explicit formulae for Kronecker coefficients.  We also use it to write down explicitly (Theorem \ref{t positivity})
all the structure coefficients for the action of the Chevalley generators on $\pNSTC$; we observe that these satisfy a certain positivity property.

Finally, in \textsection\ref{s explicit formulae for Kronecker coefficients} we show that Theorem \ref{t intro main theorem advertisement} actually produces a fairly simple positive formula for two-row Kronecker coefficients.
For example,
define the \emph{symmetric} (resp. \emph{exterior}) \emph{Kronecker generating function}
\[
g_{\varepsilon\nu}(x) := \sum_{\lambda \vdash_2 r} g_{\varepsilon\lambda\nu}x^{\lambda_1-\lambda_2}, \quad \varepsilon  = + \text{ (resp. }\varepsilon = -).
\]
Here $\nu$ is any partition of $r$ of length at most 4; let $n_i$ be the number of columns of the diagram of $\nu$ of height $i$.
For $k \in \ZZ$, define $\llbracket k\rrbracket  = x^{k} + x^{k-2} + \dots + x^{k'}$, where $k'$ is 0 (resp. $1$) if $k$ is even (resp. odd); if $k < 0$, then $\llbracket k \rrbracket := 0$.
The symmetric and exterior Kronecker generating functions are given by
\be \label{e intro Kronecker generating function}
g_{\varepsilon\nu}(x) = {\small \begin{cases}
 \llbracket n_1\rrbracket  \llbracket  n_2\rrbracket  \llbracket  n_3\rrbracket  & \text{if } \hphantom{-\ }(-1)^{n_2} = \hphantom{- }(-1)^{n_3 + n_4} \varepsilon = 1, \\
 \llbracket n_1 - 1\rrbracket  \llbracket  n_2-1\rrbracket  \llbracket  n_3\rrbracket x & \text{if } -(-1)^{n_2} = \hphantom{- }(-1)^{n_3 + n_4} \varepsilon = 1, \\
  \llbracket n_1\rrbracket  \llbracket  n_2-1\rrbracket  \llbracket  n_3-1\rrbracket x & \text{if } -(-1)^{n_2} = -(-1)^{n_3 + n_4} \varepsilon = 1, \\
   \llbracket n_1-1\rrbracket  \llbracket  n_2-2\rrbracket  \llbracket  n_3 -1\rrbracket x^2 & \text{if } \hphantom{-\ }(-1)^{n_2} = -(-1)^{n_3 + n_4} \varepsilon = 1,
\end{cases}}
\ee
where we have identified the values $+,-$ for  $\varepsilon$ with $+1,-1$.
We also easily recover a nice formula for certain two-row Kronecker coefficients from \cite{BWZ} as well as the exact conditions for two-row Kronecker coefficients to vanish, from \cite{BOR}.

\subsection{Organization}
Sections \ref{s Preliminaries and notation}--\ref{s notation for GLV GLW} are preparatory. We fix conventions for the Hecke algebra  $\H_r$ and its Kazhdan-Lusztig basis (\textsection\ref{s Canonical bases of the type A Hecke algebra}) and for the quantized enveloping algebra $\Uq$ and the quantum coordinate algebra  $\O(GL_q(V))$ (\textsection\ref{s quantum group GLq}).  Subsections \ref{ss FRT algebras}--\ref{ss compact} explain the quantum coordinate algebras $\O(M_q(V))$ and $\O(GL_q(V))$ in a way that prepares for the definitions of the corresponding nonstandard objects.
We review (\textsection\ref{s Bases for GLq modules}) global crystal bases from \cite{Kas1,Kas2}, based modules and tensoring based modules from \cite{LBook},
and projected canonical bases from \cite{BProjected}.   
Section \ref{s Quantum Schur-Weyl duality and canonical bases} contains more details about the upper canonical basis of $V^{\tsr r}$.
In \textsection\ref{s graphical calculus}, we review the  $U_q(\sl_2)$ graphical calculus from \cite{FK}.

The first part of new material in this paper (\textsection\ref{s nonstandard coordinate algebra}--\ref{s Nonstandard representation theory in the two-row case}) defines the nonstandard objects and develops their representation theory: in \textsection\ref{s nonstandard coordinate algebra}--\ref{s nonstandard quantum groups GLq} we define the nonstandard quantum group $GL_q(\nsbr{X})$ and prove Theorem \ref{t intro cmqg}.
We give explicit examples for $\O(M_q(\nsbr{X}))$ (\textsection\ref{ss examples for nonstandard O}) and for nonstandard minors  (\textsection\ref{ss nonstandard minors in the two row case}).
Then in \textsection\ref{s nonstandard Hecke algebra definition} we define the nonstandard Hecke algebra and establish some of its basic properties and representation theory. The algebra $\nsH_3$ is treated in detail in  \textsection\ref{salgb3}--\ref{scanonical}.
In  \textsection\ref{s Nonstandard Schur-Weyl duality} we prove the nonstandard analog of Schur-Weyl duality (Theorem \ref{t intro nonstandard schur-weyl duality}) and go over the two-row,  $r=3$ example in detail. In \textsection\ref{s Nonstandard representation theory in the two-row case} we discuss the approximations  $\Oqt$ and  $\Uqt$ to  $\O(GL_q(\nsbr{X}))$ and give a complete description of the representation theory of the nonstandard Hecke algebra and quantum group in the two-row case.

The second part of the new material (\textsection\ref{s A canonical basis for Yalpha}--\ref{s explicit formulae for Kronecker coefficients}) contains a proof of Theorem \ref{t intro main theorem advertisement} and consequences of this theorem.  The bulk of the proof, particularly the necessary canonical basis theory, is contained in \textsection\ref{s global crystal basis for two-row Kronecker coefficients}, and the necessary combinatorics is worked out in  \textsection\ref{s Straightened NST and semistandard tableaux}--\ref{s A Kronecker graphical calculus}.  Section \ref{s A Kronecker graphical calculus} develops the beginnings of a graphical calculus for the basis $\pNSTC$, and section \ref{s explicit formulae for Kronecker coefficients} gives explicit formulae for Kronecker coefficients.
Finally, \textsection\ref{s future work} gives more details about the conjectural basis of  $\nsbr{X}^{\tsr r}$.


\section{Basic concepts and notation}
\label{s Preliminaries and notation}
We introduce our basic notation and conventions for ground rings, tensor products, and type  $A$ combinatorics for the weight lattice, partitions, words, and tableaux. We also define cells in the general setting of modules with basis, rather than only for $W$-graphs, and recall some basic notions about comodules and Hopf algebras.

\subsection{General notation}
We work primarily over the ground rings $\field = \QQ(\u)$,  $\CC$, and $\mathbf{A} = \ZZ[\u, \ui]$. Define $\field_0$ (resp. $\field_\infty$) to be the subring of $\field$ consisting of rational functions with no pole at $\u = 0$ (resp. $\u = \infty$).  For the parts of the paper involving Gelfand-Tsetlin bases and the quantum unitary  group
$\unitary_q(V)$, we work over  the complex numbers $\CC$ and in this context $q$ is taken to be a real number not equal to  $0, \pm1$, rather than an indeterminate.

Let $\br{\cdot}$ be the involution of $\field$ determined by $\br{\u} = \ui$; it restricts to an involution of $\mathbf{A}$.
For a nonnegative integer $k$, the $\br{\cdot}$-invariant quantum integer is $[k] := \frac{\u^k - \u^{-k}}{\u - \ui} \in \mathbf{A}$ and the quantum factorial is $[k]! := [k][k-1]\dots[1]$.
If $N_\mathbf{A}$ is an  $\mathbf{A}$-module, then the  $\u = a$ \emph{specialization} $N|_{\u=a},$ $a \in \QQ$, is defined to be $\QQ \tsr_{\mathbf{A}} N_\mathbf{A}$, the map $\mathbf{A} \to \QQ$ given by $\u \mapsto a$; in a couple places we will also use this notation with $\ZZ[\frac{1}{2}]$ in place of  $\QQ$.

The notation $[k]$ also denotes the set $\{1,\ldots,k\}$ in addition to the quantum integer, but these usages should be easy to distinguish from context.
The notation $\Omega_r^n$ denotes the set of subsets of $[n]$ of size $r$.

Throughout the paper $V,W$, and $X = V \tsr W$ will denote vector spaces of dimensions $\dv,\dw,\dx$, respectively.  These will be over the field  $\field$ or $\CC$. 
For an  $R$-module  $X$, set $X^* = \hom_R(X,R)$.
See \textsection\ref{ss comodules} for important conventions about duals.

If  $R$ is a ring and  $B$ is a subset of an  $R$-module  $N$, then  $R B$ denotes the $R$-span of $B$.

Let $(W, S)$ be a Coxeter group with length function $\ell$ and Bruhat order $<$. If $\ell(vw)=\ell(v)+\ell(w)$, then $vw = v\cdot w$ is a \emph{reduced factorization}. The \emph{right descent set} of $w \in W$ is $R(w) = \{s\in S : ws < w\}$.  The type  $A_{r-1}$ Coxeter group is denoted $(\S_r, S)$, the symmetric group on $r$ letters with simple reflections $S = \{s_1,\ldots,s_{r-1}\}$.

For any $J\subseteq S$, the \emph{parabolic subgroup} $W_J$ is the subgroup of $W$ generated by $J$. Each right coset $W_Jw$ contains a unique element of minimal length called a minimal coset representative. The set of all such elements is denoted $\leftexp{J}W$.

\subsection{Tensor products}
\label{ss restitutions}
Since we will be working with complicated tensor products of many modules in this paper, we use three different symbols for tensor products depending on the context. The symbol $\tsrdual$ is used for tensor products between an object and its dual, the symbol $\tsrvw$ for tensor products of objects involving $V$ with objects involving $W$, and the symbol $\tsr$ for all other tensor products.

So, for instance, we write $V\tsrdual V^*$ for $V\tsr V^*$ and $X=V\tsrvw W$ for $V\tsr W$. We will come across expressions like
\[
(X\tsrdual X^*)^{\tsr r}\cong X^{\tsr r} \tsrdual (X^*)^{\tsr r} \cong (V \tsrdual V^*)^{\tsr r} \tsrvw (W \tsrdual W^*)^{\tsr r} \cong (U^V)^{\tsr r} \tsrvw (U^W)^{\tsr r},
\]
where  $U^V = V\tsrdual V^*$, $U^W = W \tsrdual W^*$.
This will make it more clear where different elements lie in expressions like
\[\zz{\rho(i,j)}{\rho(k,l)} = y_{\rho(i,j)}\tsrdual y^{\rho(k,l)} = x_\vw{i}{j}\tsrdual x^\vw{k}{l} = u_i^k \tsrvw u_j^l.\]

\subsection{Words and tableaux}
\label{ss type A combinatorics preliminaries}
In this paper we work almost entirely in type $A$.
The \emph{weight lattice} $\wl(\g_V)$ of the Lie algebra $\g_V := \gl(V)$ is $\ZZ^\dv$ with standard basis $\epsilon_1, \dots, \epsilon_{\dv}$. Its dual, $\wl(\g_V)^*$, has basis $\epsilon^{1}, \dots, \epsilon^{\dv}$, dual to the standard. The simple roots are $\alpha_i = \epsilon_i - \epsilon_{i+1}, i \in [\dv-1]$.

We write $\lambda \vdash_l r$ for a partition $\lambda = (\lambda_1, \ldots, \lambda_l)$ of size $r = |\lambda| := \sum_{i=1}^l \lambda_i$. A partition $\lambda \vdash_\dv r$ is identified with the weight $\lambda_1\epsilon_1 + \dots + \lambda_\dv\epsilon_\dv \in \wl(\g_V)$.
We also write $\lambda = [n_{l}, \dots, n_1]$ as an alternative notation for the partition $(n_l + \dots + n_1, n_l + \dots + n_2, \dots, n_l)$; note that
$n_i$ is the number of columns of the diagram of  $ \lambda$ of height $i$.
Let $\mathscr{P}_r$ denote the set of partitions of size $r$ and  $\mathscr{P}_{r,l}$ the set of partitions of size $r$ with at most  $l$ parts; let $\mathscr{P}'_r$ (resp.  $\mathscr{P}'_{r,l}$) be the subset of $\mathscr{P}_r$ (resp. $\mathscr{P}_{r,l}$) consisting of those partitions that are not a single row or column shape.

The partial order  $\ld, \ldneq$ on $\wl(\g_V)$ is defined by $\lambda \ld \mu$ if  $\mu - \lambda$ is a nonnegative sum of simple roots. In the case  $\lambda, \mu \vdash r$, this corresponds to the usual dominance order on partitions.
The conjugate partition $\lambda'$ of a partition $\lambda$ is the partition whose diagram is the transpose of the diagram of $\lambda$.

We let $\alpha \vDash_l^{\dx} r$ denote a composition $\alpha = (\alpha_1,\ldots,\alpha_l)$ of $r$ with $\alpha_i \in [\dx]$.
For  $\zeta = (\zeta_1,\ldots,\zeta_l)$ a weak composition of $r$ (i.e. $\zeta_i \geq 0$), let   $B_j$ be the interval $[\sum_{i=1}^{j-1}\zeta_i+1,\sum_{i=1}^{j}\zeta_i]$, $j \in [l]$.  Define  $J_\zeta = \{s_i:i, i +1\in B_j \text{ for some } j\}$ so that $(\S_r)_{J_\zeta} \cong \S_{\zeta_1} \times \dots \times \S_{\zeta_{l}}$.

%

Let $\mathbf{k} = k_1 k_2\dots k_r \in [\dv]^r$ be a word of length $r$ in the alphabet $[\dv]$. The \emph{content} of $\mathbf{k}$ is the tuple $(\zeta_1,\ldots,\zeta_\dv)$ whose  $i$-th entry $\zeta_i$ is the number of  $i$'s in $\mathbf{k}$.
The symmetric group $\S_r$ acts on $[\dv]^r$ on the right by $\mathbf{k} s_i = k_1 \dots k_{i-1} \,  k_{i+1} \, k_i \, k_{i+2} \dots k_r$.
Define $\sort(\mathbf{k})$ to be the tuple obtained by rearranging the $k_j$ in weakly increasing order.
For a word $\mathbf{k}$ of content $\zeta$, define $d(\mathbf{k})$ to be the element  $w$ of $\leftexp{J_\zeta}{\S_r}$ such that $\sort(\mathbf{k})w = \mathbf{k}$.

The set of standard Young tableaux is denoted SYT and the subset of SYT of shape $\lambda$ is denoted SYT$(\lambda)$.  The set of semistandard Young tableaux of size $r$ with entries in $[l]$ is denoted SSYT$_{l}^r$ and the subset of SSYT$_{l}^r$ of shape $\lambda \vdash r$ is SSYT$_{l}(\lambda)$.
Tableaux are drawn in English notation, so that entries of an SSYT strictly increase from north to south along columns and weakly increase from west to east along rows. For a tableau $T$, $|T|$ is the number of squares in $T$ and $\sh(T)$ its shape.  The \emph{content} of a tableau  $T$ is the content of any word with insertion tableau  $T$.

We let $P(\mathbf{k}), Q(\mathbf{k})$ denote the insertion and recording tableaux produced by the Robinson-Schensted-Knuth (RSK) algorithm applied to the word $\mathbf{k}$.  We abbreviate $\sh(P(\mathbf{k}))$ simply by $\sh(\mathbf{k})$.
Let $Z_\lambda$ be the superstandard tableau of shape and content $\lambda$---the tableau whose $i$-th row is filled with $i$'s.
Let  $Z_\lambda^*$ be the SYT of shape $\lambda$ with $1, \ldots, \lambda_1$ in the first row,  $\lambda_1+1,\ldots,\lambda_1+\lambda_2$ in the second row, etc.
The notation $\transpose{Q}$ denotes the transpose of an SYT $Q$, so that $\sh(\transpose{Q}) = \sh(Q)'$.

For an SYT $Q$, let  $\ell(Q)$ denote the distance between $Q$ and $Z_\lambda^*$ in the dual Knuth equivalence graph on SYT$(\lambda)$ (for a definition of this graph, see \cite{Sami}).
It is not hard to show that for any $P \in \text{SYT}(\lambda)$, $\ell(Q) \equiv  \ell(w)-\ell(z)\mod 2$, where $w = \text{RSK}^{-1}(P,Q),$ $z = \text{RSK}^{-1}(P,Z_\lambda^*)$.
\subsection{Cells}
\label{ss cells}
We define cells in the general setting of modules with basis, following \cite{BProjected}.
Let  $H$ be an $R$-algebra for some commutative ring $R$.
Let $M$ be a left $H$-module and $\Gamma$ an  $R$-basis of  $M$. The preorder $\klo{\Gamma}$ (also denoted $\klo{M}$) on the vertex set $\Gamma$ is generated by the relations
\be
\label{e preorder}
\delta\klocov{\Gamma}\gamma \begin{array}{c}\text{if there is an $h\in H$ such that $\delta$ appears with nonzero}\\ \text{coefficient in the expansion of $h\gamma$
in the basis $\Gamma$}. \end{array}
\ee

Equivalence classes of $\klo{\Gamma}$ are the \emph{left cells} of $(M, \Gamma)$. The preorder $\klo{M}$ induces a partial order on the left cells of $M$, which is also denoted $\klo{M}$.

A \emph{cellular submodule} of $(M, \Gamma)$ is a submodule of $M$ that is spanned by a subset of $\Gamma$ (and is necessarily a union of left cells). A \emph{cellular quotient} of $(M,\Gamma)$ is a quotient of $M$ by a cellular submodule, and a \emph{cellular subquotient} of $(M, \Gamma)$ is a cellular quotient of a cellular submodule.
We denote a cellular subquotient $R\Gamma'/ R\Gamma''$ by $R\Lambda$, where $\Gamma'' \subseteq \Gamma' \subseteq \Gamma$ span cellular submodules and $\Lambda = \Gamma' \setminus \Gamma''$.
We say that the left cells  $\Lambda$ and  $\Lambda'$ are isomorphic if $(R \Lambda, \Lambda)$ and $(R \Lambda', \Lambda')$ are isomorphic as modules with basis.

Sometimes we speak of the left cells of $M$, cellular submodules of $M$, etc. or left cells of $\Gamma$, cellular submodules of $\Gamma$, etc. if the pair $(M, \Gamma)$ is clear from context.
For a right $H$-module  $M$, the \emph{right cells}, \emph{cellular submodules}, etc. of  $M$ are defined similarly with $\gamma h$ in place of $h \gamma$ in \eqref{e preorder}. We also use the terminology $H$-cells, $H$-cellular submodules, etc. to make it clear that the algebra $H$ is acting, and we omit left and right when they are clear.

If $(M,\Gamma)$ is as above and $M \cong \bigoplus_{i \in I} M_i$ is a decomposition of $M$ as a direct sum of $H$-modules, then we say that $(M,\Gamma)$ is \emph{compatible} with the decomposition if every cellular submodule of  $M$ is of the form  $\bigoplus_{i \in J}M_i$ for some  $J \subseteq I$.

\subsection{Comodules} \label{ss comodules}
In the next two subsections, we fix some notation regarding comodules and dual pairings of Hopf algebras, mostly following \cite[Chapters 1,11]{KS}; this reference contains a good introduction to these generalities.

Let $\A$ be a coalgebra over a field $\field$, $N$ a  $\field$-vector space with basis  $e_1,\ldots,e_n$, and
$\varphi: N\rightarrow A \otimes N$ the left corepresentation of  $A$ on  $N$ given by
\begin{equation} \label{eqrightact1V}
 \varphi(e_j)=\sum_{k =1}^n  m^k_j \tsr e_k.
\end{equation}
The matrix  $(m^k_j)_{j,k \in [n]}$ is called the \emph{coefficient matrix} of  $\varphi$ (or of $N$) with respect to the basis $e_1,\ldots,e_n$ and its entries $\{m^k_j: j,k \in [n]\}$ are the \emph{matrix coefficients} of  $\varphi$ (or of $N$) with respect to $e_1,\ldots,e_n$.
Coefficient matrices and matrix coefficients are defined similarly for right corepresentations.

\begin{remark}\label{r dual convention}
In this paper we identify the endomorphism algebra $\End(V)$ with $V^* \tsr V = V^* \tsrdual V$ and the coalgebra dual to $\End(V)$ with $V \tsrdual V^*$. We adopt the convention that dual objects take upper indices and ordinary objects take lower indices. Thus for algebras, upper indices correspond to rows and lower indices to columns, and for coalgebras, lower indices correspond to rows and upper indices to columns. Dual objects will typically correspond to right corepresentations and ordinary objects to left corepresentations. In \textsection\ref{ss FRT algebras}--\ref{ss compact}, \textsection\ref{s nonstandard coordinate algebra}--\ref{s nonstandard quantum groups GLq}, and Appendices \ref{s reduction system} and \ref{s appendix Oqt}, we will work with left and right corepresentations, and we are careful to distinguish between the two. For the remainder of the paper, such care is not necessary and we will typically work with left modules and right comodules, but will write $V_\lambda$ in place of $V_\lambda^*$, $X$ in place of $X^*$, etc. to avoid extra symbols.
\end{remark}

Corresponding to the left corepresentation $\varphi$ above, there is a right corepresentation $(\varphi)_R: N^* \rightarrow N^* \tsr A$ of  $A$ on $N^* = \hom_\field(N, \field)$
given by
\begin{equation} \label{eqact2V}
 (\varphi)_R(e^k)=\sum_{j=1}^n  e^j \tsr m^k_j ,
\end{equation}
where  $e^1,\ldots,e^n$ is the basis dual to $e_1,\ldots,e_n$.
We will write  $(N)_R$ for the right comodule on  $N^*$ corresponding to $\varphi_R$.
This construction is independent of the basis  $\{e_i\}$.
The comodules $N$ and $(N)_R$ share the same coefficient matrix with respect to any basis of  $N$ and corresponding dual basis of  $(N)_R$.

Similarly, given a $\field$-vector space  $N'$ with basis  $e^1,\ldots,e^n$, and
$\varphi': N'\rightarrow N' \tsr A$ the right corepresentation of  $A$ on  $N'$ given by
\begin{equation} \label{eqact2W}
\varphi'(e^k)=\sum_{j=1}^n  e^j \tsr m^k_j,
\end{equation}
there is a left  corepresentation of  $A$ on  $N'^*$ given by
\begin{equation} \label{eqrightact1W}
 (\varphi')_L(e_j)=\sum_{k=1}^n m^k_j \tsr e_k ,
\end{equation}
where  $e_1,\ldots,e_n$ is the basis of  $N'^*$ dual to  $e^1,\ldots,e^n$.  The corresponding left comodule on  $N'^*$ is denoted $(N')_L$.
The coefficient matrix  $(m^k_j)$ of  $N'$ with respect to  $e^1,\ldots,e^n$ is the same as the coefficient matrix of $(N')_L$ with respect to $e_1,\ldots,e_n$.

\subsection{Dually paired Hopf algebras}
\label{ss dually paired  Hopf algebras}
Given two  $\field$-bialgebras  $\U$ and  $\A$, a bilinear map $\langle \cdot, \cdot \rangle : \U \times \A \to \field$
is a \emph{dual pairing of bialgebras} if
\[
\begin{array}{r@{\ = \ }lr@{\ = \ }l}
\langle \Delta_\U (f),a_1\tsr a_2 \rangle & \langle f,a_1 a_2\rangle, &\langle f_1 f_2,a\rangle & \langle f_1 \tsr f_2, \Delta_\A (a)\rangle,\\[1.3mm]
\langle f,1_\A\rangle & \epsilon_\U (f), & \langle 1_\U,a\rangle & \epsilon_\A(a)
\end{array}
\]
for all $f,f_1,f_2\in\U$ and $a,a_1,a_2\in\A$. If $\U$ and $\A$ are Hopf algebras, then compatibility with the antipode is automatic \cite[Proposition 9, Chapter 1]{KS}:
\[
\langle S_\U(f),a\rangle = \langle f, S_\A(a)\rangle,\quad f\in\U,\ a\in\A,
\]
and in this case $\langle \cdot, \cdot \rangle$ is a \emph{dual pairing of Hopf algebras}.

If $\U$ and $\A$ are dually paired bialgebras, then to any right $\A$-corepresentation $\varphi: N \to N \tsr \A$, there corresponds a left $\U$-representation   $\hat{\varphi}: \U \to \End(N)$ given by
\[\hat{\varphi}(f) x = ((\idelm \tsr f) \circ \varphi)(x) = \sum x_{(0)} \langle f, x_{(1)} \rangle, \quad f\in \U, \ x \in N, \]
where $\varphi(x) = \sum x_{(0)} \tsr x_{(1)}$ expresses $\varphi$ in Sweedler notation.
Note however that for dually paired bialgebras, a representation of one does not, in general, come from a corepresentation of the other, even if they are Hopf algebras and the pairing is nondegenerate.

\section{Hecke algebras and canonical bases}
\label{s Canonical bases of the type A Hecke algebra}
The \emph{Hecke algebra} $\H(W)$ of $(W, S)$ is the free $\mathbf{A}$-module with standard basis $\{T_w :\ w\in W\}$ and relations generated by
\be \label{e Hecke algebra def} \begin{array}{ll}T_vT_w = T_{vw} & \text{if } vw = v\cdot w\ \text{is a reduced factorization},\\
(T_s - \u)(T_s + \ui) = 0 & \text{if } s\in S.\end{array}\ee
We remark that the $q$ here is frequently $q^{1/2}$ in the literature on Hecke algebras, as it is, for instance, in \cite{KL}.
We have chosen this convention so that in quantum Schur-Weyl duality, the Hecke algebra  $q$ matches the usual notation (as in \cite{Kas1, Kas2, KS}) for the quantum group $q$.

For each $J\subseteq S$, $\H(W)_J$ denotes the subalgebra of $\H(W)$ with $\mathbf{A}$-basis $\{T_w:\ w\in W_J\}$, which is isomorphic to $\H(W_J)$.

In this section we recall the definition of the Kazhdan-Lusztig basis elements $C_w$ of \cite{KL} and some of their basic properties.
Then we specialize to type  $A$ and review the beautiful connection between cells and the RSK algorithm.

\subsection{The upper canonical basis of $\H(W)$}
\label{ss upper canonical basis of H}

The \emph{bar-involution}, $\br{\cdot}$, of $\H(W)$ is the additive map from $\H(W)$ to itself extending the $\br{\cdot}$-involution of $\mathbf{A}$ and satisfying $\br{T_w} = T_{w^{-1}}^{-1}$. Observe that $\br{T_{s}} = T_s^{-1} = T_s + \ui - \u$ for $s \in S$. Some simple $\br{\cdot}$-invariant elements of $\H(W)$ are $\C_\text{id} := T_\text{id}$, $C_s := T_s - \u = T_s^{-1} - \ui$, and $\C_s := T_s + \ui = T_s^{-1} + \u$, $s\in S$.

Define the lattice $(\H_r)_{\ZZ[\u]} := \ZZ[\u] \{ T_w : w \in W \}$ of $\H_r$.
\be
\parbox{13.4cm}{For each $w \in W$, there is  a unique element $C_w \in \H(W)$ such that $\br{C_w} = C_w$ and $C_w$ is congruent to $T_w \mod \u (\H_r)_{\ZZ[\u]}$.}
\ee
The $\mathbf{A}$-basis $\Gamma_W := \{C_w: w\in W\}$
is the \emph{upper canonical basis} of $\H(W)$ (we use this language to be consistent with that for crystal bases).

The coefficients of the upper canonical basis in terms of the standard basis are essentially the \emph{Kazhdan-Lusztig polynomials} $P_{x,w}$:
\be\label{e Kazhdan-Lusztig polynomials} C_w = \sum_{x \in W} P^-_{x,w} T_x. \ee
The  $P^-_{x,w}$ are related to the $P_{x,w}$ defined in \cite{KL} by $P^-_{x,w}(q) = \iota(q_{\text{KL}}^{(\ell(x)-\ell(w))/2}P_{x,w}(q_\text{KL}))$, where  $\iota$ is the involution of $\mathbf{A}$ defined by $\iota(\u) = -\ui$ and $q_\text{KL}$ is the $q$ used in \cite{KL}, related to ours by $q_\text{KL}^{1/2} = \u$.
Now let $\mu(x,w) \in \ZZ$ be the coefficient of $\ui$ in $\iota(P^-_{x,w})$ (resp. $\iota(P^-_{w,x})$) if $x \leq w$ (resp. $w \leq x$).
Then the right regular representation in terms of the upper canonical basis of $\H_r$ takes the following simple form:
\begin{equation}\label{e C on canbas}
C_w C_s =
\left\{\begin{array}{ll} -[2] C_w & \text{if}\ s \in R(w),\\[1mm]
\displaystyle\sum_{\substack{\{w' \in  W: s \in R(w')\}}} \mu(w',w)C_{w'} & \text{if}\ s \notin R(w).
\end{array}\right.
\end{equation}

The simplicity and sparsity of this action along with the fact that the right cells of $\Gamma_W$ often give rise to  $\CC(\u) \tsr_\mathbf{A} \H(W)$-irreducibles are among the most amazing and useful properties of canonical bases.
\subsection{Cells in type  $A$} \label{ss cell label conventions C_Q C'_Q}
Let  $\H_r = \H(\S_r)$ be the type $A$ Hecke algebra.
It is well known that $\field \H_r := \field \tsr_\mathbf{A} \H_r$ is semisimple and its irreducibles in bijection with partitions of $r$; let $M_\lambda$ and $M_\lambda^{\mathbf{A}}$ be the $\field\H_r$-irreducible and Specht module of $\H_r$ of shape $\lambda \vdash r$ (hence $M_\lambda \cong \field\tsr_\mathbf{A} M_\lambda^\mathbf{A}$).

The work of Kazhdan and Lusztig \cite{KL} shows that the decomposition of $\Gamma_{\S_r}$ into right cells is
$\Gamma_{\S_r} = \bigsqcup_{\lambda\vdash r, \, P \in \text{SYT}(\lambda)} \Gamma_P$, where $\Gamma_P := \{C_w : P(w) = P\}$.  Moreover, the right cells $\{ \Gamma_P : \sh(P) = \lambda\}$ are all isomorphic, and, denoting any of these cells by $\Gamma_\lambda$, $\mathbf{A}\Gamma_\lambda \cong M_\lambda^\mathbf{A}$.
A combinatorial discussion of left cells in type $A$ is given in \cite[\textsection 4]{B0}.

We refer to the basis $\Gamma_\lambda$ of $M_\lambda^\mathbf{A}$ as the \emph{upper canonical basis of $M_\lambda$} and denote it by $\{ C_Q : Q \in \text{SYT}(\lambda) \}$, where $C_Q$ corresponds to $C_w$ for any (every) $w \in \S_r$ with recording tableau $Q$.  Note that with these labels the action of $C_s$ on the upper canonical basis of $M_\lambda$ is similar to \eqref{e C on canbas}, with  $\mu(Q',Q):= \mu(w',w)$ for any  $w',w$ such that  $P(w')=P(w)$, $Q' = Q(w'),Q = Q(w)$, and right descent sets
\be \label{e tableau descent set}
R(C_Q) = \{ s_i : i + 1 \text{ is strictly to the south of $i$ in $Q$}\}.
\ee
\begin{example} \label{ex lambda31}
The integers $\mu(Q',Q)$ for the upper canonical basis of $M_{(3,1)}$ are given by the following graph ($\mu$ is 1 if the edge is present and 0 otherwise), and descent sets are shown below each tableau.
\setlength{\cellsize}{2.5ex}
\[
\xymatrix@R=0cm{
\ \tableau{1&2&3\\4} \ar @{-} [r] & \ \tableau{1&2&4\\3} \ar @{-} [r] & \ \tableau{1&3&4\\2} \\
\{s_3\} &   \{s_2\} & \{s_1\}
}
\]
\setlength{\cellsize}{1.5ex}
\end{example}

\section{The quantum group  $GL_q(V)$}
\label{s quantum group GLq}
The quantum group  $GL_q(V)$ is a virtual object associated to two Hopf algebras---the Drinfel$'$d-Jimbo quantized enveloping algebra $\Uq$ and the quantum coordinate algebra  $\O(GL_q(V))$.  These are dually paired Hopf algebras, and this connects the corepresentation theory of  $\O(GL_q(V))$ to the representation theory of $\Uq$.
In this section we recall the definition of $\Uq$, following \cite{Kas1,HK}, and of $\O(GL_q(V))$, following \cite{KS,RTF}.  Our treatment of  $\O(M_q(V))$ and $\O(GL_q(V))$
here will prepare us for the construction of the corresponding nonstandard objects in  \textsection\ref{s nonstandard coordinate algebra}--\ref{s nonstandard quantum groups GLq}.  In  \textsection\ref{ss modules of GLq}, we fix notation regarding representations of $GL_q(V)$.

\subsection{The quantized enveloping algebra  $\Uq$}
\label{ss quantized enveloping algebra}
The \emph{quantized universal enveloping algebra} $\Uq$ is the associative $\field$-algebra generated by
$q^h, h \in \wl(\g_V)^*$ (set $K_i = q^{\epsilon^{i}-\epsilon^{i+1}}$) and $E_i, F_i, i \in \rootset$ with relations
\be
\begin{array}{ll}
q^0 = 1, & q^hq^{h'} = q^{h + h'},   \\[.6mm]
q^hE_iq^{-h} = \u^{\langle \alpha_i, h \rangle}E_i, & q^hF_iq^{-h} = \u^{-\langle \alpha_i, h \rangle}F_i, \\[.6mm]
E_iF_j - F_jE_i = \delta_{i,j} \frac{K_i - K_i^{-1}}{\u-\ui}, & \\[.6mm]
E_iE_j - E_jE_i = F_iF_j - F_jF_i = 0 & \text{for } |i-j| > 1, \\[.6mm]
E_i^2E_j-  [2]E_iE_jE_i + E_jE^2_i = 0 & \text{for } |i-j| = 1, \\[.6mm]
F^2_iF_j - [2]F_iF_jF_i + F_jF^2_i = 0 & \text{for } |i-j| = 1. \\[.6mm]
\end{array}
\ee

The \emph{bar-involution}, $\br{\cdot}: \Uq \to \Uq$, is the  $\QQ$-linear automorphism extending the involution $\br{\cdot}$ on $\field$ and satisfying
\be
\br{q^{h}} = q^{-h}, \ \br{E}_i = E_i, \ \br{F}_i = F_i.
\ee
Let $\varphi : \Uq \to \Uq$ be the algebra antiautomorphism determined by
\[
\varphi(E_i) = F_i, \quad \varphi(F_i) = E_i, \quad \varphi(K_i) = K_i.
\]

The algebra $\Uq$ is a Hopf algebra with coproduct $\Delta$ given by
\be \label{e U_q coproduct}
\Delta(q^h) = q^h \tsr q^h, \ \ \Delta(E_i) = E_i \tsr K_i^{-1} + 1 \tsr E_i, \ \ \Delta(F_i) = F_i \tsr 1 + K_i \tsr F_i.
\ee
This is the same as the coproduct used in \cite{Brundan, Kas2, HK}; it differs from the coproduct of \cite{LBook} by $(\varphi \tsr \varphi) \circ \Delta \circ \varphi$ and from that of  \cite{KS} by $(\br{\cdot} \tsr \br{\cdot}) \circ \Delta \circ \br{\cdot}$.


\subsection{FRT-algebras}\label{ss FRT algebras}
The quantum coordinate algebra  $ \O(M_q(V))$ and nonstandard coordinate algebra $ \O(M_q(\nsbr{X}))$ will be defined in the generality of FRT-algebras \cite{RTF} (see also \cite[Chapter 9]{KS}).

Let $V$ be a  $\field$-vector space of dimension $\dv$, with standard basis $v_1,\ldots,v_\dv$.
Let $U = V \tsrdual V^*$ be the  $\field$-vector space with standard basis $\{u^j_i = v_i \tsrdual v^j : i,j \in [\dv]\}$, where  $v^1,\ldots,v^\dv$ is the basis of $V^*$ dual to
 $v_1,\ldots,v_\dv$.
We view  $U$ as the coalgebra dual to the endomorphism algebra $\End(V)$.  In terms of the standard basis,
the comultiplication and counit are given by
\[\Delta(u^j_i)=\sum_k u^k_i\otimes u^j_k, \quad \epsilon(u^j_i)=\delta_{ij}, \]
or in matrix form
\be \label{e dot tsr}
\Delta(\mb{u})=\mb{u} \dot{\tsr}  \mb{u}, \ \epsilon(\mb{u})=\mb{I},
\ee
where $\mb{u}$ is the $\dv\times \dv$ matrix  $(u^j_i)$ with entries in  $U$
and $\dot{\tsr}$ denotes matrix multiplication with tensor product in place of scalar multiplication.
For coalgebras we adopt the convention that upper indices correspond to columns and lower indices to  rows (see Remark \ref{r dual convention}).

The tensor algebra $\field\langle u^j_i \rangle = T(U)= \bigoplus_{r \geq 0} U^{\tsr r}$ is a  $\field$-bialgebra with comultiplication and counit extending those of $U$ in the unique way that makes them into algebra homomorphisms.

Let $\R=\R_{V,V} \in M_{\dv^2}(\field)$ be a nonsingular matrix, identified with an element of  $\End(V^{\tsr 2})$ via the standard basis,
and let $\hat \R_{V,V}=\hat \R=\tau \circ \R$, where $\tau$ is the flip of $V\otimes V$.
The \emph{FRT-algebra $A(\R)$} \cite{RTF} is the quotient algebra $\field\langle u^j_i \rangle/\mathcal{I}_\R$, where  $\mathcal{I}_\R$ is the two-sided ideal generated by certain degree two relations, which, in matrix form, are
\begin{equation} \label{eqhatr2}
\hat \R (\mb{u} \otimes \mb{u}) = (\mb{u} \otimes \mb{u}) \hat \R,
\end{equation}
where  $\mb{u} \otimes \mb{u}$ is the $\dv^2 \times\dv^2$ matrix with the entry $u^j_i\tsr u^l_k$ in the  $ik$-th row and  $jl$-th column.  This is to be interpreted as an equality of elements of  $M_{\dv^2}(T(U))$, i.e.,  $\dv^4$ many equations, each requiring some linear combination of
elements of $U^{\tsr 2}$ to be equal to another linear combination of elements of $U^{\tsr 2}$.  For an explicit form of these relations in the case  $\hat\R$ is given by \eqref{e hat R formula}, see \eqref{eqredclassical} below.
By \cite[Proposition 9.1]{KS},  $\mathcal{I}_\R$ is a coideal of  $T(U)$, hence  $A(\R)$ is a bialgebra with coproduct and counit given by \eqref{e dot tsr}.

\subsection{The quantum coordinate algebra $\O(M_q(V))$} \label{squantum}
The \emph{quantum coordinate algebra $\O(M_q(V))$} of the \emph{standard quantum matrix space $M_q(V)$} is the FRT-bialgebra corresponding to the $\hat \R$ given by
\be \label{e hat R formula}
\hat\R_{V,V} = \sum_{i < j} (\u - \ui) \bv^{ij}\tsrdual \bv_{ij} + \sum_{i \neq j} \bv^{ij} \tsrdual\bv_{ji} + \u \sum_{i} \bv^{ii} \tsrdual \bv_{ii},
\ee
where  $\bv_{ij} = v_i \tsr v_j$,  $\bv^{ij} = v^i \tsr v^j$.
It is known that $\hat \R$ satisfies the quadratic equation
\begin{equation} \label{eqhatR}
(\hat \R - q I)(\hat \R + q^{-1} I)=0,
\end{equation}
and has the spectral decomposition
\begin{equation}
\hat \R=q P_+-q^{-1}P_-,
\end{equation}
where the projections $P_+=P_+^V$ and $P_-=P_-^V$ are
\begin{equation} \label{eqprojV}
 P_+=\frac{1}{[2]}(\hat \R+q^{-1}I) , \quad P_-=\frac{1}{[2]}(-\hat \R+q I),
\end{equation}
so that $I=P_+ + P_-$ is the spectral decomposition of the identity.

These projections are  quantum analogs of the symmetrization and antisymmetrization
operators on $V^{\tsr 2}$.  Specifically, let the \emph{symmetric subspace}
 $\ssym{2}{V} := (V \tsr V) P_+$ be  the image of $P_+$,
and let the \emph{antisymmetric subspace} $\swedge{2}{V} := (V \tsr V)P_-$ be the image of
$P_-$.

The \emph{quantum symmetric algebra of  $V$}, denoted $\ssymalgebra{V}$, is
the quotient algebra of $T(V)$ by the two-sided ideal generated by $\swedge{2}{V}$.  Explicitly,
this is the algebra over the $v_i$'s subject to the
relations
\begin{equation} \label{eqsymspace1}
v_jv_i = \ui v_iv_j, \quad i<j.
\end{equation}
These relations can also be put in matrix form, but we have found the above two descriptions more convenient.
The \emph{quantum exterior algebra of  $V$}, denoted  $\swedgealgebra{V}$, is
the quotient algebra of $T(V)$ by the two-sided ideal generated by $\ssym{2}{V}$.  Explicitly,
this is the algebra over the $v_i$'s subject to the
relations
\begin{equation} \label{eqaltspace1}
v_i^2=0, \mbox{\ and\ } v_jv_i = -q v_iv_j, \quad i<j.
\end{equation}
Let $\ssym{r}{V}$ and $\swedge{r}{V}$ be the degree $r$-components of
$\ssymalgebra{V}$ and $\swedgealgebra{V}$, respectively.

We think of $\ssymalgebra{V}$ as the coordinate algebra of a virtual symmetric quantum space
$V_{\mbox{sym}}$ with commuting coordinates (in the quantum sense),
and $\swedgealgebra{V}$ as the coordinate algebra of
a virtual antisymmetric quantum space $V_\wedge$ with anti-commuting coordinates (in the quantum sense).

We can now give some other descriptions of $\O(M_q(V))$, which we have found to be more convenient than the matrix form \eqref{eqhatr2}.
Both $\ssymalgebra{V}$ and $\swedgealgebra{V}$ are left $\O(M_q(V))$-comodule algebras via  $v_i \mapsto \sum_j u^j_i \tsr v_j$ and
$\ssymalgebra{V^*}$ and $\swedgealgebra{V^*}$ are right $\O(M_q(V))$-comodule algebras via  $v^j \mapsto \sum_i v^i \tsr u^j_i$.
In fact, it can be shown that $\O(M_q(V))$ is the largest bialgebra quotient of  $T(U)$ such that $\ssym{2}{V}$ is a left $\O(M_q(V))$-comodule and $\ssym{2}{V}^*$ is a right $\O(M_q(V))$-comodule.
Similarly, $\O(M_q(V))$ is the largest bialgebra quotient of  $T(U)$ such that  $\swedge{2}{V}$ is a  left $\O(M_q(V))$-comodule and  $\swedge{2}{V}^*$ is a right $ \O(M_q(V))$-comodule.
This view of the standard quantum group, emphasized by Manin \cite{manin}, carries over nicely to the nonstandard setting.

Note that, by \eqref{eqprojV},
the defining relations \eqref{eqhatr2} of $\O(M_q(V))$ are equivalent
to either of
\begin{align}
P_+ (\mb{u} \otimes \mb{u})  = (\mb{u} \otimes \mb{u}) P_+, \label{eqp+2} \\
P_- (\mb{u} \otimes \mb{u}) = (\mb{u} \otimes \mb{u}) P_-. \label{eqp-2}
\end{align}
It can also be shown that the two-sided ideal $\mathcal{I}_\R$ is that generated by
\be
\label{e OMqV ideal def}
 (V \tsr V \tsrdual V^* \tsr V^*) (P_+^V \tsrdual P_-^{V^*} + P_-^V \tsrdual P_+^{V^*})   = \ssym{2}{V} \tsrdual \swedge{2}{V}^* \oplus\swedge{2}{V} \tsrdual \ssym{2}{V}^*,
\ee
where  $P_\pm^{V^*}$ are defined the same as $P_\pm^V$,  with the basis $v^1,\ldots,v^\dv$ in place of $v_1,\ldots,v_\dv$.
\subsection{The quantum determinant and the Hopf algebra $\O(GL_q(V))$}
\label{ss quantum determinant}
Let
\[
\begin{array}{l}
\varphi^R_r: \nswedge{r}{V}^* \rightarrow  \nswedge{r}{V}^* \tsr \O(M_q(V)), \\[1.7mm]
\varphi^L_r: \nswedge{r}{V} \rightarrow \O(M_q(V)) \tsr \nswedge{r}{V}
\end{array}
\]
be the right and left corepresentations corresponding to the right  $\O(M_q(V))$-comodule $\nswedge{r}{V}^*$  and left $\O(M_q(V))$-comodule $\nswedge{r}{V}$.

Recall that $\Omega_r^\dv$ denotes the set of subsets of $[\dv]$ of size $r$.
For a subset $I\in \Omega_r^\dv$, with $I=\{i_1,\ldots,i_r\}$, $i_1<i_2<\cdots$, let
$v_I=v_{i_1}\cdots v_{i_r}\in \swedge{r}{V}$ and $v^I=v^{i_1}\cdots v^{i_r}\in \swedge{r}{V}^*$. The \emph{standard monomial basis} of $\swedge{r}{V}$ (resp. $\swedge{r}{V}^*$) is $\{v_I: I\in \Omega_r^\dv\}$ (resp. $\{v^I: I\in \Omega_r^\dv\}$). It is known that the isomorphism $\swedge{r}{V}^* \cong (\swedge{r}{V})_R$ of right $\O(M_q(V))$-comodules identifies the standard monomial basis of $\swedge{r}{V}^*$ with the basis dual to the standard monomial basis of $(\swedge{r}{V})_R$  (here, $(\cdot)_R$ is the notation for dualizing comodules explained in  \textsection\ref{ss comodules}). The \emph{right quantum $r$-minors} $D^{I,R}_J(V)$ of $\O(M_q(V))$ are defined to be the
matrix coefficients of the right corepresentation $\varphi^R_r$ in the standard monomial basis. Explicitly, they are defined by
\[ \varphi^R_r(v^I)=\sum_{J\in\Omega_r^\dv} v^J \otimes D^{I,R}_J(V),\ I\in\Omega_r^\dv.\]
The \emph{left quantum $r$-minors} $D^{I,L}_J(V)$ are defined by
\[ \varphi^L_r(v_J)=\sum_{I\in\Omega_r^\dv} D^{I,L}_J(V) \otimes v_I,\ J\in\Omega_r^\dv.\]
It is known that
\begin{equation} \label{eqquantumminorstd}
D^I_J=D^{I,R}_J(V)=D^{I,L}_J(V) = \sum_{\sigma \in \S_r} (-q)^{\ell(\sigma)}u^{i_{\sigma(1)}}_{j_1}\cdots u^{i_{\sigma(r)}}_{j_r},
\end{equation}
where $\ell(\sigma)$ is the number of inversions of the permutation $\sigma$.
The quantum determinant  $D_q=D_q(V)$ of $\O(M_q(V))$
 is defined to be $D^{I}_J$, with $I=J=[\dv]$.
Explicitly,
\begin{equation} \label{eqquantumdetstd}
D_q=\sum_{\sigma \in \S_\dv} (-q)^{\ell(\sigma)}u^{{\sigma(1)}}_{1}\cdots u^{{\sigma(\dv)}}_{\dv}.
\end{equation}

The coordinate algebra $\O(GL_q(V))$
of the quantum group $GL_q(V)$ is obtained by adjoining the
inverse $D_q^{-1}$ to $\O(M_q(V))$.
By applying the corepresentation maps to the nondegenerate pairings
\begin{align*}
\swedge{\dv-1}{V}^* \otimes \swedge{1}{V}^* &\rightarrow \swedge{\dv}{V}^*,\\
\swedge{1}{V} \otimes \swedge{\dv-1}{V} &\rightarrow \swedge{\dv}{V},
\end{align*}
it can be shown that the cofactor matrix $\tilde{\mb{u}}$ with entries $\tilde{u}_{k}^{i} := (-q)^{k-i}D^{\hat{k}}_{\hat{i}}$, where $\hat{k} := [\dv]\setminus\{k\}$, satisfies
\[ \tilde{\mb{u}} \mb{u}=\mb{u} \tilde{\mb{u}} = D_q \mb{I}.\]
Then we can formally define $\mb{u}^{-1}=D_q^{-1} \tilde{\mb{u}}$. This gives the
following Hopf structure on
$\O_q(GL_q(V))$:
\begin{enumerate}
\item $\Delta(\mb{u})=\mb{u}\dot{\tsr} \mb{u}$,  $\Delta(D_q^{-1}) = D_q^{-1} \tsr D_q^{-1}$.
\item $\epsilon(\mb{u})=\mb{I}$.
\item $S(u^i_j)=\tilde u^i_j D_q^{-1}$, $S(D_q^{-1})=D_q$, where
$u^i_j$ are the entries of $\mb{u}$ and
$\tilde u^i_j$ are the entries of $\tilde{\mb{u}}$.
\end{enumerate}

\subsection{A reduction system for  $\O(M_q(V))$}
\label{ss standard reduction system}
The Poincar\'e series of $\O(M_q(V))$  coincides with the Poincar\'e series of the
commutative algebra $\CC[u^j_i]$. Because, just as in the classical case,
$\O(M_q(V))$ has a basis consisting of the standard  monomials
$(u^1_1)^{k_{11}}(u^1_2)^{k_{12}}\cdots (u^\dv_\dv)^{k_{\dv\dv}}$, $k_{ij}$ being nonnegative
integers.
To show this \cite{nym,artin}, the monomials are ordered lexicographically,
and the defining equations \eqref{eqp+2}  of $\O(M_q(V))$ are recast in the form of
a reduction system:
\begin{equation} \label{eqredclassical}
\begin{array}{lcl}
u^j_k u^i_k&\rightarrow& q^{-1}u^i_k u^j_k  \quad (i<j) \\
u^k_j u^k_i &\rightarrow& q^{-1}u^k_i u^k_j  \quad (i<j) \\
u^j_k u^i_l &\rightarrow& u^i_l u^j_k  \quad (i<j, k<l) \\
u^j_l u^i_k &\rightarrow& u^i_k u^j_l -(q-q^{-1})u^i_l u^j_k  \quad (i<j, k<l).
\end{array}
\end{equation}
Then, by the diamond lemma \cite{KS}, it suffices to show that all ambiguities in this
reduction system are resolvable. This means any term of the form $u^i_j u^k_l u^r_s $, when
reduced in any way, leads to the same result.
This has to be checked for
24 different types of configurations of the three indices $(i,j),(k,l),(r,s)$;
see \cite{artin,KS,nym} for details.

\subsection{Compactness, unitary transformations} \label{ss compact}
What sets the standard quantum group apart from other known deformations
\cite{artin,manin,RTF,reshetikhin,sudbery} of $GL(V)$
is that it has a real form that is compact. To see what this means,
we have to recall the notion of compactness due to Woronowicz  in the
quantum setting; see \cite{wor1} or \cite[Chapter 11]{KS} for details.

Let $\mathcal{A}$ be the coordinate Hopf algebra of a quantum group $G_q$.
Suppose there is an involution $*$ on $\mathcal{A}$ so that it is a Hopf $*$-algebra \cite{KS}.
We  say that $*$ defines a real form of the quantum group $G_q$.
A finite-dimensional corepresentation of $\mathcal{A}$ on a vector space $V$ with a Hermitian form
is called unitary if the matrix $m=(m_j^k)_{j,k}$ of this corepresentation with respect to an
orthonormal basis $\{e_i\}$ of $V$ satisfies $m^* m=m m^*=I$, where  $m^* := \big((m_k^j)^* \big)_{\hspace{-.5mm} j,k}$.
The algebra
$\mathcal{A}$ is called a compact matrix group algebra (CMQG) if (1) it is the linear
span of all matrix elements of finite-dimensional corepresentations of $\mathcal{A}$, and (2)
it is generated as an algebra by finitely many elements. Then

\begin{theorem}[Woronowicz \cite{wor1} (also see {\cite[Chapter 11]{KS}})]
\label{tworonowicz}
\noindent (a) A Hopf $*$-algebra $\mathcal{A}$ is a CMQG algebra if and only if there is a finite-dimensional
unitary corepresentation of $\mathcal{A}$ whose matrix elements generate $\mathcal{A}$ as an
algebra.

\noindent (b) If $\mathcal{A}$ is a CMQG algebra, then the quantum analog of the Peter-Weyl
theorem holds and any finite-dimensional corepresentation of $\mathcal{A}$ is unitarizable,
and hence, a direct sum of irreducible corepresentations.
\end{theorem}

Assume that objects are defined over $\CC$ and $q$ is a real number such that  $q \neq 0, \pm1$.
There is a unique involution $*$ on the algebra $\O(GL_q(V))$ such that
$(u^i_j)^*=S(u^j_i)$.
This involution makes $\O(GL_q(V))$ into a Hopf $*$-algebra,
denoted $\O(\unitary_q(V))$, and called the coordinate algebra of the quantum unitary group
$\unitary_q(V)$---which is, again, a virtual object.
Furthermore, $\O(\unitary_q(V))$ is a CMQG algebra.

Woronowicz \cite{wor1} has shown that the
usual results for real compact groups, such as Harmonic analysis, existence of
orthonormal bases, and so on, generalize to CMQG algebras.

\subsection{Representations of  $GL_q(V)$}
\label{ss modules of GLq}
The \emph{weight space}  $N^\zeta$ of a $\Uq$-module  $N$ for the weight $\zeta \in \wl(\g_V)$ is the $\field$-vector space $\{x \in N : q^h x = \u^{\langle \zeta, h \rangle} x\}$ (we will only consider type 1 representations of $\Uq$ in this paper).
Let $\Oint{\g_V}$ be as in \cite[Chapter 7]{HK}, the category of finite-dimensional $\Uq$-modules such that the weight of any nonzero weight space belongs to $\ZZ^\dv_{\geq 0} \subseteq \wl(\g_V)$.  It is semisimple, the irreducible objects being the highest weight modules $V_\lambda$ for partitions  $\lambda$.

Now by \cite[Corollary 54, Chapter 11]{KS}, there is a nondegenerate Hopf pairing between  $U_q(\g_V)$ and  $\O(GL_q(V))$.  So, as discussed in  \textsection\ref{ss dually paired  Hopf algebras}, any right  $\O(GL_q(V))$-comodule is also a left  $\Uq$-module.
All of the objects of  $\Oint{\g_V}$ in fact come from  $\O(GL_q(V))$-comodules;
from now on, $V_\lambda$ is understood to be both a $\Uq$-module and the corresponding $\O(GL_q(V))$-module.
By the Peter-Weyl theorem for $\O(M_q(V))$ \cite[Theorem 21, Chapter 11]{KS}, the objects of  $\Oint{\g_V}$ are exactly the $\O(M_q(V))$-comodules.

For any object $N$ of $\Oint{\g_V}$ and partition $\lambda$, let $N[\lambda]$ be the $V_\lambda$-isotypic component of $N$. Set $N[\ld\lambda] = \bigoplus_{\mu \ld\lambda} N[\mu]$, $N[\ldneq\lambda] = \bigoplus_{\mu \ldneq\lambda} N[\mu]$. Let $\varsigma_\lambda^N : N \twoheadrightarrow N[\lambda]$ be the canonical surjection and $\iota_\lambda^N : N[\lambda] \hookrightarrow N$ the canonical inclusion. Define the projector $\pi_\lambda^N: N \to N$ by $\pi_\lambda^N = \iota_\lambda^N \circ \varsigma_\lambda^N$.

\section{Bases for $GL_q(V)$ modules}
\label{s Bases for GLq modules}
We recall some facts we will need about the Gelfand-Tsetlin basis and canonical basis of $V_\lambda$.
We then recall the construction of global crystal bases in the sense of \cite{Kas1,Kas2} and of the similar notion of based modules of \cite{LBook}. We will also make use of the projected canonical basis defined in \cite{BProjected}.

\subsection{Gelfand-Tsetlin bases and Clebsch-Gordon coefficients} \label{sgelfand}
Standard results for  the unitary group $\unitary(V)$
have their analogs for $\unitary_q(V)$. In this section, we describe results of this kind
that we  need; see \cite{KS,vilenkin} for their detailed description.  As in \textsection\ref{ss compact},
we work over the field  $\CC$ and  $q$ is assumed to be a real number such that  $q \neq 0, \pm1$.

Recall that $V_\lambda$ denotes the irreducible left $U_q(\g_V)$-module of highest weight $\lambda$; this also corresponds to a right $\O(GL_q(V))$-comodule.
Let $\{ \bracket{M}\}$ denote the orthonormal
Gelfand-Tsetlin basis for $V_\lambda$, where $M$ ranges over Gelfand-Tsetlin
tableaux of shape $\lambda$.
Gelfand-Tsetlin tableaux are equivalent to semistandard Young tableaux (SSYT) and in examples, as below, we will use SSYT to label the elements of this basis.
\begin{example}
\label{ex Gelfand-Tsetlin}
The orthonormal Gelfand-Tsetlin of  $V^{\tsr r}$ is orthonormal with respect to the Hermitian form on $V^{\tsr r}$ in which the standard monomial basis is orthonormal.
The orthonormal Gelfand-Tsetlin basis of  $V^{\tsr 2}$ when  $\dv =2$ is given by
\setlength{\cellsize}{1.5ex}
\begin{equation}
\begin{array}{rl@{\qquad}rl}
\bracket{\tiny \tableau{1 & 1}}&=  \bv_{11},\\
\bracket{\tiny \tableau{1 & 2}}&= \big(\frac{q}{[2]}\big)^{1/2}(q \bv_{12}+ \bv_{21}),\\
\bracket{\tiny \tableau{2 & 2}}&=\bv_{22}, \\
\bracket{\tiny \tableau{1 \\ 2}}&=\big(\frac{q}{[2]}\big)^{1/2}(\bv_{21} - q^{-1} \bv_{12}),
\end{array}
\end{equation}
\setlength{\cellsize}{2.5ex}
where  $\bv_{ij} := v_i \tsr v_j$.
Note that in the  $\dv =2$ case, the orthonormal Gelfand-Tsetlin basis of  $V^{\tsr r}$ is proportional to the projected upper canonical basis
$\tilde{B}^r_V$ of \textsection\ref{ss upper canonical basis of bT}.
\end{example}
\begin{remark}
The orthonormal Gelfand-Tsetlin basis described in  \cite[\textsection7.3]{KS} is for a slightly larger quantized enveloping algebra $\breve{U}_q(\g_V)$, with a different coproduct than that used here.  The normalization required to make the Gelfand-Tsetlin basis orthonormal is therefore slightly different here than in this reference.
\end{remark}

The tensor product of two irreducible  $\O(GL_q(V))$-comodules decomposes as
\begin{equation} \label{eqdecomptensor}
V_\lambda \otimes V_\mu = \bigoplus_{\nu,i} (V_\nu)_i,
\end{equation}
where $i$ labels
different  copies of $V_\nu$---the number of these copies  is
the Littlewood-Richardson coefficient $c^\nu_{\lambda \, \mu}$.

The Clebsch-Gordon (Wigner) coefficients (CGCs) of this tensor product are defined by the formula
\begin{equation}\label{e CGCs}
\bracket{M}_i = \sum_{N,K} C_{N K M,i}^{\lambda,\mu,\nu}  \bracket{N} \otimes \bracket{K},
\end{equation}
where $N$ and $K$ range over Gelfand-Tsetlin tableaux of shapes $\lambda$
and $\mu$, respectively, and $\bracket{M}_i$ denotes the Gelfand-Tsetlin basis element of
$(V_\nu)_i$ in \eqref{eqdecomptensor} labeled by the Gelfand-Tsetlin tableau $M$ of
shape $\nu$.
By orthonormality, \eqref{e CGCs} can be inverted to obtain
\begin{equation}\label{e CGCs inverted}
\bracket{N} \otimes \bracket{K} = \sum_{\nu, i, M} \overline{C_{N K M,i}^{\lambda,\mu,\nu}} \bracket{M}_i,
\end{equation}
where the bar denotes complex conjugation, $\nu$ and $i$ range as in the right-hand side of \eqref{eqdecomptensor}, and $M$ ranges over Gelfand-Tsetlin tableaux of shape $\nu$.

We denote  $C_{N K M,i}^{\lambda,\mu,\nu}$ by simply $C_{N K M,i}$ if the shapes are understood.
These coefficients have been intensively studied in the literature;
 see  \cite{KS,vilenkin} and the references therein.
An explicit formula for them is known when either $V_\lambda$ or
$V_\mu$ is a  fundamental vector representation,
or more generally, a symmetric representation.
In the presence of multiplicities, the Clebsch-Gordon coefficients
are not uniquely determined, and do not have explicit formulae in general.

\subsection{Crystal bases}
\label{ss crystal bases}
An \emph{upper crystal basis at  $\u = \infty$} of $N \in \Oint{\g_V}$ is a  pair  $(\L(N),\B)$, where $\L(N)$ is a $\field_\infty$-submodule of $N$ and $\B$ is a $\QQ$-basis of $\L(N) / \ui \L(N)$ which satisfy a certain compatibility with the Kashiwara operators $\crystalusual{E_i}, \crystalusual{F_i}$ (see \cite[\textsection3.1]{Kas2}).  The lattice $\L(N)$ of the pair is called an \emph{upper crystal lattice at $\u = \infty$} of  $N$.

Kashiwara \cite{Kas2} gives a fairly explicit construction of an upper crystal basis of $V_\lambda$, which we denote by $(\L(\lambda),\B(\lambda))$. The basis  $\B(\lambda)$ is naturally labeled by $\text{SSYT}_{\dv}(\lambda)$ and we let $b_P$ denote the basis element corresponding to $P \in \text{SSYT}_{\dv}(\lambda)$ (see, for instance, \cite[Chapter 7]{HK}).
A fundamental result of \cite{Kas1,Kas2} is that an upper crystal basis is always isomorphic to a direct sum $\bigoplus_j (\L(\lambda^j),\B(\lambda^j))$, i.e., each  $N \in \Oint{\g_V}$ has a unique upper crystal basis.

The \emph{crystal graph} of an upper crystal basis $(\L,\B)$ is the colored directed graph with vertex set  $\B$, and, for each  $\flat \in \B$ such that $\crystalusual{F_i}(\flat) \neq 0$, a directed edge from $\flat$ to  $\crystalusual{F_i}(\flat)$ with color $i$. A \emph{crystal component} of a crystal graph is a connected component of the underlying undirected colorless graph.  By the uniqueness result for upper crystal bases, a crystal graph is always the disjoint union of crystal graphs of some $(\L(\lambda^j),\B(\lambda^j))$; also, it is well known that the crystal graphs of irreducibles are connected, so the decomposition of  $N$ into irreducibles is given by the decomposition of the crystal graph of  $(\L(N),\B)$  into connected components.

\subsection{Global crystal bases}
\label{ss global canonical bases}
We next define upper based $\Uq$-modules, which is similar to the based modules of \cite[Chapter 27]{LBook} (see \cite[\textsection 4.2]{BProjected}).

The \emph{$\mathbf{A}$-form $\Uq_\mathbf{A}$ of $\Uq$} is the $\mathbf{A}$-subalgebra of $\Uq$ generated by $E_i^{(m)} := \frac{E_i^m}{[m]!}, F_i^{(m)} := \frac{F_i^m}{[m]!}, q^h, \genfrac{\{}{\}}{0pt}{}{q^{h}}{m}$ for $i \in \rootset,\ m \in \ZZ_{\geq 0}$, and $h \in \wl(\g_V)^*$, where
\[\genfrac{\{}{\}}{0pt}{}{x}{m} := \prod^m_{k=1} \frac{\u^{1-k}x - \u^{k-1}x^{-1}}{\u^k-\u^{ -k}}. \]
We also define the \emph{$\QQA$-form $\Uq_\QQ$ of  $\Uq$} to be $\QQ \tsr_\ZZ \Uq_{\mathbf{A}}$.

\begin{definition}\label{d upper based}
An \emph{upper based $\Uq$-module} is a pair $(N,B)$, where $N$ is an object of $\Oint{\g_V}$
and $B$ is a $\field$-basis of $N$ such that
\begin{list}{\emph{(\alph{ctr})}} {\usecounter{ctr} \setlength{\itemsep}{1pt} \setlength{\topsep}{2pt}}
\item $B \cap N^{\zeta}$ is a basis of $N^\zeta$, for any $\zeta \in \wl(\g_V)$;
\item Define $N_\mathbf{A} := \mathbf{A} B$. The $\QQA$-submodule  $\QQ \tsr_\ZZ N_\mathbf{A}$ of $N$ is stable under  $\Uq_\QQ$;
\item the $\QQ$-linear involution $\br{\cdot} : N \to N$ defined by $\br{ab} = \br{a}b$ for all $a \in \field$ and all $b \in B$ intertwines the $\br{\cdot}$-involution of $\Uq$, i.e. $\br{fn} = \br{f}\br{n}$ for all $f \in \Uq, n \in N$;
\item Set $\L(N) = \field_\infty B$ and let $\B$ denote the image of $B$ in $\L(N)/\ui \L(N)$. Then $(\L(N), \B)$ is an upper crystal basis of $N$ at $\u = \infty$.
\end{list}
\end{definition}
The  \emph{$\br{\cdot}$-involution} of an upper based $\Uq$-module  $(N,B)$ is the involution on $N$ defined in (c), its \emph{balanced triple} is $(\QQA B, \field_0B, \field_\infty B)$, and its \emph{upper crystal basis} is that of (d).  The \emph{crystal graph}  $\G$ of  $(N,B)$ is the crystal graph of its upper crystal basis, and, as a slight abuse notation, we identify the vertex set of  $\G$ with $B$;  a \emph{crystal component} of $(N,B)$ is a crystal component of $\G$, and is identified with a subset of $B$.  Since in this paper based modules are emphasized over crystal bases, it is convenient to define the following global versions of the Kashiwara operators:
\be \label{e global Kashiwara operators}
\begin{array}{cc}
\crystal{F_i} : B \to B \sqcup \{0\}, & \crystal{F_i} := G \circ \crystalusual{F_i} \circ G^{-1},\\
\crystal{E_i} : B \to B \sqcup \{0\}, & \crystal{E_i} := G \circ \crystalusual{E_i} \circ G^{-1},
\end{array}
\ee
for any $i \in \rootset$, where $G^{-1}$ is the canonical isomorphism $B \xrightarrow{\cong} \B$.
An element  $b \in B$ is \emph{highest weight} if  $\crystal{E_i} b = 0$ for all  $i \in \rootset$.

\begin{remark}\label{r balanced triple}
In the language of Kashiwara \cite{Kas2}, the basis  $B$ in the definition above is an upper global crystal basis with respect to its balanced triple.  To define global upper crystal bases, Kashiwara first defines a balanced triple  $(\QQ \tsr_\ZZ N_\mathbf{A}, \br{\L(N)}, \L(N))$ and a basis  $\B \subseteq \L / \ui \L$ and then defines $B$ to be the inverse image of  $\B$ under the isomorphism
\[ \QQ \tsr_\ZZ N_\mathbf{A} \cap \br{\L(N)} \cap \L(N) \xrightarrow{\cong} \L / \ui \L.\]
\end{remark}

Let $\eta_\lambda$ be a highest weight vector of $V_\lambda$.  The $\br{\cdot}$-involution on $V_\lambda$ is defined by setting $\br{\eta_\lambda} = \eta_\lambda$ and requiring that it intertwines the $\br{\cdot}$-involution of $\Uq$.
The upper $\QQA$-form of  $V_\lambda$ of \cite{Kas2} is denoted $V^{\QQ \text{ up}}_\lambda$, which is a $\Uq_{\QQ}$-submodule of $V_\lambda$.
We can now state the fundamental result about the existence of global crystal bases and based modules for $V_\lambda$.
\begin{theorem}[Kashiwara \cite{Kas2}]
\label{t theorem 6.2.2 HK}
The triple $(V^{\QQ \text{ up}}_\lambda, \br{\L(\lambda)}, \L(\lambda))$ is balanced.  Then, letting $G_\lambda$ be the inverse of the canonical isomorphism
\[ V^{\QQ \text{ up}}_\lambda \cap \br{\L(\lambda)} \cap \L(\lambda)  \xrightarrow{\cong} \L(\lambda) / \ui \L(\lambda),\]
 $B(\lambda) := G_\lambda(\B(\lambda))$ is the \emph{upper global crystal basis of  $V_\lambda$} and $(V_\lambda,B(\lambda))$ is an upper based $\Uq$-module.
\end{theorem}
Note that Kashiwara proves that the triples are balanced and the conclusions about based modules follow easily (see \cite[27.1.4]{LBook} or {\cite[Theorem 6.2.2]{HK}}).  We may now define the upper integral form of $V_\lambda$ to be $V^{\mathbf{A} \text{ up}}_\lambda := \mathbf{A}B(\lambda)$.  We say that the element  $\eta_\lambda$ is the \emph{canonical highest weight vector} of $(V_\lambda,B(\lambda))$.

We will need some facts about lower based modules from \cite[Chapter 27]{LBook}, or rather, their corresponding statements for upper based $\Uq$-modules. It is shown in \cite[\textsection 5.2]{Kas2} that (see \textsection\ref{ss modules of GLq} for notation)
\be \label{e based module filtration}
\parbox{14.1cm}{
if $(N,B)$ is an upper based  $\Uq$-module, then so are $(N[\ld \lambda], B[\ld \lambda])$, $(N[\ldneq \lambda], B[\ldneq \lambda])$, and  $(N[\lambda],\varsigma_\lambda^N(B[\lambda]))$, where $B[\ld \lambda] = N[\ld \lambda] \cap B$, $B[\ldneq \lambda] = N[\ldneq \lambda] \cap B$, and  $B[\lambda] = B[\ld \lambda] \setminus B[\ldneq \lambda]$. Moreover, $(N[\lambda],\varsigma_\lambda^N(B[\lambda]))$ is isomorphic as an upper based  $\Uq$-module to a direct sum of copies of $(V_\lambda, B(\lambda))$.}
\ee
As a consequence, the  $\Uq$-cells of the module with basis $(N,B)$ coincide with its crystal components.  An additional consequence is that  $b \in B$ being highest weight is equivalent to $E_i b = 0$ for all  $i \in \rootset$ (warning: this is not true for lower based modules).
%

\subsection{Projected based modules}
\label{ss projected based modules}
We now define the projected based  $\Uq$-module  $(N,\tilde{B})$ of a based  $\Uq$-module  $(N,B)$, following \cite{BProjected}. For this we need an integral form that is different from $N_\mathbf{A}$.  Set  $\L = \L(N)$.
The upper based $\Uq$-module  $(N[\ld \lambda],B[\ld \lambda])$ from the previous subsection has balanced triple
\be \label{e lambda triple N}
(N_\mathbf{A}[\ld \lambda]/ N_\mathbf{A}[\ldneq \lambda], \br{\L[\ld \lambda]}/\br{\L[\ldneq \lambda]}, \L[\ld \lambda]/ \L[\ldneq \lambda]),
\ee
where $N_\mathbf{A}[\ld \lambda]$ and $\L[\ld \lambda]$ (resp. $N_\mathbf{A}[\ldneq \lambda]$ and $\L[\ldneq \lambda]$) are the $\mathbf{A}$- and $\field_\infty$- span of  $B[\ld \lambda]$ (resp. $B[ \ldneq \lambda]$).
Now define
\be \label{e tildeN definition upper}
\begin{array}{c@{\ :=\ }l@{\ \subseteq\ }l}
N_{\mathbf{A},\lambda} & \varsigma_\lambda^N(N_\mathbf{A}[\ld \lambda]) & N[\lambda], \\
\L_{\lambda} &\varsigma_\lambda^N(\L[\ld \lambda]) & N[\lambda], \\
\tilde{N}_\mathbf{A} & \bigoplus_\lambda N_{\mathbf{A}, \lambda} & N.
\end{array}
\ee

We will make use of the following result giving several descriptions of projections of upper based  $\Uq$-modules.  This is slightly more general than the similar result \cite[Theorem 6.1]{BProjected}, which is proved in the context of Schur-Weyl duality in type $A$.
\begin{theorem}
\label{t projected upper canonical basis of N}
Maintain the notation above and that of \textsection\ref{ss global canonical bases}.  Let $b \in B[\lambda]$ and  $\b$  its image in  $\L / \ui \L$.    The element $\varsigma_\lambda^N(b)$ belongs to a copy of  $(V_\lambda, B(\lambda))$ in  $(N[\lambda],\varsigma_\lambda^N(B[\lambda]))$ with canonical highest weight vector $\varsigma_\lambda^N(b_\text{hw})$ for some $b_\text{hw} \in B[\lambda]$.
Let $b_P \in \B(\lambda)$ be such that $\varsigma_\lambda^N(b) = G_\lambda(b_P)$ in this copy and let $V_{b_\text{hw}} = \Uq b_\text{hw} \subseteq N$.
Then the triples in (b) and (c) are balanced and the projected upper canonical basis element $\liftb$ has the following descriptions
\begin{list}{\emph{(\alph{ctr})}} {\usecounter{ctr} \setlength{\itemsep}{1pt} \setlength{\topsep}{2pt}}
\item the unique  $\br{\cdot}$-invariant element of $\tilde{N}_\mathbf{A}$ congruent to $b \mod \ui \L$,
\item $\tilde{G}(\b)$, where $\tilde{G}$ is the inverse of the canonical isomorphism
\[\QQ \tsr_\ZZ \tilde{N}_\mathbf{A} \cap \br{\L}\cap\L  \xrightarrow{\cong} \L / \ui \L, \]
\item $\tilde{G}_\lambda(\b_\lambda)$, where  $\b_\lambda$ is image of $\varsigma_\lambda^{N}(b)$ in $\L_{\lambda} / \ui \L_{\lambda}$ and  $\tilde{G}_\lambda$ is the inverse of the canonical isomorphism
\[\QQ \tsr_\ZZ N_{\mathbf{A},\lambda} \cap \br{\L_{\lambda}}\cap \L_{\lambda}  \xrightarrow{\cong} \L_{\lambda} / \ui \L_{\lambda}, \]
\item the global crystal basis element $G_\lambda(b_P)$ of $V_{b_\text{hw}}$,
\item $\pi_\lambda^N(b)$.
\end{list}
Then $(N, \tilde{B})$, with $\tilde{B} := \{\liftb: b \in B\}$, is an upper based  $\Uq$-module, referred to as the \emph{projected upper based  $\Uq$-module of $(N,B)$}.
\end{theorem}
\begin{proof}
The proof is similar to that of \cite[Theorem 6.1]{BProjected}, which follows in a straightforward way from results of \cite[\textsection 5.2]{Kas2} and the uniqueness of upper crystal bases.  The proof of   \cite[Theorem 6.1]{BProjected} goes by showing that the elements in (b)--(e) are the same and then showing that these are $\br{\cdot}$-invariant, hence equal to the element in (a).  We replace the proof of  $\br{\cdot}$-invariance in \cite{BProjected} by the following: the element in (d) is  $\br{\cdot}$-invariant because the  $\br{\cdot}$-involutions on  $N$ and $V_{b_\text{hw}}$ intertwine the bar-involution on  $\Uq$, and  $\br{b}_\text{hw} = b_\text{hw}$.
\end{proof}

Given  $(N,B)$ and  $(N,\tilde{B})$ as in theorem, let  $(m_{\tilde{b'}b})_{\{\tilde{b'}\in\tilde{B},b \in B\}}$ be the transition matrix from  $B$ to  $\tilde{B}$ (so that $b = \sum_{\tilde{b'} \in \tilde{B}} m_{\tilde{b'}b} \tilde{b'}$ for all $b \in B$).  It follows that for any partition $\mu$ and $b \in B$,
\be \label{e projection transition matrix}
\pi^N_\mu(b) = \sum_{\tilde{b'} \in \tilde{B}[\mu]} m_{\tilde{b'} b} \tilde{b'}. \ee
For later use, we record the following easy corollary.

\begin{corollary}
\label{c clifts and projections general}
Maintain the notation of the previous paragraph. For any $b \in B[\lambda]$,
\be \label{e lifts and projections general}
 \pi^N_\mu(b) \begin{cases}
=\tilde{b} & \text{ if } \mu = \lambda,\\
=0 & \text{ if } \mu \not\ld \lambda, \\
\in \ui \L \cap \u \br{\L} & \text{ if } \mu \ldneq \lambda.
\end{cases}
\ee
\end{corollary}
\begin{proof}
Theorem \ref{t projected upper canonical basis of N} (e) and \eqref{e based module filtration} yield the top and middle case of \eqref{e lifts and projections general}, respectively.  Next, note that Theorem \ref{t projected upper canonical basis of N} (b) implies $\sum_{\tilde{b'} \in \tilde{B}, b' \neq b} m_{\tilde{b'} b} \tilde{b'} = b-\tilde{b} \in \ui \L \cap \u \br{\L}$.  The bottom case then follows from \eqref{e projection transition matrix}.
\end{proof}

\subsection{Tensor products of based modules}
\label{ss Tensor products of based modules}
Let $(N,B), (N',B')$ be upper based $\Uq$-modules.  There is a basis $B \heart B'$ which makes $N \tsr N'$ into an upper based $\Uq$-module.  However, first, we need an involution on  $N \tsr N'$ that intertwines the $\br{\cdot}$-involution on $\Uq$.  This definition is not obvious and requires Lusztig's quasi-$\R$-matrix, but adapted to our coproduct as in \cite{Brundan}: let $\Theta = (\varphi \tsr \varphi)(\tilde{\Theta}^{-1})$ where $\tilde{\Theta}$ is exactly Lusztig's quasi-$\R$-matrix from \cite[\textsection4.1.2]{LBook}. It is an element of a certain completion $(\Uq \tsr \Uq)^{\wedge}$ of the algebra $\Uq \tsr \Uq$.
Then the involution $\br{\cdot}:N \tsr N' \to N \tsr N'$ is defined by $\br{n \tsr n'} = \Theta(\br{n} \tsr \br{n'})$. (This involution is denoted $\Psi$ in \cite{LBook}.)

As discussed in \cite[\textsection4.4]{BProjected}, the corresponding result for the based modules of Lusztig (\cite[Theorem 27.3.2]{LBook}) adapts to this setting:
\begin{theorem} \label{t tsr product upper canbas}
Maintain the notation above with $(N,B), (N', B')$ upper based $\Uq$-modules and set $(N \tsr N')_{\ZZ[\ui]} = \ZZ[\ui] B \tsr B'$. For any $(b,b') \in B \times B'$, there is a unique element $b \heart b' \in (N \tsr N')_{\ZZ[\ui]}$ such that $\br{b\heart b'} = b \heart b'$ and $b\heart b'-b\tsr b' \in \ui (N\tsr N')_{\ZZ[\ui]}$.

Set $B \heart B' = \{b\heart b' : b \in B, b' \in B'\}$. Then the pair $(N \tsr N', B\heart B')$ is an upper based  $\Uq$-module.
Moreover, the product $\heart$ is associative.
\end{theorem}
We define the $\heart$ product on all of $N\times N'$ by extending the product just defined $\field$-bilinearly.

We will come across the following situation in our application to the Kronecker problem.
\begin{proposition}
\label{p heart commutes with projections}
Maintain the notation of this and the previous two subsections. Let $(N_1, B_1), \ldots,$ $(N_l,B_l)$
be upper based $\Uq$-modules.  Let $b_i \in B_i$,  $i \in [l]$, be given and define $\lambda^i$ such that $b_i \in B_i[\lambda^i]$.
For each $i \in [l]$, let $\pi_i$ be either  $\pi_{\lambda^i}^{N_i}$ or the identity map on $N_i$; set $B'_i = \tilde{B}_i$ (resp.  $B_i' = B_i$) if  $\pi_i = \pi_{\lambda^i}^{N_i}$ (resp.  $\pi_i = \id^{N_i}$), where  $(N_i, \tilde{B}_i)$ is the projected upper based  $\Uq$-module of  $(N_i,B_i)$. Then
\[ \pi_1(b_1) \heartproj \cdots \heartproj \pi_l(b_l) = \pi_1 \tsr \cdots \tsr \pi_l(b_1 \heart \cdots \heart b_l),
\]
where  $\heartproj$ (resp.  $\heart$) denotes the construction of Theorem \ref{t tsr product upper canbas} for the upper based $\Uq$-modules $(N_1, B'_1),\ldots,(N_l,B'_l)$ (resp. $(N_1, B_1),\ldots,(N_l,B_l)$).
\end{proposition}
\begin{proof}
Set $b := b_1 \heart \cdots \heart b_l$ and $\pi := \pi_1 \tsr \cdots \tsr \pi_l$. It suffices to show that  $\pi(b)$ satisfies the defining properties of  $\pi_1(b_1) \heartproj \cdots \heartproj \pi_l(b_l)$. Since the elements of $\tilde{B}_i$ are $\br{\cdot}$-invariant, the transition matrix from  $B_i$ to  $\tilde{B}_i$ consists of $\br{\cdot}$-invariant elements of  $\field$.  It follows from \eqref{e projection transition matrix} that the projector  $\pi_i$ intertwines the $\br{\cdot}$-involution on $N_i$. Hence $\br{\pi(b)} = \pi(b)$.

It is evident from description (b) of Theorem \ref{t projected upper canonical basis of N} that the lattice $\L(N_i)$ is the same for the based modules $(N_i, B_i)$ and $(N_i, \tilde{B}_i)$, and that $\pi_i(\L(N_i)) = \L(N_i)$. Set $\L := \L(N_1) \tsr_{\field_\infty} \cdots \tsr_{\field_\infty} \L(N_l)$. Hence applying $\pi$ to $b - (b_1 \tsr \cdots \tsr b_l) \in \ui \L$ implies
\[
\pi(b) - \pi_1(b_1) \tsr \cdots \tsr \pi_l(b_l) \in \ui \L,
\]
so $\pi(b)$ satisfies the defining properties of $\pi_1(b_1) \heartproj \cdots \heartproj \pi_l(b_l)$.
\end{proof}

\section{Quantum Schur-Weyl duality and canonical bases}
\label{s Quantum Schur-Weyl duality and canonical bases}
Write $V=V_{\epsilon_1}$ for the natural representation of $\Uq$.
The action of  $\Uq$ on the weight basis $v_1,\ldots,v_\dv$ of  $V$ is given by $q^{\epsilon^{i}} v_j = \u^{\delta_{ij}} v_j$, $F_i v_i = v_{i+1}$, $F_i v_j = 0$ for $i \neq j$, and $E_i v_{i+1} = v_i$, $E_i v_j = 0$ for $j \neq i+1$.

In this section we describe the commuting actions of $\Uq$ and $\H_r$ on $\bT := V^{\tsr r}$ as in \cite{Jimbo,GL, Ram, Brundan} and give several characterizations of the upper canonical basis and projected upper canonical basis of $\bT$; we closely follow \cite{Brundan,BProjected} and are consistent with their conventions.  This background will be needed to construct an upper canonical basis for  $\nswedge{r}{X}$ in  \textsection\ref{s A canonical basis for Yalpha} and motivates the hypothesized basis for $\nsbr{X}^{\tsr r}$ detailed in Conjecture \ref{cj canonical basis X^r}.

\subsection{Commuting actions on $\bT = V^{\tsr r}$}
\label{ss commuting action on Vtsrr}
The action of $\Uq$ on $\bT$ is determined by the coproduct $\Delta$  \eqref{e U_q coproduct}.
The commuting action of $\H_r$ on $\bT$ is defined by sending $T_i$ to $\hat{\R}_i$, where $\hat{\R}_i$ denotes the $\Uq$-isomorphism of $V^{\tsr r}$ equal to $\hat{\R}_{V,V}$ on the  $i$ and  $i+1$-st tensor factors and the identity elsewhere. Here $\hat{\R}_{V,V}$ denotes the $\hat{\R}$-matrix defined in \textsection\ref{ss FRT algebras}.
Equation \eqref{e hat R formula} for  $\hat{\R}_{V,V}$ gives an explicit form for the $\H_r$ action, which we reformulate as follows: for a word $\mathbf{k} = k_1\dots k_r \in [\dv]^r$, let $\bv_{\mathbf{k}} = v_{k_1} \tsr v_{k_2} \tsr \dots \tsr v_{k_r}$ be the corresponding tensor monomial. Recall from  \textsection\ref{ss type A combinatorics preliminaries} the right action of  $\S_r$ on words of length  $r$. Then
\be
\label{e T inverse_i act on V}
\bv_\mathbf{k} T_i^{-1} =
\begin{cases}
\bv_{\mathbf{k} \, s_i} & \text{ if } k_i < k_{i+1}, \\
\ui \bv_\mathbf{k} & \text{ if } k_i = k_{i + 1}, \\
(\ui - \u) \bv_\mathbf{k} + \bv_{\mathbf{k} \, s_i} & \text{ if } k_i > k_{i+1}.
\end{cases}
\ee

\begin{remark}
This convention for the action of $\H_r$ on $\bT$ is consistent with that in \cite{Brundan, Ram}, but not with that in \cite{GL}. Note that  $\bv_\mathbf{k}, T_i^{-1}$ are denoted $M_\alpha, H_i$ respectively in \cite{Brundan}.
\end{remark}

Schur-Weyl duality generalizes nicely to the quantum setting:
\begin{theorem}[Jimbo \cite{Jimbo}]\label{c Schur-Weyl duality basic}
As a $(\Uq, \field\H_r)$-bimodule, $\bT$ decomposes into irreducibles as
\[
\bT \cong \bigoplus_{\lambda \vdash_\dv r}  V_\lambda \tsr M_\lambda.
\]
\end{theorem}

As an $\H_r$-module, $\bT$ decomposes into a direct sum of weight spaces: $\bT \cong \bigoplus_{\zeta \in \wl(\g_V)} \bT^\zeta$.  The weight space $\bT^\zeta$ is the $\field$-vector space spanned by $\bv_{\mathbf{k}}$ such that $\mathbf{k}$ has content $\zeta$.
Let $\epsilon_+ := M_{(r)}^\mathbf{A}$ be the trivial $\H_r$-module, i.e. the one-dimensional module identified with the map $\H_r \to \mathbf{A}$, $T_i \mapsto \u$. It is not difficult to prove using \eqref{e T inverse_i act on V} (see \cite[\textsection 4]{Brundan})
\begin{proposition}
\label{p weight space equals induced}
The map $\bT_\mathbf{A}^\zeta \to \epsilon_+ \tsr_{\H_{J_\zeta}} \H_r$ given by $\bv_\mathbf{k} \mapsto \epsilon_+ \tsr_{\H_{J_\zeta}} \br{T}_{d(\mathbf{k})}$ is an isomorphism of right $\H_r$-modules.
\end{proposition}
Here  $d(\mathbf{k})$ is as in  \textsection\ref{ss type A combinatorics preliminaries} and $\bT_\mathbf{A}$ is the integral form of $\bT$, defined below.

\subsection{Upper canonical basis of $\bT$}
\label{ss upper canonical basis of bT}
We now apply the general theory of  \textsection\ref{ss global canonical bases}, \textsection\ref{ss Tensor products of based modules} to construct a global crystal basis of $\bT$.  Recall from  \textsection\ref{ss Tensor products of based modules} that there is a $\br{\cdot}$-involution on  $\bT$ defined using the quasi-$\R$-matrix.
The $\br{\cdot}$-involution on $\H_r$ intertwines that of $\bT$, i.e., $\br{v h} = \br{v} \ \br{h}, \text{ for any } v \in \bT, \ h \in \H_r$ \cite{LBook, Brundan}.

Let $V_\mathbf{A} = \mathbf{A}\{v_i: i \in [\dv]\}$, which is the same as the integral form $V^{\mathbf{A} \text{ up}}_{\epsilon_1}$ from  \textsection\ref{ss global canonical bases}.
By Theorem \ref{t tsr product upper canbas} and associativity of the $\heart$ product, $(\bT, B^r)$ is an upper based $\Uq$-module with balanced triple $(\QQ \tsr_\ZZ \bT_\mathbf{A}, \br{\L}, \L)$, where
\be
\begin{array}{l@{\ :=\ }l}
\L & \L(\epsilon_1) \tsr_{\field_\infty} \dots \tsr_{\field_\infty} \L(\epsilon_1), \\
\bT_{\ZZ[\ui]} & \ZZ[\ui] \{\bv_\mathbf{k} : \mathbf{k} \in [\dv]^r\}, \\
\bT_\mathbf{A} & V_\mathbf{A} \tsr_\mathbf{A} \dots \tsr_\mathbf{A} V_\mathbf{A} = \mathbf{A} \tsr_\ZZ \bT_{\ZZ[\ui]}, \\
B^r & B(\epsilon_1) \heart \dots \heart B(\epsilon_1).
\end{array}
\ee
We call $B^r$ the \emph{upper canonical basis} of $\bT$ and, for each  $\mathbf{k} \in [\dv]^r$, write $c_\mathbf{k}$ for the element $v_{k_1} \heart \dots \heart v_{k_r} \in B^r$.
Figure \ref{f c_111 example upper} from the introduction gives the upper canonical basis in terms of the monomial basis for $r = 3, \ \dv =2$.

\begin{theorem}[\cite{GL, Brundan} (see {\cite[Theorem 5.6]{BProjected}})]
\label{t Schur-Weyl duality upper}
The upper canonical basis element $c_\mathbf{k}, \, \mathbf{k} \in [\dv]^r,$ has the following equivalent descriptions
\begin{list}{\emph{(\roman{ctr})}} {\usecounter{ctr} \setlength{\itemsep}{1pt} \setlength{\topsep}{2pt}}
\item the unique $\br{\cdot}$-invariant element of $\bT_{\ZZ[\ui]}$, congruent to $\bv_\mathbf{k} \mod \ui\bT_{\ZZ[\ui]}$;
\item $v_{k_1} \heart \dots \heart v_{k_r}$;
\item The image of $C_{d(\mathbf{k})}$ under the isomorphism in Proposition \ref{p weight space equals induced}.
\end{list}
\end{theorem}

The next result is a slightly more precise version of Theorem \ref{t intro standard Schur-Weyl duality}. As explained in the introduction, it connects quantum Schur-Weyl duality with the RSK correspondence and is our model for constructing a basis of $\nsbr{X}^{\tsr r}$ that solves the Kronecker problem.
\begin{theorem}[\cite{GL} (see {\cite[Corollary 5.7]{BProjected}})]
\label{t standard Schur-Weyl duality2}
\
\begin{list}{\emph{(\roman{ctr})}} {\usecounter{ctr} \setlength{\itemsep}{1pt} \setlength{\topsep}{2pt}}
\item The $ \H_r$-module with basis $(\bT, B^r)$ decomposes into $ \H_r$-cells as
\[B^r = \bigsqcup_{T \in \text{SSYT}_{\dv}^r} \Gamma_T, \quad \text{where }\ \Gamma_T := \{c_\mathbf{k} : P(\mathbf{k}) = T \}.\]
\item The $ \H_r$-cell $ \Gamma_T$ of  $\bT$ is isomorphic to $(M_{\sh(T)}, \Gamma_{\sh(T)})$ of  \textsection\ref{ss cell label conventions C_Q C'_Q}.
\item The  $\Uq$-module with basis $(\bT, B^r)$ decomposes into $\Uq$-cells as
\[B^r = \bigsqcup_{\lambda \vdash_\dv r, \ T \in \text{SYT}(\lambda)} \Lambda_T, \quad \text{where }\ \Lambda_T  = \{c_\mathbf{k} : Q(\mathbf{k}) = T \}.\]
\item The $\Uq$-cell $ \Lambda_T$ is isomorphic to $(V_{\sh(T)}, B(\sh(T)))$ of Theorem \ref{t theorem 6.2.2 HK}.
\end{list}
\end{theorem}

We conclude this subsection with an explicit description of the projected upper canonical basis $(\bT, \tilde{B}^r)$ of  $(\bT,B^r)$.
\begin{theorem}[{\cite[Theorem 6.1]{BProjected}}]
\label{t lifted upper canonical basis}
Let $(\bT, \tilde{B}^r = \{\liftc_\mathbf{k}: \mathbf{k}\in [\dv]^r\})$ be the projected upper canonical basis of  $(\bT,B^r)$ and $\tilde{\bT}_\mathbf{A} = \bigoplus_\lambda \pi_\lambda^\bT(\bT_\mathbf{A}[\ld \lambda])$ the integral form as in Theorem \ref{t projected upper canonical basis of N}.
Let $\mathbf{l} \in [\dv]^r$ and $\lambda = \sh(\mathbf{l})$. Set $\mathbf{j} = \text{RSK}^{-1}(Z_\lambda,Q(\mathbf{l}))$, where $Z_\lambda$ is the superstandard tableau of shape $\lambda$ (see  \textsection\ref{ss type A combinatorics preliminaries}).
Let $V_{Q(\mathbf{l})} = \Uq c_\mathbf{j}$.
Then the triple in (b) is  balanced and the projected upper canonical basis element $\liftc_{\mathbf{l}}$ has the following descriptions
\begin{list}{\emph{(\alph{ctr})}} {\usecounter{ctr} \setlength{\itemsep}{1pt} \setlength{\topsep}{2pt}}
\item the unique  $\br{\cdot}$-invariant element of $\tilde{\bT}_\mathbf{A}$ congruent to $\bv_\mathbf{l} \mod \ui \L$,
\item $\tilde{G}(b_{\mathbf{l}})$, where  $b_\mathbf{l}$ is the image of $c_\mathbf{l}$ in  $\L/\ui\L$ and $\tilde{G}$ is the inverse of the canonical isomorphism
\[ \QQ \tsr_\ZZ \tilde{\bT}_\mathbf{A} \cap \br{\L}\cap\L  \xrightarrow{\cong} \L / \ui \L, \]
\item the global crystal basis element $G_\lambda(b_{P(\mathbf{l})})$ of $V_{Q(\mathbf{l})}$,
\item $\pi_\lambda^\bT (c_{\mathbf{l}})$.
\end{list}
The $\Uq$- and  $\H_r$-cells of $(\bT, \tilde{B}^r)$ are given by Theorem \ref{t standard Schur-Weyl duality2} with $\liftc$  in place of $c$.
\end{theorem}
See \cite[Figure 3]{BProjected} for the example of the projected upper canonical basis corresponding to the upper canonical basis of Figure \ref{f c_111 example upper}.


\subsection{Graphical calculus for $U_q(\gl_2)$-modules}
\label{s graphical calculus}
Our study of upper based $\Uq$-modules for two-row Kronecker in \textsection\ref{s global crystal basis for two-row Kronecker coefficients}--\ref{s A Kronecker graphical calculus} depends heavily on the graphical calculus for $U_q(\gl_2)$-modules, which we now describe. Our main reference for this is \cite{FK}, though our notation differs slightly from theirs. In this subsection, fix $\dv = 2$ and let $F=F_1,\ E=E_1$.

Let  $\lambda \vdash_\dv r$, $Q \in \text{SYT}(\lambda)$, and  $\Lambda_Q$ be the  $\Uq$-cell of the upper based $\Uq$-module $(V^{\tsr r}, B^r)$ of \textsection \ref{ss upper canonical basis of bT}.  Consider the quotient map from the minimal cellular submodule  $\field \Lambda'_Q$ containing $\field\Lambda_Q$ onto  $\field\Lambda_Q$, and let $D(Q) \subseteq [\dv]^r$ be the set of  $\mathbf{k}$ such that  $c_\mathbf{k} \in \field\Lambda'_Q$. Define $e_\mathbf{k} \in \field \Lambda_Q$ to be the image of $c_\mathbf{k}$, $\mathbf{k} \in D(Q)$, under this map.  Then
\be\label{e projection V tsr r Vr}
e_\mathbf{k} =  \begin{cases}
G_{\lambda}(b_{P(\mathbf{k})}) = \frac{[r-j]!}{[r]!} F^j \eta_{\lambda} & \text{ if } \sh(\mathbf{k}) = \lambda, \\
0 & \text{ otherwise},
\end{cases}
\ee
where  $j$ is the number of 2's in the first row of   $P(\mathbf{k})$ and $\eta_{\lambda}$ is the canonical highest weight vector of $(V_{\lambda}, B(\lambda))$.  This formula follows from Theorem \ref{t standard Schur-Weyl duality2}.
Note that  $(\field\Lambda_{Q},\Lambda_{Q})$ is isomorphic to $(V_{\sh(Q)}, B(\sh(Q)))$, but it is convenient to keep the extra data of the SYT $Q$ in what follows.

\begin{figure}
\setlength{\cellsizeCol}{2.5ex}
\begin{tikzpicture}[xscale=.9]
\tikzstyle{column} = [inner sep = -4pt]
\tikzstyle{edge} = [draw,-,black]
\tikzstyle{LabelStyleH} = [text=black, fill =white, inner sep = -.8pt]

\begin{scope}
\foreach \y/\z in {-.3/-17*.25} {
    \draw[edge, bend right=70] (\z+1*.5,\y) to (\z+13*.5,\y);
    \draw[edge, bend right=70] (\z+2*.5,\y) to (\z+11*.5,\y);
    \draw[edge, bend right=70] (\z+3*.5,\y) to (\z+7*.5,\y);
    \draw[edge, bend right=70] (\z+5*.5,\y) to (\z+6*.5,\y);
    \draw[edge, bend right=70] (\z+9*.5,\y) to (\z+10*.5,\y);
    \draw[edge, bend right=70] (\z+14*.5,\y) to (\z+15*.5,\y);
    \node[column] (theNode) at (0,0) {
    $\myvcenter{\ensuremath{\mybox{\pad{2}&\pad{2}&\pad{2}} \downheart \mybox{\pad{2}& \pad{1}& \pad{1}} \downheart \mybox{\pad{2}& \pad{1} &\pad{1}} \downheart \mybox{\pad{1}& \pad{2} &\pad{1} &\pad{1}} }}$
    };
}
\end{scope}

\begin{scope}[yshift=-2.7cm]
\foreach \y/\z in {-.3/-17*.25} {
    \draw[edge, bend right=70] (\z+1*.5,\y) to (\z+10*.5,\y);
    \draw[edge, bend right=70] (\z+2*.5,\y) to (\z+6*.5,\y);
    \draw[edge, bend right=70] (\z+3*.5,\y) to (\z+5*.5,\y);
    \draw[edge, bend right=70] (\z+7*.5,\y) to (\z+9*.5,\y);
    \draw[edge, bend right=70] (\z+11*.5,\y) to (\z+13*.5,\y);
    \node[column] (theNode) at (0,0) {
    $\myvcenter{\ensuremath{\mybox{\pad{2}&\pad{2}&\pad{2}} \downheart \mybox{\pad{1}& \pad{1}&\pad{2}} \downheart \mybox{\pad{1} &\pad{1}&\pad{2}} \downheart \mybox{\pad{1} &\pad{1} &\pad{1}& \pad{2}} }}$
    };
}
\end{scope}
\setlength{\cellsizeCol}{2.1ex}
\end{tikzpicture}
\caption{The diagram corresponding to two elements of $\field(\Lambda_{Z^*_{(3)}} \tsr \Lambda_{Z^*_{(3)}} \tsr \Lambda_{Z^*_{(3)}} \tsr \Lambda_{Z^*_{(4)}})$. The top element evaluates to zero because it contains extra internal arcs. The bottom element is the canonical basis element $d_{2221121121112}$ in the notation of Theorem \ref{t FK F action on c basis}}
\label{f standard graphical calculus}
\end{figure}
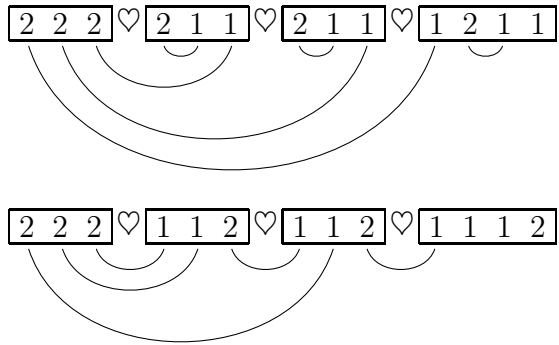

\begin{definition}
\label{d pairing internal external}
Let  $\lambda^{(1)} \vdash_\dv i_1,\ldots,\lambda^{(l)} \vdash_\dv i_l$,  $Q_j \in \text{SYT}(\lambda^{(j)})$, and $\mathbf{k} = \mathbf{k^{(1)}} \cdots \mathbf{k^{(l)}}$ such that $\mathbf{k^{(j)}} \in D(Q_j)$. The canonical basis element $e_\mathbf{k^{(1)}} \heart \dots \heart e_\mathbf{k^{(l)}} \in \field\Lambda_{Q_1} \tsr \dots \tsr \field\Lambda_{Q_l}$ is described by the \emph{diagram} of $\mathbf{k}$, denoted diagram$(\mathbf{k})$, which is the picture obtained from $\mathbf{k}$ by pairing 2's and 1's as left and right parentheses and then drawing an arc between matching pairs as shown in Figure \ref{f standard graphical calculus}.

We also record in the diagram the partitions $\lambda^{(1)}, \dots, \lambda^{(l)}$. An arc is \emph{internal} if its ends belong to the same  $\mathbf{k^{(j)}}$, and is \emph{external} otherwise.  An \emph{extra internal arc} is an internal arc with ends in $\mathbf{k^{(j)}}$ that does not occur in the diagram of  $\mathbf{l^{(j)}}$ for those  $\mathbf{l^{(j)}}$ satisfying  $Q(\mathbf{l^{(j)}}) = Q_j$ (all such diagrams have the same internal arcs).
\end{definition}
Equation  \eqref{e projection V tsr r Vr} implies the following important fact:
\be
\parbox{13.4cm}{A diagram contains an extra internal arc if and only if the corresponding basis element evaluates to zero.}
\ee
\begin{remark}
Strictly speaking, determining the extra internal arcs requires the data $Q_j$, but deciding whether there are extra internal arcs, which is what we really care about, only requires knowing the $\lambda^{(j)}$.
\end{remark}

For any upper based $U_q(\g_V)$-module $(N,B)$, define the functions  $\varphi, \varepsilon: B \to \ZZ_{\geq 0}$ by
\be \label{e phi varepsilon definition}
\varphi(b) := \max\{m: (\crystal{F})^m(b) \neq 0\}, \qquad \varepsilon(b) := \max\{m: (\crystal{E})^m(b) \neq 0\}.
\ee
These are the standard functions from crystal basis theory, but are usually defined for local rather than global crystal basis elements.
In the case $(N,B) = (V^{\tsr r},B^r)$, the statistic $\varphi(c_\mathbf{k})$ (resp.  $\varepsilon(c_\mathbf{k})$) is the number of unpaired $1$'s (resp. 2's) in the diagram of $\mathbf{k}$.  We also write $\varphi(\mathbf{k})$ (resp.  $\varepsilon(\mathbf{k})$) in place of $\varphi(c_\mathbf{k})$ (resp.  $\varepsilon(c_\mathbf{k})$).

\begin{theorem}[{\cite[\textsection 2.3]{FK}}]\label{t FK F action on c basis}
Maintain the notation of Definition \ref{d pairing internal external}.
The action of $F$ and  $E$ on the upper canonical basis $\Lambda_{Q_1} \heart \dots \heart \Lambda_{Q_l}$ is given as follows.
Let $d_{\mathbf{k}} = e_{\mathbf{k^{(1)}}} \heart \cdots \heart e_{\mathbf{k^{(l)}}} \in \field (\Lambda_{Q_1} \tsr \dots \tsr \Lambda_{Q_l})$. Then
\[F (d_{\mathbf{k}}) = \sum_{j=1}^{\varphi(\mathbf{k})} [j]d_{\F_{(j)}(\mathbf{k})}, \]
where $\F_{(j)}(\mathbf{k})$ is the word obtained by replacing the $j$-th unpaired  $1$ in  $\mathbf{k}$ with a $2$ (if the diagram of $\F_{(j)}(\mathbf{k})$ has
an extra internal arc, then $d_{\F_{(j)}(\mathbf{k})} = 0$; see \cite{FK}).
Similarly,
\[E (d_{\mathbf{k}}) = \sum_{j=1}^{\varepsilon(\mathbf{k})} [j]d_{\E_{(j)}(\mathbf{k})}, \]
where $\E_{(j)}(\mathbf{k})$ is the word obtained by replacing the $\varepsilon(\mathbf{k})-j+1$-th unpaired  $2$ in  $\mathbf{k}$ with a $1$ (so that $\E_{(1)}(\mathbf{k})$ changes the rightmost unpaired 2).
\end{theorem}
\begin{remark}
Throughout the paper we will usually only state results for  $F$ and omit the analogous statements for  $E$.
\end{remark}

In preparation for the application to the two-row Kronecker problem, we record the following corollary of Proposition \ref{p heart commutes with projections}.
Let $(V^{\tsr 2}, B^2)$ be the upper based $\Uq$-module from \textsection \ref{ss upper canonical basis of bT}. The corresponding projected basis is (where edges indicate the action of $F$)
\begin{center}
\begin{tikzpicture}[xscale =5, yscale =.82]
\tikzstyle{vertex}=[inner sep=0pt, outer sep=3pt, fill = white]
\tikzstyle{edge} = [draw, thick, ->,black]
\tikzstyle{LabelStyleH} = [text=black, anchor=north]
    \node[vertex] (11) at (2,0) {$\liftc_{11} = c_{11}$};
    \node[vertex] (12) at (1,-1) {$\liftc_{12} = c_{12} + \frac{1}{[2]}c_{21}$};
    \node[vertex] (21) at (1, 1) {$\liftc_{21}= c_{21}$};
    \node[vertex] (22) at (0, 0) {$\liftc_{22}= c_{22}$};
\draw[edge] (11) to node[LabelStyleH]{$[2]$} (12);
\draw[edge] (11) to node[LabelStyleH]{$ $} (21);
\draw[edge] (12) to node[LabelStyleH]{$ $} (22);
\end{tikzpicture}.
\end{center}
\begin{corollary}
\label{c projecting to heartproj basis}
Maintain the notation of Definition \ref{d pairing internal external} and specialize the setup of Proposition \ref{p heart commutes with projections} as follows: fix $t \in [l]$ and set $(N_t, B_t) = (V^{\tsr 2}, B^2)$ and $(N_j, B_j) = (\field \Lambda_{Q_j}, \Lambda_{Q_j})$ for $j \in [l] \setminus \{t\}$. Set $\pi_t = \pi_{(2,0)}^{N_t}$ and $\pi_j = \id$ for $j \neq t$. For convenience set $i_t = 2$ and $r = \sum_{j=1}^{l} i_j$. Then
\[
e_{\mathbf{k}^{(1)}} \heartproj \cdots \heartproj e_{\mathbf{k}^{(t-1)}} \heartproj \liftc_{\mathbf{k}^{(t)}} \heartproj e_{\mathbf{k}^{(t+1)}} \heartproj \cdots \heartproj e_{\mathbf{k}^{(l)}} = d_{\mathbf{k}} + \frac{1}{[2]} d_{\mathbf{k}} C_{m},
\]
where  $\mathbf{k}^{(t)} \in \{11,12,22\}$, $\mathbf{k} = \mathbf{k}^{(1)} \cdots \mathbf{k}^{(l)}$, $d_\mathbf{k}$ is the image of $c_{\mathbf{k}}$ under the projection
\[ \field\Lambda'_{Q_1} \tsr \dots \field\Lambda'_{Q_{t-1}} \tsr V^{\tsr 2} \tsr \field\Lambda'_{Q_{t+1}} \tsr \dots \tsr \field\Lambda'_{Q_l} \twoheadrightarrow \field\Lambda_{Q_1} \tsr \dots \field\Lambda_{Q_{t-1}} \tsr V^{\tsr 2} \tsr \field\Lambda_{Q_{t+1}} \tsr \dots \tsr \field\Lambda_{Q_l},\]
and $m = 1 + \sum_{j = 1}^{t-1} i_j$. Moreover, $d_\mathbf{k} C_{m} = d_\mathbf{k'}$, where $\mathbf{k'}$ is determined by the graphical calculus (the diagram  $\asymp$ is attached below that of $\mathbf{k}$ in position  $m$---see \cite[\textsection 2.1]{FK}), and $d_\mathbf{k'} = 0$ if the diagram of $\mathbf{k'}$ has an extra internal arc with ends in the $j$-th projector for any  $j \neq t$.
\end{corollary}
\begin{proof}
The projector $\pi_t$ is just $\frac{1}{[2]} \C_{m} = 1 + \frac{1}{[2]} C_{m}$.
\end{proof}

\section{Notation for  $GL_q(V) \times GL_q(W)$}
\label{s notation for GLV GLW}
Let $V,W$, and $X=V\tsrvw W$ be vector spaces of dimensions $\dv,\dw,\dx$, where $\tsrvw$ is our notation for tensor product between objects associated to $V$ and objects associated to $W$. As in the previous sections, $V$ is the defining $\O(GL_q(V))$-comodule and $\Uq$-module with weight basis $v_1,\cdots,v_\dv$. Similarly, $W$ is the defining $\O(GL_q(W))$-comodule and $U_q(\g_W)$-module with weight basis $w_1,\cdots,w_\dw$.
In general, notation from the previous sections for objects associated to $V$ will be used for $W$ as well, often with subscripts or superscripts to indicate whether they correspond to $V$ or $W$.

For a word $\mathbf{k} = k_1\dots k_r \in [\dv]^r$ (resp.  $\mathbf{l} = l_1 \dots l_r \in [\dw]^r$),  let $\bv_{\mathbf{k}} = v_{k_1} \tsr v_{k_2} \tsr \dots \tsr v_{k_r} \in V^{\tsr r}$ (resp. $\bw_{\mathbf{k}} = w_{l_1} \tsr w_{l_2} \tsr \dots \tsr w_{l_r} \in W^{\tsr r}$) denote the corresponding tensor monomial.  Let $x_\vw{i}{j}=v_i \tsrvw w_j \in X$ and $\bx_\vw{\mathbf{k}}{\mathbf{l}} = x_\vw{k_1}{l_1} \tsr \cdots \tsr  x_\vw{k_r}{l_r} = \bv_\mathbf{k} \tsrvw \bw_\mathbf{l} \in X^{\tsr r}$, for $\mathbf{k} \in [\dv]^r, \ \mathbf{l}\in [\dw]^r$. We sometimes identify  $[\dv] \times [\dw]$ with  $[\dx]$ via the bijection $\rho: (a, b) \mapsto (a-1)\dw+b$.  We will use the notation $y_{\rho(a,b)} = x_\vw{a}{b}$ and  $\by_\mathbf{j} = y_{j_1} \tsr \cdots \tsr y_{j_r}$,  $\mathbf{j} \in [\dx]^r$.

The weight lattice  $\wl(\g_V\oplus \g_W)$ of  $\g_V \oplus \g_W$ is equal to  $\wl(\g_V)\oplus\wl(\g_W)$.
The partial order  $\ld, \ldneq$ on weights is defined the same way as for $\wl(\g_V)$, thus $(\alpha,\beta)\ld (\gamma,\delta)$ if and only if $\alpha\ld \gamma$ and $\beta \ld \delta$.
A pair of partitions  $(\lambda,\mu)$ is identified with the weight  $(\lambda,\mu) \in \wl(\g_V \oplus \g_W)$.

Define $\Oint{\g_V \oplus \g_W}$ to be the category of finite-dimensional $\Uqvw$-modules
such that the weight of any nonzero weight space belongs to $\ZZ^\dv_{\geq 0}\oplus\ZZ^\dw_{\geq 0} \subseteq \wl(\g_V \oplus\g_W)$.
For any object  $N$ in $\Oint{\g_V \oplus \g_W}$ and partitions $\lambda, \mu$, define $\pi_{\lambda, \mu}^N: N \to N$ to be the $\Uqvw$-projector with image the $V_\lambda \tsrvw W_\mu$-isotypic component of $N$.

The definitions and results for based modules from \textsection\ref{s Bases for GLq modules} carry over in the obvious way to objects of $\Oint{\g_V \oplus \g_W}$. From now on, $B^r_V := \{c^V_{\mathbf{k}} : \mathbf{k} \in [\dv]^r\}$ denotes the upper canonical basis of $V^{\tsr r}$ constructed in \textsection \ref{ss upper canonical basis of bT} and $B^r_W := \{c^W_{\mathbf{k}} : \mathbf{k} \in [\dw]^r\}$ denotes the upper canonical basis of $W^{\tsr r}$. The basis $B^r_V \tsrvw B^r_W$ of $X^{\tsr r}$ is the \emph{upper canonical basis of $X^{\tsr r}$} and its elements are denoted $\cvw{\mathbf{k}}{\mathbf{l}} := c^V_\mathbf{k} \tsrvw c^W_\mathbf{l}$. This makes $(X^{\tsr r}, B^r_V \tsrvw B^r_W)$ an upper based $\Uqvw$-module with balanced triple  $(X^{\tsr r}_\mathbf{A}, \br{\L_V \tsrvw_{\field_\infty} \L_W},\L_V \tsrvw_{\field_\infty} \L_W)$, where  $X_\mathbf{A} = V_\mathbf{A} \tsrvw W_\mathbf{A}$.

\section{The nonstandard coordinate algebra $\O(M_q(\nsbr{X}))$}
\label{s nonstandard coordinate algebra}
Here we give the definition of the nonstandard coordinate algebra $\O(M_q(\nsbr{X}))$ as an FRT-algebra.
The theory of this object and the corresponding Hopf algebra  $\O(GL_q(\nsbr{X}))$ is developed in the next three sections,
following the treatment of their standard counterparts in \textsection\ref{ss FRT algebras}--\ref{ss compact}.
In  \textsection\ref{ss definitions ns symmetric exterior}, we construct the nonstandard symmetric and exterior  $\O(M_q(\nsbr{X}))$-comodule algebras, which are the nonstandard analogs of the quantum symmetric and exterior $\O(M_q(V))$-comodule algebras.
Subsections \ref{sexpproduct} and \ref{ss examples for nonstandard O} address explicit computations for nonstandard objects, illustrating that these are significantly more complicated than their standard counterparts.


\subsection{Definition of  $\O(M_q(\nsbr{X}))$}
\label{ss definition of OnsX}
Let  $V$ and  $W$ be  $\field$-vector spaces of dimensions  $\dv$ and  $\dw$, respectively, and let $X = V \tsrvw W$ be their tensor product (this notation for tensor product is explained in  \textsection\ref{ss restitutions}); we write  $\nsbr{X}$ in place of $X$ when this space is associated with the nonstandard objects we are about to define.
Let $\hat \R_{V,V}$ be as in  \textsection\ref{ss FRT algebras} and  $\hat \R_{W,W}$ be defined in the same way with  $W$ in place of  $V$ and standard basis $w_1,\ldots,w_\dw$ in place of  $v_1,\ldots,v_\dv$.
Define $\hat \R_{\nsbr{X},\nsbr{X}} := \hat \R_{V,V} \tsrvw  \hat \R_{W,W} \in M_{\dx^2}(\field)$.
This is different from the $\hat \R$-matrix $\hat \R_{X,X}$  obtained by thinking of
$X=V\tsrvw  W$ as a  corepresentation  of the quantum coordinate algebra $\O(GL_q(X))$.

Both $\hat \R_{V,V}$ and $\hat \R_{W,W}$ are diagonalizable with
 eigenvalues $q$ and $-q^{-1}$. Hence,
 $\hat \R_{\nsbr{X},\nsbr{X}}$ is diagonalizable with eigenvalues
$q^2,-1,q^{-2}$.
The \emph{nonstandard symmetric square} $\nssym{2}{X} \subseteq \nsbr{X} \otimes \nsbr{X}$
is defined to be the sum of the eigenspaces of  $\hat \R_{\nsbr{X},\nsbr{X}}$
corresponding to the eigenvalues $q^2$ and $q^{-2}$.
The \emph{nonstandard exterior square} $\nswedge{2}{X} \subseteq \nsbr{X} \otimes \nsbr{X}$
is defined to be the eigenspace of $\hat \R_{\nsbr{X},\nsbr{X}}$ for the eigenvalue  $-1$.
Let $P_+^\nsbr{X}: \nsbr{X}^{\tsr 2} \to \nsbr{X}^{\tsr 2}$ (resp. $P_-^\nsbr{X}$)
be the projector with image $\nssym{2}{X}$ (resp.  $\nswedge{2}{X}$).
These spaces and projectors are expressed in terms of $V$ and  $W$ as
\begin{equation} \label{e symwedge2X VW}
\begin{array}{lcl}
\nssym{2}{X}&=& \ssym{2}{V} \tsrvw \ssym{2}{W} \oplus \swedge{2}{V}\tsrvw \swedge{2}{W}, \\[1.6mm]
\nswedge{2}{X}&=& \ssym{2}{V} \tsrvw \swedge{2}{W} \oplus \swedge{2}{V}\tsrvw \ssym{2}{W},
\end{array}
\end{equation}
and,
\begin{equation} \label{eqPminusandplus}
\begin{array}{lcl}
P_+^{\nsbr{X}}&=& P_+^V \tsrvw  P_+^W + P_-^V\tsrvw  P_-^W, \\[1.6mm]
P_-^{\nsbr{X}}&=& P_-^V \tsrvw  P_+^W + P_+^V\tsrvw  P_-^W.
\end{array}
\end{equation}

Let  $\nsbr{Z} = \nsbr{X} \tsrdual \nsbr{X}^* \cong U^V \tsrvw U^W$ with standard basis $\{\zz{i}{j}: i,j \in [\dx]\}$.
Let $\nsbr{\mb{z}}=\mb{u}^V \tsrvw \mb{u}^W$ be the variable matrix  $(\zz{i}{j})$, specifying the linear functions on an endomorphism of $\nsbr{X}$.
Let $\field \langle \zz{i}{j} \rangle = T(\nsbr{Z})$ denote the free algebra over
the variable entries of $\nsbr{\mb{z}}$.
\begin{definition}
The \emph{nonstandard coordinate algebra} $\O(M_q(\nsbr{X}))$ of the virtual \emph{nonstandard matrix space}
$M_q(\nsbr{X})$ is the quotient of $\field \langle \zz{i}{j} \rangle$
by the relations
\begin{equation} \label{eqquantcommute1}
P_+^{\nsbr{X}} (\nsbr{\mb{z}} \otimes \nsbr{\mb{z}}) =
 (\nsbr{\mb{z}} \otimes \nsbr{\mb{z}}) P_+^{\nsbr{X}}.
\end{equation}
\end{definition}


We now establish some basic facts and make some remarks about the nonstandard coordinate algebra.
Since $I=P_-^{\nsbr{X}}+P_+^{\nsbr{X}}$, \eqref{eqquantcommute1} is equivalent to
\begin{equation} \label{eqquantcommute2}
P_-^{\nsbr{X}} (\nsbr{\mb{z}} \otimes \nsbr{\mb{z}}) =
 (\nsbr{\mb{z}} \otimes \nsbr{\mb{z}}) P_-^{\nsbr{X}}.
\end{equation}
Similar to the description \eqref{e OMqV ideal def} of the quantum coordinate algebra  $\O(M_q(V))$, we have:
\begin{equation} \label{eqdefnewquan no basis}
\parbox{14cm}{the nonstandard coordinate algebra  $ \O(M_q(\nsbr{X}))$ is the quotient of  $T(\nsbr{Z})$ by the two-sided ideal  $\nsbr{\mathcal{I}}$ generated by
$\nsbr{\mathcal{I}}_2 := \nssym{2}{X} \tsrdual \nswedge{2}{X}^* \oplus \nswedge{2}{X} \tsrdual \nssym{2}{X}^*.$
}
\end{equation}

It is easy to see that $\nssym{2}{X}$ (resp. $\nswedge{2}{X}$) specializes to $S^2 X$ (resp. $\Wedge^2 X$) at $q=1$.
Thus the degree 2 part of $\O(M_q(\nsbr{X}))$ coincides with the degree 2 part of $\O(M(X))$ at $q=1$. This means that the $\zz{j}{i}$'s commute at $q=1$.
These specializations are made precise and checked carefully in Appendix \ref{s reduction system}.

\begin{remark}
\label{r nonstandard FRT algebra}
The definition of  $A(\R)$ in \cite{RTF} requires  $\R$ to be nonsingular.  The relations \eqref{eqquantcommute1} or \eqref{eqquantcommute2} are
like the defining relations for an FRT-algebra except with a singular $\R$, however \eqref{eqquantcommute1} or \eqref{eqquantcommute2} is equivalent to
\begin{equation} \label{eqquantcommute3}
\hat{\mathcal{R}}_{\nsbr{X},\nsbr{X}}(a,b) (\nsbr{\mb{z}} \otimes \nsbr{\mb{z}}) =
 (\nsbr{\mb{z}} \otimes \nsbr{\mb{z}})  \hat{\mathcal{R}}_{\nsbr{X},\nsbr{X}}(a,b),
\end{equation}
where
\[ \hat{\mathcal{R}}_{\nsbr{X},\nsbr{X}}(a,b)=  a P_+^{\nsbr{X}} + b P_-^{\nsbr{X}}\]
for any distinct constants $a,b$.
Thus if  $a,b$ are distinct and nonzero, then $\O(M_q(\nsbr{X}))$ is an FRT-algebra with $\R$-matrix
$\mathcal{R}_{\nsbr{X},\nsbr{X}}(a,b) = \tau \circ \hat{\mathcal{R}}_{\nsbr{X},\nsbr{X}}(a,b)$.
\end{remark}

As explained in \textsection\ref{ss FRT algebras}, any FRT-algebra is a bialgebra, hence
$\O(M_q(\nsbr{X}))$ is a bialgebra with coproduct and counit given by
$\Delta(\nsbr{\mb{z}})=\nsbr{\mb{z}}\dot{\tsr} \nsbr{\mb{z}},$ and $\epsilon(\nsbr{\mb{z}})=\mb{I}$.

\begin{proposition} \label{phomomorph}
Let $\O(M_q(V))$  and $\O(M_q(W))$ be the quantum coordinate algebras defined in \textsection\ref{squantum}.
There is a bialgebra homomorphism
\[ \psi: \O(M_q(\nsbr{X})) \to \O(M_q(V))\tsrvw  \O(M_q(W)), \]
determined by $\nsbr{\mb{z}} \mapsto {\mathbf u^V} \tsrvw {\mathbf u^W}$.
\end{proposition}
Note that the  $\tsrvw$ in ${\mathbf u^V} \tsrvw {\mathbf u^W}$ is serving two purposes: one is as the tensor product of a $\dv \times \dv$ matrix and a $\dw \times \dw$ matrix and the other is as the $\tsrvw$ product inside the ring $\O(M_q(V))\tsrvw  \O(M_q(W))$.
\begin{proof}
One has to check that the relations obtained by substituting
$\nsbr{\mb{z}} = {\mathbf u^V} \tsrvw  {\mathbf u^W}$ in
\eqref{eqquantcommute1} defining $\O(M_q(\nsbr{X}))$
 are implied by the relations defining $\O(M_q(V))$ and $\O(M_q(W))$.

The defining relations \eqref{eqp+2} of $\O(M_q(V))$ are
\[
 P_+^V ({\mathbf u^V} \otimes {\mathbf u^V}) =
 ({\mathbf u^V} \otimes {\mathbf u^V}) P_+^V,
\]
which are equivalent to \eqref{eqp-2}:
\[
 P_-^V ({\mathbf u^V} \otimes {\mathbf u^V}) =
 ({\mathbf u^V} \otimes {\mathbf u^V}) P_-^V.
\]
Similarly, the defining relations of $\O(M_q(W))$ are either of
\[
\begin{array}{c}
 P_+^W ({\mathbf u^W} \otimes {\mathbf u^W}) =
 ({\mathbf u^W} \otimes {\mathbf u^W}) P_+^W, \\[1.2mm]
 P_-^W ({\mathbf u^W} \otimes {\mathbf u^W}) =
 ({\mathbf u^W} \otimes {\mathbf u^W}) P_-^W.
\end{array}
\]
Since $P_-^{\nsbr{X}}=P_-^V\tsrvw  P_+^W + P_+^V\tsrvw  P_-^W$ (see \eqref{eqPminusandplus}), these relations imply
\eqref{eqquantcommute1} when
$\nsbr{\mb{z}} = {\mathbf u^V} \tsrvw  {\mathbf u^W}$.

To show that $\psi$ is a bialgebra homomorphism, one has to additionally verify that
\[ \Delta \circ \psi = (\psi \otimes \psi) \circ \Delta \ \ \text{and } \epsilon=\epsilon \circ \psi, \]
which is easy.
\end{proof}

\begin{remark}\label{r no braid relation0}
Fix distinct $a,b$, and let  $\mathcal{R}=\mathcal{R}_{\nsbr{X},\nsbr{X}}(a,b)$, as in Remark \ref{r nonstandard FRT algebra}.
Given a tensor product $\nsbr{X}^{\tsr r}$, let $\mathcal{R}_i$ denote the transformation which
acts like $\mathcal{R}$ on the $i$-th and $(i+1)$-st factors, the other factors remaining unaffected.
Then, as is shown in  \textsection\ref{salgb3} (see Remark \ref{r no braid relation}),
the pairs $\hat{\mathcal{R}}_i, \hat{\mathcal{R}}_{i+1}$ do not satisfy the braid relation---equivalently, the pairs $\mathcal{R}_i, \mathcal{R}_{i+1}$ do not
satisfy the quantum Yang-Baxter equation.
Thus although $\O(M_q(\nsbr{X}))$ is an FRT-algebra, it is not coquasitriangular, hence the main theory of FRT-algebras  \cite{RTF} does not apply.
\end{remark}
%

\begin{remark}
\label{r ns coordinate algebra as large as possible}
The nonstandard coordinate algebra $\O(M_q(\nsbr{X}))$ is much smaller than $\O(M_q(X))$ as will be seen in Proposition \ref{ppoincare} and \textsection\ref{s Nonstandard representation theory in the two-row case}. However, it is the only FRT-algebra with a coalgebra homomorphism to $\O(M_q(V))\tsrvw \O(M_q(W))$ such that its degree 2 corepresentations coincide with those of $\O(M(X))$ at $q=1$. For example, another quantization of  $\O(M(X))$ we considered is the FRT-algebra $A(\R_{\nsbr{X},\nsbr{X}})$,  where $\R_{\nsbr{X},\nsbr{X}}=\tau\circ \hat \R_{\nsbr{X},\nsbr{X}}$.  However, it is smaller than $\O(M_q(\nsbr{X}))$ and even its degree 2 corepresentations are smaller than those of $\O(M(X))$, so it is not a good candidate for a quantization of $\O(M(X))$ for the Kronecker problem.
See Remark \ref{r nsH as small as possible} for a similar argument claiming that the nonstandard Hecke algebra is in some sense the only choice for a quantization of the symmetric group for the Kronecker problem.
\end{remark}

\subsection{Nonstandard symmetric and exterior algebras}
\label{ss definitions ns symmetric exterior}
\label{squantumbase}
Here we define the nonstandard symmetric and exterior algebras of $\nsbr{X}$; these play an analogous for  $\O(M_q(\nsbr{X}))$ to the role played by the quantum symmetric and exterior algebras of $V$ for  $\O(M_q(V))$, as was described in \textsection\ref{squantum}.

Maintain the notation from  \textsection\ref{s notation for GLV GLW} so that $x_\vw{i}{j}=v_i \tsrvw w_j \in \nsbr{X}$, $y_{\rho(a,b)} = x_\vw{a}{b}$, where $\rho(a, b) = (a-1)\dw+b,$ $\by_\mathbf{j} = y_{j_1} \tsr \cdots \tsr y_{j_r}$,  $\mathbf{j} \in [\dx]^r$, etc.
The \emph{standard monomial basis} of  $\nsbr{X}^{\tsr r}$ is  $\{\by_\mathbf{j} : \mathbf{j} \in [\dx]^r \}$ and its dual basis is the \emph{standard monomial basis} of  $(\nsbr{X}^*)^{\tsr r}$, denoted  $\{\by^\mathbf{j} : \mathbf{j} \in [\dx]^r \}$.
\be \label{e inner product X}
\parbox{13cm}{Define the \emph{standard bilinear form} $\langle \cdot,\cdot\rangle$ on $\nsbr{X}^{\tsr r}$ (resp. $(\nsbr{X}^*)^{\tsr r}$) to be the symmetric bilinear form for which the standard monomial basis is orthonormal (we do not want the Hermitian form here).
Both the standard bilinear form on  $\nsbr{X}^{\tsr r}$ and that on  $(\nsbr{X}^*)^{\tsr r}$ induce the isomorphism $\alpha_r:\nsbr{X}^{\tsr r} \xrightarrow{\cong} (\nsbr{X}^*)^{\tsr r}$,  $\by_\mathbf{j}\mapsto\by^\mathbf{j}$,  $\mathbf{j} \in[\dx]^r$.}
\ee

The {\em nonstandard symmetric algebra} $\nssymalgebra{}$  of $\nsbr{X}$
is the free $\field$-algebra in the $x_\vw{i}{j}$'s subject to the relations
\begin{equation}\label{eqp-}
P_-^{\nsbr{X}} (\mathbf{x} \tsr \mathbf{x})  =0,
\end{equation}
where  $\mathbf{x}$ is the column vector with entries  $x_\vw{i}{j}$.
Equivalently, $\nssymalgebra{}$ is the quotient of the tensor algebra $T(\nsbr{X}) = \bigoplus_{r \geq 0} \nsbr{X}^{\tsr r}$
by the two-sided ideal generated by  $\nswedge{2}{X}$.
It can be thought of as the coordinate ring of the  virtual  {\em  nonstandard symmetric space}  $\xsym$.

Similarly, the {\em nonstandard exterior algebra} $\nswedgealgebra{}$ of $\nsbr{X}$
is the free  $\field$-algebra in the $x_\vw{i}{j}$'s subject to the relations
\begin{equation}\label{eqp+}
 P_+^{\nsbr{X}} (\mathbf{x} \tsr \mathbf{x})=0.
\end{equation}
Equivalently, $\nswedgealgebra{}$ is the quotient of  $T(\nsbr{X})$
by the two-sided ideal generated by  $\nssym{2}{X}$.
It can be thought of as the coordinate ring of the virtual {\em nonstandard antisymmetric space} $\xwedge$.
Let $\nssym{r}{X}$ and $\nswedge{r}{X}$ be the degree $r$ components  of
$\nssymalgebra{}$ and $\nswedgealgebra{}$, respectively.

Using the standard bilinear form \eqref{e inner product X} to identify  $\nsbr{X}$ and  $\nsbr{X}^*$ defines the right-hand versions
 $\nssym{r}{X}^*$ and  $\nswedge{r}{X}^*$ of the nonstandard symmetric and exterior algebras, i.e., $\nssym{r}{X}^*$ is the quotient of $T(\nsbr{X}^*)$
 by the two-sided ideal generated by   $\nswedge{2}{X}^*$, where $\nswedge{2}{X}^* := \alpha_2(\nswedge{2}{X})$.
\begin{proposition} \label{pequidefining1}
\begin{list}{\emph{(\arabic{ctr})}} {\usecounter{ctr} \setlength{\itemsep}{1pt} \setlength{\topsep}{2pt}}
\item The nonstandard symmetric algebra $\nssymalgebra{}$ (resp.  $\nssymalgebra{^*}$) is a left (resp.  right)  $\O(M_q(\nsbr{X}))$-comodule algebra
 via  $y_i \mapsto \sum_j \zz{i}{j} \tsr y_j$ (resp.   $y^j \mapsto \sum_i y^i \tsr \zz{i}{j}$).
The nonstandard coordinate algebra $ \O(M_q(\nsbr{X}))$ is the largest bialgebra quotient of  $T(\nsbr{Z})$
such that  $\nssym{2}{X}$ is a left $\O(M_q(\nsbr{X}))$-comodule and  $\nssym{2}{X}^*$ is a right  $ \O(M_q(\nsbr{X}))$-comodule.
\item Similarly, the nonstandard exterior algebra $\nswedgealgebra{}$ (resp.  $\nswedgealgebra{^*}$) is a left (resp.  right) $\O(M_q(\nsbr{X}))$-comodule algebra.
The nonstandard coordinate algebra $ \O(M_q(\nsbr{X}))$ is the largest bialgebra quotient of  $T(\nsbr{Z})$ such that  $\nswedge{2}{X}$ is a left $\O(M_q(\nsbr{X}))$-comodule and  $\nswedge{2}{X}^*$ is a right  $ \O(M_q(\nsbr{X}))$-comodule.
\end{list}
\end{proposition}
\begin{proof} The proof is similar to the standard case. This uses the fact that the matrices $P_\pm^{\nsbr{X}}$ are symmetric, which follows from
the fact that  $P_\pm^V, \ P_\pm^W$ are symmetric.
\end{proof}


Let $\tilde{B}^V_+$ (resp. $\tilde{B}^V_-$) be a basis of $\ssym{2}{V}$ (resp.  $\swedge{2}{V}$).
For instance, we could let $\tilde{B}^2_V$  be the projected upper canonical basis of  $V^{\tsr 2}$ and $\tilde{B}^V_+ \subseteq \tilde{B}^2_V$ (resp. $\tilde{B}^V_- \subseteq \tilde{B}^2_V$) the subset  $\tilde{\Lambda}_{\tiny \tableau{1 & 2}} = \{\liftc^V_{ij} : 1 \leq i \leq j \leq \dv\}$ (resp.  $\tilde{\Lambda}_{\tiny \tableau{1\\2}} = \{\liftc^V_{ij}: 1 \leq j < i \leq \dv \}$); see \textsection\ref{ss upper canonical basis of bT}, particularly Theorems \ref{t lifted upper canonical basis} and \ref{t standard Schur-Weyl duality2}.  The  $\dv =2$ case is described explicitly in the example below.
Define $\tilde{B}^W_+$ and $\tilde{B}^W_-$ similarly.
Then by \eqref{e symwedge2X VW}, the following are bases of $\nssym{2}{X}$ and $\nswedge{2}{X}$:
\begin{equation}
\label{e basis sym wedge}
\begin{array} {lr}
\nssym{2}{X}: & \tilde{B}^V_+ \tsrvw  \tilde{B}^W_+ \sqcup \tilde{B}^V_- \tsrvw \tilde{B}^W_-, \\[1.4mm]
\nswedge{2}{X}: & \tilde{B}^V_+ \tsrvw \tilde{B}^W_- \sqcup \tilde{B}^V_- \tsrvw \tilde{B}^W_+.
\end{array}
\end{equation}
%

\begin{example}
\label{ex ns wedge sym bases}
Let $\dv=\dw=2$ and $\{v_1,v_2\}$, $\{w_1,w_2\}$ be the standard bases of $V$ and  $W$.
Then $\tilde{B}^V_+=\{\liftc^V_{11},\liftc^V_{21},\liftc^V_{22}\}$ and $\tilde{B}^V_-=\{\liftc^V_{21}\}$, where
\begin{equation}\label{eqA11}
 \begin{array}{rl}
\liftc^V_{11}&=\bv_{11}, \\
\liftc^V_{12}&= \frac{1}{[2]}(q \bv_{12}+ \bv_{21}), \\
\liftc^V_{22}&= \bv_{22}, \\
\liftc^V_{21}&= \bv_{21} - q^{-1} \bv_{12}.
\end{array}
\end{equation}
The elements $\liftc^W_\mathbf{k}$ of $\tilde{B}^W_+$ and  $\tilde{B}^W_-$ are similar with  $\bw_{\mathbf{l}}$ in place of  $\bv_\mathbf{l}$.
Set $\liftcvw{\mathbf{k}}{\mathbf{l}} = \liftc^V_{\mathbf{k}} \tsrvw \liftc^W_{\mathbf{l}}$.  The bijection $\rho$ in the $\dv=\dw=2$ case is
\[\stack{1}{1} \,\leftrightarrow 1, \stack{1}{2} \,\leftrightarrow 2, \stack{2}{1} \,\leftrightarrow 3, \stack{2}{2} \,\leftrightarrow 4.\]

Then the basis \eqref{e basis sym wedge} of $\nswedge{2}{X}$ is expressed in terms of the monomial basis of $\nsbr{X}^{\tsr 2}$ as follows.  This basis
is labeled by $\text{NST}((2))$, the set of nonstandard tabloids of shape $(2)$ (this will be explained in full generality in  \textsection\ref{s A canonical basis for Yalpha}, but for now we can take this  as the definition in the two-row case).
\begin{equation} \label{eqwedgedef2}
 \begin{array}{l@{\ :=\ }l@{\ =\ }l@{\ =\ }l}
\ctableaua{1\\2}  & \liftcvw{11}{21}&\bx_\vw{11}{21}- \ui \bx_\vw{11}{12}&\by_{21}- \ui \by_{12}, \\[2mm]
\ctableaua{3\\4}  & \liftcvw{22}{21}&\bx_\vw{22}{21}-\ui \bx_\vw{22}{12}&\by_{43}-\ui \by_{34}, \\[2mm]
\ctableaua{1\\3}  & \liftcvw{21}{11}&\bx_\vw{21}{11}-\ui \bx_\vw{12}{11}&\by_{31}-\ui \by_{13}, \\[2mm]
\ctableaua{2\\4}  & \liftcvw{21}{22}&\bx_\vw{21}{22}-\ui \bx_\vw{12}{22}&\by_{42}-\ui \by_{24}, \\[2mm]
\ctableaua{3\\2}  & \liftcvw{12}{21}&\frac{1}{[2]}(\u \bx_\vw{12}{21}-\bx_\vw{12}{12}+\bx_\vw{21}{21}- \ui \bx_\vw{21}{12}) &\frac{1}{[2]}(\u \by_{23}-\by_{14}+\by_{41}- \ui \by_{32}), \\[2mm] 
\ctableaua{2\\3}  & \liftcvw{21}{12}&\frac{1}{[2]}(\u \bx_\vw{21}{12}-\bx_\vw{12}{12}+\bx_\vw{21}{21}- \ui \bx_\vw{12}{21}) &\frac{1}{[2]}(\u \by_{32}-\by_{14}+\by_{41}- \ui \by_{23}). 
\end{array}
\end{equation}

The basis \eqref{e basis sym wedge} of $\nssym{2}{X}$ is expressed in terms of the monomial basis of $\nsbr{X}^{\tsr 2}$ as follows.  This basis
is denoted $\widetilde{\text{NST}}((1,1))$, which is defined to be the projection of  $\text{NST}((1,1))$ onto  $\nssym{2}{X}$ (the projection is needed as $\text{NST}((2)) \sqcup \text{NST}((1,1))$ is a basis for  $\nsbr{X}^{\tsr 2}$, but  $\text{NST}((1,1))$ is not a subset of $\nssym{2}{X}$).
\begin{equation} \label{eqsymdef2}
\begin{array}{l@{\ :=\ }l@{\ =\ }l@{\ =\ }l}
\widetilde{ \ctableau{1}{1}} & \liftcvw{11}{11}&\bx_\vw{11}{11}&\by_{11}, \\
 \widetilde{\ctableau{2}{2}} & \liftcvw{11}{22}&\bx_\vw{11}{22}&\by_{22}, \\
 \widetilde{\ctableau{1}{2}} & \liftcvw{11}{12}& \frac{1}{[2]}(\u \bx_\vw{11}{12} + \bx_\vw{11}{21})& \frac{1}{[2]}(\u \by_{12} + \by_{21}), \\
 \widetilde{\ctableau{3}{3}} & \liftcvw{22}{11}&\bx_\vw{22}{11}&\by_{33}, \\
 \widetilde{\ctableau{4}{4}} & \liftcvw{22}{22}&\bx_\vw{22}{22}&\by_{44}, \\
 \widetilde{\ctableau{3}{4}} & \liftcvw{22}{12}&\frac{1}{[2]}(\u \bx_\vw{22}{12}+ \bx_\vw{22}{21})&\frac{1}{[2]}(\u \by_{34}+ \by_{43}), \\
 \widetilde{\ctableau{1}{3}} & \liftcvw{12}{11}&\frac{1}{[2]}(\u \bx_\vw{12}{11}+ \bx_\vw{21}{11})&\frac{1}{[2]}(\u \by_{13}+ \by_{31}), \\
 \widetilde{\ctableau{2}{4}} & \liftcvw{12}{22}&\frac{1}{[2]}(\u \bx_\vw{12}{22}+ \bx_\vw{21}{22})&\frac{1}{[2]}(\u \by_{24}+ \by_{42}), \\
 \widetilde{\ctableau{1}{4}} & \liftcvw{12}{12}&\frac{1}{[2]^2}(\u^2 \bx_\vw{12}{12}+ \u \bx_\vw{12}{21}+\u \bx_\vw{21}{12}+ \bx_\vw{21}{21})&\frac{1}{[2]^2}(\u^2 \by_{14}+ \u \by_{23}+\u \by_{32}+ \by_{41}), \\
 \widetilde{\ctableau{4}{1}} & \liftcvw{21}{21}&\bx_\vw{21}{21} - \ui \bx_\vw{21}{12}- \ui \bx_\vw{12}{21}+ \u^{-2} \bx_\vw{12}{12}&\by_{41} - \ui \by_{32}- \ui \by_{23}+ \u^{-2} \by_{14}.
\end{array}
\end{equation}
\end{example}

\begin{proposition} \label{pbasis}
\
\begin{enumerate}
\item A basis of $\nssymalgebra{}$ is $\{y_{j_1} y_{j_2}  \cdots  y_{j_r} : 1\leq j_1\leq j_2 \leq \dots \leq j_r \leq \dx, \, r\geq 0\}$.
\item A basis of $\nswedgealgebra{}$ is $\{y_{j_1} y_{j_2}  \cdots  y_{j_r} : 1\leq j_1< j_2 < \dots < j_r\leq \dx, \, r\geq 0\}$.
\end{enumerate}
Here $y_{\rho(a,b)}=x_\vw{a}{b}$ as above, and  $y_{j_1} \cdots  y_{j_r}$ denotes the image of $y_{j_1}\tsr \cdots \tsr y_{j_r}$ in $\nssymalgebra{}$ or $\nswedgealgebra{}$.
\end{proposition}
These bases will be called {\em standard monomial bases} of $\nssymalgebra{}$ and $\nswedgealgebra{}$.
\begin{proof}
(1)  The relations \eqref{eqp-} (in the two-row case this means setting the elements in \eqref{eqwedgedef2} to 0)
can be reformulated in the form of the following reduction system:
\begin{equation} \label{eqredsym}
\begin{array}{lcl}
\bx_\vw{j}{k}\bx_\vw{i}{k}&\rightarrow& q^{-1}\bx_\vw{i}{k}\bx_\vw{j}{k} \quad (i<j) \\
\bx_\vw{k}{j}\bx_\vw{k}{i}&\rightarrow& q^{-1}\bx_\vw{k}{i}\bx_\vw{k}{j} \quad (i<j) \\
\bx_\vw{j}{k}\bx_\vw{i}{l}&\rightarrow& \bx_\vw{i}{l}\bx_\vw{j}{k} \quad (i<j, k<l) \\
\bx_\vw{j}{l}\bx_\vw{i}{k}&\rightarrow& \bx_\vw{i}{k}\bx_\vw{j}{l}-(q-q^{-1})\bx_\vw{i}{l}\bx_\vw{j}{k} \quad (i<j, k<l).
\end{array}
\end{equation}
When $\dv=\dw$, these
coincide
with the defining relations \eqref{eqredclassical} for the standard quantum
matrix space $M_q(V)$ after the change of variables $\bx_\vw{i}{j} \mapsto u^j_i $.
In this case,
the ambiguities in this reduction system can   be resolved
just as in the
case of the reduction system for
$\O(M_q(V))$   \cite{nym,artin} (see \textsection\ref{ss standard reduction system}).
This is also  so when $\dv \not = \dw$.
Hence the result follows from the diamond lemma \cite{KS}.

(2) The relations \eqref{eqp+} (in the two-row case this means setting the elements in \eqref{eqsymdef2} to 0)
can be reformulated  in the form of the following reduction
system:
\begin{equation} \label{eqredwedge}
\begin{array}{rcl}
\bx_\vw{i}{j}^2&\rightarrow& 0 \\
\bx_\vw{j}{k}\bx_\vw{i}{k}&\rightarrow& -q \bx_\vw{i}{k}\bx_\vw{j}{k} \quad (i<j) \\
\bx_\vw{k}{j}\bx_\vw{k}{i}&\rightarrow& -q \bx_\vw{k}{i}\bx_\vw{k}{j} \quad (i<j) \\
\bx_\vw{j}{l}\bx_\vw{i}{k}&\rightarrow& -\bx_\vw{i}{k}\bx_\vw{j}{l} \quad (i<j, k<l) \\
\bx_\vw{j}{k}\bx_\vw{i}{l}&\rightarrow& -\bx_\vw{i}{l}\bx_\vw{j}{k}+(q^{-1}-q)\bx_\vw{i}{k}\bx_\vw{j}{l} \quad (i<j, k<l).
\end{array}
\end{equation}
Ambiguities in this reduction system can  also be resolved just as in (1);
we omit the details.
So the result again follows from  the diamond lemma \cite{KS}.
\end{proof}

The nonstandard symmetric and exterior algebras $\nssym{r}{X}$ and  $\nswedge{r}{X}$ become $\O(M_q(V)) \tsrvw \O(M_q(W))$-comodules via the coalgebra
homomorphism $\psi$ of Proposition \ref{phomomorph}.
\begin{proposition} \label{pgelfanddecomp}
(1) As an $\O(M_q(V)) \tsrvw \O(M_q(W))$-comodule,
\[
\nssym{r}{X} \cong
\bigoplus_\stack{\lambda \vdash r}{\ell(\lambda) \leq \dv, \
\ell(\lambda) \leq \dw} V_\lambda\tsrvw W_\lambda.
\]
(2) Similarly, letting $\lambda'$ be the conjugate of $\lambda$ as in
\textsection\ref{ss type A combinatorics preliminaries},
\[
\nswedge{r}{X}
\cong \bigoplus_\stack{\lambda \vdash r}{\ell(\lambda) \leq \dv, \
\ell(\lambda') \leq \dw} V_\lambda\tsrvw W_{\lambda'}.
\]
\end{proposition}
\begin{proof}
By the proof of Proposition \ref{pbasis} (1), the $\O(M_q(V)) \tsrvw \O(M_q(W))$-comodule action on $\nssym{r}{X}$ coincides with the
two-sided coaction of  $\O(M_q(V))$ on  $\O(M_q(V))$ in the case $W = V^*$.  Thus (1)  is the $q$-analog of the Peter-Weyl theorem
for the standard quantum coordinate algebra $O(M_q(V))$ \cite[Theorem 21, Chapter 11]{KS}.

Similarly, (2) is an antisymmetric version of the $q$-analog of the Peter-Weyl theorem.  A more careful proof of (2) can be given using nonstandard Schur-Weyl duality
(Theorem \ref{t nonstandard schur-weyl duality}) and Proposition \ref{p sign in lambda tsr lambda'}.
\end{proof}

\subsection{Explicit product formulae} \label{sexpproduct}
We wish to give explicit formulae for products in
the nonstandard symmetric and exterior algebras $\nssymalgebra{}$ and
$\nswedgealgebra{}$.

Recall the Gelfand-Tsetlin basis and its notation from \textsection\ref{sgelfand} and, as there, assume that objects are over $\CC$ and $q$ is a real number such that  $q \neq 0, \pm1$. Let
\begin{equation} \label{egtbsym}
B^\text{GT}(\nssym{r}{X})=\bigsqcup_\lambda \left\{\bracket{M_\lambda} \tsrvw \bracket{N_\lambda}:M_\lambda\in \text{SSYT}_{\dv}(\lambda),N_\lambda\in \text{SSYT}_{\dw}(\lambda)\right\}
\end{equation}
be the orthonormal Gelfand-Tsetlin basis for $\nssym{r}{X}$ as per the decomposition in
Proposition \ref{pgelfanddecomp} (1)
and
\begin{equation} \label{egtbwedge}
B^\text{GT}(\nswedge{r}{X})=\bigsqcup_\lambda \left\{\bracket{M_\lambda} \tsrvw \bracket{N_{\lambda'}} : M_\lambda\in \text{SSYT}_{\dv}(\lambda),N_{\lambda'}\in \text{SSYT}_{\dw}(\lambda')\right\}
\end{equation}  that for
$\nswedge{r}{X}$ as per  Proposition \ref{pgelfanddecomp} (2).

When $W=V^*$, the basis element $\bracket{M_\lambda} \tsrvw \bracket{N_\lambda} \in
V_\lambda \tsrvw W_\lambda
\subseteq \nssymalgebra{}$ corresponds to the
matrix coefficient $u_{M_\lambda N_\lambda}$ of the comodule
$V_\lambda$ of $\O(GL_q(V))$ under the isomorphism in the proof of Proposition \ref{pbasis} (1).

It is of interest to know explicit transformation matrices connecting the Gelfand-Tsetlin
bases of $\nssym{r}{X}$ and $\nswedge{r}{X}$
with their standard monomial bases in Proposition \ref{pbasis}.
In other words, we want to know the decompositions in Proposition \ref{pgelfanddecomp} (1) and
(2) in terms of the monomial bases. When $W=V^*$, this amounts to finding explicit formulae for the
matrix coefficients of irreducible representations of $GL_q(V)$. This problem
has been studied intensively in the literature. When $\dv=2$, explicit formulae
for matrix coefficients in terms little $q$-Jacobi polynomials are known. In general,
the problem is not completely understood at present; see the survey \cite{vilenkin} and the
references therein.

The advantage of working with the Gelfand-Tsetlin bases of
$\nssymalgebra{}, \nswedgealgebra{}$, instead of the standard monomial bases in
Proposition \ref{pbasis} is that
multiplication is simpler in terms of the former, and has explicit formulae
in terms of Clebsch-Gordon coefficients.
We now state these formulae.

When $W=V^*$, the following
multiplication formula  for matrix coefficients can be deduced from \eqref{e CGCs}, \eqref{e CGCs inverted}, and the bialgebra structure of $\O(M_q(V))$ (see \cite[\textsection 7.2.2]{KS}):
\begin{equation}
u_{N_\lambda R_\lambda} u_{K_\mu S_\mu}
= \sum_{\nu,M_\nu,L_\nu} \left(\sum_i C_{N_\lambda K_\mu M_\nu, i} \,
\overline{C_{R_\lambda S_\mu L_\nu, i}}\right) u_{M_\nu L_\nu},
\end{equation}
where the bar denotes complex conjugation.


It follows that multiplication in the Gelfand-Tsetlin basis $\bigsqcup_{r \geq 0} B^\text{GT}(\nssym{r}{X})$ of $\nssymalgebra{}$ is given by
\begin{equation}
(\bracket{N_\lambda} \tsrvw \bracket{R_\lambda})
(\bracket{K_\mu} \tsrvw \bracket{S_\mu})
= \sum_{\nu,M_\nu,L_\nu} \left(\sum_i \overline{C_{N_\lambda K_\mu M_\nu, i}} \
\overline{C_{R_\lambda S_\mu L_\nu, i}}\right) \bracket{M_\nu} \tsrvw \bracket{L_\nu},
\end{equation}
where $\lambda,\mu$ are partitions with at most $\min(\dv,\dw)$ parts.

Similarly, multiplication in the basis $\bigsqcup_{r \geq 0} B^\text{GT}(\nswedge{r}{X})$ of $\nswedgealgebra{}$ is given by
\begin{equation} \label{eqwedgeprod1}
(\bracket{N_\lambda} \tsrvw \bracket{R_{\lambda'}})
(\bracket{K_\mu} \tsrvw \bracket{S_{\mu'}})
= \sum_{\nu,M_\nu,L_\nu} \left(\sum_i \overline{C_{N_\lambda K_\mu M_\nu, i}} \
\overline{C_{R_{\lambda'} S_{\mu'} L_{\nu'}, i}}\right) \bracket{M_\nu} \tsrvw \bracket{L_{\nu'}},
\end{equation}
where $\lambda,\mu$ are partitions with $\leq\dv$ parts and largest part $\leq\dw$.


\subsection{Examples and computations for $\O(M_q(\nsbr{X}))$}
\label{ss examples for nonstandard O}
Here we give some flavor of the defining relations of $\O(M_q(\nsbr{X}))$ in terms of the bases defined in  \textsection\ref{ss definitions ns symmetric exterior}.
We also show that the reduction system \eqref{eqredclassical} for  $\O(M_q(V))$ does not have an analog (in any way we can determine) in the nonstandard case.  This means that computations in  $\O(M_q(\nsbr{X}))$ as well as the corepresentation theory of  $\O(M_q(\nsbr{X}))$ are significantly more difficult than in the standard case.

In this section and in examples later on, we use the notation
\[\zz{\rho(i,j)}{\rho(k,l)} = y_{\rho(i,j)}\tsrdual y^{\rho(k,l)} = x_\vw{i}{j}\tsrdual x^\vw{k}{l},\]
where $y$ and $\rho$ are as in \textsection\ref{s notation for GLV GLW}, and $\tsrdual$ is our symbol for tensor product in this setting as explained in \textsection\ref{ss restitutions}.
We  also drop the $\otimes$ symbol for elements of $\nsbr{Z}^{\tsr r}$ so that $\zz{a}{a'}\zz{b}{b'}$ means $\zz{a}{a'}\tsr\zz{b}{b'}$. For example,
\[\zz{2}{1}\zz{3}{4} = \by_{23}\tsrdual\by^{14} = \bx_\vw{12}{21} \tsrdual x^\vw{12}{12}.\]

Let $\nsbr{B}^{\nsbr{X}}_+, \nsbr{B}^{\nsbr{X}}_-$ be the  bases of $\nssym{2}{X}$ and
$\nswedge{2}{X}$ from \eqref{e basis sym wedge}. Let $\nsbr{B}^{\nsbr{X}^*}_+$ and
$\nsbr{B}^{\nsbr{X}^*}_-$ be defined similarly, with the elements $\by^{ij}$ in place of  $\by_{ij}$, i.e. $\nsbr{B}^{\nsbr{X}^*}_\pm := \alpha_2 (\nsbr{B}^{\nsbr{X}}_\pm)$ where  $\alpha_2$ is as in  \eqref{e inner product X}.
With these bases, the defining relations of $\O(M_q(\nsbr{X}))$ take the form:
\begin{equation} \label{eqdefnewquan}
\begin{array}{l}
b'_+\tsrdual b_-=0, \quad  b'_+ \in \nsbr{B}^{\nsbr{X}}_+, b_- \in \nsbr{B}^{\nsbr{X}^*}_-, \\[2.2mm]
b'_-\tsrdual b_+=0, \quad  b'_- \in \nsbr{B}^{\nsbr{X}}_-, b_+ \in \nsbr{B}^{\nsbr{X}^*}_+.
\end{array}
\end{equation}

\begin{example} \label{sdiamondvia}
Let $\dv=\dw=2$.  Let
\[
\begin{array}{ccl}
 \text{NST}((2))&=&\left\{\ctableaua{1 \\ 2},\ctableaua{3 \\ 4},\ctableaua{1 \\ 3},\ctableaua{2 \\ 4},\ctableaua{3 \\ 2},\ctableaua{2 \\ 3}\right\}, \\[2.5mm]
 \widetilde{\text{NST}}((1,1))&=&\big\{ \widetilde{\ctableau{1}{1}},\widetilde{\ctableau{2}{2}},\widetilde{\ctableau{1}{2}},\widetilde{\ctableau{3}{3}},\widetilde{\ctableau{4}{4}}, \widetilde{\ctableau{3}{4}},\widetilde{\ctableau{1}{3}},\widetilde{\ctableau{2}{4}}, \widetilde{\ctableau{1}{4}},\widetilde{\ctableau{4}{1}}\big\},
\end{array}
\]
be as in \eqref{eqwedgedef2} and \eqref{eqsymdef2}.

The defining relations \eqref{eqdefnewquan} of
$\O(M_q(\nsbr{X}))$ are now 120 in number.
We show one such typical relation below (to avoid extra notation, we use the same symbol for an element of $\nsbr{B}^{\nsbr{X}^*}_\pm$ as its corresponding element of $\nsbr{B}^{\nsbr{X}}_\pm$):
\begin{equation}
\begin{array}{lc@{\, }c@{\hspace{-.2mm}}l}
0= \widetilde{\ctableau{1}{4}} \tsrdual \ctableaua{3 \\ 2}
 &=&& \frac{1}{[2]^2}(q^2 \bx_\vw{12}{12} + q \bx_\vw{12}{21} + q \bx_\vw{21}{12} + \bx_\vw{21}{21}) \tsrdual  \frac{1}{[2]}(q \bx^\vw{12}{21}-\bx^\vw{12}{12} + \bx^\vw{21}{21} - \ui \bx^\vw{21}{12})\\[3mm]
& =&& \frac{1}{[2]^2}(q^2 \by_{14} + q \by_{23} + q \by_{32} + \by_{41}) \tsrdual  \frac{1}{[2]}(q \by^{23}- \by^{14} +  \by^{41} - \ui\by^{32})\\[3mm]
&=&\frac{1}{[2]^3}\Big(&+ q^3 \zz{1}{2}  \zz{4}{3}  +  q^2 \zz{2}{2}  \zz{3}{3}  + q^2 \zz{3}{2}  \zz{2}{3}  + q \zz{4}{2}  \zz{1}{3}\\[3mm]
&& &- q^2 \zz{1}{1}  \zz{4}{4}  - q \zz{2}{1}  \zz{3}{4}  - q \zz{3}{1}  \zz{2}{4}  -  \zz{4}{1}  \zz{1}{4}  \\[3mm]
&& &+ q^2\zz{1}{4}  \zz{4}{1}  +q  \zz{2}{4}  \zz{3}{1}  +q \zz{3}{4}  \zz{2}{1}  + \zz{4}{4}  \zz{1}{1} \\[3mm]
&& &- q \zz{1}{3}  \zz{4}{2}  -  \zz{2}{3}  \zz{3}{2}  - \zz{3}{3}  \zz{2}{2}  -\ui \zz{4}{3}  \zz{1}{2} \Big).
\end{array}
\end{equation}
\end{example}

The $2\binom{\dx+1}{2}\binom{\dx}{2}$ relations of \eqref{eqdefnewquan}, after taking appropriate linear combinations,
can be recast in the form of a
reduction system---just as the relations \eqref{eqp+2} were recast in the form of a
reduction system \eqref{eqredclassical}---where each reduction rule is of the form
\[ \zz{a}{a'} \zz{b}{b'} = \sum_{i} \alpha_i \zz{a_i}{a_i'} \zz{b_i}{b_i'}, \]
each $\zz{a_i}{a_i'} \zz{b_i}{b_i'}$ being {\em descending}, meaning that $(a_i,a'_i) \geq (b_i,b'_i)$ (say, lexicographically).
The resulting reduction system is described in Appendix \ref{s reduction system}.
It turns out that this system does not satisfy the diamond property.
For example, the monomial
$\zz{1}{1} \zz{1}{2} \zz{2}{3}$, when reduced in two different ways,
yields the following two distinct linear combinations of descending monomials:
\begin{align*}
 l_{121}&=(-1+{q}^{2}) \cdot \zz{2}{1}\zz{1}{3}\zz{1}{2}+{\frac {{q}^{2}-1}{[2]}} \cdot \zz{2}{1}\zz{1}{4}\zz{1}{1}+{\frac {{q}-{q}^{-1}}{[2]}} \cdot \zz{2}{2}\zz{1}{3}\zz{1}{1}\\
 &+ {\frac {2{q}}{[2]}} \cdot \zz{2}{3}\zz{1}{2}\zz{1}{1}+{\frac {1-{q}^{2}}{[2]}} \cdot \zz{2}{4}\zz{1}{1}\zz{1}{1}, \\
 l_{212}&={\frac {q^3 +q -3{q}^{-1}+{q}^{-3}}{[2]}}
\cdot \zz{2}{1}\zz{1}{3}\zz{1}{2}+
{\frac {2{q}-2q^{-1}}{[2]^{2}}} \cdot \zz{2}{1}\zz{1}{4}\zz{1}{1}+ {\frac { 2-2{q}^{-2}}{[2]^{2}}} \cdot \zz{2}{2} \zz{1}{3} \zz{1}{1}\\
&+{\frac {{q}^{2}+4-{q}^{-2}}{[2]^{2}}} \cdot \zz{2}{3}\zz{1}{2}\zz{1}{1} -  {\frac { 2q-2{q}^{-1}}{[2]^{2}}}\cdot \zz{2}{4}\zz{1}{1}\zz{1}{1}.
\end{align*}
This means we have the following nontrivial relation among descending monomials:
\[ l_{121}-l_{212}=0.\]
See Appendix \ref{s reduction system} for details.

This failure of the diamond property has the following consequence:
\begin{proposition} \label{ppoincare}
The Poincar\'e series of $\O(M_q(\nsbr{X}))$ does not, in general, coincide with
the Poincar\'e series of the classical $\O(M( X ))$.
\end{proposition}
Here by Poincar\'e series of $\O(M_q(\nsbr{X}))$ we mean the series
\[\sum_{r \geq 0} \dim(\O(M_q(\nsbr{X}))_r) \, t^r,\]
where $\O(M_q(\nsbr{X}))_r$ denotes the degree $r$ component of
$\O(M_q(\nsbr{X}))$.
As an example, when $\dv=\dw=2$,
$\dim(\O(M_q(\nsbr{X}))_3)= 688$, whereas the classical $\dim(\O(M(X))_3)=816$;
this example will be explained thoroughly in  \textsection\ref{sex3}.

Proposition \ref{ppoincare}  has important consequences. In the standard case,
the Poincar\'e series of $\O(M_q(V))$ coincides with
the Poincar\'e series of the classical $\O(M(V))$. By the Peter-Weyl theorem,
this implies that the irreducible representations of $GL_q(V)$ are in one-to-one correspondence with those of $GL(V)$, and dimensions agree under this correspondence.
Intuitively, this is  why the
irreducible representations of $GL_q(V)$ turn out to be  deformations of the irreducible
representations of $GL(V)$.
Proposition \ref{ppoincare} implies that this is no longer true for the nonstandard quantum group.

\section{Nonstandard determinant and minors} \label{squantumdet}
Here we define the left and right nonstandard determinant and minors of  $M_q(\nsbr{X})$.  After some examples in  \textsection\ref{ss nonstandard minors in the two row case}, we show (\textsection\ref{ssymmetry}) that the left and right determinants and minors with respect to the orthonormal Gelfand-Tsetlin basis agree.   Finally in  \textsection\ref{sexpformulaedet}, we give explicit formulae for certain nonstandard minors and present an intriguing conjecture about lengths of canonical basis elements related to these minors.

\subsection{Definitions}
Recall from Proposition \ref{pequidefining1} (2) that $\nswedge{r}{X}^*$  (resp. $\nswedge{r}{X}$) is  a right (resp.  left)  $\O(M_q(\nsbr{X}))$-comodule and
let
\begin{equation}
\begin{array}{l}
\varphi^R_r: \nswedge{r}{X}^* \rightarrow  \nswedge{r}{X}^* \tsr \O(M_q(\nsbr{X})), \\[1.7mm]
\varphi^L_r: \nswedge{r}{X} \rightarrow \O(M_q(\nsbr{X})) \tsr \nswedge{r}{X}
\end{array}
\end{equation}
be the corresponding right and left corepresentations.

Recall that $\Omega_r^\dx$ is the set of subsets of $[\dx] = [\dv \dw]$ of size $r$.
For a subset $I\in \Omega_r^\dx$, with $I=\{i_1,\ldots,i_r\}$, $i_1<i_2<\cdots < i_r$, let
$\by_I$ be the monomial $y_{i_1}\cdots y_{i_r}$ in the notation of Proposition \ref{pbasis}.
By this proposition, the set of standard monomials $\{\by_I\}_{I \in \Omega_r^\dx}$ is a basis of $\nswedge{r}{X}$.

We define the \emph{right nonstandard determinant} $\nsbr{D}^R$ to be the matrix coefficient of
the right comodule $\nswedge{\dx}{X}^*$, which is independent of the choice of basis as  $\nswedge{\dx}{X}^*$ is one-dimensional.
The \emph{left nonstandard determinant} $\nsbr{D}^L$ is the matrix coefficient of the left comodule
$\nswedge{\dx}{X}$.
The nonstandard determinants are nonzero since $\epsilon(\nsbr{D}^R) = \epsilon(\nsbr{D}^L) = 1$, where  $\epsilon$ is the counit.

More generally, the \emph{right nonstandard $r$-minors} of $M_q(\nsbr{X})$ in the standard monomial basis are defined to be the
matrix coefficients of the right corepresentation  $\varphi^R_r$ in the standard
monomial basis of $\nswedge{r}{X}^*$:
for $I \in \Omega_r^\dx$, define the right   nonstandard $r$-minors $\nsbr{D}^{I,R}_J$ by
\[ \varphi^R_r(\by^I)=\sum_{J \in \Omega_r^\dx}  \by^J \tsr \nsbr{D}^{I,R}_J.\]
The \emph{left  nonstandard $r$-minors} $\nsbr{D}^{I,L}_J$ are defined by
\[ \varphi^L_r(\by_J)=\sum_{I \in \Omega_r^\dx}  \nsbr{D}^{I,L}_J \tsr \by_I.\]
Let  $M^{r,R}_\wedge = (\nsbr{D}^{I,R}_J)$ and $M^{r,L}_\wedge = (\nsbr{D}^{I,L}_J)$ denote the corresponding coefficient
matrices.

Similarly, define the right and left nonstandard minors of $M_q(\nsbr{X})$ in the
orthonormal Gelfand-Tsetlin basis to be
the matrix coefficients of the right and left corepresentations $\varphi^R_r$ and
$\varphi^L_r$ in the
orthonormal Gelfand-Tsetlin bases $B^{\prime\text{GT}}(\nswedge{r}{X}^*)$ and $B^{\prime\text{GT}}(\nswedge{r}{X})$ of $\nswedge{r}{X}^*$ and $\nswedge{r}{X}$.
Here $B^{\prime\text{GT}}(\nswedge{r}{X})$ is like the basis $B^\text{GT}(\nswedge{r}{X})$ of \eqref{egtbwedge}, except may differ from it by a diagonal transformation;
orthonormal must be interpreted in a certain way here, which is explained in the proof of Proposition-Definition \ref{pdeterminant}.
Also, the bases $B^\text{$\prime$GT}(\nswedge{r}{X}^*)$ and $B^\text{$\prime$GT}(\nswedge{r}{X})$ are related by  $B^\text{$\prime$GT}(\nswedge{r}{X}^*)=\alpha_r (B^\text{$\prime$GT}(\nswedge{r}{X}))$, where $\alpha_r$ is as in \eqref{e inner product X}; the same notation  $\bracket{M_\lambda} \tsrvw \bracket{N_{\lambda'}}$ will be used for both bases.

These minors are defined explicitly as follows:
for  $\bracket{M_\lambda} \tsrvw \bracket{N_{\lambda'}} \in B^\text{$\prime$GT}(\nswedge{r}{X}^*)$,
 $\lambda \vdash r$,
define the \emph{right   nonstandard $r$-minors} $\nsbr{D}^{M_\lambda,N_{\lambda'},R}_{K_\mu,L_{\mu'}}$ by
\[ \varphi^R_r(\bracket{M_\lambda}\tsrvw \bracket{N_{\lambda'}})=\sum_{\bracket{K_\mu} \tsrvw \bracket{L_{\mu'}} \in B^\text{$\prime$GT}(\nswedge{r}{X}^*)}
\bracket{K_\mu} \tsrvw \bracket{L_{\mu'}} \tsr \nsbr{D}^{M_\lambda,N_{\lambda'},R}_{K_\mu,L_{\mu'}}. \]
The \emph{left  nonstandard $r$-minors} $\nsbr{D}^{M_\lambda,N_{\lambda'},L}_{K_\mu,L_{\mu'}}$ are defined similarly.
Let  $\tilde M^{r,R}_\wedge = \big(\nsbr{D}^{M_\lambda,N_{\lambda'},R}_{K_\mu,L_{\mu'}}\big)$ and $\tilde M^{r,L}_\wedge = \big(\nsbr{D}^{M_\lambda,N_{\lambda'},L}_{K_\mu,L_{\mu'}}\big)$ denote the corresponding coefficient
matrices.

\subsection{Nonstandard minors in the two-row case}
\label{ss nonstandard minors in the two row case}

\begin{example}
Let us first give an  explicit formula for $\nsbr{D}^L$ and $\nsbr{D}^R$ when
$\dv=\dw=2$.
The nonstandard exterior algebra  $\nswedgealgebra{}$ is the quotient of  $T(\nsbr{X})=\bigoplus_{r \geq 0} \nsbr{X}^{\otimes r}$ by the two-sided ideal  $\nsbr{\mathcal{I}}_\wedge$ generated by the elements
\[\widetilde{\ctableau{1}{1}},\, \widetilde{\ctableau{2}{2}},\, \widetilde{\ctableau{1}{2}},\, \widetilde{\ctableau{3}{3}},\, \widetilde{\ctableau{4}{4}},\, \widetilde{\ctableau{3}{4}},\, \widetilde{\ctableau{1}{3}},\, \widetilde{\ctableau{2}{4}},\, \widetilde{\ctableau{1}{4}},\, \widetilde{\ctableau{4}{1}}\]
from  \eqref{eqsymdef2}.
The degree $r$ component $\nswedge{r}{X}$ has standard monomial basis
$\{y_{i_1} \cdots y_{i_r} : 1\leq i_1<i_2< \cdots < i_r \leq \dx\}$, where
\[y_1=  x_\vw{1}{1},y_2= x_\vw{1}{2},y_3= x_\vw{2}{1},y_4= x_\vw{2}{2}.\]

The relations  $\widetilde{\ctableausmall{1}{4}},\, \widetilde{\ctableausmall{4}{1}} = 0$
imply that $y_4y_1=-y_1y_4$. Since
 $y_1$ and $y_4$ quasicommute with all the $y_i$'s, and
$y_i^2=0$ for all $i$, it is easy to show that
$y_iy_jy_ky_l$ is zero modulo $\nsbr{\mathcal{I}}_\wedge$,
unless it is of the form $\prod y_{\sigma(i)}$, for some permutation
$\sigma$, or is either $y_2y_3y_2y_3$ or
$y_3y_2y_3y_2$. Furthermore, we have
\[\prod y_{\sigma(i)}=(-1)^{\ell(\sigma)}q^{\iota(\sigma)} y_1y_2y_3y_4,\]
where $\ell(\sigma)$ is the number of inversions in $\sigma$, and
$\iota(\sigma)$ is the number of inversions in $\sigma$ not involving
$(2,3)$ or $(1,4)$. Also
\[\begin{array}{lcl}
y_2y_3y_2y_3&=&(\ui-q)q^2 y_1y_2y_3y_4 \\
y_3y_2y_3y_2&=&(q-\ui)q^2 y_1y_2y_3y_4.
\end{array}
\]

The right (resp.  left) determinant $\nsbr{D}^R$ (resp.  $\nsbr{D}^L$) is the  the matrix coefficient
of the right comodule $\nswedge{4}{X}^*$ (resp. left comodule $\nswedge{4}{X}$).
From the preceding remarks, it easily follows that (see  \textsection\ref{ss examples for nonstandard O} for notation)
\begin{align*}
\nsbr{D}^R = \Big(\sum_\sigma (-1)^{\ell(\sigma)}q^{\iota(\sigma)} \zz{\sigma(i)}{i}\Big)
+ (\ui-q)q^2 \zz{2}{1} \zz{3}{2} \zz{2}{3} \zz{3}{4}
+ (q-\ui)q^2 \zz{3}{1} \zz{2}{2} \zz{3}{3} \zz{2}{4}, \\
\nsbr{D}^L= \Big(\sum_\sigma (-1)^{\ell(\sigma)}q^{\iota(\sigma)} \zz{i}{\sigma(i)}\Big)
+ (\ui-q)q^2 \zz{1}{2} \zz{2}{3} \zz{3}{2} \zz{4}{3}
+ (q-\ui)q^2 \zz{1}{3} \zz{2}{2} \zz{3}{3} \zz{4}{2}.
\end{align*}
These expressions are equal in $\O(M_q(\nsbr{X}))$, which can be checked from the relations \eqref{eqdefnewquan}, or we can appeal to the more abstract argument given in  \textsection\ref{ssymmetry}.
Compare these with the formula \eqref{eqquantumdetstd} for the standard quantum determinant.

We will also give another formula for the nonstandard determinant in Proposition \ref{p ns minor formula} in terms of the upper canonical basis $B^\dx_V \tsrvw B^\dx_W$. In this case it is
\[
\nsbr{D}^L = \nsbr{D}^R = \left(\cvw{2121}{2211} - \cvw{2211}{2121}\right) \tsrdual \bx^\vw{2121}{2211}.
\]
\end{example}

\begin{example}
The nonstandard minors in the two-row,  $r=2$, case are as follows (see Example \ref{ex ns wedge sym bases} for notation):
\[\begin{array}{c}
\matb{c}{-q \ctableaua{1\\2} \\[3mm]
-q \ctableaua{1\\3} \\[3mm]
-\ui \ctableaua{3\\2} - q\ctableaua{2\\3} \\[3mm]
\ctableaua{3\\2} - \ctableaua{2\\3} \\[3mm]
-q \ctableaua{2\\4} \\[3mm]
-q \ctableaua{3\\4}
}
\dot{\tsrdual}
-\frac{1}{[2]}\matb{c@{\hspace{7mm}}c@{\hspace{7mm}}c@{\hspace{7mm}}c@{\hspace{7mm}}c@{\hspace{7mm}}c}{\ctableaua{1\\2} &
\ctableaua{1\\3} &
\ctableaua{3\\2} + \ctableaua{2\\3} &
-q \ctableaua{3\\2} +\ui \ctableaua{2\\3} &
\ctableaua{2\\4} &
\ctableaua{3\\4}
},\\[1mm]
\text{The coefficient matrix } M^{2,R}_\wedge\\[6mm]
-\frac{1}{[2]}\matb{c}{\ctableaua{1\\2} \\[3mm]
\ctableaua{1\\3} \\[3mm]
\ctableaua{3\\2} + \ctableaua{2\\3} \\[3mm]
-q \ctableaua{3\\2} +\ui \ctableaua{2\\3} \\[3mm]
\ctableaua{2\\4} \\[3mm]
\ctableaua{3\\4}
}
\dot{\tsrdual}
\matb{c@{\hspace{7mm}}c@{\hspace{7mm}}c@{\hspace{7mm}}c@{\hspace{7mm}}c@{\hspace{7mm}}c}{-q \ctableaua{1\\2} &
-q \ctableaua{1\\3} &
-\ui \ctableaua{3\\2} - q\ctableaua{2\\3} &
\ctableaua{3\\2} - \ctableaua{2\\3} &
-q \ctableaua{2\\4} &
-q \ctableaua{3\\4}
},\\[1mm]
\text{The coefficient matrix } M^{2,L}_\wedge
\end{array}\]
where  $\dot{\tsrdual}$ is like $\dot{\tsr}$ of \eqref{e dot tsr} using the tensor product $\tsrdual$.  Here the matrix  $M^{2,R}_\wedge$ is with respect to the ordered basis  $(\by^{12},\by^{13},\by^{14},\by^{23},\by^{24},\by^{34})$ of  $\nswedge{2}{X}^*$ so that, for instance, its third column gives the entries in the second tensor factor of  $\varphi^R_2(\by^{14}) = \sum_{J \in \Omega_2^4}  \by^J \tsr \nsbr{D}^{14, R}_J$.  The matrix  $M^{2,L}_\wedge$ is with respect to the ordered basis  $(\by_{12},\by_{13},\by_{14},\by_{23},\by_{24},\by_{34})$ of  $\nswedge{2}{X}$ so that, for instance, its fourth row gives the entries in the second tensor factor of  $\varphi^L_2(\by_{23}) = \sum_{I \in \Omega_2^4}  \nsbr{D}^{I,L}_{23} \tsr \by_I$.

In the two-row case, the orthonormal Gelfand-Tsetlin basis $B^{\prime\text{GT}}(\nswedge{r}{X})$ and the NSC basis (as defined in  \textsection\ref{s A canonical basis for Yalpha}) of  $\nswedge{r}{X}$ differ by a diagonal transformation; this follows from the fact that the Gelfand-Tsetlin basis and projected upper canonical basis of $V^{\tsr r}$ differ by a diagonal transformation---see \eqref{eqA11} and Example \ref{ex Gelfand-Tsetlin}.  For the example at hand, $B^{\prime\text{GT}}(\nswedge{2}{X})$ is given by
\[
\ctableaua{1\\2}^{\text{GT}} \hspace{-1.4mm} = a\ctableaua{1\\2}, \ \ \
\ctableaua{1\\3}^{\text{GT}} \hspace{-1.4mm} = a\ctableaua{1\\3}, \ \ \
\ctableaua{3\\2}^{\text{GT}} \hspace{-1.4mm} = \ctableaua{3\\2}, \ \ \
\ctableaua{2\\3}^{\text{GT}} \hspace{-1.4mm} = \ctableaua{2\\3}, \ \ \
\ctableaua{2\\4}^{\text{GT}} \hspace{-1.4mm} = a\ctableaua{2\\4}, \ \ \
\ctableaua{3\\4}^{\text{GT}} \hspace{-1.4mm} = a\ctableaua{3\\4}, \ \ \
\]
where $a=\left(\frac{q}{[2]}\right)^{1/2}$. The coefficient matrices $\tilde M^{2,R}_\wedge$ and $\tilde M^{2,L}_\wedge$ with respect to the Gelfand-Tsetlin basis are
\[
\begin{array}{c}
\matb{c}{
\ctableaua{1\\2}^{\text{GT}} \\[3mm]
\ctableaua{1\\3}^{\text{GT}} \\[3mm]
\ctableaua{3\\2}^{\text{GT}} \\[3mm]
\ctableaua{2\\3}^{\text{GT}} \\[3mm]
\ctableaua{2\\4}^{\text{GT}} \\[3mm]
\ctableaua{3\\4}^{\text{GT}}
}
\dot{\tsrdual}
\matb{c@{\hspace{7mm}}c@{\hspace{7mm}}c@{\hspace{7mm}}c@{\hspace{7mm}}c@{\hspace{7mm}}c}{
\ctableaua{1\\2}^{\text{GT}} &
\ctableaua{1\\3}^{\text{GT}} &
\ctableaua{3\\2}^{\text{GT}} &
\ctableaua{2\\3}^{\text{GT}} &
\ctableaua{2\\4}^{\text{GT}} &
\ctableaua{3\\4}^{\text{GT}}
}.\\[1mm]
\text{The coefficient matrix } \tilde M^{2,R}_\wedge = \tilde M^{2,L}_\wedge
\end{array}
\]
It will be shown in Proposition \ref{p ns minor formula} that for any  $r \in [\dx]$, the coefficient matrix $\tilde M^{r,R}_\wedge = \tilde M^{r,L}_\wedge$ has a similar form as an outer product.
\end{example}

\subsection{Symmetry of the determinants and minors}\label{ssymmetry}
A basic property of the standard quantum minors  is the agreement of the left and right-handed versions \eqref{eqquantumminorstd}.  We now show that the same holds in the nonstandard case.  This will be important for defining the Hopf algebra $\O(GL_q(\nsbr{X}))$.

\begin{propdef} \label{pdeterminant}
The left and right nonstandard determinants agree, so we can define
\[\nsbr{D} := \nsbr{D}^L=\nsbr{D}^R. \]
More generally,
assuming that all objects are over  $\CC$ and $q$ is transcendental,
the left and right nonstandard minors in the bases $B^{\text{$\prime$GT}}(\nswedge{r}{X})$ and $B^{\text{$\prime$GT}}(\nswedge{r}{X}^*)$ agree, so we can define
\[\nsbr{D}^{M_\lambda,N_{\lambda'}}_{K_\mu,L_{\mu'}} := \nsbr{D}^{M_\lambda,N_{\lambda'},L}_{K_\mu,L_{\mu'}}=
\nsbr{D}^{M_\lambda,N_{\lambda'},R}_{K_\mu,L_{\mu'}}.\]
Equivalently, $\tilde M^{r,L}_\wedge= \tilde M^{r,R}_\wedge$.
\end{propdef}


In what follows, the tensor symbol $\tsrdual$ will be treated as an ``outer tensor'' in the sense that we are only using the coalgebra structure of  $\O(M_q(\nsbr{X}))$ for this tensor, not its bialgebra structure: if  $N'$ is a left  $\O(M_q(\nsbr{X}))$-comodule and  $N$ is a right  $\O(M_q(\nsbr{X}))$-comodule, then we consider $N' \tsrdual N$ as an $\O(M_q(\nsbr{X}))$-bicomodule.


To prove Proposition-Definition \ref{pdeterminant}, we will first show that $\nswedge{r}{X}^*$ and $(\nswedge{r}{X})_R$ are isomorphic as right $\O(M_q(\nsbr{X}))$-comodules, where
$(\cdot)_R$ is the notation for dualizing comodules explained in  \textsection\ref{ss comodules}.
Recall that $\nsbr{Z}= \nsbr{X} \tsr \nsbr{X}^*$ and by \eqref{eqdefnewquan no basis}, $\O(M_q(\nsbr{X}))$ is the quotient of the tensor algebra $T(\nsbr{Z})$ by the two-sided ideal  $\nsbr{\mathcal{I}}$ generated by $\nsbr{\mathcal{I}}_2 = \nssym{2}{X} \tsrdual \nswedge{2}{X}^*  \oplus \nswedge{2}{X} \tsrdual \nssym{2}{X}^*$.  Let  $\nsbr{\mathcal{J}}$ be the two-sided ideal of $T(\nsbr{Z})$ generated by $\nssym{2}{X} \tsrdual \nssym{2}{X}^*$ and set
\[ \mathcal{R} := T(\nsbr{Z})/(\nsbr{\mathcal{I}}+\nsbr{\mathcal{J}}) \cong \O(M_q(\nsbr{X}))/\br{\nsbr{\mathcal{J}}},\]
where  $\br{\nsbr{\mathcal{J}}}$ denotes the image of  $\nsbr{\mathcal{J}}$ in  $\O(M_q(\nsbr{X}))$.  The quotient coalgebra $\mathcal{R}$ turns out to be cosimple, as we now show, and will help us understand the nonstandard exterior algebra.

\begin{lemma} \label{llrdet2}
Let $\mathcal{R}$,  $\nsbr{\mathcal{I}}$, $\nsbr{\mathcal{J}}$ be as above and $\mathcal{R}_r, \nsbr{\mathcal{I}}_r, \nsbr{\mathcal{J}}_r$ denote their degree $r$ parts.
\begin{enumerate}
\item There is an isomorphism $\mathcal{R}_r \cong \nswedge{r}{X} \tsrdual \nswedge{r}{X}^*$ of  $\O(M_q(\nsbr{X}))$-bicomodules.
\item The coalgebra $\mathcal{R}_r$ is cosimple.
\item There is an isomorphism $\nswedge{r}{X}^* \cong (\nswedge{r}{X})_R$ (resp. $\nswedge{r}{X} \cong (\nswedge{r}{X}^*)_L$) of right (resp. left) $\O(M_q(\nsbr{X}))$-comodules.
\end{enumerate}
\end{lemma}
\begin{proof}
To prove (1), define
\begin{align*}
 Y^2 &:= \nssym{2}{X} \tsrdual (\nsbr{X}^*)^{\tsr 2},\\
 Y'^2 &:= X^{\tsr 2} \tsrdual \nssym{2}{X}^*,
\end{align*}
\begin{align*}
 M^r &:= \sum_{i = 1}^{r-1} \nsbr{X}^{\tsr i-1} \tsr \nssym{2}{X} \tsr (\nsbr{X})^{\tsr r-i-1}, \ r >2, \\
 M'^r &:= \sum_{i = 1}^{r-1} (\nsbr{X}^*)^{\tsr i-1} \tsr \nssym{2}{X}^* \tsr (\nsbr{X}^*)^{\tsr r-i-1}, \ r >2, \\
 Y^r &:= \sum_{i = 1}^{r-1} \nsbr{Z}^{\tsr i-1} \tsr Y^2 \tsr \nsbr{Z}^{\tsr r-i-1} = M^r \tsrdual (\nsbr{X}^*)^{\tsr r},\ r >2,\\
 Y'^r &:= \sum_{i = 1}^{r-1} \nsbr{Z}^{\tsr i-1} \tsr Y'^2 \tsr \nsbr{Z}^{\tsr r-i-1} = \nsbr{X}^{\tsr r} \tsrdual M'^r, \ r >2.
\end{align*}
All of the  $Y$'s are  $\O(M_q(\nsbr{X}))$-bicomodules, and $M^r$ and  $M'^r$ are left and right $\O(M_q(\nsbr{X}))$-comodules, respectively.
The bicomodule $\mathcal{R}$ is the quotient of $T(\nsbr{Z})$ by the two-sided ideal generated by  $\nsbr{\mathcal{I}}_2 + \nsbr{\mathcal{J}}_2 = Y^2+Y'^2$
, hence we have the following isomorphisms of $\O(M_q(\nsbr{X}))$-bicomodules
\begin{align}
\mathcal{R}_r &\cong \nsbr{Z}^{\tsr r} / (Y^r + Y'^r) \\
&\cong \nsbr{X}^{\tsr r} \tsrdual (\nsbr{X}^*)^{\tsr r} / \big(M^r \tsrdual (\nsbr{X}^*)^{\tsr r}+ \nsbr{X}^{\tsr r} \tsrdual M'^r \big) \\
&\cong  \big(\nsbr{X}^{\tsr r} / M^r \big)  \tsrdual \big((\nsbr{X}^*)^{\tsr r} / M'^r \big)  \\
&\cong  \nswedge{r}{X} \tsrdual \nswedge{r}{X}^*,
\end{align}
where the last isomorphism is by the definition of $\nswedge{r}{X}$, and the second to last is just an application of the general fact that
\[
\parbox{14cm}{for vector spaces $A \subseteq B$ and $A' \subseteq B'$,\ \,  $B \tsr B' /(A \tsr B' + B \tsr A') \cong B/A \tsr B'/A'$.}
\]

Now statement (2) follows from (1) and by applying the following claim to the algebra  $\mathcal{R}_r^*$ dual to the coalgebra  $\mathcal{R}_r$.
\[
\parbox{14cm}{Suppose $H$ is a finite-dimensional algebra over a field and  $M$ (resp.  $N$) is a left (resp.  right)  $H$-module.  If $H \cong M \tsr N$ as  $H$-bimodules and  $\dim(M) = \dim(N)$, then  $H$ is split simple.}
\]
The claim holds because  $H \cong M \tsr N$ implies that $M$ is a faithful left  $H$-module, hence  $H \hookrightarrow \End(M)$ is an inclusion of algebras.  Counting dimensions shows that this inclusion is an isomorphism.

Statement (3) follows from (2).
\end{proof}

\begin{proof}[Proof of Proposition-Definition \ref{pdeterminant}]
Lemma \ref{llrdet2} (3) implies that the coefficient
 matrices of $\nswedge{r}{X}^*$ and
$(\nswedge{r}{X})_R$ are similar, i.e., there exists a nonsingular
similarity matrix $Q$ such that
$\tilde M^{r,L}_\wedge=Q^{-1} \tilde M^{r,R}_\wedge Q$.
We want to show that $Q$ is the identity matrix.
Since the decomposition of  $\nswedge{r}{X}^*$ as an  $\O(M_q(V)) \tsrvw \O(M_q(W))$-comodule is multiplicity-free (Proposition \ref{pgelfanddecomp} (2)), a
Gelfand-Tsetlin basis for $\nswedge{r}{X}^*$ is uniquely determined up to a diagonal transformation.  This means that the  $\O(M_q(V)) \tsrvw \O(M_q(W))$-comodule isomorphism
$\nswedge{r}{X}^* \xrightarrow{\cong} (\nswedge{r}{X})_R$ must take $B^{\text{$\prime$GT}}(\nswedge{r}{X}^*)$ to a basis of $(\nswedge{r}{X})_R$ that differs from the basis dual to
$B^{\text{$\prime$GT}}(\nswedge{r}{X})$ by a diagonal transformation, i.e., $Q$ is diagonal.

If  $Q$ were not the identity, the basis elements could be normalized by square roots of the entries of  $Q$ to fix this (we are assuming all objects are defined over $\CC$ and $q$ is transcendental).   See the discussion below for a better explanation of this normalization.
\end{proof}
A good way to say how the Gelfand-Tsetlin bases must be normalized uses the realization of $\nswedge{r}{X}$ as a subset of $\nsbr{X}^{\tsr r}$ described in \textsection\ref{s A canonical basis for Yalpha}.
Recall from \eqref{e inner product X} that $\alpha_r:\nsbr{X}^{\tsr r} \xrightarrow{\cong} (\nsbr{X}^*)^{\tsr r}$ is the isomorphism induced by the standard bilinear form on  $\nsbr{X}^{\tsr r}$.
Restricting  $\alpha_r$ to  $\nswedge{r}{X}$ yields an isomorphism  $\nswedge{r}{X} \xrightarrow{\alpha_r} \nswedge{r}{X}^*$ of vector spaces.
Also, the composition  $(\nswedge{r}{X})_R \hookrightarrow (\nsbr{X}^{\tsr r})_R \xrightarrow{\cong} (\nsbr{X}^*)^{\tsr r}$
has image  $\nswedge{r}{X}^*$, so restricting the codomain yields  an  $\O(M_q(\nsbr{X}))$-comodule isomorphism  $\beta : (\nswedge{r}{X})_R \xrightarrow{\cong} \nswedge{r}{X}^*$.
This follows, for instance, from Lemma \ref{llrdet2} (3) and Theorem \ref{t nonstandard schur-weyl duality}.

Now, given a basis $B$ of  $\nswedge{r}{X}$, there are two ways to obtain a basis of  $\nswedge{r}{X}^*$: one is to take the image  $\alpha_r(B)$, and the other is to take the basis  $B^*$ dual to  $B$ and apply $\beta$ to obtain $\beta(B^*)$.  We want to know when  these bases agree.  This is equivalent to  $\beta^{-1}(\alpha_r (B))$ and  $B$ being dual bases, which exactly means that  $B$ is orthonormal with respect to the standard bilinear form restricted to $\nswedge{r}{X}$.
In the present situation, $B = B^\text{$\prime$GT}(\nswedge{r}{X})$ and, by definition,  $B^{\prime\text{GT}}(\nswedge{r}{X}^*) = \alpha_r(B)$.
The matrix  $Q$ is the transition matrix between $\beta(B^*)$ and  $\alpha_r(B)$, so  $Q$ being the identity matrix is equivalent to $B^\text{$\prime$GT}(\nswedge{r}{X})$ being orthonormal.



\subsection{Formulae for nonstandard minors}\label{sexpformulaedet}
Here we give as explicit as possible formulae for the nonstandard determinant
$\nsbr{D}$ and for certain nonstandard minors in the orthonormal Gelfand-Tsetlin basis.

\begin{proposition}\label{p ns minor formula}
In terms of the bases  $B^{\prime\text{GT}}(\nswedge{r}{X}) \subseteq \nswedge{r}{X} \subseteq \nsbr{X}^{\tsr r}$ and
$B^{\prime\text{GT}}(\nswedge{r}{X}^*) \subseteq \nswedge{r}{X}^* \subseteq (\nsbr{X}^*)^{\tsr r}$, there holds
\be \label{e nonstandard minor}
\nsbr{D}^{M_\lambda,N_{\lambda'}}_{K_\mu,L_{\mu'}} = \bracket{K_\mu} \tsrvw \bracket{L_{\mu'}} \tsrdual \bracket{M_\lambda} \tsrvw \bracket{N_{\lambda'}}.
\ee
In terms of the NSC basis of  \textsection\ref{s A canonical basis for Yalpha}, the highest weight nonstandard minors are given by
\be\label{e hw nonstandard minor2}
\nsbr{D}^{Z_\lambda,Z_{\lambda'}}_{Z_\lambda,Z_{\lambda'}} =
\text{NSC}_{Z_\lambda,Z_{\lambda'}}\tsrdual\bx^\vw{\mathbf{k}}{\mathbf{l}} :=
\sum_{Q \in SYT(\lambda)} (-1)^{\ell(\transpose{Q})} \cvw{RSK^{-1}(Z_\lambda,Q)}{RSK^{-1}(Z_{\lambda'},\transpose{Q})} \tsrdual\bx^\vw{\mathbf{k}}{\mathbf{l}},
\ee
where $\mathbf{k} = \text{RSK}^{-1}(Z_\lambda,\transpose{(Z_{\lambda'}^*)})$, $\mathbf{l} = \text{RSK}^{-1}(Z_{\lambda'},Z_{\lambda'}^*)$,
and $Z_\lambda$, $Z_{\lambda'}^*$, and  $\ell(\transpose{Q})$ are as in \textsection\ref{ss type A combinatorics preliminaries}.
\end{proposition}
\begin{proof}
The first formula \eqref{e nonstandard minor} follows from the proof of Proposition-Definition \ref{pdeterminant} and the discussion following it.  Maintain the notation of this discussion.  The coefficient matrix of  $\nsbr{X}^{\tsr r}$ in the standard monomial basis is $\big( \by_\mathbf{j} \tsrdual \by^{\mathbf{j}'}\big)_{\mathbf{j},\mathbf{j}' \in [\dx]^r}$.  In general, for any orthonormal basis  $B$ of  $\nsbr{X}^{\tsr r}$ with respect to the standard bilinear form, the coefficient matrix with respect to  $B$ has the similar form  $\big(b \tsrdual \alpha_r(b')\big)_{b,b' \in B}$.  The coefficient matrix of any  $ \O(M_q(\nsbr{X}))$-subcomodule  $N$ of  $\nsbr{X}^{\tsr r}$ with respect to an orthonormal basis of $N$ has a similar form.  In the present situation, the subcomodule is  $\nswedge{r}{X}$ and the orthonormal basis is $B^{\prime\text{GT}}(\nswedge{r}{X})$.

The right-hand side of \eqref{e hw nonstandard minor2} lies in $\nswedge{r}{X}\tsrdual (\nsbr{X}^*)^{\tsr r}$. By Lemma \ref{llrdet2},  $\nswedge{r}{X}$ is irreducible, and thus by Theorem \ref{t intro cmqg}(d), the image of $\nswedge{r}{X}\tsrdual (\nsbr{X}^*)^{\tsr r}$ in $\O(M_q(\nsbr{X}))$ is equal to the coefficient coalgebra $\nswedge{r}{X}\tsrdual \nswedge{r}{X}^* \subseteq \O(M_q(\nsbr{X}))$.  Considering $\nswedge{r}{X}\tsrdual \nswedge{r}{X}^* $ as an $\O(M_q(V))\tsr \O(M_q(W))$-bicomodule, the highest weight space of left and right weight $(\lambda,\lambda')$ is one-dimensional, so must be spanned by the right-hand side of \eqref{e hw nonstandard minor2} provided it is nonzero.

Since the nonstandard minors are matrix coefficients, we have $\epsilon(\nsbr{D}^{Z_\lambda,Z_{\lambda'}}_{Z_\lambda,Z_{\lambda'}})=1$.
Then to prove \eqref{e hw nonstandard minor2}, it remains to check that the counit evaluates to 1 on this quantity. This amounts to checking that the coefficient of $\bx_\vw{\mathbf{k}}{\mathbf{l}}$ in $\cvw{\mathbf{k}'}{\mathbf{l}'}$, for $\mathbf{k}',\mathbf{l}'$ such that $P(\mathbf{k}')=Z_\lambda, P(\mathbf{l}')=Z_{\lambda'}$, is 1 if $\mathbf{k}=\mathbf{k}', \mathbf{l}=\mathbf{l}'$ and 0 otherwise. By theorem \ref{t Schur-Weyl duality upper} (iii),  the coefficient of $\bv_\mathbf{l}$ in $c_{\mathbf{l}'}$ is just an adjusted Kazhdan-Lusztig polynomial $\br{P^-_{d(\mathbf{l}),d(\mathbf{l}')}}$, where $P^-_{d(\mathbf{l}),d(\mathbf{l}')}$ is as in \eqref{e Kazhdan-Lusztig polynomials}.
Noting that the $\mathbf{l}$ defined in the proposition is the maximal in Bruhat order element of the minimal coset representatives $\leftexp{J_{\lambda'}}{W}$, the desired result follows from the fact that $P'_{x,w}$ is 0 unless $x\leq w$.
\end{proof}

The \emph{length squared} $\|x\|^2$ of an element $x\in\nsbr{X}^{\tsr r}$ is defined to be $\langle x,x\rangle$ using the standard bilinear form of \eqref{e inner product X}.
Let $[k]_{-}$ be the $q$-analog $q^{-k+1}[k]=1+q^{-2}+\dots+q^{-2k+2}$ of $k$. The following computation for $\lambda=(r)$ should be mostly familiar:
\begin{align}
\|\text{NSC}_{Z_\lambda,Z_{\lambda'}}\|^2 = \|\cvw{1\,2\,\cdots\, r\ \ }{r\,r-1\cdots 1}\|^2 = \big\|\sum_{\sigma\in\S_r}(-q)^{-\ell(\sigma)}\bx_\vw{1\,2\,\cdots\, r\ \ \ \ }{(r\,r-1\cdots 1)\sigma} \big\|^2 = \sum_{\sigma\in\S_r}q^{-2\ell(\sigma)} = [r]_-!.
\end{align}
We conjecture the following generalization:
\begin{conjecture}\label{cj minor hook}
The length squares of the highest weight nonstandard columns are given by the following $q$-analogs of $r!$:
\[
\|\text{NSC}_{Z_\lambda,Z_{\lambda'}}\|^2 := \Big\|\sum_{Q \in SYT(\lambda)} (-1)^{\ell(\transpose{Q})} \cvw{RSK^{-1}(Z_\lambda,Q)}{RSK^{-1}(Z_{\lambda'},\transpose{Q})} \Big\|^2
= |\text{SYT}(\lambda)|\prod_{b\,\in\, \lambda}[h(b)]_-,
\]
where $\lambda \vdash_\dv r$,  $\ell(\lambda') \leq \dw$, the product ranges over the squares $b$ of the diagram of $\lambda$, and $h(b)$ denotes the hook length of $b$. Here $Z_\lambda$ is the superstandard tableau of shape and content $\lambda$ and $\cvw{\mathbf{k}}{\mathbf{l}}$ is the product $c^V_\mathbf{k} \tsrvw c^W_\mathbf{l}$ of upper canonical basis elements (see \textsection\ref{s notation for GLV GLW}).
\end{conjecture}
This conjecture has been checked for all $\lambda$ of size less than or equal to 6.  We suspect that these coefficients reflect something inherent in the integral form $\nswedge{r}{X}^\mathbf{A}$ of  $\nswedge{r}{X}$ (see  \textsection\ref{ss nonstandard columns labeled a canonical basis for Lambda r X}), rather than something specifically about canonical bases.

\begin{remark}
An interesting and, as far as we know, unstudied problem is to compute the lengths of canonical basis elements $c_\mathbf{k}$ (here the symmetric bilinear form is that for which the monomial basis of $V^{\tsr r}$ is orthonormal). These lengths are polynomials in $q^{-1}$, and, by the nonnegativity of type  $A$ Kazhdan-Lusztig polynomials, have nonnegative integer coefficients.   Finding a combinatorial interpretation for these lengths seems like a difficult but tractable problem, or at least much easier than understanding Kazhdan-Lusztig polynomials combinatorially.
\end{remark}

\section{The nonstandard quantum groups  $GL_q(\nsbr{X})$ and  $\unitary_q(\nsbr{X})$}
\label{s nonstandard quantum groups GLq}
Here we define the nonstandard coordinate Hopf algebra
$\O(GL_q(\nsbr{X}))$ of the (virtual) \emph{nonstandard quantum group $GL_q(\nsbr{X})$} by inverting the determinant of  $\O(M_q(\nsbr{X}))$
and using the results of the previous section to put a Hopf structure on the resulting bialgebra.
A natural $*$-structure is put on
$\O(GL_q(\nsbr{X}))$, and $\unitary_q(\nsbr{X})$ is defined to be the virtual object corresponding to this  $*$-Hopf algebra---this is
the analog of the unitary group in this setting.
Finally, we restate our main theorem about $GL_q(\nsbr{X})$ and  $\unitary_q(\nsbr{X})$ (Theorem \ref{t intro cmqg}, restated as Theorem \ref{tcmqg}) and assemble its proof.

\subsection{Hopf structure}
\label{shopf}
To define  a cofactor matrix of $\nsbr{\mb{z}}$ we need the following.
\begin{proposition} \label{ppairing}
The left $\O(M_q(\nsbr{X}))$-comodule homomorphism
\[\nswedge{r}{X} \otimes \nswedge{\dx-r}{X} \rightarrow \nswedge{\dx}{X},\ \,  \by_I \tsr \by_J \mapsto \by_I \by_J \]
is a nondegenerate pairing.  A similar statement holds for the corresponding right comodules.
\end{proposition}
\begin{proof}
Note that the given map is a left $\O(M_q(\nsbr{X}))$-comodule homomorphism because  $\nswedgealgebra{}$ is a left  $\O(M_q(\nsbr{X}))$-comodule algebra.
The nondegeneracy follows from Proposition \ref{pbasis}, the reduction system \eqref{eqredwedge},  and the nondegeneracy of the
pairing in the $q=1$ case.
\end{proof}

\begin{proposition} \label{pcofactor}
There exists a  cofactor matrix $\tilde{\nsbr{\mb{z}}}$
so that
\[ \tilde{\nsbr{\mb{z}}} \nsbr{\mb{z}}=\nsbr{\mb{z}} \tilde{\nsbr{\mb{z}}} = \nsbr{D} \mb{I}.\]
\end{proposition}
\begin{proof}
The matrix form of the nondegenerate pairing in Proposition \ref{ppairing} yields
a $q$-analog of Laplace expansion for $\O(M_q(\nsbr{X}))$ in the present context.
In particular, we have  nondegenerate pairings
\begin{equation}\label{eqpairing1}
\nswedge{\dx-1}{X}^* \otimes \nswedge{1}{X}^*  \xrightarrow{m^*_{\dx-1,1}} \nswedge{\dx}{X}^*,
\end{equation}
\begin{equation}\label{eqpairing2}
\nswedge{1}{X} \otimes \nswedge{\dx-1}{X} \xrightarrow{m_{1,\dx-1}} \nswedge{\dx}{X}.
\end{equation}
The homomorphism in \eqref{eqpairing1} (resp. \eqref{eqpairing2}) is a right (resp.  left) $\O(M_q(\nsbr{X}))$-comodule homomorphism.
Note that $\nswedge{1}{X}=\nsbr{X}$ is the fundamental vector representation.

Let $B^{\prime\text{GT}}(\nswedge{1}{X})=\{x_\vw{i}{j}\}$ and  $B^{\prime\text{GT}}(\nswedge{\dx-1}{X}) = \{\dual{x}_\vw{i}{j}\}$ be the Gelfand-Tsetlin bases of $\nsbr{X}$ and $\nswedge{{\dx-1}}{X}$ as in the proof of Proposition-Definition \ref{pdeterminant}; in terms of our previous notation for Gelfand-Tsetlin basis elements, $\dual{x}_\vw{i}{j} := \bracket{M_\lambda} \tsrvw \bracket{N_{\lambda'}}$, where  $\lambda = (\dw^{\dv-1},\dw-1)$ and  $M_\lambda$ (resp.  $N_{\lambda'}$) is superstandard in the first  $\dw-1$ (resp.  $\dv-1$) columns and its last column has entries  $[\dv]\setminus \{i\}$ (resp. $[\dw]\setminus \{j\}$). Let  $\nsbr{d} = x_\vw{1}{1} \dual{x}_\vw{1}{1}$ be our chosen basis element for $\nswedge{\dx}{X}$.  Then since
$\nswedge{\dx-1}{X} \cong \Wedge_q^{\dv-1}V \tsr(\Wedge_q^\dv V)^{\tsr\dw-1} \tsrvw
\Wedge_q^{\dw-1}W \tsr(\Wedge_q^\dw W)^{\tsr\dv-1}$
as an  $\O(M_q(V))\tsrvw \O(M_q(W))$-comodule, just as in the standard case (see the proof of \cite[Proposition 8, Chapter 9]{KS}), there holds
\begin{equation}\label{e pairing elements2}
x_\vw{i}{j}\tsr \dual{x}_\vw{k}{l} \xrightarrow{m_{1,\dx-1}}  \delta_{ik} \delta_{jl} (-q)^{i-1+j-1}\nsbr{d},
\end{equation}
Applying the left corepresentation maps corresponding to the comodules in \eqref{eqpairing2} to both sides of \eqref{e pairing elements2} implies that
\[\sum_{k,l}(-q)^{k+l-i-j}\, \zz{\rho(r,s)}{\rho(k,l)}\, \nsbr{D}^{\hat{k}\hat{l},L}_{\hat{i}\hat{j}} = \delta_{ri}\delta_{sj}\nsbr{D}^L,\]
where the $\nsbr{D}^{\hat{k}\hat{l},L}_{\hat{i}\hat{j}}$ is an abbreviated notation for the entries of $\tilde M^{\dx-1,L}_\wedge$.  Thus the cofactor matrix $\tilde {\nsbr{\mb{z}}}$ with entries $\tzz{\rho(k,l)}{\rho(i,j)} := (-q)^{k+l-i-j}\nsbr{D}^{\hat{k}\hat{l},L}_{\hat{i}\hat{j}}$
satisfies
\[ \nsbr{\mb{z}} \tilde{\nsbr{\mb{z}}} = \nsbr{D} \mb{I}.\]
(Recall our convention that lower indices correspond to rows and upper indices to columns.)
By Proposition-Definition \ref{pdeterminant},  $\nsbr{D}^{\hat{k}\hat{l},L}_{\hat{i}\hat{j}} = \nsbr{D}^{\hat{k}\hat{l},R}_{\hat{i}\hat{j}}$ and  $\nsbr{D}^L=\nsbr{ D}^R$.
A similar computation using \eqref{eqpairing1} then yields
\[ \tilde{\nsbr{\mb{z}}} \nsbr{\mb{z}}= \nsbr{D} \mb{I}.\]
\end{proof}

This result implies, just as in the standard case \cite[\textsection9.2.2]{KS}, that $\nsbr{D}$ belongs to the center
of $\O(M_q(\nsbr{X}))$.
The coordinate algebra $\O(GL_q(\nsbr{X}))$
of the sought quantum group $GL_q(\nsbr{X})$ is obtained by adjoining the
inverse $\nsbr{D}^{-1}$ to $\O(M_q(\nsbr{X}))$.
We formally define $\nsbr{\mb{z}}^{-1}=\nsbr{D}^{-1} \tilde{\nsbr{\mb{z}}}$.
This allows us to define a Hopf structure on $\O(GL_q(\nsbr{X}))$
just as in the standard case (\textsection\ref{ss quantum determinant}).
\begin{proposition} \label{phopf}
There is a unique  Hopf algebra structure on
$\O(GL_q(\nsbr{X}))$  so that
\begin{enumerate}
\item $\Delta(\nsbr{\mb{z}})=\nsbr{\mb{z}}\dot{\tsr} \nsbr{\mb{z}}$,  $\Delta(\nsbr{D}^{-1})=\nsbr{D}^{-1} \tsr \nsbr{D}^{-1}$.
\item $\epsilon(\nsbr{\mb{z}})=\mb{I}$.
\item $S(\zz{j}{i})=\tzz{j}{i} \nsbr{D}^{-1}$, $S(\nsbr{D}^{-1})=\nsbr{D}$, where
$\zz{j}{i}$ are the entries of $\nsbr{\mb{z}}$ and
$\tzz{j}{i}$ are the entries of $\tilde{\nsbr{\mb{z}}}$.
\end{enumerate}
\end{proposition}
\begin{proof}
Most of the work has been done in Proposition \ref{pcofactor}.  The remaining details are similar to the standard case
\cite[Proposition 10, Chapter 9]{KS}.
\end{proof}

\subsection{Compact real form} \label{sstar}

\begin{proposition} \label{phopfstar}
The algebra $\O(GL_q(\nsbr{X}))$ is a Hopf $*$-algebra with the involution
$*$ determined by $(\zz{j}{i})^*=S(\zz{i}{j})$.
\end{proposition}
\begin{proof} This follows from \cite[Proposition 3, Chapter 9]{KS}.
This requires that the rule
$\nsbr{\mb{z}} \mapsto \transpose{\nsbr{\mb{z}}}$ determines an algebra automorphism of $ \O(GL_q(\nsbr{X}))$, which follows from the fact that $P_+^\nsbr{X}$ is symmetric.
\end{proof}

\begin{proposition}
\label{phomomorph GL}
Let  $\psi: \O(M_q(\nsbr{X})) \to \O(M_q(V))\tsrvw  \O(M_q(W))$ be as in Proposition \ref{phomomorph}.  There holds
\be \label{e psi nsD}
\psi(\nsbr{D}) = D_q(V)^\dw D_q(W)^\dv,
\ee
where  $D_q(V), D_q(W)$ are the quantum determinants of  $\O(M_q(V))$ and $\O(M_q(W))$, respectively (see  \textsection\ref{ss quantum determinant}).
There is a unique Hopf  $*$-algebra homomorphism
\[ \tilde{\psi}: \O(GL_q(\nsbr{X})) \to \O(GL_q(V))\tsrvw  \O(GL_q(W)) \]
extending  $\psi$.
\end{proposition}
\begin{proof}
The composition
\[\nswedge{\dx}{X}\xrightarrow{\varphi^L_\dx}\O(M_q(\nsbr{X}))\tsr \nswedge{\dx}{X} \xrightarrow{\psi\tsr \idelm} \O(M_q(V))\tsrvw  \O(M_q(W)) \tsr \nswedge{\dx}{X}\]
takes $y_1\cdots y_\dx$ to $\psi(\nsbr{D})\tsr y_1\cdots y_\dx.$
Thus $\psi(\nsbr{D})$ is the matrix coefficient of $\nswedge{\dx}{X}$ considered as a left $\O(M_q(V))\tsrvw  \O(M_q(W))$-comodule. By Proposition \ref{pgelfanddecomp} (2), $\nswedge{\dx}{X} \cong V_{(\dw^\dv)}\tsrvw W_{(\dv^\dw)}$, where $(\dw^\dv)$ (resp. $(\dv^\dw)$) is the rectangular partition with $\dv$ rows and $\dw$ columns (resp. $\dw$ rows and $\dv$ columns). Since $V_{(\dw^\dv)}\cong (\Wedge_q^\dv V)^{\tsr\dw}$  (as  $\O(M_q(V))$-comodules), the identity \eqref{e psi nsD} follows.

Any algebra homomorphism extending $\psi$ must satisfy $\nsbr{D}^{-1} \mapsto D_q(V)^{-\dw} D_q(W)^{-\dv}$, hence the uniqueness of $\tilde{\psi}$. It is easy to check that $\tilde{\psi}$ is a bialgebra homomorphism and a bialgebra homomorphism of Hopf algebras is always a Hopf algebra homomorphism \cite[\textsection1.2.4]{KS}.
That $\tilde{\psi}$ intertwines the $*$-involutions, i.e.  $\tilde{\psi}(z^*) = \tilde{\psi}(z)^*$, follows from the fact that it intertwines the antipodes and it intertwines the algebra automorphisms of  $\O(GL_q(\nsbr{X}))$ and  $ \O(GL_q(V))\tsrvw  \O(GL_q(W))$ determined by $\nsbr{\mb{z}} \mapsto \transpose{\nsbr{\mb{z}}}$ and  $\mb{u}^V \mapsto\transpose{(\mb{u}^V)}, \ \mb{u}^W \mapsto\transpose{(\mb{u}^W)}$, respectively.
 \end{proof}

\begin{proposition} \label{pcmqg}
The Hopf $*$-algebra $\O(GL_q(\nsbr{X}))$ is a CMQG algebra (see \textsection\ref{ss compact}).
\end{proposition}
\begin{proof}
The fundamental corepresentation $\nsbr{X}$ of $\O(GL_q(\nsbr{X}))$ is unitary by
Proposition \ref{pcofactor}, and  $\nsbr{D}^{-1}$ is a unitary element of
$\O(GL_q(\nsbr{X}))$. Furthermore, $\O(GL_q(\nsbr{X}))$ is generated by the
matrix elements of the unitary corepresentation $\nsbr{\mb{z}} \oplus (\nsbr{D}^{-1})$.
Hence, the result follows from Theorem \ref{tworonowicz} (a).
\end{proof}

\subsection{Complete reducibility}
We conclude this section by restating our main theorem about  $GL_q(\nsbr{X})$ and  $ \unitary_q(\nsbr{X})$ and collecting its proof.
\begin{theorem} \label{tcmqg}
Assume that all objects are over the field $\CC$ and $q$ is real and transcendental. Then

\noindent (a)
The Hopf algebra $\O(GL_q(\nsbr{X}))$ can be made into a Hopf $*$-algebra.
This is considered to be the coordinate ring of the compact real form of the nonstandard quantum group $GL_q(\nsbr{X})$. This virtual compact real form is
denoted $\unitary_q(\nsbr{X})$, which is a compact quantum group
in the sense of Woronowicz \cite{wor1}.

\noindent (b) There is a Hopf  $*$-algebra homomorphism
\[
\tilde{\psi}: \O(GL_q(\nsbr{X})) \rightarrow \O(GL_q(V)) \otimes \O(GL_q(W)),
\]

\noindent (c) Every finite-dimensional representation of $\unitary_q(\nsbr{X})$ (meaning a corepresentation of
$\O(GL_q(\nsbr{X}))$) is unitarizable, and
hence, is a direct sum of irreducible representations.

\noindent (d) An analog of the Peter-Weyl theorem holds:
\[
\O(GL_q(\nsbr{X})) = \bigoplus_{\alpha \in \nsP} (\nsbr{\X}_\alpha)_L \otimes \nsbr{\X}_\alpha,
\]
where $\nsP$ is an index set for the irreducible right comodules of $\O(GL_q(\nsbr{X}))$ and $\nsbr{\X}_\alpha$
is the comodule labeled by $\alpha$.
\end{theorem}
\begin{proof}
Part (a) is Proposition \ref{phopfstar} and Proposition \ref{pcmqg}, part (b) is contained in Proposition \ref{phomomorph GL}, and parts (c) and (d) follow from Proposition \ref{pcmqg} and Theorem \ref{tworonowicz}.
\end{proof}

\section{The nonstandard Hecke algebra $\nsH_r$} \label{s nonstandard Hecke algebra definition}
We now turn to the nonstandard Hecke algebra, which plays the role of the symmetric group in the nonstandard setting.
The group algebra  $\QQ \S_r$ is a Hopf algebra with coproduct $\Delta_{\QQ \S_r} :\QQ \S_r \to \QQ \S_r \tsrvw \QQ \S_r$,  $w \mapsto w \tsrvw w$.
This makes the tensor product of  $\QQ \S_r$-modules into a  $\QQ \S_r$-module, and allows us to define Kronecker coefficients.  The Hecke algebra $\H_r$ is not a Hopf algebra in a natural way.  The nonstandard Hecke algebra $\nsH_r$, which  we soon define, approximates the Hopf algebra $\QQ \S_r$ in the smallest possible way in a certain sense.  Despite its being as small as possible, $\nsH_r$ has dimension much larger than that of $\S_r$.

In this section we show that  $\nsH_r$  is semisimple and describe some of its basic properties and representation theory.
This section is mostly a summary of results from the older unpublished version of this paper \cite{GCT4} and \cite{Bnsbraid, B4}.
%

\subsection{Definition of  $\nsH_r$ and basic properties}
\label{ss definition of nsH}
Let  $S = \{s_1,\ldots,s_{r-1}\}$ be the set of simple reflections of the Coxeter group $\S_r$.  Let $\H_r = \H(\S_r)$ be the type $A$ Hecke algebra as introduced in \textsection\ref{ss cell label conventions C_Q C'_Q}.
Let $\C_s = T_s + \u$ and  $C_s = T_s - \ui$ for each  $s \in S$.  These are the simplest lower and upper Kazhdan-Lusztig basis elements (see  \textsection\ref{ss upper canonical basis of H} and \cite{BProjected}).  They are also proportional to the primitive central idempotents of  $\Ap (\H_r)_{\{s\}} \cong \Ap \H_2$, where  $\Ap := \mathbf{A}[\frac{1}{[2]}]$.  Specifically,
$\frac{1}{[2]} \C_{s_1}$ (resp. $- \frac{1}{[2]} C_{s_1}$) is the  idempotent corresponding to the trivial (resp. sign) representation of $\Ap \H_2$.

\begin{definition}
\label{d nonstandard Hecke algebra}
The \emph{type $A$ nonstandard Hecke algebra}  $\nsH_r$ is the subalgebra of $\H_r \tsrvw \H_r$ generated by the elements
\be \label{e sP definition}
\sP_s := \C_s \tsrvw \C_s + C_s \tsrvw C_s, \ s \in S.
\ee
We let $\nsbr{\Delta}:\nsH_r \hookrightarrow   \H_r \tsrvw \H_r$ denote the canonical inclusion, which we think of as a deformation of the coproduct
$\Delta_{\ZZ \S_r} :\ZZ \S_r \to \ZZ \S_r \tsrvw \ZZ \S_r$,  $w \mapsto w \tsrvw w$.
\end{definition}
The nonstandard Hecke algebra is also the subalgebra of $\H_r \tsrvw \H_r$ generated by
\[ \sQ_s := [2]^2 - \sP_s = - \C_s \tsrvw C_s - C_s \tsrvw \C_s, \ s \in S. \]
We will write $\sP_{i_1 i_2 \ldots i_l}$ as shorthand for  $\sP_{s_{i_1}} \sP_{s_{i_2}} \cdots \sP_{s_{i_l}}$.
For a ring homomorphism $\field \to \mathbf{A}$, we have the specialization $\field \nsH_r := \field \tsr_{\mathbf{A}} \nsH_r$ of the nonstandard Hecke algebra.

\begin{remark}
The notation $\tsrvw$ is our notation for tensor product when tensoring objects associated to $V$ with objects associated to $W$. We use this notation in the present context because of the natural action of $\H_r \tsrvw \H_r$ on $V^{\tsr r}\tsrvw W^{\tsr r}$ that will be discussed in \textsection\ref{s Nonstandard Schur-Weyl duality}.
\end{remark}

Write $\epsilon_+ = M_{(r)}^\mathbf{A}, \epsilon_- = M_{(1^r)}^\mathbf{A}$ for the one-dimensional trivial and sign representations of $\H_r$, which are defined by
\[ \begin{array}{cccc}
  \epsilon_+ : \C_s \mapsto \two, && \epsilon_- : \C_s \mapsto 0, & s \in S.
\end{array} \]
We identify these algebra homomorphisms $\epsilon_+, \epsilon_-: \H_r \to \mathbf{A}$ with right $\H_r$-modules in the usual way.

There are also one-dimensional trivial and sign representations of $\nsH_r$, which we denote by $\nsbr{\epsilon}_{+}$ and $\nsbr{\epsilon}_{-}$:
\[ \begin{array}{cccc}
  \nsbr{\epsilon}_{+} : \sP_{s} \mapsto \two^{2}, && \nsbr{\epsilon}_{-} : \sP_s \mapsto 0, & s \in S.
\end{array} \]

\medskip

There is an $\mathbf{A}$-algebra automorphism $\theta : \H_r \to \H_r$ defined by $\theta(T_s) = - T_s^{-1},\ s \in S$.  Note that  $\theta(\C_s) = -C_s$,  $\theta(C_s) = -\C_s$.
Let $1^{\text{op}}$ be the $\mathbf{A}$-anti-automorphism of $\H_r$ given by $1^{\text{op}}(T_w) = T_{w^{-1}}$.
Let $\theta^{\text{op}}$ be the $\mathbf{A}$-anti-automorphism of $\H_r$ given by $\theta^{\text{op}} = \theta \circ 1^{\text{op}} = 1^{\text{op}} \circ \theta$.
Let  $\eta$ be the unique $\mathbf{A}$-algebra homomorphism from  $\mathbf{A}$ to $\H_r$.
At  $\u=1$, the maps $\eta, \epsilon_+, 1^\text{op}$ specialize to the unit, counit, and antipode of the Hopf algebra  $\ZZ \S_r$.

\begin{proposition}[\cite{Bnsbraid}]
\label{p H2 Hopf algebra}
Set  $\Ap = \mathbf{A}[\frac{1}{[2]}]$ and  $A'_1 = \ZZ[\frac{1}{2}]$.  We have $\Ap \nsH_2 \cong \Ap\H_2$ by $\sP_1 \mapsto [2] \C_{s_1}$. Then
\begin{list}{\emph{(\roman{ctr})}} {\usecounter{ctr} \setlength{\itemsep}{1pt} \setlength{\topsep}{2pt}}
\item $\Ap \H_2$ is a Hopf algebra with coproduct $\Delta =\nsbr{\Delta}$, antipode $1^{\text{op}}$, counit $\epsilon_+$, and unit $\eta$.
\item the Hopf algebra $\Ap \H_2|_{\u = 1}$, with Hopf algebra structure coming from (i), is isomorphic to the group algebra $A'_1 \S_2$ with its usual Hopf algebra structure.
\end{list}

Moreover, the Hopf algebra structure of (i) is the unique way to make the algebra $\Ap\H_2$ into a Hopf algebra so that (ii) is satisfied.
\end{proposition}
\begin{remark}
\label{r nsH as small as possible}
If we want to construct a subalgebra  $H$ of $\field (\H_r \tsrvw \H_r)$ that is compatible with the coproducts  $\field (\H_r)_{\{s_i\}} \hookrightarrow \field((\H_r)_{\{s_i\}} \tsrvw (\H_r)_{\{s_i\}})$, then the nonstandard Hecke algebra  $\field \nsH_r$ is the smallest possible choice and the choice making  $H$ and $H \hookrightarrow \field (\H_r \tsrvw \H_r)$ as close as possible to  $\QQ \S_r$ and $\Delta_{\QQ \S_r}$ at  $\u =1$.
\end{remark}

The next proposition gives another way that $\nsH_r$ is like a Hopf algebra.
\begin{proposition}[\cite{Bnsbraid}] \label{p hecke algebra antipode}
The involutions $1^{\text{op}}$ and $\theta^{\text{op}}$ are antipodes in the following sense:
\begin{flalign}
\label{e antipode 1}
\mu \circ (1^{\text{op}} \tsr 1) \circ \nsbr{\Delta} &= \eta \circ \nsbr{\epsilon}_+,  \\
\label{e antipode 2}
\mu \circ (\theta^{\text{op}} \tsr 1) \circ \nsbr{\Delta} &= \eta \circ \nsbr{\epsilon}_-,
\end{flalign}
where these are equalities of maps from $\nsH_r$ to $\H_r$ and  $\mu$ is the multiplication map for  $\H_r$.
\end{proposition}

We next study some algebra involutions of  $\H_r \tsrvw \H_r$ and their restrictions to  $\nsH_r$.  This will help us understand the representation theory of  $\nsH_r$.
Let  $\tau : \H_r \tsrvw \H_r \to \H_r \tsrvw \H_r$, be the algebra involution given by $\tau(a \tsrvw b) \mapsto b \tsrvw a$.
Let $\ttaut$ be the subgroup of the group of algebra automorphisms of $\H_r \tsrvw \H_r$ generated by  $\theta \tsrvw 1$, $1 \tsrvw \theta$, and  $\tau$.  Let  $\A_{\theta, \tau}$ be the subgroup of $\ttaut$ generated by $\theta \tsrvw \theta$ and $\tau$.  Note that $\ttaut \cong \A_{\theta, \tau} \rtimes \S_2$ is isomorphic to the dihedral group of order eight.
\begin{proposition}
\label{p theta on nsH}
The elements of $\ttaut$ restrict nicely to $\nsH_r$:
\begin{list}{\emph{(\roman{ctr})}} {\usecounter{ctr} \setlength{\itemsep}{1pt} \setlength{\topsep}{2pt}}
\item $\alpha(\sP_s) =
\begin{cases} \sP_{s} & \text{if } \alpha \in \A_{\theta,\tau} \\
\sQ_{s} & \text{if } \alpha \in (\theta \tsrvw 1) \A_{\theta,\tau}
\end{cases}
\text{ for all } s \in S.$
\item $\nsH_r$ is left stable by the elements of $\ttaut$.
\item There is an  $\mathbf{A}$-algebra involution $\Theta: \nsH_r \to \nsH_r$ determined by  $\Theta(\sP_s) = \sQ_s$, $s \in S$.
\item  The restriction of an element of $\ttaut$ to $\nsH_r$ corresponds to the map $\ttaut \to \aut(\nsH_r)$ given by
$\theta \tsrvw 1 \mapsto \Theta,  \ 1 \tsrvw \theta \mapsto\Theta, \ \tau \mapsto 1.$
\item The nonstandard Hecke algebra is at the beginning of the chain of subalgebras $\nsH_r \subseteq (S^2 \H_r)^\theta  \subseteq S^2 \H_r \subseteq \H_r \tsrvw \H_r$, where
$(S^2 \H_r)^\theta$ is the subalgebra of  $\H_r \tsrvw \H_r$ fixed by the elements of $\A_{\theta,\tau}$.
\end{list}
\end{proposition}
\begin{proof}
Statement (i) follows from the definition of $\sP_s$ \eqref{e sP definition}.  The remaining statements follow easily from (i).
\end{proof}

\subsection{Semisimplicity of $\field\nsH_r$}
\begin{lemma}
\label{l transpose semisimple}
Let  $\field = \QQ(\u)$ and let $H \subseteq M_{m}(\field)$ be a $\field$-subalgebra of the matrix algebra $M_m(\field)$. If for every $M \in H$, the transpose $\transpose{M}$ is also in $H$, then $H$ is semisimple.
\end{lemma}
\begin{proof}
Since  $H$ is a finite-dimensional algebra over a field, its Jacobson radical $J(H)$ is nilpotent, i.e., $J(H)^k = 0$ for some $k$.  Thus  $H$ is semisimple (equivalently,  $J(H) = 0$) if and only if there does not exist a nonzero two-sided ideal  $I$ of  $H$ such that  $I^2=0$. Now suppose  $I$ is a two-sided ideal of  $H$ such that  $I^2 = 0$ and let $M \in I$. Then  $N := \transpose{M} M \in I$ implies  $\transpose{N} N = N^2 = 0$.  Taking the trace of both sides of this equation and letting  $N_{ij}$ denote the entries of  $N$ yields  $\sum N_{i j}^2 = 0$.  It follows that $N = 0$.  The same argument then shows $M =0$, thus  $I=0$.
\end{proof}

\label{ssemisimple}
\begin{proposition} \label{p nsH semisimple}
The nonstandard Hecke algebra $\field\nsH_r$ is semisimple, where $\field=\QQ(\u)$. There is a finite extension $\field'$ of $\field$ such that $\field'\nsH_r$ is split semisimple.
\end{proposition}
\begin{proof}
We know that there is a right action of $\field \H_r$
on $V^{\otimes r}$, which is faithful when $\dv = \dim_\field V \geq r$.  It is easy to see from \eqref{e T inverse_i act on V} that the action of each $\C_s$, when expressed in the standard monomials basis of  $V^{\tsr r}$, is a symmetric matrix.  Consequently there is a faithful
representation of $\field (\H_r \tsrvw \H_r)$ on $(V \tsrvw W)^{\otimes r}$ such that the matrix corresponding to each $\sP_s = \C_s \tsrvw \C_s + C_s \tsrvw C_s$ is symmetric.

Next, let us check that $\field \nsH_r \xrightarrow{\field\nsbr{\Delta}} \field (\H_r \tsrvw \H_r)$ is an inclusion (this deserves some care since it fails for the specialization $\mathbf{A} \to \ZZ$,  $\u \mapsto 1$). We have the inclusion $\nsH_r \xrightarrow{\nsbr{\Delta}} \H_r \tsrvw \H_r$ of  $\mathbf{A}$-modules.
Since localizations are flat, $\field$ is a flat $\mathbf{A}$-module, and thus  $\field\nsbr{\Delta}$ is also an inclusion.
Then $\field \nsH_r$ is a subalgebra of  $M_{(r^2)^r}(\field)$ generated by symmetric matrices, so by Lemma \ref{l transpose semisimple} $\field \nsH_r$ is semisimple.

The second fact follows from the general fact that any finite-dimensional associative algebra over a field becomes split after a finite field extension \cite[Proposition 7.13]{CR}.
\end{proof}

\begin{remark}
The specialization  $\nsH_r|_{q=1} := \QQ \tsr_\mathbf{A} \nsH_r$, the map  $\mathbf{A} \to \QQ$ given by  $\u \mapsto 1$, has  $\QQ$ dimension equal to  $\dim_\field \field \nsH_r$.  This is because $\QQ \tsr_\ZZ \nsH_r$ is a free $\QQA$-module since it is a submodule of a free $\QQA$-module and  $\nsH_r|_{q=1} = \QQ \tsr_\QQA \QQ \tsr_\ZZ \nsH_r$.
It can then be shown that the algebra $\nsH_r|_{q=1}$ is not semisimple for  $r > 2$. It has Jacobson radical  $J = \ker(\nsbr{\Delta}|_{q=1})$ and the quotient $\nsH_r|_{q=1}/J \cong \im(\nsbr{\Delta}|_{q=1})$ is equal to $\im(\Delta) \cong \QQ \S_r$, where  $\Delta : \QQ\S_r \to \QQ \S_r \tsrvw \QQ S_r$ is the usual coproduct.

\end{remark}

\subsection{Representation theory of $S^2\H_r$}
\label{ss representation theory of S2Hr}
We briefly discuss the representation theory of $S^2\H_r$ because this is close to the representation theory of $\nsH_r$, especially in the two-row case. We will return to this again in \textsection\ref{s Nonstandard representation theory in the two-row case}.

First note that we have the following commutativity property for any $\H_r$-modules  $M$ and  $M'$:
\be
\label{e commutativity S2H}
\Res_{S^2\H_r} M \tsrvw M' \cong \Res_{S^2\H_r} M' \tsrvw M,
\ee
where the isomorphism is given by  $\tau$.

Recall from  \textsection\ref{ss type A combinatorics preliminaries} that $\mathscr{P}_{r}$ denotes the set of partitions of size $r$ and $\mathscr{P}'_{r}$ the set of  partitions of  $r$ that are not a single row or column shape.
Recall from \textsection\ref{ss cell label conventions C_Q C'_Q} that $M^\mathbf{A}_\lambda$ denotes the Specht module of $\H_r$ so that $M_\lambda \cong \field\tsr_\mathbf{A} M_\lambda^\mathbf{A}$.
\begin{propdef}
\label{p S2Hr representations}
Define the following  $S^2 \H_r$-modules.  After tensoring these with $\field$, this is the list of distinct $\field S^2\H_r$-irreducibles
\begin{list}{\emph{(\arabic{ctr})}} {\usecounter{ctr} \setlength{\itemsep}{1pt} \setlength{\topsep}{2pt}}
\item $M^\mathbf{A}_{\{\lambda,\mu\}} := \Res_{S^2\H_r} M^\mathbf{A}_\lambda \tsrvw M^\mathbf{A}_\mu$, $\{\lambda,\mu\}\subseteq \mathscr{P}_r$, $\lambda\neq\mu$,
\item $S^2 M^\mathbf{A}_\lambda := \Res_{ S^2\H_r} S^2 M^\mathbf{A}_\lambda$, $\lambda\in\mathscr{P}_r$,
\item $\Wedge^2 M^\mathbf{A}_\lambda := \Res_{ S^2\H_r} \Wedge^2 M^\mathbf{A}_\lambda$, $\lambda\in\mathscr{P}'_r$.
\end{list}
Let $ M_{\{\lambda,\mu\}}$, $ S^2 M_\lambda$,
 $ \Wedge^2 M_\lambda$ denote the corresponding $\field S^2 \H_r$-modules.
\end{propdef}
\begin{proof}
This follows from a general result about the structure of  $S^2 H$ for  $H$ any split semisimple algebra over a field  $R$ of characteristic zero.  Such a result is proved in \cite[\textsection5]{LarsenRowell} in the case   $R = \CC$.  The proof goes by using the result for the case  $H = \End(V)$,  $V = R^{\oplus k}$ to deduce the general case.  And this special case follows from
\[S^2 (\End(V)) \cong S^2 (V^* \tsr V) \cong S^2 V^* \tsr S^2 V \oplus \Wedge^2 V^* \tsr \Wedge^2 V \cong \End(S^2 V) \oplus \End(\Wedge^2 V). \]
On the level of vector spaces, this is just the degree 2 part of Proposition \ref{pgelfanddecomp} (1) in the case  $V = V^*, \ W= V,\  q=1$.  One checks that this is in fact an isomorphism of algebras that holds over any field $R$ of characteristic zero.
\end{proof}

\subsection{Some representation theory of  $\nsH_r$}
\label{ss some representation theory of  nsH}
In this subsection, we give some flavor for the representation theory of $\nsH_r$. This is far from being a thorough treatment as our knowledge is limited outside the two-row case (see Theorem \ref{t nsH irreducibles two row case} for a complete answer in this case).
From computations we have done up to $r=6$, we suspect that most of the $\nsH_r$-irreducibles are restrictions of $(S^2 \H_r)^\theta$-irreducibles (see Proposition \ref{p theta on nsH}), except for the trivial and sign representations.

We have already defined the trivial and sign representations  $ \nsbr{\epsilon}_+$,  $\nsbr{\epsilon}_-$ of  $\nsH_r$.
In addition, there are the following six types of $\nsH_r$-modules:
\be  \label{e some nsH modules}
\parbox{14cm}{
\begin{list}{(\arabic{ctr})} {\usecounter{ctr} \setlength{\itemsep}{1pt} \setlength{\topsep}{2pt}}
\item $\nsbr{M}^\mathbf{A}_{\{\lambda, \mu\}} :=  \Res_{\nsH_r} M^\mathbf{A}_{\{\lambda,\mu\}},$
\item $S^2\nsbr{M}^\mathbf{A}_\lambda  :=  \Res_{\nsH_r} S^2 M^\mathbf{A}_\lambda, $
\item $\Wedge^2 \nsbr{M}^\mathbf{A}_\lambda  :=  \Res_{\nsH_r} \Wedge^2 M^\mathbf{A}_\lambda, $
\item $S' \nsbr{M}^\mathbf{A}_\lambda := \ker(\trace) \cap S^2 \nsbr{M}^\mathbf{A}_\lambda, $
\item $(S' \nsbr{M}^\mathbf{A}_\lambda)^\dualtheta, $
\item $(\Wedge^2 \nsbr{M}^\mathbf{A}_\lambda)^\dualtheta,$
\end{list}}
\ee
where the right-hand sides of the first three lines are restrictions of the  $S^2\H_r$-modules of Proposition \ref{p S2Hr representations}.
The last three modules will be explained below, after we discuss contragradients of  $\nsH_r$-modules. Let $ \nsbr{M}_{\{\lambda,\mu\}}$, $ S^2 \nsbr{M}_\lambda$, etc. denote the corresponding $\field \nsH_r$-modules.


Any anti-automorphism $S$ of an $\mathbf{A}$-algebra $H$ allows us to define contragradients of $H$-modules: let $\langle \cdot, \cdot \rangle: M \tsr M^* \to \mathbf{A}$ be the canonical pairing, where $M^*$ is the $\mathbf{A}$-module  $\hom_{\mathbf{A}}(M,\mathbf{A})$. Then the $H$-module structure on  $M^*$ is defined by
\[ \langle m,h m'\rangle = \langle S(h)m,m' \rangle \text{ for any } h \in H,\ m \in M, m' \in M^*.\]

Recall the anti-automorphisms $1^\text{op}$,  $\theta^\text{op}$ of  $\H_r$ defined in  \textsection\ref{ss definition of nsH}.
There are also anti-automorphisms $1^{\text{op}} := 1^{\text{op}} \tsrvw 1^{\text{op}}$ and $(\Theta)^\text{op} := 1^{\text{op}} \circ \Theta$ of $\nsH_r$, where $\Theta$ is defined in Proposition \ref{p theta on nsH} (iii).
For an $\H_r$-module $M$, we write $M^\dualone$ (resp.  $M^\dualtheta$) for the contragradient of $M$ corresponding to the anti-automorphism  $1^\text{op}$ (resp.  $\theta^\text{op}$).
For an $\nsH_r$-module $\nsbr{M}$, we also write $\nsbr{M}^\dualone$ (resp.  $\nsbr{M}^\dualtheta$) for the contragradient of $\nsbr{M}$ corresponding to the anti-automorphism  $1^\text{op}$ (resp.  $\Theta^\text{op}$).

\begin{proposition}[\cite{B4} (see also {\cite[Exercises 2.7, 3.14]{Mathas}})]
\label{p dual basis to C' S = 1 theta op}
The contragradients of the Specht module  $M^\mathbf{A}_\lambda$ of  $\H_r$ are given by
\[
\begin{array}{lcr}
 (M^\mathbf{A}_\lambda)^\dualone \cong M^\mathbf{A}_{\lambda} &  \text{and} & (M^\mathbf{A}_\lambda)^\dualtheta \cong M^\mathbf{A}_{\lambda'}.
\end{array}
\]
\end{proposition}

We now explain (4)--(6) of \eqref{e some nsH modules}.
Let $\mathbf{A} \xrightarrow{I} (M^\mathbf{A}_{\lambda})^\dualone \tsrvw M^\mathbf{A}_{\lambda}$
be the canonical inclusion given by sending  $1 \in \mathbf{A}$ to  $I \in \End(M^\mathbf{A}_\lambda) \cong (M^\mathbf{A}_{\lambda})^\dualone \tsrvw M^\mathbf{A}_{\lambda}$.
Let
$ M^\mathbf{A}_{\lambda} \tsrvw (M^\mathbf{A}_{\lambda})^\dualone \xrightarrow{\trace} \mathbf{A}$
be the canonical surjection.
It follows from Proposition \ref{p hecke algebra antipode} (see, e.g., \cite{B4}) and the  $\H_r$-module isomorphism $(M^\mathbf{A}_\lambda)^\dualone \cong M^\mathbf{A}_\lambda$ that there are the following $\nsH_r$-module homomorphisms
\begin{align*}
\nsbr{\epsilon}_+ &\xrightarrow{I} \nsbr{M}^\mathbf{A}_{\{\lambda,\lambda\}}, \\
\ker(\trace) &\hookrightarrow  \nsbr{M}^\mathbf{A}_{\{\lambda,\lambda\}} \xrightarrow{\trace} \nsbr{\epsilon}_+.
\end{align*}
Since $\frac{1}{|\text{SYT}(\lambda)|} I$ is a splitting of $\trace$, we obtain the decomposition
\be  \label{e nsH trace}
\ker(\trace) \oplus \nsbr{\epsilon}_+ \cong \nsbr{M}^\mathbf{A}_{\{\lambda,\lambda\}}
\ee
of $\nsH_r$-modules.

Define  $S' \nsbr{M}^\mathbf{A}_\lambda := \ker(\trace) \cap S^2 \nsbr{M}^\mathbf{A}_\lambda$.  The decomposition \eqref{e nsH trace} yields the decomposition
\be  \label{e nsH S' iso}
S^2 \nsbr{M}^\mathbf{A}_\lambda \cong S' \nsbr{M}^\mathbf{A}_\lambda \oplus \nsbr{\epsilon}_+.
\ee
Applying  $\dualtheta$ to \eqref{e nsH trace} and \eqref{e nsH S' iso} yields
\begin{align}
(\nsbr{M}^\mathbf{A}_{\{\lambda, \lambda\}})^\dualtheta &\cong (\ker(\trace))^\dualtheta \oplus \nsbr{\epsilon}_-, \label{e ns sign decomposition} \\
(S^2 \nsbr{M}^\mathbf{A}_\lambda)^\dualtheta &\cong (S' \nsbr{M}^\mathbf{A}_\lambda)^\dualtheta \oplus \nsbr{\epsilon}_-. \notag
\end{align}



\begin{proposition}
\label{p 1 op theta op representation facts}
There hold the following isomorphisms of $\nsH_r$-modules
\be \label{e nsH duals1}
\begin{array}{ccccc}
(\Res_{\nsH_r} M^\mathbf{A}_l \tsr M^\mathbf{A}_r)^\dualone &\cong & \Res_{\nsH_r} (M^\mathbf{A}_l)^\dualone \tsr (M^\mathbf{A}_r)^\dualone &\cong & \Res_{\nsH_r} (M^\mathbf{A}_l)^\dualtheta \tsr (M^\mathbf{A}_r)^\dualtheta, \\
(\Res_{\nsH_r} M^\mathbf{A}_l \tsr M^\mathbf{A}_r)^\dualtheta &\cong & \Res_{\nsH_r} (M^\mathbf{A}_l)^\dualone \tsr (M^\mathbf{A}_r)^\dualtheta &\cong & \Res_{\nsH_r} (M^\mathbf{A}_l)^\dualtheta \tsr (M^\mathbf{A}_r)^\dualone,
\end{array}
\ee
for any  $\H_r$-modules  $M^\mathbf{A}_l, M^\mathbf{A}_r$.
Hence the following  $\nsH_r$-modules are isomorphic
\begin{align}
\nsbr{\epsilon}_{\pm}^\dualone &\cong  \nsbr{\epsilon}_\pm, \label{e nsHdual iso1}\\
\nsbr{\epsilon}_\pm^\dualtheta &\cong \nsbr{\epsilon}_\mp, \\
 (\nsbr{M}^\mathbf{A}_{\{\lambda,\mu\}})^\dualone &\cong \nsbr{M}^\mathbf{A}_{\{\lambda,\mu\}}\cong  \nsbr{M}^\mathbf{A}_{\{\lambda',\mu'\}}, \\
 (\nsbr{M}^\mathbf{A}_{\{\lambda,\mu\}})^\dualtheta &\cong \nsbr{M}^\mathbf{A}_{\{\lambda,\mu'\}} \cong  \nsbr{M}^\mathbf{A}_{\{\lambda',\mu\}},\\
 (S^2\nsbr{M}^\mathbf{A}_{\lambda})^\dualone &\cong S^2\nsbr{M}^\mathbf{A}_{\lambda}\cong  S^2\nsbr{M}^\mathbf{A}_{\lambda'}, \label{e nsHdual iso5}\\
 (\Wedge^2 \nsbr{M}^\mathbf{A}_{\lambda})^\dualone &\cong \Wedge^2 \nsbr{M}^\mathbf{A}_{\lambda}\cong \Wedge^2 \nsbr{M}^\mathbf{A}_{\lambda'}, \label{e nsHdual iso6}\\
 (S'\nsbr{M}^\mathbf{A}_{\lambda})^\dualone &\cong S'\nsbr{M}^\mathbf{A}_{\lambda}\cong  S'\nsbr{M}^\mathbf{A}_{\lambda'}. \label{e nsHdual iso7}
\end{align}
\end{proposition}
\begin{proof}
The isomorphisms in \eqref{e nsH duals1} follow from  Proposition \ref{p theta on nsH} (iv) (see also \cite[Proposition 2.9]{Bnsbraid}). The isomorphisms \eqref{e nsHdual iso1}--\eqref{e nsHdual iso6} are straightforward from \eqref{e nsH duals1} and Proposition \ref{p dual basis to C' S = 1 theta op}. The isomorphisms in \eqref{e nsHdual iso7} follow from \eqref{e nsHdual iso5}, \eqref{e nsHdual iso1}, and \eqref{e nsH S' iso}.
\end{proof}

\subsection{The sign representation in the canonical basis}
For future reference, we record the right action of $\frac{\sQ_s}{[2]^2}$ on  $\H_r \tsrvw \H_r$ in terms of the basis $\Gamma_{\S_r} \tsrvw \Gamma_{\S_r}$.
\refstepcounter{equation}
\begin{align*}
&\quad C_v \tsrvw C_w \frac{\sQ_s}{[2]^2}= \tag{\theequation}\label{e sQ on CC}\\
&\begin{cases}
0 & \text{if } s \in R(v), s \in R(w), \\
C_v \tsrvw C_w  + \frac{1}{[2]} \displaystyle \sum_{s \in R(w')} \mu(w', w) C_{v} \tsrvw C_{w'} & \text {if } s \in R(v) \text{ and } s \notin R(w), \\
C_v \tsrvw C_w + \frac{1}{[2]} \displaystyle \sum_{s \in R(v')} \mu(v', v) C_{v'} \tsrvw C_w & \text {if } s \notin R(v) \text{ and } s \in R(w), \\
-\frac{1}{[2]} \Big( \displaystyle \sum_{s \in R(v')} \mu(v', v) C_{v'} \tsrvw C_w + \sum_{s \in R(w')} \mu(w', w) C_{v} \tsrvw C_{w'} \Big) +\\
-\frac{2}{[2]^2} \displaystyle \sum_{s \in R(v'), s \in R(w')} \mu(v', v) \mu(w', w) C_{v'} \tsrvw C_{w'} & \text{if } s \notin R(w) \text{ and } s \notin R(v). \\
\end{cases}
\end{align*}
This is immediate from \eqref{e C on canbas}.  This also gives the action on any cellular subquotient  $\mathbf{A} \Gamma$ of $\mathbf{A}\Gamma_{\S_r} \tsrvw \Gamma_{\S_r}$ by restricting $\mu$ to  $\Gamma$.

The next proposition expresses the decomposition \eqref{e ns sign decomposition} in terms of canonical bases.  This will allow us to construct a canonical basis for $\nswedgealgebra{}$ explicitly as a subset of  $T(\nsbr{X})$ in \textsection\ref{ss nonstandard columns labeled a canonical basis for Lambda r X}.

\begin{proposition}[\cite{B4}]
\label{p sign in lambda tsr lambda'}
The sign representation $\field\nsbr{\epsilon}_-$ of $\field\nsH_r$ occurs with multiplicity one in $M_\lambda \tsrvw M_{\lambda'}$ and not at all in $M_\lambda \tsrvw M_\mu$, $\mu \neq \lambda'$. Moreover, the inclusion $\nsbr{i}_-: \field\nsbr{\epsilon}_- \hookrightarrow M_\lambda \tsrvw M_{\lambda'}$ can be expressed in terms of the basis $\Gamma_\lambda \tsrvw \Gamma_{\lambda'}$ as
\[
1 \mapsto \sum_{Q \in \text{SYT}(\lambda)} (-1)^{\ell(\transpose{Q})} C_Q \tsrvw C_{\transpose{Q}},
\]
and the surjection $\nsbr{s}_- : M_\lambda \tsrvw M_{\lambda'} \twoheadrightarrow \field\nsbr{\epsilon}_-$ by
\[
\sum_{Q', Q \in \text{SYT}(\lambda)} a^{Q'\transpose{Q}}C_{Q'} \tsrvw C_{\transpose{Q}} \mapsto \frac{1}{|\text{SYT}(\lambda)|} \sum_{Q \in \text{SYT}(\lambda)} (-1)^{\ell(\transpose {Q})}a^{Q\transpose{Q}} \ \ (a^{Q'\transpose{Q}} \in \field),
\]
where $\ell(\transpose{Q})$ is as in \textsection\ref{ss type A combinatorics preliminaries}.
\end{proposition}

\subsection{The algebra $\nsH_3$} \label{salgb3}
In \cite{GCT4}, the example $\nsH_3$ is described in detail.  This example is generalized in \cite{Bnsbraid}, and we follow this reference to recall some of the main results.

Let us first describe the irreducible representations of $\field \nsH_3$.
\begin{proposition}
\label{p nsH S3 2D reps}
For $\field = \QQ(\u),$ the irreducible representations of $\field \nsH_3$ consist of the trivial and sign representations
\[ \nsbr{\epsilon}_+, \ \nsbr{\epsilon}_-, \]
and the two two-dimensional representations
\[\nsbr{M}_{\{(3),(2,1)\}},\ \ S'\nsbr{M}_{(2,1)}.\]
\end{proposition}
This is shown in \cite{GCT4} and is a special case of \cite[Theorem 3.4]{Bnsbraid}.
Note that the two-dimensional representations both specialize to the representation $M_{(2,1)}|_{q=1}$  of $\QQ\S_3$ at $q=1$. Thus the algebra $\field\nsH_3$ is similar to $\field\H_3$ except with two two-dimensional representations instead of one. Let us now see how these two-dimensional representations differ.

These two-dimensional representations are both of the form $\nsbr{N}(a)$, where $\nsbr{N}(a) \cong \field^{\oplus 2}$, $a \in \field$,  is the representation of  $\nsH_3$ determined by the following matrices giving the action of $\sP_i$ on $\field^{\oplus 2}$:
\be
\label{e bar N action}
\sP_1 \mapsto \mat{[2]^{2} & 0 \\ a & 0},\ \sP_2 \mapsto \mat{0 & a \\ 0 & [2]^{2} }.
\ee
Here we have specified a basis $(e_1, e_2)$ for  $\nsbr{N}(a)$ and are thinking of matrices as acting on the right on row vectors, so that the $j$-th row of these matrices gives the coefficients of $e_j \sP_i$ in the basis $(e_1, e_2)$. Those $a$ for which  $ \nsbr{N}(a)$ defines a representation of $\field \nsH_3$ that is irreducible are
\[a_1=[2],\ a_2=[2]^2-2,\]
and
\[\nsbr{M}_{\{(3),(2,1)\}}=\nsbr{N}(a_1),\ \ S'\nsbr{M}_{(2,1)}=\nsbr{N}(a_2).\]

\begin{remark}
In view of the generalization of the nonstandard Hecke algebra in \cite{Bnsbraid}, the constants $a_1, a_2$ have a nice explanation: $a_1, a_2$ is the beginning of the sequence $T_1(\frac{1}{[2]}), T_2(\frac{1}{[2]}),\ldots$ (up to a certain normalization by factors of $\pm[2]^j$), where the $T_k(x)$ are the Chebyshev polynomials of the first kind.
\end{remark}

The algebra $\nsH_3$ has a nice presentation using these coefficients.
\begin{theorem}[\cite{GCT4},\cite{Bnsbraid}]
\label{t S3 coprod relations}
The algebra $\nsH_3$ is the associative $\mathbf{A}$-algebra generated by $\sP_s, s \in S=\{s_1,s_2\},$ with quadratic relations
\be
\label{e nsH quadratic relation}
(\sP_s)^2 = [2]^2 \sP_s, \ \ \ s \in S,
\ee
and nonstandard braid relation
\be
\label{e nsH braid relation}
\sP_1 (\sP_{21}-a_1^2)(\sP_{21}-a_2^2) = \sP_2 (\sP_{12}-a_1^2)(\sP_{12}-a_2^2).
\ee
Moreover, $\nsH_3$ is free as an $\mathbf{A}$-module.
\end{theorem}

\begin{remark}\label{r no braid relation}
Since $\sP_1,\sP_2$ do not satisfy the relation
\[
\C_1\C_2\C_1-\C_1=\C_2\C_1\C_2-\C_2
\]
satisfied by $\C_1,\C_2$ (which is equivalent to the braid relation for $T_1,T_2$), the  $\R$-matrices $\mathcal{R}_1, \mathcal{R}_2$ (in the notation of Remark \ref{r no braid relation0}) do not satisfy the quantum Yang-Baxter equation. Hence the Hopf algebra $\O(GL_q(\nsbr{X}))$ is not coquasitriangular.
\end{remark}


\subsection{A canonical basis of $\nsH_3$}
\label{scanonical}
We now  ask if $\nsH_r$ has a canonical basis $\nsbr{\Cbasis}^r$ akin to the
Kazhdan-Lusztig basis of $\H_r$.
There are two properties we would like such a basis to satisfy.

The first is that  for each  $J \subseteq S$, the nonstandard right (resp. left) $J$-descent space is spanned by a subset of $ \nsbr{\Cbasis}^r$, where
the  \emph{nonstandard  right (resp. left) $J$-descent space} of  $\nsH_r$ is defined to be
\begin{align*}
& \{h \in \nsH_r: h \sQ_{s} = [2]^2 h \text{ for all } s\in J\}, \\
(\text{resp. } & \{h \in \nsH_r: \sQ_{s} h = [2]^2 h \text{ for all } s\in J\}).
\end{align*}

The second property is that  $\nsbr{\Cbasis}^r$ be a cellular basis in the sense of Graham-Lehrer \cite{GrahamLehrer}.
Graham and Lehrer's theory of cellular algebras \cite{GrahamLehrer} was in fact made to abstract and generalize some of the nice properties
satisfied by the Kazhdan-Lusztig basis of $\H_r$.
We now review the definition of a cellular basis and recall the cellular basis of $R \nsH_3$ (for suitable $R$) from \cite{GCT4,Bnsbraid}.


Let $H$ be an algebra over a commutative ring $R$.
\begin{definition}
\label{d cellular algebra}
Suppose that $(\Lambda, \geq)$ is a (finite) poset and that for each $\lambda \in \Lambda$ there is a finite indexing set $\Tab(\lambda)$ and distinct elements $C^\lambda_{ST} \in H$ for all $S,T \in \Tab(\lambda)$ such that
\[ \Cbasis = \{ C^\lambda_{S T} : \lambda \in \Lambda \text{ and } S,T \in \Tab(\lambda) \} \]
is a (free)  $R$-basis of $H$. For $\lambda \in \Lambda$, let $H_{< \lambda}$ be the $R$-submodule of $H$ with basis $\{ C^\mu_{S T} : \mu < \lambda \text{ and } S,T \in \Tab(\mu)\}$.

The triple $(\Cbasis, \Lambda, \Tab)$ is a \emph{cellular basis} of $H$ if
\begin{list}{(\roman{ctr})} {\usecounter{ctr} \setlength{\itemsep}{1pt} \setlength{\topsep}{2pt}}
\item the $R$-linear map $* : H \to H$ determined by $(C^{\lambda}_{S T})^* = C^\lambda_{T S}$, for all $\lambda \in \Lambda$ and all $S$ and $T$ in $\Tab(\lambda)$, is an algebra anti-isomorphism of $H$,
\item for any $\lambda \in \Lambda$ and $h \in H$ there exist $r_{S',S} \in R$, for  $S',S \in \Tab(\lambda)$, such that for all $T \in \Tab(\lambda)$
\[ h C^\lambda_{S T} \equiv \sum_{S' \in \Tab(\lambda)} r_{S', S} C^\lambda_{S' T} \mod H_{< \lambda}. \]
\end{list}
\end{definition}
For each  $\lambda \in \Lambda$, the \emph{cell representation} corresponding to  $\lambda$ is the left $H$-module that is the submodule of $H_{\leq \lambda} / H_{<\lambda}$ with $R$-basis $\{ C^{\lambda}_{ST} : S \in \Tab (\lambda) \}$ for some $T \in \Tab (\lambda)$; this basis is independent of $T$.

Let $R$ be a commutative ring with a map $\mathbf{A} \to R$ such that the images of $a_1$ and  $a_2$ are invertible.
There are several cellular bases for  $R \nsH_3$.  We define one such cellular basis $(\nsbr{\Cbasis}^3, \nsbr{\Lambda}, \nsbr{\Tab})$ for the poset $\nsbr{\Lambda}$ given by
\be
\label{e lambda poset definition}
\xymatrix@R=10pt@C=8pt{
\nsbr{\epsilon}_{+} \ar @{-} [d]\\
\nsbr{N}(a_1) \ar @{-} [d]\\
\nsbr{N}(a_2) \ar @{-} [d]\\
\nsbr{\epsilon}_{-}
}
\ee
Set  $\nsbr{\Tab}(\nsbr{\epsilon}_\pm) = \{\pm\}$,  $\nsbr{\Tab}(\nsbr{N}(a_i)) = \{1,2\}$.
The basis $\nsbr{\Cbasis}^3$ consists of
\be
\label{e canonical basis definition m odd}
\begin{array}{l@{\ \ :=  \ \ }l}
C^{\nsbr{\epsilon}_+} &  1, \\
C^{\nsbr{N}(a_1)}_{1 1} & a_1 \sQ_1, \\
C^{\nsbr{N}(a_1)}_{2 1} & \sQ_{21}, \\
C^{\nsbr{N}(a_1)}_{2 2} & a_1 \sQ_2, \\
C^{\nsbr{N}(a_1)}_{1 2} & \sQ_{12}, \\
C^{\nsbr{N}(a_2)}_{1 1} & a_2 \sQ_1 (\sQ_{21} - a_1^{2}), \\
C^{\nsbr{N}(a_2)}_{2 1} & \sQ_{21} (\sQ_{21} - a_1^{2}), \\
C^{\nsbr{N}(a_2)}_{2 2} & a_2 \sQ_2 (\sQ_{12} - a_1^{2}), \\
C^{\nsbr{N}(a_2)}_{1 2} & \sQ_{12} (\sQ_{12} - a_1^{2}), \\
C^{\nsbr{\epsilon}_-} & \sQ_1 (\sQ_{21}-a_1^2)(\sQ_{21}-a_2^2).
\end{array}
\ee
It is is not hard to check, given Theorem \ref{t S3 coprod relations},  that  $(\nsbr{\Cbasis}^3, \nsbr{\Lambda}, \nsbr{\Tab})$ is a cellular basis; see \cite[Proposition 5.3]{Bnsbraid} for a careful proof.  In the case  $R = \field$, the radical \cite[Definition 3.1]{GrahamLehrer}  of each cell representation is 0.  Thus we recover the four absolutely irreducible $\field \nsH_3$-modules of Proposition \ref{p nsH S3 2D reps} as cell representations.

\begin{remark}
It is shown in the older version of this paper \cite{GCT4} that a similar basis $B$ of  $\nsH_3$ has the following positivity property: the coefficients of the expansion of any $b \in B$ in terms of the basis $\{C_v \tsrvw C_w : v,w \in \S_3\}$
are  $\br{\cdot}$-invariant Laurent polynomials in $q$ with nonnegative integer coefficients.
We are uncertain if a similar, perhaps slightly weaker, form of positivity holds for $\nsH_4$ and beyond.
In Example \ref{ex ns Schur-Weyl duality canonical basis r4}, we construct a nice basis of each $\nsH_4$-irreducible.  The coefficients for the action of  $\sQ_i$ on these bases
are $\br{\cdot}$-invariant Laurent polynomials in $q$ and, after factoring out powers of $\u-\ui$, have all nonnegative or all nonpositive integer coefficients.
Similarly, Theorem \ref{t positivity} and Example \ref{ex not positive} show that a weak form of positivity holds for the basis  $\pNSTC(\nu)$ of  $\nsbr{X}_\nu$, but that this cannot be strengthened.
\end{remark}

\subsection{The algebra $\nsH_4$} \label{sb4}
The algebra $\nsH_4 $ turns out to be  considerably more complicated
and is of dimension $114$.
The irreducible representations of $\field \nsH_4$ are
\[
\field\nsbr{\epsilon}_-,\,
S'\nsbr{M}_{(2,2)},\,
\nsbr{\Wedge}^2\nsbr{M}_{(3,1)},\,
\nsbr{M}_{\{(3,1),(2,2)\}},\,
S'\nsbr{M}_{(3,1)},\,
(S'\nsbr{M}_{(3,1)})^\dualtheta,\,
\nsbr{M}_{\{(4),(2,2)\}},\,
\nsbr{M}_{\{(4),(3,1)\}},\,
\field\nsbr{\epsilon}_+.
\]
The dimension count corresponding to expressing $\field \nsH_4$ as the sum of its minimal two-sided ideals is
\[
1^2+2^2+3^2+6^2+5^2+5^2+2^2+3^2+1^2=114.
\]

We have not been able to determine a presentation for $\nsH_4$ akin to the presentation \eqref{e Hecke algebra def} for the
Hecke algebra or the presentation of  $\nsH_3$ from Theorem \ref{t S3 coprod relations}.

The ideal of relations expressing  $\nsH_4$ as a quotient of the free  $\mathbf{A}$-algebra in the $\sP_i$'s is not generated by the quadratic relations \eqref{e nsH quadratic relation} and the nonstandard braid relations \eqref{e nsH braid relation} for the parabolic subalgebras $\langle \sP_i,\sP_{i+1}\rangle \subseteq \nsH_4$.
We determined by computer the ideal of relations by a simple procedure of generating
monomials in the  $\sP_i$'s systematically and of increasing degree while retaining only those
which were not linear combinations of earlier monomials. The top degree
obtained thus was $9$. In other words, every monomial of degree $10$ and
above is a linear combination of some smaller monomials. However, these
linear combinations seem fairly complicated.
To give an idea of the difficulties involved,
the simplest  relation among the generators of $\nsH_4$ that cannot be deduced
from the quadratic and nonstandard braid relations
is a linear combination of $74$ monomials of degrees $\leq 7$;
it is reported in the older version of this paper \cite{GCT4}.


\section{Nonstandard Schur-Weyl duality}\label{s Nonstandard Schur-Weyl duality}
We prove a nonstandard analog of quantum Schur-Weyl duality for the coaction of  $\O(M_q(\nsbr{X}))$ and action of $\nsH_r$ on $\nsbr{X}^{\tsr r}$.
This allows us to relate the representation theory of $\nsH_r$ to the corepresentation theory of $ \O(M_q(\nsbr{X}))$.  We illustrate this for the two-row,  $r=3$ case.

\subsection{Nonstandard Schur-Weyl duality}
As explained in \textsection\ref{ss commuting action on Vtsrr}, $\H_r$ acts  on $V^{\tsr r}$ on the right by sending  $T_i$
to the endomorphism of  $V^{\tsr r}$ given by $\hat{\R}_{V,V}$ acting on the  $i$ and  $i+1$-st tensor factors.
Let  $\field\mathscr{S}(V,r)$ be the \emph{$q$-Schur algebra} over  $\field$, which is defined to be the endomorphism algebra $\End_{\field \H_r}(V^{\tsr r})$.  It is known that this is also equal to the algebra dual to the coalgebra $\O(M_q(V))_r$.
By definition, the right action of  $\H_r$ on  $V^{\tsr r}$ commutes with the action of
$\field \mathscr{S}(V,r)$ on  $V^{\tsr r}$, hence  $V^{\tsr r}$ is a $(\field \mathscr{S}(V,r), \field \H_r)$-bimodule.
Quantum Schur-Weyl duality now takes the same form
$V^{\tsr r} \cong \bigoplus_{\lambda}  V_\lambda \tsr M_\lambda$
as Theorem \ref{c Schur-Weyl duality basic}, except this is considered as an isomorphism of $(\field \mathscr{S}(V,r), \field \H_r)$-bimodules rather of $(U_q(\g_V), \field \H_r)$-bimodules.

Next, consider the commuting actions of $\H_r \tsrvw \H_r$ and $\field\mathscr{S}(V,r) \tsrvw \field\mathscr{S}(W,r)$ on  $X^{\tsr r}$.
The nonstandard Hecke algebra $\nsH_r$ acts on $\nsbr{X}^{\tsr r}$ by sending $\sP_i$ to the endomorphism of  $\nsbr{X}^{\tsr r}$ given by
$[2]^2 P_+^{\nsbr{X}}$ acting on the  $i$ and  $i+1$-st tensor factors.
This is a well-defined action of  $\nsH_r$ on  $\nsbr{X}^{\tsr r}$ by the definition of the action of $\H_r$ on  $V^{\tsr r}$ and the similar forms of
$\sP_i$ \eqref{e sP definition} and  $P_+^{\nsbr{X}}$ \eqref{eqPminusandplus}.
Define the \emph{nonstandard Schur algebra}, denoted $\field \nsSchur{r}$, to be the algebra dual to the coalgebra $\O(M_q(\nsbr{X}))_r$.
By the definition of $\O(M_q(\nsbr{X}))$, the action of $\nsH_r$ on  $\nsbr{X}^{\tsr r}$ commutes with the action of $\field \nsSchur{r}$---this will be shown carefully in the proof below.

We have the following nonstandard analog of quantum Schur-Weyl duality:
\begin{theorem} \label{t nonstandard schur-weyl duality}
As a $(\field \nsSchur{r}, \field \nsH_r)$-bimodule,  $ \nsbr{X}^{\tsr r}$ decomposes into irreducibles as
\be \label{e nonstandard Schur-Weyl duality}
\nsbr{X}^{\tsr r} \cong \bigoplus_{\alpha \in \nsP_r} \nsbr{\X}_\alpha \tsr \nsbr{M}_\alpha,
\ee
where $\nsP_r$ is an index set so that $\nsbr{\X}_\alpha$ ranges over irreducible $\O(M_q(\nsbr{X}))_r$-comodules (which are the same as irreducible $\field \nsSchur{r}$-modules) and $\nsbr{M}_\alpha$ ranges over $\field \nsH_r$-irreducibles.
\end{theorem}
\begin{remark}
This theorem should use $\nsbr{X}^*$ in place of $\nsbr{X}$ to be consistent with our conventions in \textsection\ref{s nonstandard coordinate algebra}--\ref{s nonstandard quantum groups GLq} (see Remark \ref{r dual convention}). The proof uses $\nsbr{X}^*$ in place of $\nsbr{X}$ and pays careful attention to duals.  The left $\field \nsSchur{r}$-module $(\nsbr{X}^*)^{\tsr r}$ gives rise to an injection $\field \nsSchur{r} \hookrightarrow \End(\nsbr{X}^{\tsr r})$ and this is our starting point in the proof below.
\end{remark}
\begin{proof}
Let $\End_{\field \nsH_r}(\nsbr{X}^{\tsr r})$ be the algebra of endomorphisms of  $\nsbr{X}^{\tsr r}$ intertwining the action of $\field \nsH_r$.  By well-known algebraic generalities used to prove classical Schur-Weyl duality (see, e.g., \cite[Lemma 6.22]{fultonrepr}; the algebras in this lemma are over  $\CC$, but it is not hard to check that it extends to the present setting) and the semisimplicity of $\field \nsH_r$, it suffices to show that
\[
\field \nsSchur{r} = \End_{\field \nsH_r}(\nsbr{X}^{\tsr r}).
\]
Clearly,  $f \in \End(\nsbr{X}^{\tsr r})$ lies in $\End_{\field \nsH_r}(\nsbr{X}^{\tsr r})$ if and only if  $f$ commutes with each  $\frac{\sP_i}{[2]^2}$.  Note that, in general, an endomorphism $g$ commutes with a projector  $p$ if and only if  $g(\im(p)) \subseteq \im(p)$ and  $g(\ker(p)) \subseteq \ker(p)$.  Hence  $f$ commutes with $\frac{\sP_i}{[2]^2}$ if and only if
\begin{align*}
f\big(\nsbr{X}^{\tsr i-1} \tsr \nssym{2}{X} \tsr \nsbr{X}^{\tsr r-i-1} \big) &\subseteq \nsbr{X}^{\tsr i-1} \tsr \nssym{2}{X} \tsr \nsbr{X}^{\tsr r-i-1}, \ \text{and} \\
f\big(\nsbr{X}^{\tsr i-1} \tsr \nswedge{2}{X} \tsr \nsbr{X}^{\tsr r-i-1} \big) &\subseteq \nsbr{X}^{\tsr i-1} \tsr \nswedge{2}{X} \tsr \nsbr{X}^{\tsr r-i-1}.
\end{align*}
This is equivalent to
\[f \in (\nsbr{Z}^*)^{\tsr i-1} \tsr\big(\nssym{2}{X}^*\tsrdual\nssym{2}{X}\oplus \nswedge{2}{X}^* \tsrdual \nswedge{2}{X}\big) \tsr (\nsbr{Z}^*)^{\tsr r-i-1},\]
where  $ \nsbr{Z} = \nsbr{X} \tsr \nsbr{X}^*$.
Regarding  $f$ as an element of $\hom(\nsbr{Z}^{\tsr r}, \field)$, this is equivalent to  $f$ vanishing on
\[\nsbr{\mathcal{I}}_{r,i}:= \nsbr{Z}^{\tsr i-1} \tsr\big(\nssym{2}{X}\tsrdual\nswedge{2}{X}^*\oplus \nswedge{2}{X} \tsrdual \nssym{2}{X}^*\big) \tsr \nsbr{Z}^{\tsr r-i-1}.\]

Since by \eqref{eqdefnewquan no basis}, $\O(M_q(\nsbr{X}))_r$ is the quotient of $\nsbr{Z}^{\tsr r}$ by  $\sum_{i =1}^{r-1} \nsbr{\mathcal{I}}_{r,i}$,  $f \in \field \nsSchur{r}$ if and only if   $f$ vanishes on $\sum_{i =1}^{r-1} \nsbr{\mathcal{I}}_{r,i}$ if and only if $f$ commutes with each $\frac{\sP_i}{[2]^2}$.
\end{proof}

\subsection{Consequences for the corepresentation theory of  $\O(M_q(\nsbr{X}))$}
As a corollary to nonstandard Schur-Weyl duality, we obtain a result similar to Theorem \ref{tcmqg} (d) (but without the assumption that the field is  $\CC$).
\begin{corollary}
As a coalgebra over  $\field = \QQ(\u)$,  $\O(M_q(\nsbr{X}))$ is cosemisimple.
\end{corollary}

Note that, just as in the classical case, nonstandard Schur-Weyl duality allows us to describe the irreducible  $\O(M_q(\nsbr{X}))$-comodule $\nsbr{\X}_\alpha$ if a primitive idempotent  $\nsbr{e}_\alpha$
corresponding to $\nsbr{M}_\alpha$ is known: $\nsbr{\X}_\alpha$ is equal to $\nsbr{X}^{\tsr r} \nsbr{e}_\alpha$.
Computing such idempotents explicitly is difficult, but there is a similar and easier way to relate the corepresentation theory of  $ \O(M_q(\nsbr{X}))$ to the representation theory of $\nsH_r$:
combining standard quantum Schur-Weyl duality applied to  $V^{\tsr r}$ and  $W^{\tsr r}$ with \eqref{e nonstandard Schur-Weyl duality} yields
\be
\bigoplus_{\alpha \in \nsP_r} \nsbr{\X}_\alpha \tsr \nsbr{M}_\alpha \cong \bigoplus_{\lambda \vdash_\dv r,\ \mu \vdash_\dw r}  V_\lambda \tsrvw W_\mu \tsr M_\lambda \tsrvw M_\mu.
\ee
Viewing this as an isomorphism of $(\field\mathscr{S}(V,r) \tsrvw \field\mathscr{S}(W,r), \nsH_r)$-bimodules shows that there are nonnegative integers $n^{\lambda,\mu}_\alpha= n_{\lambda,\mu}^\alpha$ that correspond to the multiplicities in the following two decomposition problems:
\be \label{e n lambda mu multiplicities}
\nsbr{\X}_\alpha \cong \bigoplus_{\lambda, \mu}(V_\lambda \tsrvw W_\mu)^{\oplus n^{\lambda,\mu}_\alpha}, \quad  \Res_{\nsH_r} M_\lambda \tsrvw M_\mu \cong \bigoplus_{\alpha} \nsbr{M}_\alpha^{\oplus n_{\lambda,\mu}^\alpha}.
\ee

Applying \eqref{e n lambda mu multiplicities} to $\alpha$ corresponding to the trivial or sign representation of $\nsH_r$ together with Proposition \ref{p sign in lambda tsr lambda'} (and the analogous statement for $\nsbr{\epsilon}_+$) yields
\begin{corollary}
The $\O(M_q(\nsbr{X}))$-comodules $\nswedge{r}{X}$ and  $\nssym{r}{X}$ are irreducible and decompose into irreducible $\O(M_q(V))\tsrvw  \O(M_q(W))$-comodules as in Proposition \ref{pgelfanddecomp}.
\end{corollary}
This was already known from Lemma \ref{llrdet2} and Proposition \ref{pgelfanddecomp}, but this gives another proof.

\subsection{The two-row, $r=3$ case} \label{sex3}
In  the next section, we will give a complete description of nonstandard Schur-Weyl duality in the two-row case, but for now let us give some feel for the result by working out the  two-row ($\dv = \dw =2$),  $r=3$ case.  This will make use of our understanding of  $\nsH_3$ from  \textsection\ref{salgb3}.

With the given assumptions, \eqref{e nonstandard Schur-Weyl duality} takes the form
\be \label{e ns Schur-Weyl r3}
\nsbr{X}^{\tsr 3}\cong \nssym{3}{X}\dualcnv \tsr \field \nsbr{\epsilon}_+ \oplus \nsbr{\X}_{\{(3), (2,1)\}} \tsr \nsbr{M}_{\{(3), (2,1)\}} \oplus \nsbr{\X}_{+(2,1)} \tsr S' \nsbr{M}_{(2,1)} \oplus \nswedge{3}{X}\dualcnv \tsr \field \nsbr{\epsilon}_-.
\ee
The $\O(M_q(\nsbr{X}))$-comodules  $\nsbr{\X}_{\{(3),(2,1)\}}, \ \nsbr{\X}_{+(2,1)}$ have not yet been defined, but we know from \eqref{e n lambda mu multiplicities}, and the fact that $\Res_{\nsH_3} M_{(2,1)} \tsrvw M_{(2,1)} \cong \nsbr{\epsilon}_+ \oplus S' \nsbr{M}_{(2,1)} \oplus \nsbr{\epsilon}_-$, that
they decompose into irreducible $\O(M_q(V))\tsrvw  \O(M_q(W))$-comodules as
\be\label{e ns comodules r=3}
\begin{array}{lll}
\nssym{3}{X}\dualcnv&\cong& V_{(3)} \tsrvw W_{(3)} \oplus V_{(2,1)}\tsrvw W_{(2,1)}, \\[1.4mm]
\nsbr{\X}_{\{(3),(2,1)\}}&\cong& V_{(3)}\tsrvw W_{(2,1)} \oplus V_{(2,1)} \tsrvw W_{(3)} \\[1.4mm]
\nsbr{\X}_{+(2,1)} &\cong& V_{(2,1)} \tsrvw W_{(2,1)}, \\[1.4mm]
\nswedge{3}{X}\dualcnv &\cong& V_{(2,1)} \tsrvw W_{(2,1)}.
\end{array}
\ee
(The first and last lines are already known from Proposition \ref{pgelfanddecomp}.)

As a check, the dimension count for \eqref{e ns Schur-Weyl r3} is
\[
4^3 = 20\cdot1+16\cdot2+4\cdot2+4\cdot1.
\]

See Example \ref{ex ns Schur-Weyl duality canonical basis r3} for a nice basis of  $\nsbr{X}^{\tsr 3}$ that realizes nonstandard Schur-Weyl duality in this case.

We can also compare this to standard quantum Schur-Weyl duality for  $X^{\tsr 3}$ as a  $\O(GL_q(X))$-comodule (this has the same form as Schur-Weyl duality between $GL(X)$ and  $\S_3$):
\[ X^{\tsr 3} \cong X_{(3)} \tsr M_{(3)} \oplus X_{(2,1)} \tsr M_{(2,1)} \oplus X_{(1,1,1)} \tsr M_{(1,1,1)},
\]
where  $X_\lambda$ denotes the $\O(GL_q(X))$-comodule of highest weight $\lambda$.
The dimension count here is
\[ 4^3 = 20 \cdot 1 + 20 \cdot 2 + 4 \cdot 1.\]
By comparing this to Schur-Weyl duality for $V^{\tsr 3}$ and $W^{\tsr 3}$ and setting $q=1$ (we are just computing Kronecker coefficients for partitions of size 3) we obtain the left-hand isomorphisms below.

\[
\begin{array}{r@{\,\cong\,}l@{\,\cong\,}l}
X_{(3)}|_{q=1} & \left(V_{(3)} \tsrvw W_{(3)} \oplus V_{(2,1)}\tsrvw W_{(2,1)}\right)|_{q=1} & \nssym{3}{X}\dualcnv|_{q=1}, \\[1.4mm]
X_{(2,1)}|_{q=1} &   \left(V_{(3)}\tsrvw W_{(2,1)} \oplus V_{(2,1)} \tsrvw W_{(3)} \oplus V_{(2,1)} \tsrvw W_{(2,1)}\right)|_{q=1} & \nsbr{\X}_{\{(3),(2,1)\}}|_{q=1} \oplus \nsbr{\X}_{+(2,1)}|_{q=1}, \\[1.4mm]
X_{(1,1,1)}|_{q=1} & \left(V_{(2,1)} \tsrvw W_{(2,1)}\right)|_{q=1} &  \nswedge{3}{X}\dualcnv|_{q=1}.
\end{array}
\]
The right-hand isomorphisms are by \eqref{e ns comodules r=3}, all objects here being thought of as $\O(GL(V))\tsrvw \O(GL(W))$-comodules. Thus $\nsbr{\X}_{\{(3),(2,1)\}} \oplus \nsbr{\X}_{+(2,1)}$ is some quantization of $X_{(2,1)}|_{q=1}$, which we believe to be a better quantization than $X_{(2,1)}$ for the Kronecker problem.

We can also use the knowledge just gained about  $\O(M_q(\nsbr{X}))_3$-comodules to understand
the Peter-Weyl theorem for $\O(M_q(\nsbr{X}))$ (Theorem \ref{tcmqg} (d)) and  explain the dimension count
$\dim(\O(M_q(\nsbr{X}))_3)=688$.
By \eqref{e ns comodules r=3} and Theorem \ref{tcmqg} (d),
\[ \O(M_q(\nsbr{X}))_3 \cong \nssym{3}{X} \tsr \nssym{3}{X}^* \oplus (\nsbr{\X}_{\{(3),(2,1)\}})_L \tsr \nsbr{\X}_{\{(3),(2,1)\}}\oplus (\nsbr{\X}_{+(2,1)})_L \tsr \nsbr{\X}_{+(2,1)}\oplus \nswedge{3}{X} \tsr \nswedge{3}{X}^*, \]
with corresponding dimensions $688=20^2+16^2+4^2+4^2$.  Compare this to the dimension count
$\dim(\O(M_q(X))_3)= 816= 20^2+20^2+4^2$ for the standard Peter-Weyl theorem.

\section{Nonstandard representation theory in the two-row case}\label{s Nonstandard representation theory in the two-row case}
It turns out that in the two-row ($\dv=\dw =2$) case, the nonstandard Hecke algebra $\nsH_r$ is quite close to  $S^2 \H_r$ (by the ``two-row case'' of  $\nsH_r$, we mean the nonstandard Temperley-Lieb quotient $\nsH_{r,2}$, defined in  \textsection\ref{ss nonstandard two-row case}) and the nonstandard coordinate algebra  $ \O(GL_q(\nsbr{X}))$ is close to
the smash coproduct $\Oqt := \O(GL_q(V))\tsrvw\O(GL_q(W)) \rtimes \F(\S_2)$ (as defined in Appendix \ref{s appendix Oqt}).  To develop a theory of crystal bases for $\O(GL_q(\nsbr{X}))$ and its comodules, we would prefer to work with a ``nonstandard enveloping algebra'' that is Hopf dual to $\O(GL_q(\nsbr{X}))$. However, we have not been able to construct such a nonstandard enveloping algebra explicitly.  For the two-row case however, the approximation $\Uqt := \Uqvw \rtimes \S_2$ has proved to be close enough, and this is the object we work with in the detailed study of two-row Kronecker in  \textsection\ref{s global crystal basis for two-row Kronecker coefficients}--\ref{s A Kronecker graphical calculus}.

Throughout this section, assume $\dv=\dw$.  We discuss the (co)representation theory of  $\Uqt$ and  $\Oqt$ and a Schur-Weyl duality between these and  $S^2 \H_r$.  A complete description of nonstandard Schur-Weyl duality between $\nsH_{r,2}$ and $\O(M_q(\nsbr{X}))$ is then given in \textsection\ref{ss nonstandard two-row case}.  We also extend the theory of upper based  $\Uqvw$-modules to  $\Uqt$-modules (\textsection\ref{ss upper based Uqt}).

\subsection{The Hopf algebra  $\Uqt$}
Define  $\Uqt$ to be the wreath product $U_q(\gl_\dv) \wr \S_2$, also equal to $\Uqvw \rtimes \S_2$ since we are assuming $\dv =\dw$.  Explicitly,  $\Uqt$ is the algebra containing  $\Uqvw$ and an element $\tau$ such that  $\tau g_V \tau = g_W$, for any  $g \in U_q(\gl_\dv)$, where $g_V$,  $g_W$ denote the corresponding elements of  $U_q(\g_V)$ and  $U_q(\g_W)$, respectively.

The algebra $\Uqt$ is a Hopf algebra containing $\Uqvw$ as a Hopf subalgebra.  Its coproduct $\Delta : \Uqt \to \Uqt \tsr \Uqt$ is the unique algebra homomorphism extending the coproduct of $\Uqvw$ (see \eqref{e U_q coproduct}) and satisfying
\be \label{e Uq tau coproduct}
\Delta(\tau)= \tau \tsr \tau.
\ee
The counit  $\varepsilon$ of  $\Uqt$ is the algebra homomorphism determined by $\varepsilon(q^h) = \varepsilon(\tau)= 1$,  $\varepsilon(E_i) = \varepsilon(F_i) = 0$, and the antipode $S$ is the algebra anti-homomorphism determined by
\be \label{e Uqt antipode}
S(q^h) = q^{-h},  \ \ S(E_i) = -E_i K_i, \ \ S(F_i) = - K_i^{-1} F_i, \ \ S(\tau) = \tau.
\ee

The bar-involution on  $\Uqvw$ extends to a  $\QQ$-linear automorphism of  $\Uqt$ by $\br{\tau} = \tau$, also called the \emph{bar-involution} and denoted $\br{\cdot}$.
\subsection{The Hopf algebra  $\Oqt$}
\label{ss Hopf algebra Oqt}
Let  $\F(\S_2)$ denote the Hopf algebra of functions on the group  $\S_2$ taking values in  $\field$.
Dual to $\Uqt$, there is an object
\[\Oqt := \O(GL_q(V))\tsrvw\O(GL_q(W)) \rtimes \F(\S_2). \]
Here,  $\rtimes$ is a ``semidirect coproduct,'' often called a smash coproduct in this setting.  Since this is a less familiar operation than the semidirect product, we  give it a careful treatment, but leave this to Appendix \ref{s appendix Oqt} since the details are somewhat technical.  The object  $\Oqt$ can be given the structure of a Hopf algebra (Proposition \ref{p Oqt Hopf algebra technical}) such that the pairing between $\Uqt$ and  $\Oqt$ coming from the pairing between $U_q(\g_V)$ and  $\O(GL_q(V))$ is a nondegenerate Hopf-pairing (Corollary \ref{c Oqt Uqt pairing}).  Moreover, we show (Proposition \ref{phomomorph tau}) that
 there is a Hopf algebra homomorphism $\tilde{\psi}^\tau$ such that the composition
\[ \O(GL_q(\nsbr{X})) \xrightarrow{\ \tilde{\psi}^\tau}  \Oqt \xrightarrow{\ \pi} \O(GL_q(V))\tsrvw \O(GL_q(W)) \]
is equal to $\tilde{\psi}$ of Proposition \ref{phomomorph GL}; here $\pi$ is the canonical surjection.

\subsection{Representation theory of  $\Uqt$ and  $\Oqt$}
\label{ss representation theory Uqt Oqt}
Let  $\Ointtau$ be the full subcategory of $\Uqt$-modules with objects
\[\{\X \in \Uqt\text{-}\Mod: \Res_{\Uqvw}\X \in \Oint{\g_V \oplus \g_W}\}.\]
All such modules are completely reducible and the irreducibles are described below.  This follows from general results about the representation theory of the wreath product of a universal enveloping algebra (or an algebra with similar properties) and a finite group, as treated in \cite{khare}.

Let $\eta_{\lambda, \mu}$ denote the canonical highest weight vector of $V_\lambda \tsrvw W_\mu$.
Recall that $\mathscr{P}_{r,l}$ denotes the set of partitions of size $r$ with at most  $l$ parts.
The irreducible objects of $\Ointtau$ are
\begin{list}{(\arabic{ctr})} {\usecounter{ctr} \setlength{\itemsep}{1pt} \setlength{\topsep}{2pt}}
\item $\X_{\{\lambda,\mu\}} := V_\lambda \tsrvw W_\mu \oplus V_\mu \tsrvw W_\lambda$,  $\lambda \in \mathscr{P}_{r_1,\dv},\ \mu \in \mathscr{P}_{r_2,\dv}, \ r_1,r_2 \geq 0$,  $\lambda \neq \mu$, with the action of $\tau$ determined by $\tau(\eta_{\lambda,\mu}) = \eta_{\mu,\lambda}$,
\item $\X_{+ \lambda} := V_\lambda \tsrvw W_\lambda$,  $\lambda \in \mathscr{P}_{r,\dv}, \ r \geq 0$, with the action of $\tau$ determined by $\tau(\eta_{\lambda,\lambda}) = \eta_{\lambda,\lambda}$,
\item $\X_{- \lambda} := V_\lambda \tsrvw W_\lambda$, $\lambda \in \mathscr{P}_{r,\dv}, \ r \geq 0$, with the action of $\tau$ determined by $\tau(\eta_{\lambda,\lambda}) = -\eta_{\lambda,\lambda}$.
\end{list}

Since  $\Uqt$ and  $\Oqt$ are dually paired Hopf algebras, any right $\Oqt$-comodule is also a left $\Uqt$-module (see  \textsection\ref{ss dually paired  Hopf algebras}).
All of the objects of  $\Ointtau$ in fact come from  $\Oqt$-comodules and the irreducibles are the same:
let  $\X$ be any of the $\Uqt$-modules (1)--(3).
First of all, given the forms of (1)--(3), it follows from  \textsection\ref{ss modules of GLq} that there is a corepresentation $\varphi: \X \to \X \tsr \O(GL_q(V)) \tsrvw \O(GL_q(W)), \ x \mapsto \sum x_{(0)} \tsr x_{(1)}$  corresponding to  $\Res_\Uqvw \X$.
The corepresentation  $\varphi^\tau: \X \to \X \tsr \Oqt$ giving rise to the  $\Uqt$-module  $\X$ is then given by
\[
\begin{cases}
\varphi^\tau(x) = \sum x_{(0)}  \tsr x_{(1)} \sharp \dual{e} + \tau(x_{(0)}) \tsr \tau(x_{(1)}) \sharp \dual{\tau}, & \text{if  $\X = \X_{\{\lambda,\mu\}}$ or  $\X = \X_{+\lambda}$},   \\
\varphi^\tau(x) = \sum x_{(0)}  \tsr x_{(1)} \sharp \dual{e} - \tau(x_{(0)}) \tsr \tau(x_{(1)}) \sharp \dual{\tau}, & \text{if  $\X = \X_{-\lambda}$},
\end{cases}
\]
where $\tau(\cdot)$ denotes both  the involution of  $\X$ and of $\O(GL_q(V)) \tsrvw \O(GL_q(W))$ given by  $\tau(f \tsrvw g) = g \tsrvw f$; see Appendix \ref{s appendix Oqt} for notation.
Moreover, by the Peter-Weyl theorem for  $\O(M_q(V))$, the comodules corresponding to (1)--(3) are all the irreducible comodules of  $\Oqt$ up to tensoring with powers of the determinant.

\subsection{Schur-Weyl duality between  $\Uqt$ and  $S^2 \H_r$}
\label{ss Schur-Weyl duality Uqt S2Hr}
Recall from  \textsection\ref{ss representation theory of S2Hr} the irreducible $\field S^2 \H_r$-modules $ M_{\{\lambda,\mu\}}$, $ S^2 M_\lambda$,
 $ \Wedge^2 M_\lambda$.
Recall that $\mathscr{P}'_{r,l}$ is the subset of $\mathscr{P}_{r,l}$ consisting of those partitions that are not a single row or column shape.
\begin{proposition}\label{p schur-weyl duality tau}
As a $(\Uqt,  S^2\H_r)$-bimodule, $X^{\tsr r}$ decomposes into irreducibles as
\[
X^{\tsr r} \cong \bigoplus_{\stackrel{\{\lambda, \mu\} \subseteq \mathscr{P}_{r,\dv}}{\lambda \neq \mu}} \X_{\{\lambda, \mu\}} \tsr  M_{\{\lambda, \mu\}} \oplus
\bigoplus_{\lambda \in \mathscr{P}_{r,\dv}} \X_{+\lambda} \tsr  S^2 M_\lambda\oplus \bigoplus_{\lambda \in \mathscr{P}'_{r,\dv}} \X_{-\lambda} \tsr  \Wedge^2 M_\lambda.
\]
\end{proposition}
This follows easily from Proposition \ref{p S2Hr representations} and the decomposition of  $V^{\tsr r} \tsrvw W^{\tsr r}$ as a  $(\Uqvw, \H_r \tsrvw \H_r)$-bimodule.

Now we can define the symmetric (resp. exterior) Kronecker coefficient $g_{+  \lambda\nu}$ (resp. $g_{-  \lambda\nu}$) to be either of the following quantities
\begin{itemize}
\item the multiplicity of $M_\nu|_{q=1}$ in $S^2 M_\lambda|_{q=1}$ (resp. $\Wedge^2 M_\lambda|_{q=1}$),
\item the multiplicity of $\X_{+  \lambda}|_{\u =1}$ (resp. $\X_{-  \lambda}|_{\u = 1}$) in $\Res_{U^\tau} (X_\nu|_{\u = 1})$,
\end{itemize}
where  $U^\tau := U(\g_V \oplus \g_W) \rtimes \S_2$. These multiplicities are the same by Proposition \ref{p schur-weyl duality tau} and standard Schur-Weyl duality for $X^{\tsr r}$  $\big(X^{\tsr r} \cong \bigoplus_{\nu \vdash_\dx r} X_\nu \tsr M_\nu \big)$.

\subsection{Upper based  $\Uqt$-modules}
\label{ss upper based Uqt}
For the detailed study of two-row Kronecker in  \textsection\ref{s global crystal basis for two-row Kronecker coefficients}--\ref{s A Kronecker graphical calculus}, we need some theory of canonical bases for $\Uqt$-modules.

\begin{definition}
A \emph{weak upper based $\Uqt$-module} is a pair $(N,B)$, where $N$ is an object of $\Ointtau$ and $B$ is a $\field$-basis of $N$ such that
$(\Res_\Uqvw N,B)$ is an upper based  $\Uqvw$-module  and  $\tau(b) \in \pm B$ for all $b \in B$.

A weak upper based  $\Uqt$-module  $(N,B)$ is a \emph{upper based $\Uqt$-module} if for any highest weight $b \in B$ of weight  $(\lambda,\lambda)$,  $\tau(b) = \pm b$.
\end{definition}

In order to check that the tensor product of weak upper based  $\Uqt$-modules is a weak upper based $\Uqt$-module, we must first check that
\be \label{e tau and bar involution tsr}
\parbox{13.4cm}{if the  $\br{\cdot}$-involution on $\Uqt$ intertwines that of  $N_1$ and  $N_2$, then it intertwines the  $\br{\cdot}$-involution on  $N_1 \tsr N_2$.}
\ee
This amounts to checking that $\Theta \ \br{\cdot} \tsr \br{\cdot}(\Delta(\tau)) = \Delta(\br{\tau}) \Theta$, which follows from $\br{\tau} = \tau$,  $\Delta(\tau)= \tau \tsr \tau$, and  $\Theta = \Theta_V \Theta_W$, where  $\Theta_V$ (resp.  $\Theta_W$) denotes the  quasi-$\R$-matrix for  $U_q(\g_V)$ (resp. $U_q(\g_W)$). Here \br{\cdot} \tsr \br{\cdot} denotes the map  $\Uqt \tsr \Uqt \to \Uqt \tsr \Uqt, \ x \tsr x' \mapsto \br{x} \tsr \br{x'}$.

\begin{proposition}
\label{p upper based tau}
If $(N,B), (N', B')$ are weak upper based $\Uqt$-modules, then $(N \tsr N', B\heart B')$ is a weak upper based  $\Uqt$-module with  $\tau(b \heart b') = \tau(b) \heart \tau(b')$.
\end{proposition}
\begin{proof}
Note that by the definition of weak upper based $\Uqt$-module, the  $\br{\cdot}$-involution on  $\Uqt$ intertwines that of  $N$ and  $N'$.  Then by \eqref{e tau and bar involution tsr}, there holds $\br{\tau(b \heart b')}= \br{\tau}(\br{b \heart b'}) = \tau (b \heart b')$.  Moreover, $\tau(b\heart b')-\tau(b)\tsr \tau(b') \in \ui (N\tsr N')_{\ZZ[\ui]}$.  Hence $\tau(b \heart b')$ satisfies the defining properties of $\tau(b) \heart \tau(b')$.  The hypotheses then imply $\tau(b \heart b') = \tau(b) \heart \tau(b') \in \pm B \heart B'$, hence $(N \tsr N', B\heart B')$ is a weak upper based  $\Uqt$-module.
\end{proof}


An upper based  $\Uqt$-module, unlike a weak upper based  $\Uqt$-module, has irreducible  $\Uqt$-cells.
The are four types of isomorphism classes of upper based  $\Uqt$-modules  $(N,B)$ for which  $N$ is irreducible (numbered to match Proposition-Definition \ref{p S2Hr representations}):
\be \label{e Uqt canonical bases Z lambda mu}
\parbox{14cm}{
\begin{list}{(\arabic{ctr})} {\usecounter{ctr} \setlength{\itemsep}{1pt} \setlength{\topsep}{2pt}}
\item[($1_+$)] $(\X_{\{\lambda,\mu\}}, B_V(\lambda) \tsrvw B_W(\mu) \sqcup B_V(\mu) \tsrvw B_W(\lambda))_+$,
\item[($1_-$)] $(\X_{\{\lambda,\mu\}}, B_V(\lambda) \tsrvw B_W(\mu) \sqcup B_V(\mu) \tsrvw B_W(\lambda))_-$,
\item[(2)\ ] $(\X_{+ \lambda}, B_V(\lambda) \tsrvw B_W(\lambda)),$
\item[(3)\ ] $(\X_{- \lambda}, B_V(\lambda) \tsrvw B_W(\lambda)),$
\end{list}}
\ee
where the action of $\tau$ is given by
\[
\begin{array}{rrr}
 \tau(G_\lambda(b^V_{P_V}) \tsrvw G_\mu(b^W_{P_W})) &= & \ G_\mu(b^V_{P_W}) \tsrvw G_\lambda(b^W_{P_V}), \\[1.3mm]
 \tau(G_\lambda(b^V_{P_V}) \tsrvw G_\mu(b^W_{P_W})) &= & -G_\mu(b^V_{P_W}) \tsrvw G_\lambda(b^W_{P_V}), \\[1.3mm]
 \tau(G_\lambda(b^V_{P_V}) \tsrvw G_\lambda(b^W_{P_W})) &= & G_\lambda(b^V_{P_W}) \tsrvw G_\lambda(b^W_{P_V}),\\[1.3mm]
 \tau(G_\lambda(b^V_{P_V}) \tsrvw G_\lambda(b^W_{P_W})) &= & -G_\lambda(b^V_{P_W}) \tsrvw G_\lambda(b^W_{P_V}).
\end{array}
\]

\subsection{The nonstandard two-row case}
\label{ss nonstandard two-row case}
We know that  $V_\mathbf{A}^{\tsr r}$ is a right  $\H_r$-module, where $V_\mathbf{A} = \mathbf{A}\{v_i: i \in [\dv]\}$ is the integral form of $V$.  This defines an  $\mathbf{A}$-algebra homomorphism  $\H_r \to \End_\mathbf{A}(V_\mathbf{A}^{\tsr r})$.
Define the \emph{Temperley-Lieb} quotient $\H_{r,\dv}$ of $\H_r$ to be the image of this homomorphism for  $\dv$ equal to the dimension of $V$.
Equivalently, this is the quotient of $\H_r$ by the two-sided ideal
\[
\bigoplus_{\stackrel{\lambda \vdash r,\ \ell(\lambda) > d,}{P\in \text{SYT}(\lambda)}} \mathbf{A}\Gamma_P = \mathbf{A}\{C_w:\ell(\sh(w)) > d\}.
\]
Here  $\Gamma_P$ denotes a right Kazhdan-Lusztig cell of  $\S_r$ (see  \textsection\ref{ss cell label conventions C_Q C'_Q}).

Next, define the \emph{nonstandard Temperley-Lieb} quotient $\nsH_{r,d}$ of $\nsH_r$ to be the subalgebra of $\H_{r,d}\tsrvw \H_{r,d}$ generated by the elements
\[
\sP_{s} := \C_s \tsrvw \C_s + C_s \tsrvw C_s, \ s \in S.
\]

Recall that $\mathscr{P}_{r,2}$ is the set of partitions of size $r$ with at most two parts and $\mathscr{P}'_{r,2}$ is the subset of $\mathscr{P}_{r,2}$ consisting of those partitions that are not a single row or column shape. Define the index set $\nsP_{r,2}$ for the $\field \nsH_{r,2}$-irreducibles as follows:
\be\label{e definition nsp r2}
\nsP_{r,2} =  \{ \{\lambda,\mu\}: \lambda, \mu \in \mathscr{P}_{r,2}, \, \lambda \neq \mu\} \sqcup\{+\lambda: \lambda \in \mathscr{P}'_{r,2}\} \sqcup \{-\lambda: \lambda \in \mathscr{P}'_{r,2}\} \sqcup \{\nsbr{\epsilon}_+\}.
\ee
\begin{theorem}[\cite{B4}] \label{t nsH irreducibles two row case}
The algebra $\field\nsH_{r,2}$ is split semisimple and the list of distinct irreducibles is
\begin{list}{\emph{(\arabic{ctr})}} {\usecounter{ctr} \setlength{\itemsep}{1pt} \setlength{\topsep}{2pt}}
\item $\nsbr{M}_\alpha := \Res_{\nsH_{r,2}}M_\lambda \tsr M_\mu$, for $\alpha = \{\lambda, \mu\} \in \nsP_{r,2}$,
\item $\nsbr{M}_\alpha := S' \nsbr{M}_\lambda$, for $\alpha = +\lambda \in \nsP_{r,2}$,
\item $\nsbr{M}_\alpha := \Wedge^2 \nsbr{M}_\lambda$, for $\alpha = -\lambda \in \nsP_{r,2}$,
\item $\nsbr{M}_\alpha := \field\nsbr{\epsilon}_+$, for $\alpha = \nsbr{\epsilon}_+\in \nsP_{r,2}$.
\end{list}
\end{theorem}

Nonstandard Schur-Weyl duality in the two-row case thus takes the form
\be
\nsbr{X}\dualcnv^{\tsr r}\cong \nssym{r}{X} \tsr \field \nsbr{\epsilon}_+ \oplus \bigoplus_{\stackrel{\{\lambda, \mu\} \subseteq \mathscr{P}_{r,2}}{\lambda \neq \mu}} \nsbr{\X}_{\{\lambda, \mu\}} \tsr \nsbr{M}_{\{\lambda, \mu\}}  \oplus \bigoplus_{\lambda \in \mathscr{P}'_{r,2}} \nsbr{\X}_{+\lambda} \tsr  S' \nsbr{M}_\lambda \oplus \bigoplus_{\lambda \in \mathscr{P}'_{r,2}} \nsbr{\X}_{-\lambda} \tsr \Wedge^2 \nsbr{M}_\lambda.
\ee
Here $\nsbr{\X}_{\{\lambda, \mu\}}, \nsbr{\X}_{\pm\lambda}$ are irreducible  $\O(M_q(\nsbr{X}))$-comodules (for  $\dv = \dw =2$) such that the corresponding $\Oqt$-modules obtained via $\tilde{\psi}^\tau$ are isomorphic to $\X_{\{\lambda, \mu\}}, \X_{\pm\lambda}$, respectively.
This follows from Proposition \ref{p schur-weyl duality tau},  \textsection\ref{ss representation theory Uqt Oqt}, and arguments similar to those producing \eqref{e n lambda mu multiplicities}.  Then, setting $\nsbr{\X}_{\nsbr{\epsilon}_+} := \nssym{r}{X}$, we have
\begin{corollary}
\label{c irreducible nsOM comodules}
The distinct irreducible $\O(M_q(\nsbr{X}))$-comodules in the two-row case are the $\nsbr{\X}_\alpha$, as $r$ ranges over nonnegative integers and $\alpha$ ranges over $\nsP_{r,2}$.
\end{corollary}

\section{A canonical basis for $\nsbr{Y}_\alpha$}
\label{s A canonical basis for Yalpha}
Throughout this section, let $\alpha \vDash_l^{\dx} r$, set $ \nsbr{\bT} = \nsbr{X}^{\tsr r}$, and define
\be \label{e Y alpha definition}
\nsbr{Y}_\alpha := \nswedge{{\alpha_1}}{X} \tsr \nswedge{{\alpha_2}}{X} \tsr \dots \tsr \nswedge{{\alpha_l}}{X} \subseteq \nsbr{\bT}.
\ee
We define a canonical basis of $\nsbr{Y}_\alpha$ by first defining a canonical basis of $\nswedge{r }{X}$ and then putting these together with Lusztig's construction for tensoring based modules (\textsection\ref{ss Tensor products of based modules}).  The bases of $\nswedge{r }{X}$ and  $\nsbr{Y}_\alpha$ are labeled by what we call nonstandard columns and nonstandard tabloids, respectively.  Almost all the results in this section hold for general $\dv, \dw$, though a few things are only made explicit  in the two-row $(\dv = \dw = 2)$ case.

\begin{remark}
For convenience, we require  $\alpha_i > 0$, but this is not essential as  $\nswedge{0}{X} = \field$.
\end{remark}

\subsection{Nonstandard columns label a canonical basis for $\nswedge{r }{X}$}
\label{ss nonstandard columns labeled a canonical basis for Lambda r X}
Here we define a basis  $\text{NSC}^{\, r}$ of $\nswedge{r }{X}$ making  $(\nswedge{r }{X}, \text{NSC}^{\, r})$ into an upper based $\Uqt$-module.  Since $\Res_{\Uqvw}\nswedge{r }{X} \cong \bigoplus_{\lambda \vdash r} V_\lambda \tsrvw W_{\lambda'}$ and the weights  $\{(\lambda, \lambda'):\lambda \vdash r\}$ are pairwise incomparable, it follows from
\eqref{e based module filtration} that $\Res_{\Uqvw}\nswedge{r }{X}$ can be made into an upper based  $\Uqvw$-module in a unique way (this is not so for  $\nssym{r}{X}$).  The main content of this subsection is to realize the basis  $ \text{NSC}^{\, r}$ explicitly as a subset of  $ \nsbr{\bT}$, which will turn out to be important later on.

We define the elements NSC$_{P_V, P_W}$ of the basis $\text{NSC}^{\, r}$ by the following formula;
Proposition \ref{p NSC facts} will explain this formula and establish some important facts about this basis.
\begin{definition}
Let  $\lambda$ be a partition and $P_V \in \text{SSYT}_{\dv}(\lambda)$,  $P_W \in \text{SSYT}_{\dw}(\lambda')$. Define the element NSC$_{P_V, P_W}$ of $ \nsbr{\bT}$ by
\be
NSC_{P_V,P_W} := \sum_{Q \in SYT(\lambda)} (-1)^{\ell(\transpose{Q})} \liftcvw{RSK^{-1}(P_V,Q)}{RSK^{-1}(P_W,\transpose{Q})},
\ee
where $\liftcvw{\mathbf{k}}{\mathbf{l}} := \liftc^V_{\mathbf{k}} \tsrvw \liftc^W_{\mathbf{l}}$ and  $\liftc^V_\mathbf{k}$ (resp. $\liftc^W_\mathbf{k}$) is the projected upper canonical basis element of $V^{\tsr r}$ (resp. $W^{\tsr r}$) from Theorem \ref{t lifted upper canonical basis}, and  $\ell(\transpose{Q})$ is as in \textsection\ref{ss type A combinatorics preliminaries}.
Also define the sets
\be
\begin{array}{rl}
\text{NSC}(\lambda) := &\{ \text{NSC}_{P_V,P_W}: P_V \in \text{ SSYT}_{\dv}(\lambda),  P_W \in \text{SSYT}_{\dw}(\lambda')\}, \\[1mm]
\text{NSC}^{\, r} := &\bigsqcup_{\lambda \vdash r} \text{NSC}(\lambda).
\end{array}
\ee
\end{definition}

\begin{remark}
Recall that $Z_\lambda$ is the superstandard tableau of shape and content $\lambda$ (\textsection\ref{ss type A combinatorics preliminaries}).
For the highest weight NSC, i.e. those of the form NSC$_{Z_\lambda,Z_{\lambda'}}$, we do not need the projected basis:
\[NSC_{Z_\lambda, Z_{\lambda'}} = \sum_{Q \in SYT(\lambda)} (-1)^{\ell(\transpose{Q})} \cvw{RSK^{-1}(Z_\lambda,Q)}{RSK^{-1}(Z_{\lambda'},\transpose{Q})}.\]
This is immediate from Theorem \ref{t lifted upper canonical basis}.
\end{remark}

For each NSC$_{P_V,P_W} \in \text{NSC}(\lambda)$, choose some  $Q \in \text{SYT}(\lambda)$ and set  $\mathbf{k} = RSK^{-1}(P_V,Q), \mathbf{l} = RSK^{-1}(P_W,\transpose{Q})$ and label this element by the column $\ctableausmalla{j_r \\ {\tiny \cdot\hspace{-.5mm}\cdot\hspace{-.5mm}\cdot} \\ j_2 \\ j_1}$,  where  $j_i := \rho(k_i, l_i)$ and $\rho(a, b) = (a-1)\dw+b$ as in \textsection\ref{s notation for GLV GLW}.
The set of columns obtained in this way from $ \text{NSC}^{\, r}$ is the set of \emph{nonstandard columns} of \emph{height} $r$.
In the case $\dv = \dw = 2$, the nonstandard columns of height $r$ and their identifications with  $\text{NSC}^{\, r}$, $r \in [\dx]$, are shown in Figure \ref{f nonstandard columns label basis}. The choices for the SYT $Q$ above are implicit in these identifications, as the following example clarifies.

\begin{example}\label{e nonstandard columns label basis}
For $P_V = {\tiny \tableau{1 & 1 \\ 2}}, P_W = {\tiny \tableau{1 & 2\\ 2}}$, let us explain the identification of NSC$_{P_V,P_W} = \cvw{211}{221} - \cvw{121}{212}$ with the nonstandard column $\ctableausmalla{1 \\ 2 \\ 4}$.   This corresponds to the choice $Q = {\tiny \tableau{1 & 3 \\ 2}}$ so that $\mathbf{k} = 211, \mathbf{l} = 221$; hence $\rho(\mathbf{k}, \mathbf{l}) = 421$.
\end{example}
\begin{figure}
$\begin{array}{llllll}
  \myvcenter{\ensuremath{\column{1}}} = \cvw{1}{1} & \myvcenter{\ensuremath{\column{2}}} = \cvw{1}{2} & \myvcenter{\ensuremath{\column{3}}} = \cvw{2}{1} & \myvcenter{\ensuremath{\column{4}}} = \cvw{2}{2} && \vspace{0.35cm}\\ \vspace{0.35cm}
  \myvcenter{\ensuremath{\column{1 \\ 2}}} = \cvw{11}{21} & \myvcenter{\ensuremath{\column{1 \\ 3}}} = \cvw{21}{11} & \myvcenter{\ensuremath{\column{3 \\ 2}}} = \cvw{12}{21} + \frac{1}{[2]}\cvw{21}{21} & \myvcenter{\ensuremath{\column{2 \\ 3}}} = \cvw{21}{12} + \frac{1}{[2]}\cvw{21}{21} & \myvcenter{\ensuremath{\column{3 \\ 4}}} = \cvw{22}{21} & \myvcenter{\ensuremath{\column{2 \\ 4}}} = \cvw{21}{22}\\
  \vspace{0.35cm}
  \myvcenter{\ensuremath{\column{1 \\ 2 \\ 3}}} = \cvw{211}{121} - \cvw{121}{211} & \myvcenter{\ensuremath{\column{1 \\ 2 \\ 4}}} = \cvw{211}{221} - \cvw{121}{212} & \myvcenter{\ensuremath{\column{1 \\ 3 \\ 4}}} = \cvw{212}{121} - \cvw{221}{211} & \myvcenter{\ensuremath{\column{2 \\ 3 \\ 4}}} = \cvw{212}{221} - \cvw{221}{212} && \\
  \vspace{0.35cm}
  \myvcenter{\ensuremath{\column{1 \\ 2 \\ 3 \\ 4}}} =  \cvw{2121}{2211}-\cvw{2211}{2121} &&&&&
\end{array}$
\caption{Nonstandard columns of height $r$ are identified with NSC$^{\, r}$.  These are a basis of $\nswedge{r }{X}$.}
\label{f nonstandard columns label basis}
\end{figure}

\begin{remark}
In \cite{QuantumDeformations}, it is shown how to put a $\Uqvw$-crystal structure on
the set  SSYT$_\dx((1^r))$.  This gives a natural way to identify NSC$^{\, r}$ with  SSYT$_\dx((1^r))$ for general $\dv, \dw$.
\end{remark}

Define the integral form   $\nswedge{r }{X}^\mathbf{A}$ and the $\field_\infty$-lattice  $\L_{(r)}$ of   $\nswedge{r }{X}$ by
\be
\begin{array}{ccl}
\nswedge{r }{X}^\mathbf{A} &=& \mathbf{A} \text{NSC}^{\, r} \\
\L_{(r)} &=& \field_\infty \text{NSC}^{\, r}.
\end{array}
\ee

Let $\nsbr{p}_{(r)} : \nsbr{\bT} \to \nsbr{\bT}$ be the  $\field \nsH_r$-module projector with image the  $\field\nsbr{\epsilon}_-$-isotypic component of $ \nsbr{\bT}$, which is equal to  $\nswedge{r}{X}$.  Define the projector $\nsbr{\rho}_-^{\lambda, \lambda'}: M_\lambda \tsrvw M_{\lambda'} \to M_\lambda \tsrvw M_{\lambda'}$ by $\nsbr{\rho}_-^{\lambda,\lambda'} = \nsbr{i}_- \circ \nsbr{s}_-$, where  $\nsbr{i}_-, \nsbr{s}_-$ are as in Proposition \ref{p sign in lambda tsr lambda'}.  Extend this to a projector $\nsbr{p}_-^{\lambda, \lambda'}:\nsbr{\bT}\to\nsbr{\bT}$ that acts as  $\nsbr{\rho}_-^{\lambda, \lambda'}$ on each  $M_\lambda \tsrvw M_{\lambda'}$-isotypic component for the  $\H_r \tsrvw \H_r$-action and 0 on the other isotypic components.  Recall from \textsection\ref{s notation for GLV GLW} that $\pi_{\lambda,\lambda'}^{\nsbr{\bT}} :\nsbr{\bT}\to\nsbr{\bT} $ is the  $\Uqvw$-projector with image the  $V_\lambda\tsrvw W_{\lambda'}$-isotypic component of  $\nsbr{\bT}$.
Then by Proposition \ref{pgelfanddecomp} (2),
\be \label{e epsilon - projectors}
\nsbr{p}_{(r)} = \sum_\lambda \nsbr{p}_{(r)} \pi_{\lambda,\lambda'}^{\nsbr{\bT}} = \sum_\lambda \nsbr{p}_-^{\lambda, \lambda'} \pi_{\lambda,\lambda'}^{\nsbr{\bT}}.
\ee

\begin{proposition}\label{p NSC facts}
Maintain the notation above and that of Proposition \ref{p sign in lambda tsr lambda'}.
\begin{list}{\emph{(\alph{ctr})}} {\usecounter{ctr} \setlength{\itemsep}{1pt} \setlength{\topsep}{2pt}}
\item The set NSC$(\lambda)$ is an  $\mathbf{A}$-basis of $V_\lambda^\mathbf{A} \tsrvw W_{\lambda'}^\mathbf{A} \subseteq \nswedge{r }{X}^\mathbf{A}$, and
\[(V_\lambda \tsrvw W_{\lambda'} , B_V(\lambda) \tsrvw B_W(\lambda')) \to (\nswedge{r }{X}, \text{NSC}^{\, r}), \qquad
  G_\lambda(b^V_{P_V})\tsrvw G_{\lambda'}(b^W_{P_W}) \mapsto \text{NSC}_{P_V, P_W} \]
is an inclusion of upper based $\Uqvw$-modules (where $G_\lambda, b^V, b^W$ are as in \textsection\ref{ss crystal bases}--\ref{ss global canonical bases}).  These inclusions combine to give an isomorphism of upper based  $\Uqt$-modules
\[\bigoplus_{\stackrel{\lambda \vdash_{\dv} r,}{\lambda' \vdash_{\dw} r}}(V_\lambda \tsrvw W_{\lambda'} , B_V(\lambda) \tsrvw B_W(\lambda')) \xrightarrow{\cong} (\nswedge{r }{X}, \text{NSC}^{\, r}).\]
\item
\[\nsbr{p}_-^{\lambda, \lambda'}  \pi_{\lambda,\lambda'}^{\nsbr{\bT}}(\cvw{\mathbf{k}}{\mathbf{l}}) \begin{cases}
= (-1)^{\ell(\transpose{Q(\mathbf{l})})} \text{NSC}_{P(\mathbf{k}),P(\mathbf{l})} & \text{ if } Q(\mathbf{k}) = \transpose{Q(\mathbf{l})} \text{ has shape } \lambda, \\
= 0 & \text{ if } \sh(\mathbf{k}) = \sh(\mathbf{l})' = \lambda \text{ and } Q(\mathbf{k}) \neq \transpose{Q(\mathbf{l})}, \\
= 0 & \text{ if } \sh(\mathbf{k}) \not\gd \lambda \text{ or } \sh(\mathbf{l}) \not\gd \lambda', \\
\in \ui \L_{(r)} \cap \u \br{\L}_{(r)} & \text{ otherwise,}
\end{cases}\]
\item
\[\nsbr{p}_{(r)}(\cvw{\mathbf{k}}{\mathbf{l}}) \begin{cases}
= (-1)^{\ell(\transpose{Q(\mathbf{l})})} \text{NSC}_{P(\mathbf{k}),P(\mathbf{l})} & \text{ if } Q(\mathbf{k}) = \transpose{Q(\mathbf{l})}, \\
= 0 & \text{ if } \sh(\mathbf{k}) = \sh(\mathbf{l})' \text{ and } Q(\mathbf{k}) \neq \transpose{Q(\mathbf{l})}, \\
\in \ui \L_{(r)} \cap \u \br{\L}_{(r)} & \text{ otherwise,}
\end{cases}\]
\item $\nsbr{p}_{(r)}(\L_V \tsrvw_{\field_\infty} \L_W) = \L_{(r)}.$
\end{list}
\end{proposition}
\begin{proof}
We prove (b) first.  Assume we are in one of the top two cases of (b).  By Theorem \ref{t lifted upper canonical basis} (d), $\pi_{\lambda,\lambda'}^{\nsbr{\bT}} (\cvw{\mathbf{k}}{\mathbf{l}}) = \liftcvw{\mathbf{k}}{\mathbf{l}}$.  Theorem \ref{t standard Schur-Weyl duality2} (ii) for  $\tilde{B}$ shows that
\[\{\liftcvw{\mathbf{k'}}{\mathbf{l'}}: P(\mathbf{k'}) = P(\mathbf{k}), P(\mathbf{l'})= P(\mathbf{l})\} \xrightarrow{\cong} \Gamma_\lambda \tsr \Gamma_{\lambda'}, \quad  \liftcvw{\mathbf{k'}}{\mathbf{l'}} \mapsto  C_{Q(\mathbf{k'})} \tsr C_{Q(\mathbf{l'})}\]
is an isomorphism of $\H_r\tsr \H_r$-cells.
Now applying  $\nsbr{p}_-^{\lambda, \lambda'}$ to  $\liftcvw{\mathbf{k}}{\mathbf{l}}$, using Proposition \ref{p sign in lambda tsr lambda'}, yields the top two cases of (b).
For the bottom two cases of (b), we use Corollary \ref{c clifts and projections general} to apply the projector $\pi_{\lambda,\lambda'}^{\nsbr{\bT}}$ and then the easy fact that $\nsbr{p}_-^{\lambda, \lambda'} (\field_\infty(\Gamma_\lambda \tsr \Gamma_{\lambda'})) \subseteq \field_\infty( \Gamma_\lambda \tsr \Gamma_{\lambda'})$.

Next, we use \eqref{e epsilon - projectors} to show (b) implies (c). This is straightforward, noting that the third case of (b) applies if  $\sh(\mathbf{k}) = \sh(\mathbf{l})' \neq \lambda$.  Statement (d) is immediate from (c) and definitions.

Finally, we prove (a).
By Theorem \ref{t lifted upper canonical basis}, the following bijection of the $\Uqvw$-cells of $(\nsbr{\bT} , \tilde{B}_V \tsrvw \tilde{B}_W)$ on the  left-hand side with the right-hand side gives rise to an isomorphism of upper based $\Uqt$-modules:
\[
\begin{array}{ccc}
\displaystyle \bigsqcup_{\stackrel{\lambda \vdash_{\dv} r,}{\lambda' \vdash_{\dw} r}} \{ \liftcvw{\mathbf{k'}}{\mathbf{l'}} : Q(\mathbf{k'}) = \transpose{(Z_{\lambda'}^*)}, Q(\mathbf{l'}) = Z_{\lambda'}^*\} & \xrightarrow{\cong}& \displaystyle \bigsqcup_{\stackrel{\lambda \vdash_{\dv} r,}{\lambda' \vdash_{\dw} r}} B_V(\lambda) \tsrvw B_W(\lambda'), \\[1.6mm]
\ \ \qquad \liftcvw{\mathbf{k'}}{\mathbf{l'}} & \mapsto & G_\lambda(b^V_{P(\mathbf{k'})})\tsr G_{\lambda'}(b^W_{P(\mathbf{l'})}).
\end{array}
\]
Composing the inverse of this with $ \nsbr{p}_{(r)}$, using (c), yields (a).
\end{proof}

\begin{remark}
In this section and onward, we mostly work with $\Uqt$-modules instead of $\O(GL_q(\nsbr{X}))$-comodules because we have a theory of based modules for the former.  Regarding the previous proposition, we know that $\nswedge{r}{X}$ is an irreducible  $ \O(GL_q(\nsbr{X}))$-comodule, but it would be desirable to see this explicitly by describing the action of generators of the hypothetical nonstandard enveloping algebra on the basis $\text{NSC}^{\, r}$.
\end{remark}
\subsection{Nonstandard tabloids label a canonical basis of  $ \nsbr{Y}_\alpha$}
We define two products  $\heartp$ and  $\heartl$ that we use to construct a canonical basis of $\nsbr{Y}_\alpha$ from the canonical bases $\text{NSC}^{\alpha_j}$ of  $\nswedge{{\alpha_j} }{ X}$.  These products turn out to agree, so we believe the resulting canonical basis to be the ``correct'' basis for $\nsbr{Y}_\alpha$. Before defining these products, we first introduce nonstandard tabloids, which will label this basis of $\nsbr{Y}_\alpha$.

\begin{definition}
A \emph{column-diagram} of shape  $\alpha \vDash_l^{\dx} r$ is a sequence of columns of heights $\alpha_1,\ldots,\alpha_l$ with their tops aligned as in Example \ref{ex column diagram}.
A \emph{nonstandard tabloid} (NST) of shape  $\alpha$ is a column-diagram whose columns are filled from the set of nonstandard columns (see Figure \ref{f nonstandard columns label basis} for the two-row case).
The set of all NST of shape $\alpha$ is denoted NST$(\alpha)$.

The \emph{column reading word} of an NST $T$ is the word $\mathbf{j}$ obtained by reading its columns from bottom to top and then left to right. The \emph{$V$-word} (resp. \emph{$W$-word}) of $T$ is $\mathbf{k}$ (resp.  $\mathbf{l}$), where $(\mathbf{k}, \mathbf{l}) = \rho^{-1}(\mathbf{j})$ (see the beginning of  \textsection\ref{ss nonstandard columns labeled a canonical basis for Lambda r X}).

For $T \in \text{NST}(\alpha)$, let $T|_c$ denote the $c$-th column of  $T$ ($T|_1$ is the leftmost column).  For a sequence of integers $1 \leq c_1 < c_2 < \dots < c_k \leq l$, let  $T|_{\{c_1,\ldots,c_k\}}$ be the NST $T|_{c_1} T|_{c_2} \cdots T|_{c_k}$, i.e., the NST consisting of the specified columns of  $T$, in the same order as they occur in  $T$.  Such an NST is a \emph{subtabloid} of  $T$ and, if $\{c_1,\ldots,c_k\}$ is equal to the interval $[c_1, c_k]$,  it is \emph{contiguous}.
\end{definition}

\begin{example}
\label{ex column diagram}
A column-diagram and nonstandard tabloid of shape  $(2, 4, 3, 3, 1, 2)$:
\[\ctableauf{ \\ }{ \\  \\  \\ }{ \\  \\ }{ \\  \\ }{}{ \\ } \qquad \ctableauf{2 \\ 3}{1 \\ 2 \\ 3 \\ 4}{1 \\ 2 \\ 4}{2 \\ 3 \\ 4}{2}{3 \\ 2}  \]
The column reading word (top),  $V$-word (middle), and $W$-word (bottom) of the NST are
\[{\footnotesize
\begin{array}{c}
32\, 4321\, 421\, 432\, 2\, 23 \\
21 \, 2211 \, 211 \, 221 \, 1 \, 12 \vspace{-1mm} \\
12 \, 2121 \, 221 \, 212 \, 2 \, 21
\end{array}
}.\]
If  $T$ is the NST above, then $T|_{[3,5]} = \ctableausmallc{1 \\ 2 \\ 4}{2 \\ 3 \\ 4}{2}$ is a contiguous subtabloid of  $ T$.
\end{example}

Let $\nsbr{p}_\alpha: \nsbr{\bT} \to \nsbr{\bT}$ be the
$\field(\nsH_{\alpha_1} \tsr \cdots \tsr \nsH_{\alpha_l})$
-module projector with image $\nsbr{ Y}_\alpha$. There holds  $\nsbr{p}_\alpha = \nsbr{p}_{\alpha_1} \tsr \dots \tsr \nsbr{p}_{\alpha_l}$.

If  $x_i \in V^{\tsr r_i} \tsrvw W^{\tsr r_i}, i = 1,2$, then let $x_1 \heartvw x_2$ be the element of $V^{\tsr r} \tsrvw W^{\tsr r}$,  $r = r_1 + r_2$,  defined using the $\heart$ product from  \textsection\ref{ss upper canonical basis of bT} for  $V$ and $W$; equivalently, we can define  $\heartvw$ by setting $\cvw{\mathbf{k^1}}{\mathbf{l^1}} \heartvw \cvw{\mathbf{k^2}}{\mathbf{l^2}} = \cvw{\mathbf{k^1}\mathbf{k^2}}{\mathbf{l^1}\mathbf{l^2}}, \mathbf{k^i} \in [\dv]^{r_i}, \mathbf{l^i} \in [\dw]^{r_i}$, and extending bilinearly.  Now for $T$ an NST of shape $\alpha$, define the element $T_\heartvw \in \nsbr{\bT}$  to be  $T|_1 \heartvw \cdots \heartvw T|_l$.
\be
\parbox{13cm}{We identify  $T$ with the element $T|_1 \heartp \cdots\heartp T|_l := \nsbr{p}_\alpha (T_\heartvw)$ of $\nsbr{Y}_\alpha$.}
\ee


\begin{remark}
In the two-row case, the $\heartp$ product can be computed as follows:
let  $J_\alpha$ be as in \textsection \ref{ss type A combinatorics preliminaries} and let $J_2 \subseteq J_\alpha$ be the subset of $J_\alpha$ corresponding to the parts of $\alpha$ equal to  $2$, i.e. $J_2 = \{j \in J_\alpha: j-1,j+1 \notin J_\alpha\}$.
Then for any $T \in  \text{ NST}(\alpha)$ there holds
\be
T = \nsbr{p}_{J_\alpha}(T_\heartvw) = \nsbr{p}_{J_2}(T_\heartvw) = T_\heartvw \big(\prod_{j \in J_2} \frac{\sQ_j}{[2]^2} \big),
\ee
where the second equality holds because  $\Res_{U_q(\sl(V)\oplus\sl(W))} \nswedge{3 }{X} \cong \Res_{U_q(\sl(V)\oplus\sl(W))} \nswedge{1 }{X}$.
\end{remark}

\begin{example}
\label{ex heartp}
For the NST  $T$ shown below, we compute $T_\heartvw$ and $T$ in terms of the upper canonical basis of $X^{\tsr r}$:
\[\myvcenter{\ensuremath{\pad{\column{2 \\ 3}}\newpad{\raisebox{0pt}{$\heartvw$}}\pad{\column{1\\3}}}} = \big(\cvw{21}{12} + \frac{1}{[2]}\cvw{21}{21}\big) \heartvw \cvw{21}{11} = \cvw{2121}{1211} + \frac{1}{[2]}\cvw{2121}{2111},\]
\[\myvcenter{\ensuremath{\ctableau{2\\3}{1\\3}}} = \myvcenter{\ensuremath{\column{2 \\ 3}\raisebox{-15pt}{$\heartp$}\column{1\\3}}} = \big(\cvw{21}{12} + \frac{1}{[2]}\cvw{21}{21}\big) \heartp \cvw{21}{11} = \big(\cvw{2121}{1211} + \frac{1}{[2]} \cvw{2121}{2111}\big) \frac{\sQ_1}{[2]^2} \frac{\sQ_3}{[2]^2} = \cvw{2121}{1211} + \frac{1}{[2]}\cvw{2121}{1121} + \frac{1}{[2]}\cvw{2121}{2111}. \]
The last equality can be computed using \eqref{e sQ on CC}.
\end{example}

The $\heartp$ product was the first way we computed a nice basis for $\nsbr{Y}_\alpha$ and is well-suited for explicit computation in the two-row case. Later, we realized that this is a special case of Lusztig's construction (Theorem \ref{t tsr product upper canbas}), which is theoretically cleaner.  We now explain Lusztig's construction in detail in this context and denote the product by $\heartl$.

Define the following $\field_\infty$-lattice of $\nsbr{Y}_\alpha$
\be \label{e L alpha}
\L_\alpha := \L_{(\alpha_1)} \tsr_{\field_\infty} \dots \tsr_{\field_\infty} \L_{(\alpha_l)} = \nsbr{p}_\alpha(\L_V \tsrvw_{\field_\infty} \L_W),
\ee
where the equality is by Proposition \ref{p NSC facts} (d).
Define the integral form
\be
\nsbr{Y}^\mathbf{A}_{\alpha} := \nswedge{{\alpha_1} }{X}^\mathbf{A} \tsr_{\mathbf{A}} \dots \tsr_{\mathbf{A}} \nswedge{{\alpha_l} }{X}^\mathbf{A}.
\ee

It follows from Theorem  \ref{t tsr product upper canbas} that
\begin{enumerate}
\item[] there is a unique  $\br{\cdot}$-invariant element  $T|_1 \heartl T|_2 \heartl \cdots \heartl T|_l$ of  $ \nsbr{Y}^\mathbf{A}_\alpha$ congruent to \\ $T|_1 \tsr T|_2 \tsr \cdots \tsr T|_l \mod \ui \L_\alpha$,  for any $T \in \text{NST}(\alpha)$.
\end{enumerate}
As will be justified by the next proposition, we may identify the element  $T|_1 \heartl T|_2 \heartl \cdots \heartl T|_l$ with  $T$. Then, with this identification and by Proposition \ref{p upper based tau}, $(\nsbr{Y}_\alpha, \text{NST}(\alpha))$ is a weak upper based $\Uqt$-module with balanced triple
$(\nsbr{Y}^\mathbf{A}_\alpha,\br{\L}_\alpha,\L_\alpha)$.

\begin{proposition}\label{p definitions of heart_sQ product agree}
The products $\heartp$ and $\heartl$ agree.
\end{proposition}
\begin{proof}
It suffices to show that the $\heartp$ product satisfies the characterizing conditions of the $\heartl$ product. Let $\alpha = (\beta,\gamma) \vDash_l^{\dx} r$ and $T \in \nsbr{Y}_\beta^{\mathbf{A}}, T' \in \nsbr{Y}_\gamma^{\mathbf{A}}$ and assume by induction that we have shown that  $T$ and  $T'$ are equal to the $\heartp$ and the $\heartl$  products of their columns.
It follows from the $\br{\cdot}$-invariance of the $\liftcvw{\mathbf{k}}{\mathbf{l}}$ and the proof of Proposition \ref{p NSC facts} (c) that $\nsbr{p}_{(r)}$ intertwines the $\br{\cdot}$-involution on $ \nsbr{\bT}$.  Since $\nsH_r$ acts faithfully on $\nsbr{\bT}$ when $\dv, \dw \geq r$, this implies that the minimal central idempotent of $\field \nsH_r$ corresponding to  $\nsbr{\epsilon}_-$ is $\br{\cdot}$-invariant.  Thus  $\nsbr{p}_\alpha$ intertwines the $\br{\cdot}$-involution, hence $\nsbr{p}_\alpha(T \heartvw T')$ is $\br{\cdot}$-invariant.

Next, we can write $T \heartvw T' \in T \tsr T' + \ui \L_V \tsrvw_{\field_\infty} \L_W$, which implies
\[ \nsbr{p}_\alpha(T \heartvw T') \in \nsbr{p}_\alpha(T \tsr T') + \ui \nsbr{p}_\alpha (\L_V \tsrvw_{\field_\infty} \L_W)= \nsbr{p}_\alpha(T \tsr T') + \ui \L_\alpha = T \tsr T' + \ui \L_\alpha, \]
where the first equality is by \eqref{e L alpha} and the last equality is simply because $T \in \nsbr{Y}_\beta, T' \in \nsbr{Y}_\gamma$, so  $\nsbr{p}_\alpha(T \tsr T') = \nsbr{p}_\beta(T) \tsr \nsbr{p}_\gamma(T') = T \tsr T'$.
Thus $T \heartp T' = \nsbr{p}_\alpha(T \heartvw T') = T \heartl T'$.
\end{proof}

\begin{remark}
This proposition is very similar to Proposition \ref{p heart commutes with projections}, the difference being that the projector  $\nsbr{p}_\alpha = \nsbr{p}_{\alpha_1} \tsr \dots \tsr \nsbr{p}_{\alpha_l}$ is more complicated than $\pi = \pi_1 \tsr \cdots \tsr \pi_l$.  A little extra care is needed to check that $\nsbr{p}_\alpha$ intertwines the $\br{\cdot}$-involution, but otherwise the proofs are essentially the same.
\end{remark}

\begin{remark}
We believe that the $\mathbf{A}$-module $\nsbr{p}_\alpha(\nsbr{\bT}_{\mathbf{A}})$ is not a good choice  for an integral form of $ \nsbr{Y}_\alpha$.  It can be strictly larger than the integral form $\nsbr{Y}^{\mathbf{A}}_\alpha$.  For example,
\[\nsbr{p}_{(2)}(\cvw{12}{12}) = \cvw{12}{12} \frac{\sQ_1}{[2]^2} = -\frac{1}{[2]}\cvw{21}{12} - \frac{1}{[2]}\cvw{12}{21} - \frac{2}{[2]^2}\cvw{21}{21} = -\frac{1}{[2]}\Big(\ctableaua{2 \\ 3} + \ctableaua{3 \\ 2}\Big) \notin \nsbr{Y}^\mathbf{A}_{(2)}\]
(the second equality can be computed using \eqref{e sQ on CC}).
\end{remark}

\begin{example}
Continuing Example \ref{ex heartp}, we compute the corresponding  $\heartns$ product of nonstandard columns:
from $c_{12} \tsr c_{11} = c_{1211} + \ui c_{1121} + \u^{-2} c_{1112}$ we deduce
\[\myvcenter{\ensuremath{\column{2 \\ 3}\downtsr\column{1\\3}}} := \big(\cvw{21}{12} + \frac{1}{[2]}\cvw{21}{21}\big) \tsr \cvw{21}{11} = \cvw{2121}{1211} + \ui\cvw{2121}{1121} + \u^{-2}\cvw{2121}{1112} + \frac{1}{[2]}\cvw{2121}{2111} =\]
\[\cvw{2121}{1211} + \frac{1}{[2]}\cvw{2121}{1121} + \frac{1}{[2]}\cvw{2121}{2111} + \u^{-2}\big( \cvw{2121}{1112}+ \frac{1}{[2]}\cvw{2121}{1121} \big) \equiv \cvw{2121}{1211} + \frac{1}{[2]}\cvw{2121}{1121} + \frac{1}{[2]}\cvw{2121}{2111} \mod \ui \L_{(2,2)}.\]
This last  equivalence follows from
\[ \cvw{2121}{1112}+ \frac{1}{[2]}\cvw{2121}{1121} = \cvw{21}{11} \heartvw \big(\cvw{21}{12} + \frac{1}{[2]}\cvw{21}{21}\big) = \cvw{21}{11} \tsr \big(\cvw{21}{12} + \frac{1}{[2]}\cvw{21}{21}\big) = \myvcenter{\ensuremath{\column{1 \\ 3}\newpad{\raisebox{-10pt}{\heartl}}\column{2\\3}}} = \ctableau{1\\3}{2\\3}. \]
To summarize,
\be \label{e 2121 1211 computation}
\cvw{2121}{1211} + \frac{1}{[2]}\cvw{2121}{1121} + \frac{1}{[2]}\cvw{2121}{2111} = \myvcenter{\ensuremath{\column{2 \\ 3}\downtsr\column{1\\3}}} - \u^{-2}\ctableau{1\\3}{2\\3} \equiv \myvcenter{\ensuremath{\column{2 \\ 3}\downtsr\column{1\\3}}} \mod \ui \L_{(2,2)}.
\ee
The left-hand quantity of \eqref{e 2121 1211 computation} is  $\br{\cdot}$-invariant, so it must be $\myvcenter{\ensuremath{\column{2 \\ 3}\newpad{\raisebox{-10pt}{\heartl}}\column{1\\3}}}$, in agreement with the computation of the $\heartp$ product in Example \ref{ex heartp}.
\end{example}
\begin{example}
Here is the result of a similar computation for $\alpha=(2,1)$:
\[\ctableau{2 \\ 4}{1} = \myvcenter{\ensuremath{\column{2 \\ 4}\downtsr\column{1}}} - \u^{-2}\ctableau{2\\3}{2},\quad  \ctableau{2 \\ 4}{1} = \cvw{211}{221} + \frac{1}{[2]}\cvw{211}{212}, \ \text{and } \ \ctableau{2 \\ 3}{2} = \cvw{211}{122} + \frac{1}{[2]}\cvw{211}{212}.\]
\end{example}
%

\begin{remark}
By  \textsection\ref{ss Tensor products of based modules}, $\heartns$ is associative: if $(\beta, \gamma,\delta) \vDash_l^{\dx} r$ and if $A,B,C$ are NST of shape $ \beta, \gamma,\delta$ respectively, then $A \heartns (B \heartns C) = (A \heartns B) \heartns C$ as elements of $\nsbr{Y}_{(\beta, \gamma,\delta)}$.  It is important to keep in mind that the product $\heartns$ depends implicitly on $(\beta, \gamma,\delta)$, even though this is not included in the notation.
\end{remark}

\subsection{The action of the Kashiwara operators and $\tau$ on NST}
We have shown that $(\nsbr{Y}_\alpha, \text{NST}(\alpha))$ is a weak upper based $\Uqt$-module; let  $(\L_\alpha, \mathcal{NST}(\alpha))$ be its upper crystal basis.  The action of $\tau$ on NST is deduced easily and the action of the Kashiwara operators  $\crystalusual{(F_i)}_V, \crystalusual{(F_j)}_W$,  $i \in [\dv-1], j \in [\dw-1]$ on  $\mathcal{NST}$ is given by the well-known rule for tensoring $U_q(\g_V)$ crystal bases (see e.g. \cite[Chapter 7]{HK}).  We now describe these actions explicitly in the two-row case.

\begin{proposition}\label{p tau on NST}
The action of $\tau$ on NSC$^{\, r}$ is given by
\be\label{e tau on NSC}
\tau(\text{NSC}_{P_V, P_W}) = (-1)^{\ell(\transpose{(Z_\lambda^{*})})} \text{NSC}_{P_W, P_V}, \text{ where } \lambda = \sh(P_V).
\ee
This can be made more explicit using
\be\label{e length of Z lambda star}
\ell(\transpose{(Z_\lambda^{*})}) \equiv \binom{n}{2} + \sum_i \binom{\lambda_i}{2} + i\binom{\lambda'_i}{2} \mod 2.
\ee
Moreover, for an NST $T$ with $l$ columns, $\tau(T) = \tau(T|_1) \heartns \cdots \heartns \, \tau(T|_l)$.
\end{proposition}
\begin{proof}
Formula \eqref{e tau on NSC} is straightforward from definitions and the fact $\tau(\liftcvw{\mathbf{k}}{\mathbf{l}}) = \liftcvw{\mathbf{l}}{\mathbf{k}}$. The length computation \eqref{e length of Z lambda star} is slightly involved and we omit the proof. The last statement is exactly Proposition \ref{p upper based tau} in this setting.
\end{proof}

For the remainder of this section set $\dv = \dw=2.$

\begin{definition}\label{d NST diagram}
The $V$-diagram (resp. $W$-diagram) of an NST  $T$ of shape $\alpha$ is the diagram obtained from its $V$-word (resp. $W$-word) according to the rule in \textsection\ref{s graphical calculus}.  The  \emph{$V$-arcs} (resp.  \emph{$W$-arcs}) of  $T$ are the arcs of the $V$-diagram (resp.  $W$-diagram) of  $T$.  Internal and external $V$-arcs (resp. $W$-arcs) are defined as in Definition \ref{d pairing internal external} with  $i_j = \alpha_j$ and  $\lambda^{(j)} = \sh(P_V)$ (resp. $\lambda^{(j)} = \sh(P_W)$) for all $j \in [l]$, where $P_V,P_W$ are defined by  $T|_j = \text{NSC}_{P_V,P_W}$.  An \emph{arc} of $T$ is either a $V$-arc or a $W$-arc of  $T$.
\end{definition}
Let  $\varphi_V, \varphi_W$ be as in \eqref{e phi varepsilon definition}, defined using $\crystal{F}_V$ and $\crystal{F}_W$, respectively ($\crystal{F}$ is the global Kashiwara operator defined in \eqref{e global Kashiwara operators}).  Then for any NST  $T$, the statistic $\varphi_V(T)$ (resp. $\varphi_W(T)$) is the number of unpaired 1's in the $V$-diagram (resp. $W$-diagram) of $T$.

For an NST $T$ with $V$-word $\mathbf{k}$, let $\F_{(j)V}(T)$ be the NST corresponding to $\F_{(j)}(\mathbf{k})$ (as defined in Theorem \ref{t FK F action on c basis}), defined precisely as follows: if $\F_{(j)}(\mathbf{k})$ has an extra internal arc, then $\F_{(j)V}(T) = 0$, and otherwise $\F_{(j)V}(T)$ is the NST obtained from  $T$ by replacing the column $T|_c$ containing the $j$-th unpaired $1$ in  $\mathbf{k}$ with  $\crystal{F}_V(T|_c)$ if the  $j$-th unpaired $1$ lies in the $c$-th column of $V$-diagram$(T)$. The NST $\F_{(j)W}(T)$,  $\E_{(j)V}(T)$, and $\E_{(j)W}(T)$ are defined in a similar way using $W$-word in place of $V$-word and $\E_{(j)}(\mathbf{k})$ in place of $\F_{(j)}(\mathbf{k})$ as appropriate.

\begin{proposition} \label{p crystal action on Y}
The pair $(\nsbr{Y}_\alpha, \text{NST}(\alpha))$ is a weak upper based $\Uqt$-module with global Kashiwara operators given by
\be
\begin{array}{ll}
\crystal{F}_V(T) &= \F_{(\varphi_V(T))V}(T), \\
\crystal{F}_W(T) &= \F_{(\varphi_W(T))W}(T).
\end{array}
\ee
Thus the highest weight  $ \text{NST}(\alpha)$ are those whose $V$-diagram and $W$-diagram have no unpaired 2's.
The action of $\tau$ on NST$(\alpha)$ is given by $\tau(T) = (-1)^j T'$, where $T'$ is obtained from $T$ by changing 2's to 3's and 3's to 2's with the exception that columns $\ctableausmalla{1 \\ 2 \\ 3},$ $\ctableausmalla{2 \\ 3 \\ 4}$ do not change, and $j$ is the number of parts of $\alpha$ equal to 3 or 4.
\end{proposition}

The following stronger result for the action of $F_V$ and $F_W$ is also useful.
\begin{proposition}\label{p F action on heartsQ basis}
For an NST $T$ 
there holds
\[
\begin{array}{ll}
F_V T &= \sum_{j=1}^{\varphi_V(T)} [j]\F_{(j)V}(T), \\
F_W T &= \sum_{j=1}^{\varphi_W(T)} [j]\F_{(j)W}(T).
\end{array}
\]
\end{proposition}
\begin{proof}
This is a special case of Proposition \ref{t FK F action on c basis} since  $\nswedge{r }{X}$, considered as an upper based  $U_q(\g_V)$-module, is isomorphic to a direct sum of $(V_{(r-i,i)}, B((r-i,i)))$.
\end{proof}

\section{A global crystal basis for two-row Kronecker coefficients}
\label{s global crystal basis for two-row Kronecker coefficients}
For the remainder of this paper, set  $\dv =\dw =2$ (the two-row case). We now come to our main result on the two-row Kronecker problem and  the deepest canonical basis theory of this paper.

We show that for any $\nu \vdash r$,  the $ \O(GL_q(\nsbr{X}))$-comodule quotient $\nsbr{X}_\nu$ of $\nsbr{Y}_{\nu'}$
defined in  \textsection\ref{ss the approach of Adsul, Sohoni, and Subrahmanyam} satisfies
$\Res_{U^\tau} (\nsbr{X}_\nu|_{\u = 1}) \cong \Res_{U^\tau} (X_\nu|_{\u=1})$, where $X_\nu|_{\u=1}$ is the  $U(\g_X)$-module of highest weight $\nu$ and  $U^\tau = U(\gl_2) \wr \S_2$.
Recall from \textsection\ref{ss the approach of Adsul, Sohoni, and Subrahmanyam} that $\nsbr{X}_\nu$ is defined to be  $ \nsbr{Y}_{\nu'} / \nsbr{Y}_{\gdneq \nu'}$, where the submodule $\nsbr{Y}_{\gdneq \alpha}$ of  $\nsbr{Y}_{\alpha}$ (for simplicity,  $ \nsbr{Y}_{\gdneq \alpha}$ was discussed only for partitions $\alpha$ in  \textsection\ref{ss the approach of Adsul, Sohoni, and Subrahmanyam}, but it can be defined for any composition $\alpha \vDash^\dx r$) is defined ``by hand'' for  $\ell(\alpha)=2$ and for  $\ell(\alpha) > 2$ is defined to be the (generally, not direct) sum over all  $i \in [l-1]$ of
\[\nsbr{Y}_{\gdneq^i \alpha}:= \nsbr{Y}_{(\alpha_1,\ldots,\alpha_{i-1})} \tsr \nsbr{Y}_{\gdneq (\alpha_{i},\alpha_{i+1})} \tsr \nsbr{Y}_{(\alpha_{i+2},\ldots,\alpha_l)}. \]

We define a subset $\pNSTC(\nu) \subseteq \nsbr{X}_\nu$ to be the image of (a rescaled version of) a certain subset of NST$(\nu')$, and we show that  $\pNSTC(\nu)$ is
a global crystal basis of $\nsbr{X}_\nu$.
This gives an elegant solution to the two-row Kronecker problem: the Kronecker coefficient $g_{\lambda \mu \nu}$ is equal to the number of highest weight $\pNSTC(\nu)$ of weight $(\lambda, \mu)$.
This section is devoted to the algebraic portion of the proof of this as well as the verification that  $\nsbr{X}_\nu$ behaves correctly at $\u =1$.  One of the main difficulties is that  $\nsbr{Y}_{\gdneq^i \alpha}$ is not easily expressed in terms of the basis NST$(\alpha)$.  We develop some tools  to remedy this:  a grading on $\nsbr{Y}_{\alpha}$ (\textsection\ref{ss invariants}) and a canonical basis for $\nsbr{Y}_{\gdneq^i \alpha}$ (\textsection\ref{ss agrees in associated graded}).

\subsection{Invariants}
\label{ss invariants}
As defined explicitly below, an invariant is a minimal NST that is killed by $F_V, F_W, E_V$, and $E_W$.  This allows us to define a grading on $\nsbr{Y}_{\alpha}$ corresponding to how many invariants an NST contains.  This will help us organize the relations satisfied by the image of NST$(\nu')$ in $\nsbr{X}_\nu$.

An \emph{invariant} is an NST equal to one of:
\[\ctableau{4}{1}, \ctableau{2 \\ 4}{1\\3}, \ctableau{3 \\ 4}{1 \\ 2}, \ctableau{2 \\ 3 \\ 4}{1 \\ 2 \\ 3}, \ctableau{2 \\ 3 \\4}{1}, \ctableau{4}{1 \\ 2\\3}\text{ or } \ctableaua{1 \\ 2 \\ 3 \\ 4}.\]
If the columns of an invariant have the same height $j$, then  $j$ is the \emph{height} of the invariant.  We will not be too interested in the invariants  $\ctableausmall{2 \\ 3 \\4}{1}$ and $\ctableausmall{4}{1\\2\\3}$ because they belong to $\nsbr{Y}_{\gdneq (3,1)}$ and $\nsbr{Y}_{\gdneq (1,3)}$, respectively.
\begin{definition}\label{d invariants}
An \emph{invariant column pair} of an NST  $T$ is a pair of columns of $T$ that are paired by two arcs (see  Definition \ref{d NST diagram}).

The \emph{invariant record} of an NST $T$ is the tuple $(i_4, i_3, i_2, i_1)$, where
\[
\begin{array}{l}
i_4 \text{ is the number of height-4 columns}, \\
i_j \text{ is the number of invariant column pairs of height-$j$ of  $T$, \  ($j = 1,2,3$).}
\end{array}
\]
The \emph{degree} of $T$, denoted  $\deg(T)$, is  $i_4$ plus the number of invariant column pairs. 
The  \emph{invariant-free part} of an NST $T$ is the (possibly empty) NST obtained by removing all invariant column pairs and all height-4 columns.
\end{definition}
The columns of an invariant column pair have no $V$- or  $W$-arcs with an end outside the pair, so the definition of the invariant-free part is sound in that after all invariant column pairs are removed, the resulting NST has no invariant column pairs. It is easy to check that the invariants listed above, except the height-4 column, are invariant column pairs, and all invariant column pairs are of this form (after removing the columns not in the pair).


Before introducing an associated graded of $\nsbr{Y}_\alpha$, we recall some algebraic generalities.
Let $0 \subseteq X_r \subseteq X_{r-1} \subseteq \dots \subseteq X_0 = X$ be a filtered $R$-module.  The associated graded of $X$ is  the graded $R$-module $\gr(X) := \bigoplus_{i=0}^{r-1} X_i/X_{i + 1}$.  The rule  $X \mapsto \gr(X)$ is a functor from filtered $R$-$\Mod$ to graded $R$-$\Mod$.
A submodule (resp. quotient module) $M$ of $X$ inherits a filtration from that of  $X$, so we write $\gr(M)$ for the associated graded module; it is a submodule (resp. quotient module) of $\gr(X)$.

For $x \in X$, the \emph{degree of $x$}, denoted $\deg(x)$, is the largest integer $h$ such that $x \in X_h$.
For $x \in X_h$, let $\grin_h(x)$ denote the image of $x$ under the composition $X_h \twoheadrightarrow X_h/ X_{h+1} \hookrightarrow \gr(X)$.
And for $x \in X$, set $\grin(x) = \grin_{\deg(x)}(x)$.


\begin{propdef} \label{pd gr Y alpha}
Let $\alpha \vDash^{\dx}_l r$, as usual.  Set
\[
\begin{array}{ccl}
\text{NST}(\alpha)_{\geq h} &:=& \{T \in \text{NST}(\alpha): \deg(T) \geq h \}, \\
\text{NST}(\alpha)_{h} &:=& \{T \in \text{NST}(\alpha): \deg(T) = h \},
\end{array}
\]
for  $h \geq 0$, and let
$(\nsbr{Y}_\alpha)_h$, $(\L_\alpha)_h$, $(\nsbr{Y}_\alpha^\mathbf{A})_h$ be the $\field, \field_\infty,$ and $\mathbf{A}$ span of  $\text{NST}(\alpha)_{\geq h}$, respectively.
The pair $((\nsbr{Y}_\alpha)_h, \text{NST}(\alpha)_{\geq h})$ is a weak upper based $\Uqt$-module with balanced triple  $((\nsbr{Y}_\alpha^\mathbf{A})_h,\br{(\L_\alpha)}_h,(\L_\alpha)_h)$.
The filtration
\[0\subseteq ((\nsbr{Y}_\alpha)_l, \text{NST}(\alpha)_{\geq l}) \subseteq \ldots \subseteq ((\nsbr{Y}_\alpha)_1, \text{NST}(\alpha)_{\geq 1}) \subseteq ((\nsbr{Y}_\alpha)_0, \text{NST}(\alpha)_{\geq 0}) = (\nsbr{Y}_\alpha,\text{NST}(\alpha))\]
is a filtration of weak upper based $\Uqt$-modules, and hence  $(\gr(\nsbr{Y}_\alpha), \grin(\text{NST}(\alpha)))$ is a weak upper based  $\Uqt$-module.
\end{propdef}
\begin{proof}
The inclusion  $(\nsbr{Y}_\alpha)_h \hookrightarrow \nsbr{Y}_\alpha$ is a
$\Uqt$-module homomorphism because applying a Chevalley generator
or  $\tau$ to an NST  $T$ yields a linear combination  $\sum_i c_i T^i$ of NST such that every invariant column pair of  $T^i$ is an invariant column pair of $T$; see Proposition \ref{p F action on heartsQ basis}.
\end{proof}
Note that  $(\gr(\nsbr{Y}_\alpha), \grin(\text{NST}(\alpha)))$ and  $(\nsbr{Y}_\alpha, \text{NST}(\alpha))$ are not isomorphic as weak upper based $\Uqt$-modules.


\subsection{Two-column moves}
\label{ss two-column relations}
We now define $\Uqt$-submodules $\nsbr{Y}_{\gdneq\gamma} \subseteq \nsbr{Y}_\gamma$ for  $\gamma$ a composition of length 2.  Let $\gamma'$ be the conjugate of the partition obtained by sorting the parts of $\gamma$ in weakly decreasing order.
We are in luck: it turns out that $\Res_{U^\tau} (X_{\gamma'}|_{\u=1})$ is multiplicity-free, so in order for $\Res_{\Uqt} \nsbr{Y}_\gamma/\nsbr{Y}_{\gdneq \gamma}$ to be a  $q$-analog of $\Res_{U^\tau} (X_{\gamma'}|_{\u=1})$, the submodule $\nsbr{Y}_{\gdneq\gamma}$ must be a direct sum of certain $\Uqt$-irreducibles of  $\nsbr{Y}_\gamma$, which are easily computed.  In other words, there is only one way to define $\nsbr{Y}_{\gdneq\gamma}$ so that  $\nsbr{Y}_\gamma/\nsbr{Y}_{\gdneq \gamma}$ is what it is supposed to be at $q=1$.

The Figures \ref{f straightening11}--\ref{f straightening33} below serve several purposes: they give, for each partition $\gamma$ of length 2 such that $\gamma_1 \leq 3$, an explicit description of $\nsbr{Y}_{\gdneq \gamma}$ by depicting a basis $\text{NST}(\gdneq \gamma)$ for this space, and these bases will play an important role in subsequent arguments; they give examples of the action of $F_V$, $F_W$, and $\tau$ on NST as determined by Propositions \ref{p crystal action on Y} and \ref{p F action on heartsQ basis} (the horizontal (resp. vertical) arrows give the action of $F_V$ (resp. $F_W$) on the basis elements and the labels on the arrows indicate the coefficient;
the action of $\tau$ is only given for highest weight basis elements to avoid cluttering the diagrams); in some cases, the basis elements agree with NST of a different shape by thinking of all the $\nsbr{Y}_\gamma$ with  $\gamma \vDash^\dx_2 r$ as subspaces of $\nsbr{X}^{\tsr r}$, and we indicate this in the figures.

\begin{figure}[H]
\begin{tikzpicture}[xscale =  5.5,yscale = 4]
\tikzstyle{vertex}=[inner sep=-1pt, outer sep=5.3pt, fill = white]
\tikzstyle{edge} = [draw, thick, ->,black]
\tikzstyle{aedge} = [draw, thick, <->,black]
\tikzstyle{LabelStyleH} = [text=black, anchor=south]
\tikzstyle{LabelStyleV} = [text=black, anchor=east]
\tikzstyle{LabelStyleR} = [text=black, anchor=west]
\tikzstyle{LabelStyleB} = [text=black, anchor=north]
\foreach \x / \y in {.3 / -.4}{
    \node[vertex] (21) at (1+\x,0+\y) {$\myvcenter{\ensuremath{\column{2}\column{1}}} = \myvcenter{\ensuremath{\column{1 \\ 2}}}$};
    \node[vertex] (23) at (0+\x,0+\y) {${\myvcenter{\ensuremath{\column{2}\column{3}}} \ + \ \frac{1}{[2]} \myvcenter{\ensuremath{\column{4}\column{1}}} \atop
    = \ \ \myvcenter{\ensuremath{\column{3 \\ 2}}} }$};
    \node[vertex] (43) at (-1+\x,0+\y) {$\myvcenter{\ensuremath{\column{4}\column{3}}} = \myvcenter{\ensuremath{\column{3 \\ 4}}}$};
}
    \node[vertex] (31)  at (0,1) {$\myvcenter{\ensuremath{\column{3}\column{1}}} = \myvcenter{\ensuremath{\column{1 \\ 3}}}$};
    \node[vertex] (32) at (0,0) {${\myvcenter{\ensuremath{\column{3}\column{2}}} \ + \ \frac{1}{[2]} \myvcenter{\ensuremath{\column{4}\column{1}}} \atop
    = \ \ \myvcenter{\ensuremath{\column{2 \\ 3}}} }$};
    \node[vertex] (42) at (0,-1) {$\myvcenter{\ensuremath{\column{4}\column{2}}} = \myvcenter{\ensuremath{\column{2 \\ 4}}}$};

\draw[edge] (21) to node[LabelStyleH]{$[2]$} (23);
\draw[edge] (31) to node[LabelStyleV]{$[2]$} (32);
\draw[edge] (32) to node[LabelStyleV]{$1$} (42);
\draw[edge] (23) to node[LabelStyleH]{$1$} (43);
\draw[aedge] (21) to node[LabelStyleH]{$\tau$} (31);
\draw[edge] (21) to node[LabelStyleB]{$F_V $} (23);
\draw[edge] (31) to node[LabelStyleR]{$F_W $} (32);
\foreach \x / \y in {.3 / -.4}{
    \node[vertex] (21) at (1+\x,0+\y) {$\myvcenter{\ensuremath{\column{2}\column{1}}} = \myvcenter{\ensuremath{\column{1 \\ 2}}}$};
    \node[vertex] (23) at (0+\x,0+\y) {${\myvcenter{\ensuremath{\column{2}\column{3}}} \ + \ \frac{1}{[2]} \myvcenter{\ensuremath{\column{4}\column{1}}} \atop
    = \ \ \myvcenter{\ensuremath{\column{3 \\ 2}}} }$};
    \node[vertex] (43) at (-1+\x,0+\y) {$\myvcenter{\ensuremath{\column{4}\column{3}}} = \myvcenter{\ensuremath{\column{3 \\ 4}}}$};
}
    \node[vertex] (31)  at (0,1) {$\myvcenter{\ensuremath{\column{3}\column{1}}} = \myvcenter{\ensuremath{\column{1 \\ 3}}}$};
    \node[vertex] (32) at (0,0) {${\myvcenter{\ensuremath{\column{3}\column{2}}} \ + \ \frac{1}{[2]} \myvcenter{\ensuremath{\column{4}\column{1}}} \atop
    = \ \ \myvcenter{\ensuremath{\column{2 \\ 3}}} }$};
    \node[vertex] (42) at (0,-1) {$\myvcenter{\ensuremath{\column{4}\column{2}}} = \myvcenter{\ensuremath{\column{2 \\ 4}}}$};

\end{tikzpicture}
\caption{The basis NST$(\gdneq (1,1))$ of $\nsbr{Y}_{\gdneq (1,1)}$, which consists of graded and nonintegral $\gdneq \text{NST}$.}
\label{f straightening11}
\end{figure}

\begin{figure}[H]
\begin{tikzpicture}[xscale = 2.7, yscale= 2.2]
\tikzstyle{vertex}=[inner sep=0pt, outer sep=3pt]
\tikzstyle{edge} = [draw, thick, ->,black]
\tikzstyle{aedge} = [draw, thick, <->,black]
\tikzstyle{LabelStyleH} = [text=black, anchor=south]
\tikzstyle{LabelStyleV} = [text=black, anchor=east]
\tikzstyle{LabelStyleVw} = [text=black, anchor=west]

\node[vertex] (v1) at (1,1){$  \myvcenter{\ensuremath{\column{2 \\ 3}\column{1}}} - \myvcenter{\ensuremath{\column{3 \\ 2}\column{1}}} = \myvcenter{\ensuremath{\column{1 \\ 2 \\ 3}}} $};
\node[vertex] (v2) at (-1,1){$\myvcenter{\ensuremath{\column{2 \\ 3}\column{3}}} - \myvcenter{\ensuremath{\column{3 \\ 4}\column{1}}} = \myvcenter{\ensuremath{\column{1 \\ 3 \\ 4}}}  $};
\node[vertex] (v3) at (-1,-1){$\myvcenter{\ensuremath{\column{2 \\ 4}\column{3}}} - \myvcenter{\ensuremath{\column{3 \\ 4}\column{2}}} = \myvcenter{\ensuremath{\column{2 \\ 3 \\ 4}}}  $};
\node[vertex] (v4) at (1,-1){$\myvcenter{\ensuremath{\column{2 \\ 4}\column{1}}} - \myvcenter{\ensuremath{\column{3 \\ 2}\column{2}}} = \myvcenter{\ensuremath{\column{1 \\ 2 \\ 4}}}   $};
\draw[edge] (v1) to node[LabelStyleV]{$1$} (v4);
\draw[edge] (v1) to node[LabelStyleH]{$1$} (v2);
\draw[edge] (v2) to node[LabelStyleV]{$1$} (v3);
\draw[edge] (v4) to node[LabelStyleH]{$1$} (v3);
\draw[aedge, loop right, looseness=3] (v1) to node[LabelStyleVw]{-$\tau$} (v1);
\end{tikzpicture}
\caption{The basis NST$(\gdneq (2,1))$ of $\nsbr{Y}_{\gdneq (2,1)}$, which consists of degree-preserving and integral $\gdneq \text{NST}$.}
\label{f straightening21}
\end{figure}

\begin{figure}[H]
\begin{tikzpicture}[xscale=5, yscale = 4]
\tikzstyle{vertex}=[inner sep=0pt, outer sep=3pt]
\tikzstyle{edge} = [draw, thick, ->,black]
\tikzstyle{aedge} = [draw, thick, <->,black]
\tikzstyle{LabelStyleH} = [text=black, anchor=south]
\tikzstyle{LabelStyleV} = [text=black, anchor=east]

\foreach \x / \y in {.3 / -.3}{
    \node[vertex] (21) at (1+\x,0+\y) {$\myvcenter{\ensuremath{\column{3 \\ 2}\column{1 \\ 2}}}$};
    \node[vertex] (23) at (.1+\x,0+\y) {$\myvcenter{\ensuremath{\column{3 \\ 2}\column{3 \\ 2}}} + \frac{1}{[2]}\myvcenter{\ensuremath{\column{3\\4}\column{1\\2}}}$};
    \node[vertex] (43) at (-1+\x*.5,0+\y) {$\myvcenter{\ensuremath{\column{3 \\ 4}\column{3 \\ 2}}}$};
}
    \node[vertex] (31)  at (0,1) {$\myvcenter{\ensuremath{\column{2 \\ 3}\column{1 \\ 3}}}$};
    \node[vertex] (32) at (0,0) {$\myvcenter{\ensuremath{\column{2 \\ 3}\column{2 \\ 3}}} + \frac{1}{[2]}\myvcenter{\ensuremath{\column{2\\4}\column{1\\3}}}$};
    \node[vertex] (42) at (0,-1) {$\myvcenter{\ensuremath{\column{2 \\ 4}\column{2 \\ 3}}}$};
\draw[edge] (21) to node[LabelStyleH]{$[2]$} (23);
\draw[edge] (31) to node[LabelStyleV]{$[2]$} (32);
\draw[edge] (32) to node[LabelStyleV]{$1$} (42);
\draw[edge] (23) to node[LabelStyleH]{$1$} (43);
\draw[aedge] (21) to node[LabelStyleH]{$\tau$} (31);
\end{tikzpicture}
\caption{The graded elements of the basis NST$(\gdneq (2,2))$ of $\nsbr{Y}_{\gdneq (2,2)}$, which are all nonintegral.}
\label{f straightening22a}
\end{figure}


\begin{figure}[H]
\begin{tikzpicture}[xscale = 2, yscale=1.5]

\tikzstyle{vertex}=[inner sep=0pt, outer sep=3pt]

\tikzstyle{edge} = [draw, thick, ->,black]
\tikzstyle{aedge} = [draw, thick, <->,black]

\tikzstyle{LabelStyleH} = [text=black, anchor=south]

\tikzstyle{LabelStyleV} = [text=black, anchor=east]
\tikzstyle{LabelStyleVw} = [text=black, anchor=west]

\node[vertex] (v1) at (0,-2){$\myvcenter{\ensuremath{\column{3 \\ 2}\column{2 \\ 4}}} - \myvcenter{\ensuremath{\column{2 \\ 4}\column{3 \\ 2}}}$};
\node[vertex] (v2) at (2,2){$\myvcenter{\ensuremath{\column{1 \\ 2}\column{1 \\ 3}}} - \myvcenter{\ensuremath{\column{1 \\ 3}\column{1 \\ 2}}}$};
\node[vertex] (v3) at (-2,-2){$\myvcenter{\ensuremath{\column{3 \\ 4}\column{2 \\ 4}}} - \myvcenter{\ensuremath{\column{2 \\ 4}\column{3 \\ 4}}}$};
\node[vertex] (v4) at (2,0){$\myvcenter{\ensuremath{\column{1 \\ 2}\column{2 \\ 3}}} - \myvcenter{\ensuremath{\column{2 \\ 3}\column{1 \\ 2}}}$};
\node[vertex] (v5) at (0,2){$ \myvcenter{\ensuremath{\column{3 \\ 2}\column{1 \\ 3}}} - \myvcenter{\ensuremath{\column{1 \\ 3}\column{3 \\ 2}}}$};
\node[vertex] (v6) at (-2,2){$ \myvcenter{\ensuremath{\column{3 \\ 4}\column{1 \\ 3}}} - \myvcenter{\ensuremath{\column{1 \\ 3}\column{3 \\ 4}}}$};
\node[vertex] (v7) at (-2,0){$  \myvcenter{\ensuremath{\column{3 \\ 4}\column{2 \\ 3}}} - \myvcenter{\ensuremath{\column{2 \\ 3}\column{3 \\ 4}}}$};
\node[vertex] (v8) at (2,-2){$\myvcenter{\ensuremath{\column{1 \\ 2}\column{2 \\ 4}}} - \myvcenter{\ensuremath{\column{2 \\ 4}\column{1 \\ 2}}}$};
\node[vertex] (v9) at (0,0){$ \myvcenter{\ensuremath{\column{3 \\ 2}\column{2 \\ 3}}} - \myvcenter{\ensuremath{\column{2 \\ 3}\column{3 \\ 2}}}$};

\draw[edge] (v1) to node[LabelStyleH]{$1$} (v3);
\draw[edge] (v2) to node[LabelStyleH]{$[2] $} (v5);
\draw[edge] (v2) to node[LabelStyleV]{$[2] $} (v4);
\draw[aedge, loop right, looseness=2] (v2) to node[LabelStyleVw]{-$\tau$} (v2);
\draw[edge] (v4) to node[LabelStyleH]{$[2] $} (v9);
\draw[edge] (v4) to node[LabelStyleV]{$1$} (v8);
\draw[edge] (v5) to node[LabelStyleH]{$1$} (v6);
\draw[edge] (v5) to node[LabelStyleV]{$[2] $} (v9);
\draw[edge] (v6) to node[LabelStyleV]{$[2] $} (v7);
\draw[edge] (v7) to node[LabelStyleV]{$1$} (v3);
\draw[edge] (v8) to node[LabelStyleH]{$[2] $} (v1);
\draw[edge] (v9) to node[LabelStyleV]{$1$} (v1);
\draw[edge] (v9) to node[LabelStyleH]{$1$} (v7);

\end{tikzpicture}
\caption{Some of the degree-preserving elements of the basis NST$(\gdneq (2,2))$ of $\nsbr{Y}_{\gdneq (2,2)}$, which are all integral.}
\label{f straightening22b}
\end{figure}

\begin{figure}[H]
\begin{tikzpicture}[scale = 4]
\tikzstyle{vertex}=[inner sep=0pt, outer sep=3pt]
\tikzstyle{edge} = [draw, thick, ->,black]
\tikzstyle{aedge} = [draw, thick, <->, black]
\tikzstyle{LabelStyleH} = [text=black, anchor=south]
\tikzstyle{LabelStyleV} = [text=black, anchor=east]
\tikzstyle{LabelStyleVw} = [text=black, anchor=west]
\node[vertex] (32) at (0,0) {$\myvcenter{\ensuremath{\ctableau{2\\4}{1\\3} - \ctableau{3\\4}{1\\2}}} = \myvcenter{\ensuremath{\column{2 \\ 3 \\ 4}\column{1}}}$};
\draw[aedge, loop right, looseness=2] (32) to node[LabelStyleVw]{-$\tau$} (32);
\end{tikzpicture}
\caption{A degree-preserving and integral element of the basis NST$(\gdneq (2,2))$ of $\nsbr{Y}_{\gdneq (2,2)}$.}
\label{f straightening22c}
\end{figure}

\begin{figure}[H]
\begin{tikzpicture}[scale = 4]
\tikzstyle{vertex}=[inner sep=0pt, outer sep=3pt]
\tikzstyle{edge} = [draw, thick, ->,black]
\tikzstyle{aedge} = [draw, thick, <->,black]
\tikzstyle{LabelStyleH} = [text=black, anchor=south]
\tikzstyle{LabelStyleV} = [text=black, anchor=east]
\tikzstyle{LabelStyleVw} = [text=black, anchor=west]
\node[vertex] (32) at (0,0) {$\myvcenter{\ensuremath{\column{2 \\ 3 \\ 4}\column{1}}} = \myvcenter{\ensuremath{\column{1 \\ 2\\ 3 \\ 4}}} $};
\draw[aedge, loop right, looseness=2] (32) to node[LabelStyleVw]{-$\tau$} (32);
\end{tikzpicture}
\caption{The element of the basis NST$(\gdneq (3,1))$ of $\nsbr{Y}_{\gdneq (3,1)}$, which is graded and integral.}
\label{f straightening31}
\end{figure}

\begin{figure}[H]
\begin{tikzpicture}[xscale = 2.4,yscale=2]

\tikzstyle{vertex}=[inner sep=0pt, outer sep=3pt]

\tikzstyle{edge} = [draw, thick, ->,black]
\tikzstyle{aedge} = [draw, thick, <->,black]
\tikzstyle{LabelStyleH} = [text=black, anchor=south]

\tikzstyle{LabelStyleV} = [text=black, anchor=east]
\tikzstyle{LabelStyleVw} = [text=black, anchor=west]

\node[vertex] (v1) at (1,1){$\myvcenter{\ensuremath{\column{1 \\ 2 \\ 4}\column{1 \\ 3}}} + \myvcenter{\ensuremath{\column{1 \\ 3 \\ 4}\column{1 \\ 2}}}$};
\node[vertex] (v2) at (-1,-1){$ \myvcenter{\ensuremath{\column{2 \\ 3 \\ 4}\column{2 \\ 3}}} + \myvcenter{\ensuremath{\column{2 \\ 3 \\ 4}\column{3 \\ 2}}}$};
\node[vertex] (v3) at (-1,1){$\myvcenter{\ensuremath{\column{2 \\ 3 \\ 4}\column{1 \\ 3}}}  + \myvcenter{\ensuremath{\column{1 \\ 3 \\ 4}\column{3 \\ 2}}}$};
\node[vertex] (v4) at (1,-1){$\myvcenter{\ensuremath{\column{1 \\ 2 \\ 4}\column{2 \\ 3}}} + \myvcenter{\ensuremath{\column{2 \\ 3 \\ 4}\column{1 \\ 2}}}$};
\draw[edge] (v1) to node[LabelStyleV]{$1$} (v4);
\draw[edge] (v1) to node[LabelStyleH]{$1$} (v3);
\draw[edge] (v3) to node[LabelStyleV]{$1$} (v2);
\draw[edge] (v4) to node[LabelStyleH]{$1$} (v2);
\draw[aedge, loop right, looseness=3] (v1) to node[LabelStyleVw]{-$\tau$} (v1);

\end{tikzpicture}
\caption{The basis NST$(\gdneq (3,2))$ of $\nsbr{Y}_{\gdneq (3,2)}$, which consists of degree-preserving and integral $\gdneq \text{NST}$.}
\label{f straightening32}
\end{figure}
\begin{figure}[H]
\begin{tikzpicture}[xscale = 5, yscale=4]
\tikzstyle{vertex}=[inner sep=-1pt, outer sep=5.3pt]
\tikzstyle{edge} = [draw, thick, ->,black]
\tikzstyle{aedge} = [draw, thick, <->,black]
\tikzstyle{LabelStyleH} = [text=black, anchor=south]
\tikzstyle{LabelStyleV} = [text=black, anchor=east]

\foreach \x / \y in {.43 / -.43}{
    \node[vertex] (21) at (1+\x,0+\y) {$\myvcenter{\ensuremath{\column{1 \\ 2 \\ 4}\column{1 \\ 2 \\ 3}}}$};
    \node[vertex] (23) at (0+\x,0+\y) {$\myvcenter{\ensuremath{\column{1\\ 2\\ 4}\column{1\\ 3 \\4}}} \ + \ \frac{1}{[2]} \myvcenter{\ensuremath{\column{2\\ 3\\4}\column{1\\2\\3}}}$};
    \node[vertex] (43) at (-1+\x*.5,0+\y) {$\myvcenter{\ensuremath{\column{2 \\ 3 \\ 4 }\column{1 \\ 3 \\ 4}}}$};
}
    \node[vertex] (31)  at (0,1) {$\myvcenter{\ensuremath{\column{1 \\ 3 \\ 4 }\column{1 \\ 2 \\ 3}}}$};
    \node[vertex] (32) at (0,0) {$\myvcenter{\ensuremath{\column{1\\ 3\\ 4}\column{1\\ 2 \\4}}} \ + \ \frac{1}{[2]} \myvcenter{\ensuremath{\column{2\\ 3\\4}\column{1\\2\\3}}}$};
    \node[vertex] (42) at (0,-1.2) {$\myvcenter{\ensuremath{\column{2 \\ 3 \\ 4 }\column{1 \\ 2 \\ 4}}}$};
\draw[edge] (21) to node[LabelStyleH]{$[2]$} (23);
\draw[edge] (31) to node[LabelStyleV]{$[2]$} (32);
\draw[edge] (32) to node[LabelStyleV]{$1$} (42);
\draw[edge] (23) to node[LabelStyleH]{$1$} (43);
\draw[aedge] (21) to node[LabelStyleH]{$\tau$} (31);

\end{tikzpicture}
\caption{The basis NST$(\gdneq (3,3))$ of $\nsbr{Y}_{\gdneq (3,3)}$, which consists of graded and nonintegral $\gdneq \text{NST}$.}
\label{f straightening33}
\end{figure}
\subsection{Invariant moves}
\label{ss invariant moves}
\begin{figure}[H]
\begin{tikzpicture}[scale = 3]
\tikzstyle{vertex}=[inner sep=0pt, outer sep=3pt]
\tikzstyle{edge} = [draw, thick, ->,black]
\tikzstyle{aedge} = [draw, thick, <->,black]
\tikzstyle{LabelStyleH} = [text=black, anchor=south]
\tikzstyle{LabelStyleV} = [text=black, anchor=east]
\tikzstyle{LabelStyleVw} = [text=black, anchor=west]

\node[vertex] (v1) at (1,1){$\myvcenter{\ensuremath{\column{1}\column{4}\column{1}}} -
 \myvcenter{\ensuremath{\column{4}\column{1}\column{1}}}$};
\node[vertex] (v2) at (-1,1){$\myvcenter{\ensuremath{\column{3}\column{4}\column{1}}} -
 \myvcenter{\ensuremath{\column{4}\column{1}\column{3}}}$};
\node[vertex] (v3) at (-1,-1){$\myvcenter{\ensuremath{\column{4}\column{4}\column{1}}} -
 \myvcenter{\ensuremath{\column{4}\column{1}\column{4}}}$};
\node[vertex] (v4) at (1,-1){$\myvcenter{\ensuremath{\column{2}\column{4}\column{1}}} -
 \myvcenter{\ensuremath{\column{4}\column{1}\column{2}}}$};
\draw[edge] (v1) to node[LabelStyleV]{$1$} (v4);
\draw[edge] (v1) to node[LabelStyleH]{$1$} (v2);
\draw[edge] (v2) to node[LabelStyleV]{$1$} (v3);
\draw[edge] (v4) to node[LabelStyleH]{$1$} (v3);
\draw[aedge, loop right, looseness=4] (v1) to node[LabelStyleVw]{$\tau$} (v1);
\end{tikzpicture}
\caption{The elements of  NST$(\gdneq (1,1,1))$, which span a $\Uqt$-submodule of $\nsbr{Y}_{\gdneq (1,1,1)}$ and are all degree-preserving and integral.  This shows that height-1 invariants commute with height-1 columns in $\nsbr{X}_{(1,1,1)}$.}
\label{f straightening111}
\end{figure}
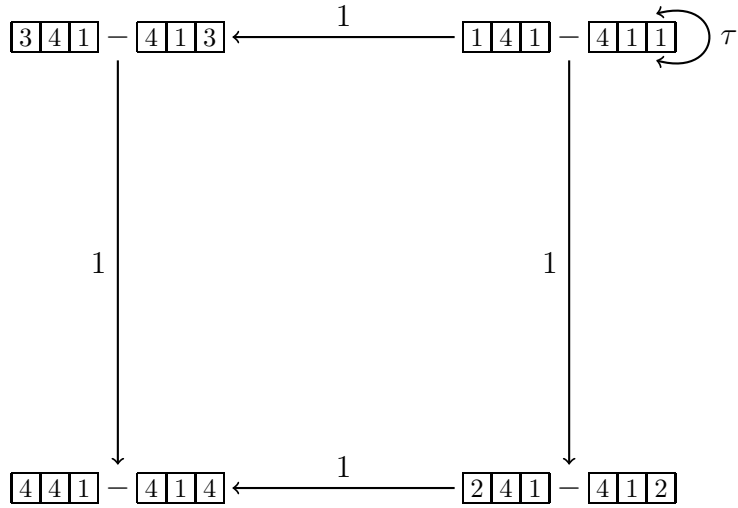
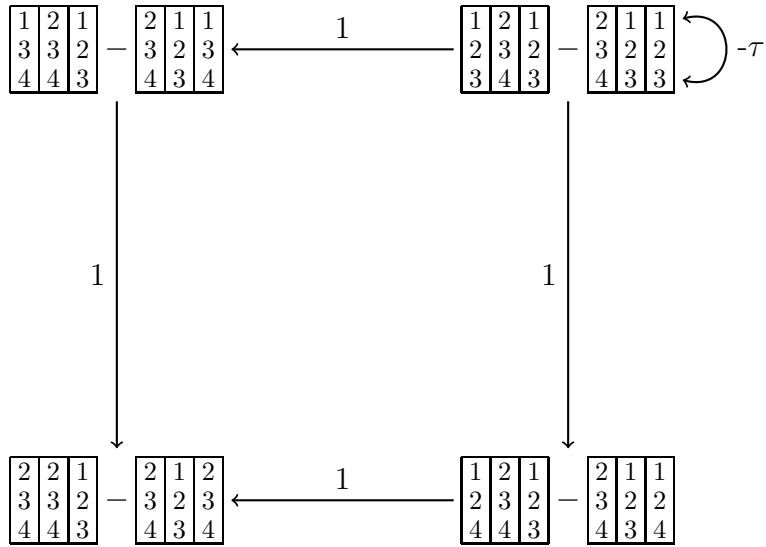
\begin{figure}[H]
\begin{tikzpicture}[scale = 3]
\tikzstyle{vertex}=[inner sep=0pt, outer sep=3pt]
\tikzstyle{edge} = [draw, thick, ->,black]
\tikzstyle{aedge} = [draw, thick, <->,black]
\tikzstyle{LabelStyleH} = [text=black, anchor=south]
\tikzstyle{LabelStyleV} = [text=black, anchor=east]
\tikzstyle{LabelStyleVw} = [text=black, anchor=west]

\node[vertex] (v1) at (1,1){$\myvcenter{\ensuremath{\column{1 \\ 2 \\ 3}\column{2\\3\\4}\column{1\\2\\3}}} -
 \myvcenter{\ensuremath{\column{2\\3\\4}\column{1\\2\\3}\column{1\\2\\3}}}$};
\node[vertex] (v2) at (-1,1){$\myvcenter{\ensuremath{\column{1\\3\\4}\column{2\\3\\4}\column{1\\2\\3}}} -
 \myvcenter{\ensuremath{\column{2\\3\\4}\column{1\\2\\3}\column{1\\3\\4}}}$};
\node[vertex] (v3) at (-1,-1){$\myvcenter{\ensuremath{\column{2\\3\\4}\column{2\\3\\4}\column{1\\2\\3}}} -
 \myvcenter{\ensuremath{\column{2\\3\\4}\column{1\\2\\3}\column{2\\3\\4}}}$};
\node[vertex] (v4) at (1,-1){$\myvcenter{\ensuremath{\column{1\\2\\4}\column{2\\3\\4}\column{1\\2\\3}}} -
 \myvcenter{\ensuremath{\column{2\\3\\4}\column{1\\2\\3}\column{1\\2\\4}}}$};
\draw[edge] (v1) to node[LabelStyleV]{$1$} (v4);
\draw[edge] (v1) to node[LabelStyleH]{$1$} (v2);
\draw[edge] (v2) to node[LabelStyleV]{$1$} (v3);
\draw[edge] (v4) to node[LabelStyleH]{$1$} (v3);
\draw[aedge, loop right, looseness=2.5] (v1) to node[LabelStyleVw]{-$\tau$} (v1);
\end{tikzpicture}
\caption{The elements of  NST$(\gdneq (3,3,3))$, which span a $\Uqt$-submodule of $\nsbr{Y}_{\gdneq (3,3,3)}$ and are all degree-preserving and integral.  This shows that height-3 invariants commute with height-3 columns in $\nsbr{X}_{(3,3,3)}$.}
\label{f straightening333}
\end{figure}

\begin{figure}[H]
\begin{tikzpicture}[scale = 3]
\tikzstyle{vertex}=[inner sep=0pt, outer sep=3pt]
\tikzstyle{edge} = [draw, thick, ->,black]
\tikzstyle{aedge} = [draw, thick, <->,black]
\tikzstyle{LabelStyleH} = [text=black, anchor=south]
\tikzstyle{LabelStyleV} = [text=black, anchor=east]
\tikzstyle{LabelStyleVw} = [text=black, anchor=west]
\node[vertex] (3) at (0,0) {$\myvcenter{\ensuremath{\column{2 \\ 3 \\ 4} \column{2 \\ 4} \column{1 \\ 3} \downdots \column{1}}}$};
\node[vertex] (2) at (1,0) {$\myvcenter{\ensuremath{\column{2 \\ 3 \\ 4} \column{3 \\ 4} \column{1 \\ 2} \downdots \column{1}}}$};
\draw[aedge] (2) to node[LabelStyleH]{$-\tau$} (3);
\end{tikzpicture}
\caption{The elements of  NST$(\gdneq (3,2^{2t},1))$, $t \geq 1$, which span a $\Uqt$-submodule of $\nsbr{Y}_{\gdneq (3,2^{2t},1)}$ and are graded and integral; the dots represent $t-1$ height-2 invariants (so that this picture represents $2^t$ NST).
}
\label{f straightening3221}
\end{figure}

For  $\gamma$ equal to $(1, 1, 1)$, $(3, 3, 3)$, or $(3,2^{2t},1)$ ($t \geq 1$), let NST$(\gdneq \gamma)$ be the elements of $\nsbr{Y}_{\gdneq \gamma}$ shown in Figure \ref{f straightening111}, \ref{f straightening333}, or \ref{f straightening3221}, respectively.  The notation  NST$(\gdneq \gamma)$ is somewhat misleading in this case as this set is not a basis for $\nsbr{Y}_{\gdneq \gamma}$, but it allows many of the definitions and results below to be stated uniformly.

Let $\gdneq \text{NST}$ be the union of $\text{NST}(\gdneq \gamma)$ over all $\gamma$ for which this is defined.
A $\gdneq \text{NST}$ corresponding to Figure \ref{f straightening11}, \ref{f straightening22a}, or \ref{f straightening33} is a \emph{nonintegral $\gdneq \text{NST}$} and an \emph{integral $\gdneq \text{NST}$} otherwise.
A $\gdneq \text{NST}$ corresponding to Figure \ref{f straightening11}, \ref{f straightening22a}, \ref{f straightening33}, \ref{f straightening31}, or \ref{f straightening3221} is a \emph{graded $\gdneq \text{NST}$} and a \emph{degree-preserving $\gdneq \text{NST}$} otherwise.
A $\gdneq \text{NST}$ corresponding to Figure \ref{f straightening111}, \ref{f straightening333}, or \ref{f straightening3221} is an \emph{invariant $\gdneq \text{NST}$} and a \emph{two-column $\gdneq \text{NST}$} otherwise.  The reasons for this terminology will be explained shortly.

\begin{proposition}
\label{p invariant moves}
The elements shown in Figures \ref{f straightening111}, \ref{f straightening333}, and \ref{f straightening3221} belong to $\nsbr{Y}_{\gdneq (1,1,1)}$, $\nsbr{Y}_{\gdneq (3, 3, 3)}$, and $\nsbr{Y}_{\gdneq (3,2^{2t},1)}$, respectively.
\end{proposition}
Part of this proposition can be rephrased as saying that for $j =1,3$, an invariant of height $j$ commutes with columns of height $j$ in $\nsbr{X}_{(j,j,j)'}$.  This can be proved by directly calculating the elements in the figures in terms of tensor products of NSC, however we postpone the proof to  \textsection\ref{ss explicit formulae for nonintegral}, where we  establish a result that makes the proof easy.

The importance of the invariant  $ \gdneq \text{NST}$ will become more clear in \textsection\ref{s Straightened NST and semistandard tableaux}.  Essentially, they are needed because, though the corresponding relations in $ \nsbr{X}_{\gamma'}$ are consequences of the  relations corresponding to two-column $\gdneq \text{NST}$, they are not consequences in an easy, combinatorial way.

\subsection{Nonstandard tabloid classes}
Here we introduce nonstandard tabloid classes (NSTC) and the subset  $\pNSTC$ of NSTC.  The  $\pNSTC$ of shape $\nu$ are the combinatorial objects that will eventually be identified with a basis of $\nsbr{X}_\nu$.
We begin by defining directed graphs  $\tgraph(\nu)$ on (a rescaled version of) NST$(\nu)$.

\label{ss NSTC}
\begin{definition}
A \emph{scaled nonstandard tabloid} (SNST) is an element of
\[\bigsqcup_{T \in \text{NST}}\Big\{({\textstyle -\frac{1}{[2]}})^{\deg(T)}T, -({\textstyle -\frac{1}{[2]}})^{\deg(T)}T\Big\} \subseteq \field\text{NST}, \]
where  $\field \text{NST}$ denotes the  $\field$-vector space with basis $\text{NST}$.
The shape of an SNST  $T$ is the shape of the NST in $\field T$, and SNST$(\alpha)$ denotes the set of SNST of shape $\alpha$.
\end{definition}
The notions of subtabloid, invariant record, degree, etc. for nonstandard tabloids extend in the obvious way to scaled nonstandard tabloids.  For instance, the invariant-free part of $a T$, for an NST $T$ and  $a \in \field $, is $a T'$, where $T'$ is the invariant-free part of $T$. Also, $(aT)|_{[i,j]} := a(T|_{[i,j]})$, for any NST $T$ and $a \in \field $.

\begin{figure}[h]
\begin{tikzpicture}[xscale = 3.64, yscale = 2]
\tikzstyle{vertex}=[inner sep=0pt, outer sep=3pt]
\tikzstyle{edge} = [draw, very thick, -,black]
\tikzstyle{Aedge} = [draw, very thick, ->,black]
\tikzstyle{LabelStyleH} = [text=black, anchor=south]
\tikzstyle{LabelStyleV} = [text=black, anchor=east]

\node[vertex] (v7) at (.5,-.5){$+\myvcenter{\ensuremath{\column{1 \\ 2 \\ 4}\column{3 \\
2}\column{1 \\ 3}\column{1}} }$};
\node[vertex] (v25) at (1.5,-.5){$+\myvcenter{\ensuremath{\column{1 \\ 2 \\ 4}\column{1 \\
3}\column{3 \\ 2}\column{1}} }$};
\node[vertex] (v29) at (2.5,0){$+\myvcenter{\ensuremath{\column{1 \\ 2 \\ 4}\column{1 \\
3}\column{2 \\ 3}\column{1}} }$};
\node[vertex] (v10) at (2.5,-1){$-\myvcenter{\ensuremath{\column{1 \\ 3 \\ 4}\column{1 \\
2}\column{3 \\ 2}\column{1}} }$};
\node[vertex] (v14) at (3.5,-.5){$-\myvcenter{\ensuremath{\column{1 \\ 3 \\ 4}\column{1 \\
2}\column{2 \\ 3}\column{1}} }$};
\node[vertex] (v20) at (4.5,-.5){$-\myvcenter{\ensuremath{\column{1 \\ 3 \\ 4}\column{2 \\
3}\column{1 \\ 2}\column{1}} }$};

\node[vertex] (v1) at (1,-2){$+\myvcenter{\ensuremath{\column{1 \\ 2 \\ 3}\column{3 \\
2}\column{3 \\ 2}\column{1}} }$};
\node[vertex] (v5) at (2,-2){$+\myvcenter{\ensuremath{\column{1 \\ 2 \\ 3}\column{3 \\
2}\column{2 \\ 3}\column{1}} }$};
\node[vertex] (v17) at (3,-2){$+\myvcenter{\ensuremath{\column{1 \\ 2 \\ 3}\column{2 \\
3}\column{3 \\ 2}\column{1}} }$};
\node[vertex] (v21) at (4,-2){$+\myvcenter{\ensuremath{\column{1 \\ 2 \\ 3}\column{2 \\
3}\column{2 \\ 3}\column{1}} }$};

\node[vertex] (v3) at (2,-3){$-\frac{1}{[2]}\myvcenter{\ensuremath{\column{1 \\ 2 \\ 3}\column{3 \\
4}\column{1 \\ 2}\column{1}} }$};
\node[vertex] (v23) at (3,-3){$-\frac{1}{[2]}\myvcenter{\ensuremath{\column{1 \\ 2 \\ 3}\column{2 \\
4}\column{1 \\ 3}\column{1}} }$};

\node[vertex] (v11) at (1,-4){$+\myvcenter{\ensuremath{\column{1 \\ 3 \\ 4}\column{3 \\
2}\column{1 \\ 2}\column{1}} }$};
\node[vertex] (v28) at (2,-4){$-\myvcenter{\ensuremath{\column{2 \\ 3 \\ 4}\column{1 \\
3}\column{1 \\ 2}\column{1}} }$};
\node[vertex] (v16) at (3,-4){$-\myvcenter{\ensuremath{\column{2 \\ 3 \\ 4}\column{1 \\
2}\column{1 \\ 3}\column{1}} }$};
\node[vertex] (v31) at (4,-4){$+\myvcenter{\ensuremath{\column{1 \\ 2 \\ 4}\column{2 \\
3}\column{1 \\ 3}\column{1}} }$};

\node[vertex] (v0) at (2.5,-5){$0$};

\draw[edge] (v1) to (v5);
\draw[edge] (v3) to (v23);
\draw[edge] (v5) to (v17);
\draw[edge] (v7) to (v25);
\draw[edge] (v10) to (v25);
\draw[edge] (v10) to (v14);
\draw[edge] (v11) to (v28);
\draw[edge] (v14) to (v29);
\draw[edge] (v14) to (v20);
\draw[edge] (v16) to (v28);
\draw[edge] (v16) to (v31);
\draw[edge] (v17) to (v21);
\draw[edge] (v25) to (v29);

\draw[Aedge] (v11) to (v0);
\draw[Aedge] (v31) to (v0);

\draw[Aedge] (v1) to (v3);
\draw[Aedge] (v21) to (v23);

\end{tikzpicture}
\caption{\small The graph $\tgraph((3, 2, 2, 1)')$ restricted to  highest weight SNST of weight $((5, 3),(5, 3))$,  and, for each pair $\{T,-T\}$ of SNST, we have only drawn one of the pair.   Edges without arrows indicate a directed edge in both directions and are degree-preserving moves; edges with arrows are graded moves. There are two strong components that are honest NSTC (the one of size 6 and the one of size 2) 
corresponding to the fact that the Kronecker coefficient $g_{(5, 3),(5, 3),(3, 2, 2, 1)'}=2$.}
\label{f tgraph example}
\end{figure}
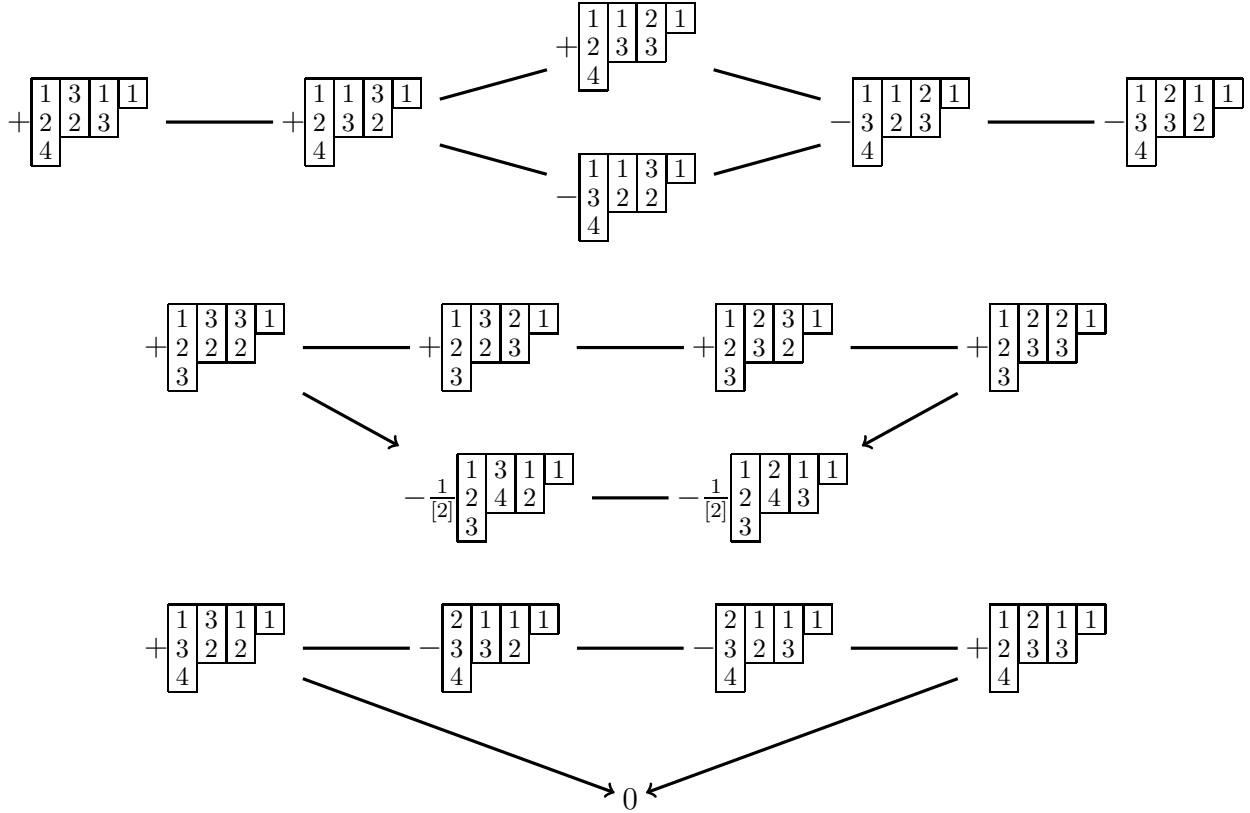

\begin{definition}
\label{d tgraph definition}
For each  $\nu \vdash r$, let $\tgraph(\nu)$ be the directed graph with vertex set $\text{SNST}(\nu') \sqcup \{0\}$ and edge set given by $T \to T'$ if the following conditions are satisfied for some  $t$ and $i \in [\nu_1 - t +1]$:
\begin{itemize}
\item $cT|_{[i,i+t-1]} - c'T'|_{[i,i+t-1]} \in \text{NST}(\gdneq (\nu'_i,\ldots,\nu'_{i+t-1})),$ for some $c,c'$ both in $\{(-[2])^j \ : \ j \in \ZZ_{\geq 0} \}$ or both in $\{-(-[2])^j \ : \ j \in \ZZ_{\geq 0} \}$;
\item $T$ and  $T'$ agree outside columns $i, \ldots, i+t-1$, i.e.  $\field T|_{[i-1]} = \field T'|_{[i-1]}$ and $\field T|_{[i+t,l]} = \field T'|_{[i+t,l]}$;
\item $\deg(T) \leq \deg(T')$
\end{itemize}
(we also allow $T' = 0$ and define $\deg(0) = \infty$, i.e.  $T \to 0$ if $cT|_{[i,i+t-1]} \in \text{NST}(\gdneq (\nu'_i,\ldots,\nu'_{i+t-1}))$ for some $c \in \{[2]^{\deg(T)}, -[2]^{\deg(T)}\}$, $i \in [\nu_1 - t +1]$).

See Figure \ref{f tgraph example} for the example $\tgraph((3,2,2,1)')$.

A directed edge is a \emph{graded move} (resp. \emph{degree-preserving move, integral, nonintegral, invariant move, two-column move}) if the corresponding $\gdneq \text{NST}$ is graded (resp. degree-preserving, integral, nonintegral, invariant, two-column); a directed edge is a \emph{move defined by Figure i} if the corresponding $\gdneq \text{NST}$ appear in this figure.  Also, if the directed edge corresponds to  $t, i$ in the definition above, then we say that it is a (graded, degree-preserving, etc.) \emph{move at $[i,i+t-1]$}.
\end{definition}

\begin{definition}
\label{d NSTC}
A \emph{nonstandard tabloid class} (NSTC) $\mathbf{T}$ of shape $\nu$ is a strong component of $\tgraph(\nu)$ (we will often identify a strong component with its vertex set).
The set of NSTC of shape  $\nu$  is denoted  NSTC$(\nu)$. An NSTC is \emph{nonorientable} if it contains $T$ and $-T$ for some SNST $T$. An NSTC $\mathbf{T}$ is \emph{dishonest} if it is the vertex 0 itself, it is nonorientable, or it has a directed edge to some other NSTC. Otherwise, we say $\mathbf{T}$ is \emph{honest}.

A SNST is  \emph{honest} (resp. \emph{dishonest}) if it belongs to an honest  (resp.  dishonest) NSTC.
If two SNST $T, T'$ lie the same NSTC, then we say that $T$ and  $T'$ are \emph{equivalent} and also denote this by $T \equiv T'$.
\end{definition}
It is easy to check directly (and is done in the proof of Theorem \ref{t main canonical basis} (iv)) that if $T \equiv T'$, then  $T$ is highest weight if and only if $T'$ is.
We then define an NSTC to be \emph{highest weight} if every SNST in its class is highest weight.

The next proposition shows that the strong components of $\tgraph(\nu)$ are quite easy to describe and justifies our terminology degree-preserving and graded.
\begin{proposition}
\label{p connected component of degree-preserving moves}
\
\begin{list}{\emph{(\alph{ctr})}} {\usecounter{ctr} \setlength{\itemsep}{1pt} \setlength{\topsep}{2pt}}
\item If  $T \to T'$ is a degree-preserving move, then the invariant records of  $T$ and $ T'$ agree.
\item If  $T \to T'$ is a graded move, then $\deg(T) < \deg(T')$.
\item An NSTC is a connected component in the undirected graph on SNST consisting of degree-preserving moves.
\end{list}
\end{proposition}
\begin{proof}
The key point here is that modifying part of an NST only affects arcs having one or both ends in the modified part. With this in mind, one can check (a)
directly for each degree-preserving move.  For instance, if  $T'$ is obtained from  $T$ by replacing a contiguous subtabloid of $T$ equal to $\ctableausmall{1 \\ 2}{1\\3}$  with  $\ctableausmall{1\\3}{1\\2}$, then the  $V$- and  $W$-words of  $T$ and  $T'$ look like
\[{\cdots {\substack{1121 \\2111}} \cdots \atop T} \ \longleftrightarrow \  {\cdots {\substack{2111 \\1121}} \cdots \atop T'} \]
The column $\ctableausmalla{1 \\2}$ of this contiguous subtabloid of  $T$ is paired by two arcs to another column of  $T$ if and only if the column  $\ctableausmalla{1\\2}$ of  $T'$ is paired by two arcs to another column of  $T'$.    A similar statement holds for $\ctableausmalla{1\\3}$.

Statement (b) holds by the definition $\deg(0) = \infty$ except in the following case:
$T \to T'$ is a graded move at  $[i,i+1]$ and $T|_{[i,i+1]}=  C, T'|_{[i, i+1]} = C'$, for $C + \frac{1}{[2]}C'$ a graded $\gdneq \text{NST}$ in the center of Figure \ref{f straightening11}, \ref{f straightening22a}, or \ref{f straightening33}.
To see that $\deg(T) < \deg(T')$ in this case, note that $T'$ contains an invariant column pair at columns $i$ and  $i+1$, while an invariant column pair of  $T$ cannot contain column $i$ or $i+1$.  Since the arcs of  $T$ and  $T'$ not involving columns $i$ and  $i+1$ are the same, any invariant column pair of  $T$ is an invariant column pair of $T'$.

Statement (c) follows from (a) and (b).
\end{proof}
The \emph{invariant record} (resp. \emph{degree}) of an honest NSTC 
is the invariant record (resp. \emph{degree}) of any SNST in its class.
An honest NSTC is \emph{invariant-free} if its invariant record is $(0,0,0,0)$.

Define the set of \emph{positive honest nonstandard tabloid classes} ($\pNSTC$) as follows: for each honest NSTC $\mathbf{T}$, declare either $+\mathbf{T}$ or $-\mathbf{T}$ to be positive. It does not really matter how these choices are made, but for computations in \textsection\ref{s A Kronecker graphical calculus} we have found it convenient to adopt the following convention.
A 3-2 arc of an SNST is an arc between a height-3 column and a height-2 column (Definition \ref{d 2-1 arc}).
\be \label{e sign convention pNSTC}
\parbox{14cm}{If every (equivalently, any) $T \in \mathbf{T}$ has no 3-2 arc, then declare $\mathbf{T}$ to be positive if it contains $(-\frac{1}{[2]})^{\deg(T)}T$ for some NST $T$. Otherwise, declare $\mathbf{T}$ to be positive if it contains $(-\frac{1}{[2]})^{\deg(T)}T$ for some NST $T$ whose 3-2 arc is a 3-2 $W$-arc.}
\ee
Let $\pSNST$ be the set of SNST that belong to some $\pNSTC$. The notions of invariant-free, $\text{NSTC}(\nu)$, $\text{SNST}(\nu)$, etc. carry over in the obvious way to  $\pNSTC$ and $\pSNST$.
Let  $\pNSTC(\nu)_{h}$ (resp. $ \pNSTC(\nu)_{\geq h}$) be the subset of $\pNSTC$ of shape $\nu$ and degree  $h$ (resp. at least $h$).


In the next subsection, we identify  $\pNSTC(\nu)$ with a subset of  $\nsbr{X}_\nu$.
We will show (as part of Theorem \ref{t main canonical basis}) that $\pNSTC(\nu)$ is a basis for $\nsbr{X}_\nu$ and
the number of highest weight elements of $\pNSTC(\nu)$ of weight  $(\lambda,\mu)$ is the Kronecker coefficient $g_{\lambda \mu \nu}$.

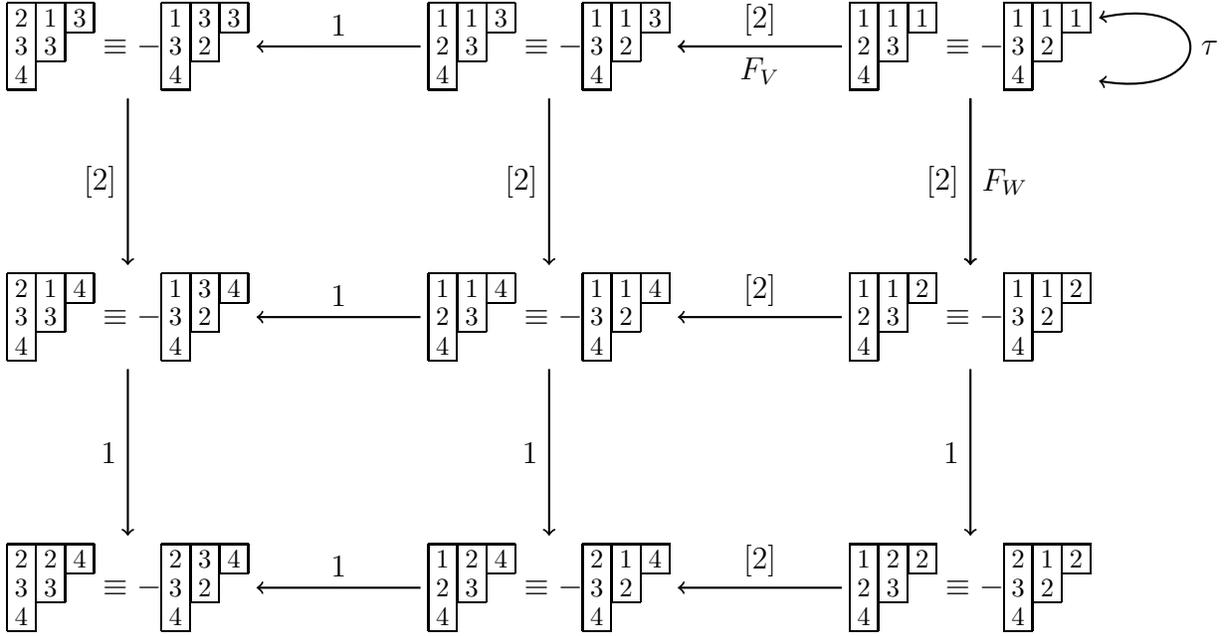
\begin{figure}[h]
\begin{tikzpicture}[yscale = 1.8, xscale=2.8]
\tikzstyle{vertex}=[inner sep=0pt, outer sep=3pt]
\tikzstyle{edge} = [draw, thick, ->,black]
\tikzstyle{aedge} = [draw, thick, <->,black]
\tikzstyle{LabelStyleH} = [text=black, anchor=south]
\tikzstyle{LabelStyleV} = [text=black, anchor=east]
\tikzstyle{LabelStyleR} = [text=black, anchor=west]
\tikzstyle{LabelStyleB} = [text=black, anchor=north]
\tikzstyle{LabelStyleVw} = [text=black, anchor=west]

\node[vertex] (v3) at (2,2){$\myvcenter{\ensuremath{\column{1 \\ 2 \\ 4}\column{1 \\ 3}\column{1}}}  \equiv -\myvcenter{\ensuremath{\column{1 \\ 3 \\ 4}\column{1 \\ 2}\column{1}}}   $};
\node[vertex] (v1) at (2,-2){$\myvcenter{\ensuremath{\column{1 \\ 2 \\ 4}\column{2 \\ 3}\column{2}}} \equiv - \myvcenter{\ensuremath{\column{2 \\ 3 \\ 4}\column{1 \\ 2}\column{2}}}   $};
\node[vertex] (v2) at (0,2){$\myvcenter{\ensuremath{\column{1 \\ 2 \\ 4}\column{1 \\ 3}\column{3}}}  \equiv -\myvcenter{\ensuremath{\column{1 \\ 3 \\ 4}\column{1 \\ 2}\column{3}}}   $};

\node[vertex] (v4) at (-2,2){$\myvcenter{\ensuremath{\column{2 \\ 3 \\ 4}\column{1 \\ 3}\column{3}}} \equiv -\myvcenter{\ensuremath{\column{1 \\ 3 \\ 4}\column{3 \\ 2}\column{3}}}     $};
\node[vertex] (v5) at (-2,-2){$\myvcenter{\ensuremath{\column{2 \\ 3 \\ 4}\column{2 \\ 3}\column{4}}} \equiv -\myvcenter{\ensuremath{\column{2 \\ 3 \\ 4}\column{3 \\ 2}\column{4}}}    $};
\node[vertex] (v6) at (0,0){$\myvcenter{\ensuremath{\column{1 \\ 2 \\ 4}\column{1 \\ 3}\column{4}}}  \equiv -\myvcenter{\ensuremath{\column{1 \\ 3 \\ 4}\column{1 \\ 2}\column{4}}}   $};
\node[vertex] (v7) at (2,0){$\myvcenter{\ensuremath{\column{1 \\ 2 \\ 4}\column{1 \\ 3}\column{2}}}  \equiv -\myvcenter{\ensuremath{\column{1 \\ 3 \\ 4}\column{1 \\ 2}\column{2}}}   $};
\node[vertex] (v8) at (0,-2){$\myvcenter{\ensuremath{\column{1 \\ 2 \\ 4}\column{2 \\ 3}\column{4}}}  \equiv -\myvcenter{\ensuremath{\column{2 \\ 3 \\ 4}\column{1 \\ 2}\column{4}}}   $};
\node[vertex] (v9) at (-2,0){$\myvcenter{\ensuremath{\column{2 \\ 3 \\ 4}\column{1 \\ 3}\column{4}}}  \equiv -\myvcenter{\ensuremath{\column{1 \\ 3 \\ 4}\column{3 \\ 2}\column{4}}}   $};
\draw[edge] (v1) to node[LabelStyleH]{$[2] $} (v8);
\draw[edge] (v2) to node[LabelStyleH]{$1$} (v4);
\draw[edge] (v2) to node[LabelStyleV]{$[2] $} (v6);
\draw[edge] (v3) to node[LabelStyleV]{$[2] $} (v7);
\draw[edge] (v3) to node[LabelStyleR]{$F_W $} (v7);
\draw[edge] (v3) to node[LabelStyleH]{$[2] $} (v2);
\draw[edge] (v3) to node[LabelStyleB]{$F_V $} (v2);
\draw[edge] (v4) to node[LabelStyleV]{$[2] $} (v9);
\draw[edge] (v6) to node[LabelStyleH]{$1$} (v9);
\draw[edge] (v6) to node[LabelStyleV]{$1$} (v8);
\draw[edge] (v7) to node[LabelStyleV]{$1$} (v1);
\draw[edge] (v7) to node[LabelStyleH]{$[2] $} (v6);
\draw[edge] (v8) to node[LabelStyleH]{$1$} (v5);
\draw[edge] (v9) to node[LabelStyleV]{$1$} (v5);
\draw[aedge, loop right, looseness=3] (v3) to node[LabelStyleVw]{$\tau$} (v3);
\end{tikzpicture}
\caption{A  $\Uqt$-cell of $\pNSTC((3, 2, 1))$; all SNST belonging to each $\pNSTC$ in this cell are shown.}
\label{f crystal component 321}
\end{figure}

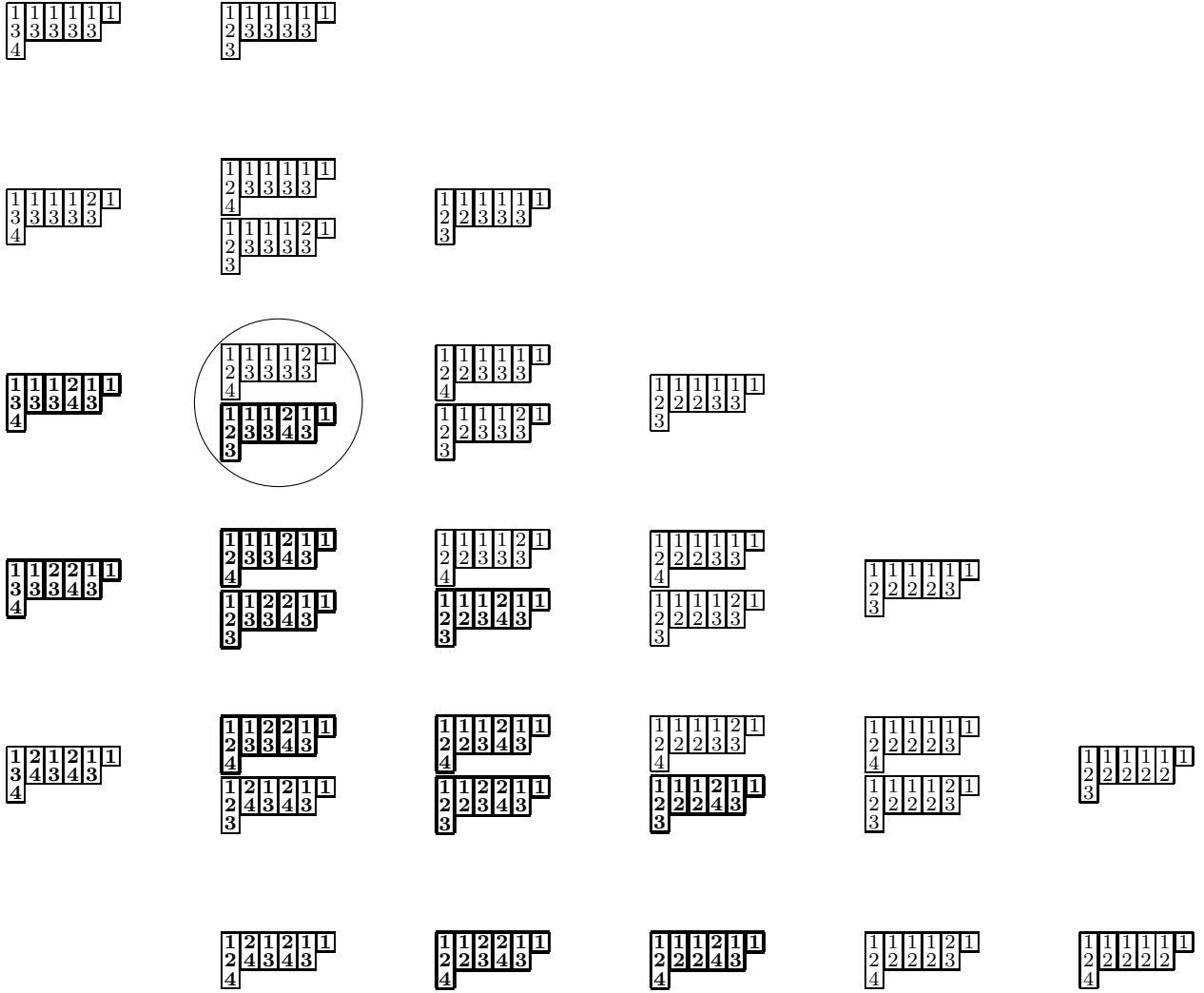
\begin{figure}[h]
{\tiny
\cellsizeCol=2.1ex
\begin{tikzpicture}[xscale = 1.5, yscale = 1.3, xshift = -1cm]
\tikzstyle{vertex}=[inner sep=0pt, outer sep=3pt]
\tikzstyle{vcircle}=[draw, circle, inner sep=0pt, outer sep=3pt]
\tikzstyle{edge} = [draw, thick, ->,black]
\tikzstyle{LabelStyleH} = [text=black, anchor=south]
\tikzstyle{LabelStyleV} = [text=black, anchor=east]

\node[vertex] (v1) at (5,5){$\ensuremath{\columnL{1 \\ 2 \\ 4}\columnL{1 \\ 2}\columnL{1 \\ 3}\columnL{1 \\ 3}\columnL{2 \\ 3}\columnL{1}}\atop \ensuremath{\columnB{1 \\ 2 \\ 3}\columnB{1 \\ 2}\columnB{1 \\ 3}\columnB{2 \\ 4}\columnB{1 \\ 3}\columnB{1}} $};
\node[vertex] (v2) at (1,11){$\ensuremath{\columnL{1 \\ 3 \\ 4}\columnL{1 \\ 3}\columnL{1 \\ 3}\columnL{1 \\ 3}\columnL{1 \\ 3}\columnL{1}} $};
\node[vertex] (v3) at (11,3){$\ensuremath{\columnL{1 \\ 2 \\ 3}\columnL{1 \\ 2}\columnL{1 \\ 2}\columnL{1 \\ 2}\columnL{1 \\ 2}\columnL{1}} $};
\node[vertex] (v4) at (3,1){$\ensuremath{\columnBB{1 \\ 2 \\ 4}\columnBB{2 \\ 4}\columnBB{1 \\ 3}\columnBB{2 \\ 4}\columnBB{1 \\ 3}\columnBB{1}} $};
\node[vertex] (v5) at (5,7){$\ensuremath{\columnL{1 \\ 2 \\ 4}\columnL{1 \\ 2}\columnL{1 \\ 3}\columnL{1 \\ 3}\columnL{1 \\ 3}\columnL{1}}\atop \ensuremath{\columnL{1 \\ 2 \\ 3}\columnL{1 \\ 2}\columnL{1 \\ 3}\columnL{1 \\ 3}\columnL{2 \\ 3}\columnL{1}} $};
\node[vertex] (v6) at (1,5){$\ensuremath{\columnB{1 \\ 3 \\ 4}\columnB{1 \\ 3}\columnB{2 \\ 3}\columnB{2 \\ 4}\columnB{1 \\ 3}\columnB{1}} $};
\node[vertex] (v7) at (11,1){$\ensuremath{\columnL{1 \\ 2 \\ 4}\columnL{1 \\ 2}\columnL{1 \\ 2}\columnL{1 \\ 2}\columnL{1 \\ 2}\columnL{1}} $};
\node[vertex] (v8) at (9,5){$\ensuremath{\columnL{1 \\ 2 \\ 3}\columnL{1 \\ 2}\columnL{1 \\ 2}\columnL{1 \\ 2}\columnL{1 \\ 3}\columnL{1}} $};
\node[vertex] (v9) at (5,9){$\ensuremath{\columnL{1 \\ 2 \\ 3}\columnL{1 \\ 2}\columnL{1 \\ 3}\columnL{1 \\ 3}\columnL{1 \\ 3}\columnL{1}} $};
\node[vertex] (v10) at (1,7){$\ensuremath{\columnB{1 \\ 3 \\ 4}\columnB{1 \\ 3}\columnB{1 \\ 3}\columnB{2 \\ 4}\columnB{1 \\ 3}\columnB{1}} $};
\node[vertex] (v11) at (3,11){$\ensuremath{\columnL{1 \\ 2 \\ 3}\columnL{1 \\ 3}\columnL{1 \\ 3}\columnL{1 \\ 3}\columnL{1 \\ 3}\columnL{1}} $};
\node[vertex] (v12) at (9,1){$\ensuremath{\columnL{1 \\ 2 \\ 4}\columnL{1 \\ 2}\columnL{1 \\ 2}\columnL{1 \\ 2}\columnL{2 \\ 3}\columnL{1}} $};
\node[vertex] (v13) at (1,3){$\ensuremath{\columnBB{1 \\ 3 \\ 4}\columnBB{2 \\ 4}\columnBB{1 \\ 3}\columnBB{2 \\ 4}\columnBB{1 \\ 3}\columnBB{1}} $};
\node[vertex] (v14) at (3,9){$\ensuremath{\columnL{1 \\ 2 \\ 4}\columnL{1 \\ 3}\columnL{1 \\ 3}\columnL{1 \\ 3}\columnL{1 \\ 3}\columnL{1}}\atop \ensuremath{\columnL{1 \\ 2 \\ 3}\columnL{1 \\ 3}\columnL{1 \\ 3}\columnL{1 \\ 3}\columnL{2 \\ 3}\columnL{1}} $};
\node[vertex] (v15) at (9,3){$\ensuremath{\columnL{1 \\ 2 \\ 4}\columnL{1 \\ 2}\columnL{1 \\ 2}\columnL{1 \\ 2}\columnL{1 \\ 3}\columnL{1}}\atop \ensuremath{\columnL{1 \\ 2 \\ 3}\columnL{1 \\ 2}\columnL{1 \\ 2}\columnL{1 \\ 2}\columnL{2 \\ 3}\columnL{1}} $};
\node[vertex] (v16) at (7,3){$\ensuremath{\columnL{1 \\ 2 \\ 4}\columnL{1 \\ 2}\columnL{1 \\ 2}\columnL{1 \\ 3}\columnL{2 \\ 3}\columnL{1}}\atop \ensuremath{\columnB{1 \\ 2 \\ 3}\columnB{1 \\ 2}\columnB{1 \\ 2}\columnB{2 \\ 4}\columnB{1 \\ 3}\columnB{1}} $};
\node[vertex] (v17) at (7,1){$\ensuremath{\columnB{1 \\ 2 \\ 4}\columnB{1 \\ 2}\columnB{1 \\ 2}\columnB{2 \\ 4}\columnB{1 \\ 3}\columnB{1}} $};
\node[vertex] (v18) at (7,7){$\ensuremath{\columnL{1 \\ 2 \\ 3}\columnL{1 \\ 2}\columnL{1 \\ 2}\columnL{1 \\ 3}\columnL{1 \\ 3}\columnL{1}} $};
\node[vertex] (v19) at (7,5){$\ensuremath{\columnL{1 \\ 2 \\ 4}\columnL{1 \\ 2}\columnL{1 \\ 2}\columnL{1 \\ 3}\columnL{1 \\ 3}\columnL{1}}\atop \ensuremath{\columnL{1 \\ 2 \\ 3}\columnL{1 \\ 2}\columnL{1 \\ 2}\columnL{1 \\ 3}\columnL{2 \\ 3}\columnL{1}} $};
\node[vcircle] (v20) at (3,7){$\ensuremath{\columnL{1 \\ 2 \\ 4}\columnL{1 \\ 3}\columnL{1 \\ 3}\columnL{1 \\ 3}\columnL{2 \\ 3}\columnL{1}}\atop \ensuremath{\columnB{1 \\ 2 \\ 3}\columnB{1 \\ 3}\columnB{1 \\ 3}\columnB{2 \\ 4}\columnB{1 \\ 3}\columnB{1}} $};
\node[vertex] (v21) at (5,1){$\ensuremath{\columnB{1 \\ 2 \\ 4}\columnB{1 \\ 2}\columnB{2 \\ 3}\columnB{2 \\ 4}\columnB{1 \\ 3}\columnB{1}} $};
\node[vertex] (v22) at (3,5){$\ensuremath{\columnB{1 \\ 2 \\ 4}\columnB{1 \\ 3}\columnB{1 \\ 3}\columnB{2 \\ 4}\columnB{1 \\ 3}\columnB{1}}\atop \ensuremath{\columnB{1 \\ 2 \\ 3}\columnB{1 \\ 3}\columnB{2 \\ 3}\columnB{2 \\ 4}\columnB{1 \\ 3}\columnB{1}} $};
\node[vertex] (v23) at (5,3){$\ensuremath{\columnB{1 \\ 2 \\ 4}\columnB{1 \\ 2}\columnB{1 \\ 3}\columnB{2 \\ 4}\columnB{1 \\ 3}\columnB{1}}\atop \ensuremath{\columnB{1 \\ 2 \\ 3}\columnB{1 \\ 2}\columnB{2 \\ 3}\columnB{2 \\ 4}\columnB{1 \\ 3}\columnB{1}} $};
\node[vertex] (v24) at (1,9){$\ensuremath{\columnL{1 \\ 3 \\ 4}\columnL{1 \\ 3}\columnL{1 \\ 3}\columnL{1 \\ 3}\columnL{2 \\ 3}\columnL{1}} $};
\node[vertex] (v25) at (3,3){$\ensuremath{\columnB{1 \\ 2 \\ 4}\columnB{1 \\ 3}\columnB{2 \\ 3}\columnB{2 \\ 4}\columnB{1 \\ 3}\columnB{1}}\atop \ensuremath{\columnBB{1 \\ 2 \\ 3}\columnBB{2 \\ 4}\columnBB{1 \\ 3}\columnBB{2 \\ 4}\columnBB{1 \\ 3}\columnBB{1}} $};
\end{tikzpicture}}
\caption{\small An NST representative of each highest weight $\pNSTC$ of shape  $\nu = (3, 2, 2, 2, 2, 1)'$ (these are the straightened representatives, defined in  \textsection\ref{s Straightened NST and semistandard tableaux}).  The NST of weight $(\lambda,\mu) = ([l_2,l_1],[m_2,m_1])$ are drawn at position $(\frac{l_1} {2},\frac{m_1}{2})$ so, for instance, the two NST circled at position (1,3) corresponds to $g_{(7,5) \, (9,3) \, \nu} = 2$.
The bold borders and numbers make it easier to read off the NST of fixed degree.}
\label{f highest weight NSTC}
\end{figure}

A $\Uqt$-cell of $(\nsbr{X}_{(3,2,1)}, \pNSTC((3, 2, 1)))$ is shown in Figure \ref{f crystal component 321}.
An NST representative of each highest weight $\pNSTC$ of shape $(3, 2, 2, 2,2, 1)'$ is shown in Figure \ref{f highest weight NSTC}.

\subsection{Justifying the combinatorics}
\label{ss agrees in associated graded}
Here we establish the precise relationship between the combinatorial NSTC$(\nu)$ of the previous subsection and the relations satisfied by the image of $\text{SNST}(\nu')$ in  $\nsbr{X}_\nu$.

Let us recall some of the notation introduced at the beginning of the section and introduce some new notation:  let $\alpha \vDash^{\dx}_l r$ and suppose $\alpha= (\beta,\gamma,\delta)$ with  $\beta = (\alpha_1,\ldots,\alpha_{i-1})$, $\gamma = (\alpha_i,\ldots,\alpha_{i + t-1})$, $\delta= (\alpha_{i+t},\ldots,\alpha_l)$, and $\gamma$ is such that
$\text{NST}(\gdneq\gamma)$ is defined.
Set
\be
\label{e gdneq alpha def}
\begin{array}{lll}
\nsbr{Y}_{\gdneq \gamma} &:=& \field \, \text{NST}(\gdneq\gamma), \\
\nsbr{Y}_{(\beta, \gdneq\gamma,\delta)} &:=& \nsbr{Y}_{\beta} \tsr \nsbr{Y}_{\gdneq \gamma} \tsr \nsbr{Y}_{\delta} \subseteq \nsbr{Y}_\alpha,\\[1.4mm]
\nsbr{Y}_{\gdneq^i \alpha} &:=& \nsbr{Y}_{(\beta, \gdneq\gamma,\delta)}, \ \text{ if  $\ell(\gamma) = 2$,} \\[1.2mm]
\nsbr{Y}_{\gdneq\alpha}&:=& \sum_{i = 1}^{l-1} \nsbr{Y}_{\gdneq^i \alpha},\\[1mm]
\nsbr{X}_{\nu} &:=& \nsbr{Y}_{\nu'}/\nsbr{Y}_{\gdneq\nu'}, \ \text{for $\nu \vdash r$},\\[1mm]
\gr(\nsbr{X}_\nu) &:=& \gr(\nsbr{Y}_{\nu'})/\gr(\nsbr{Y}_{\gdneq\nu'}).
\end{array}
\ee
Let $\varpi_{\nu}$ denote the projection $\nsbr{Y}_{\nu'} \twoheadrightarrow \nsbr{X}_{\nu}$.
It is sometimes convenient to use
\be \label{e alternative expression grXnu}
\gr(\nsbr{X}_\nu) = \bigoplus_{h \geq 0} (\nsbr{Y}_{\nu'})_h/((\nsbr{Y}_{\gdneq\nu'})_h + (\nsbr{Y}_{\nu'})_{h+1}).
\ee

The difficulty in understanding the image of $\text{SNST}(\alpha)$ in  $\nsbr{Y}_{\alpha} / \nsbr{Y}_{\gdneq\alpha}$ is that $\nsbr{Y}_{(\beta, \gdneq \gamma,\delta)}$ is not easily expressed in terms of the basis  $\text{SNST}(\alpha)$. This is mostly because of the nonintegral $\gdneq \text{NST}$. To remedy this, we define a basis  $\text{NST}(\beta, \gdneq \gamma,\delta)$ of $\nsbr{Y}_{(\beta, \gdneq \gamma,\delta)}$, which is analogous to the basis $B'_1 \heartproj \cdots \heartproj B'_l$ of Proposition \ref{p heart commutes with projections}.
Let  $(\nsbr{Y}_{(\beta, \gdneq \gamma,\delta)},\text{NST}(\beta, \gdneq \gamma,\delta))$ be the weak upper based $\Uqt$-module obtained by tensoring the weak upper based $\Uqt$-modules $(\nsbr{Y}_{\beta}, \text{NST}(\beta)), (\nsbr{Y}_{\gdneq \gamma}, \text{NST}(\gdneq \gamma)), (\nsbr{Y}_{\delta}, \text{NST}(\delta))$ (see \textsection\ref{ss Tensor products of based modules} and \textsection\ref{ss upper based Uqt}).  Let $\heartg$ denote the $\heart$ product in this setting, so that
\begin{align*}
\text{NST}(\beta, \gdneq \gamma,\delta) &:=  \text{NST}(\beta) \heartg \text{NST}(\gdneq \gamma) \heartg \text{NST}(\delta) \\
&= \Big\{ B \heartg \tilde{C} \heartg D : B \in \text{NST}(\beta), \tilde{C} \in \text{NST}(\gdneq \gamma), D \in \text{NST}(\delta) \Big\}.
\end{align*}




The main result of this subsection will follow from the following two general propositions about based modules.

\begin{proposition}\label{p hearts agree with same integral form}
Let  $(N_1',B_1'), (N_1,B_1), (N_2,B_2)$  be upper based  $\Uqvw$-modules and  $N_1'$ a submodule of  $N_1$.  Suppose that the canonical inclusion $\iota:N_1' \hookrightarrow N_1$ induces a morphism of balanced triples, i.e.,  $\iota$ restricts to maps
\[ \QQA B_1' \to \QQA B_1,\quad \field_0 B_1' \to \field_0 B_1,\quad \field_\infty B_1' \to \field_\infty B_1 .\]
Then  $b_1' \heart' b_2 = b_1' \heart b_2$ for all  $b_1' \in B_1', \ b_2 \in B_2$, where  $\heart'$ (resp.  $\heart$)  denotes the product of  \textsection\ref{ss Tensor products of based modules} for tensoring $(N_1',B_1')$ and $(N_2,B_2)$ (resp. $(N_1,B_1)$ and $(N_2,B_2)$).
\end{proposition}
\begin{proof}
Let $\L'=\field_\infty (B_1'\tsr B_2)$ (resp. $\L=\field_\infty (B_1 \tsr B_2)$) be the crystal lattice of $(N_1' \tsr N_2,B_1'\heart' B_2)$ (resp. ($N_1 \tsr N_2,B_1\heart B_2$)).
This follows simply from the fact that the canonical isomorphism
\[ \QQA (B_1 \tsr B_2) \cap \br{\L} \cap \L \xrightarrow{\cong} \L/\ui \L\]
restricts to the canonical isomorphism
\[ \QQA (B_1' \tsr B_2) \cap \br{\L'} \cap \L' \xrightarrow{\cong} \L'/\ui \L'.\]
\end{proof}

Recall that $\Uqvw$-irreducibles are parameterized by pairs of partitions and that  $\ldneq$ denotes dominance order on pairs of partitions as well as dominance order on partitions, as discussed in  \textsection\ref{s notation for GLV GLW}.
\begin{proposition}\label{p hearts agree mod lower cells}
Let  $(N_1',B_1'), (N_1,B_1), (N_2,B_2)$  be upper based  $\Uqvw$-modules and  $N_1'$ a  submodule of  $N_1$. Let  $\heart'$ and  $\heart$ be as in the previous proposition.  Then  for all pairs of partitions $\theta$, $\ b_1' \in B_1'[\theta],$ and $b_2 \in B_2$,
\[b_1' \heart' b_2 - b_1' \heart b_2 \in N_1[\ldneq \theta] \tsr N_2.\]
\end{proposition}
\begin{proof}
First assume that $N_1=N_1[\theta]$ for some pair of partitions $\theta = (\lambda,\mu)$. The result in this case follows from the fact that  $(N_1',B_1')$ and  $(N_1,B_1)$ are both isomorphic to a direct sum of $(V_\lambda \tsrvw W_\mu,B(\lambda) \tsrvw B(\mu))$ (the result follows despite the fact that the inclusion  $N_1' \hookrightarrow N_1$ need not induce a morphism of upper based $\Uqvw$-modules or even a map  $\field_\infty B_1' \to \field_\infty B_1$).
Now for the general case, let $\theta, b_1', b_2$ be as in the statement. Recall that $\varsigma_\theta^{N_1}(b_1')$ is the image of $b_1'$ in $N_1[\theta]$. Then
\[
b_1' \heart' b_2 \equiv \varsigma_\theta^{N_1'}(b_1')\tilde {\heart}' b_2 = \varsigma_\theta^{N_1}(b_1')\tilde{\heart} b_2 \equiv b_1' \heart b_2,
\]
where the equivalences are mod $N_1[\ldneq \theta] \tsr N_2$ and the equality is by the result for the $N_1=N_1[\theta]$ case; the product $\tilde{\heart}'$ (resp.  $\tilde{\heart}$) is for the tensor product of $(N_1'[\theta],\varsigma_\theta^{N_1'}(B_1'[\theta]))$ (resp. $(N_1[\theta],\varsigma_\theta^{N_1}(B_1[\theta]))$) and  $(N_2, B_2)$.
\end{proof}


\begin{theorem}\label{t heartsQ good for submodules}
Let  $\alpha = (\beta,\gamma,\delta)$ as above.  For any $B \in \text{NST}(\beta), \tilde{C} \in \text{NST}(\gdneq \gamma), D \in \text{NST}(\delta)$, there holds
\be \label{e two cases ABC-ABC}
 B \heartns \tilde{C} \heartns D - B \heartg \tilde{C} \heartg D \begin{cases}
 = 0 & \text{if  $\tilde{C}$ is integral}, \\
 \in (\nsbr{Y}_{\alpha})_{h+1} & \text{if  $\tilde{C}$ is nonintegral},
 \end{cases}
 \ee
where $h = \deg(B \heartns \tilde{C} \heartns D)$ ($\deg(x)$ is defined for any  $x \in \nsbr{Y}_{\alpha}$ in \textsection\ref{ss invariants}).  Hence
\be \label{e heartssQ hearts}
\gr(\nsbr{Y}_{(\beta, \gdneq \gamma,\delta)}) = \gr(\nsbr{Y}_{\beta} \heartns \nsbr{Y}_{\gdneq \gamma} \heartns \nsbr{Y}_{\delta}).
\ee
\end{theorem}
\begin{proof}
The top case of \eqref{e two cases ABC-ABC} follows from Proposition \ref{p hearts agree with same integral form} and the bottom from Proposition \ref{p hearts agree mod lower cells}.
The propositions are applied with $(N_1',B_1') = (\nsbr{Y}_{\gdneq \gamma},\text{NST}(\gdneq \gamma))$ and  $(N_1,B_1) = (\nsbr{Y}_\gamma, \text{NST}(\gamma))$, except the application of Proposition \ref{p hearts agree with same integral form} in the case $\gamma = (2,2)$. For this case, since nonintegral $\text{NST}(\gdneq (2,2))$ span a submodule of  $\nsbr{Y}_{\gdneq (2,2)}$, we instead apply Proposition \ref{p hearts agree with same integral form} with  $(N_1',B_1')$ equal to the based module corresponding to this submodule.

For the bottom case, the necessary connection between the $\ldneq$ of Proposition \ref{p hearts agree mod lower cells} and degree is made as follows:
put $\gamma = (j,j)$, $j =1,2$, or  $3$.  The $\theta$ of the proposition is equal $((j+1,j-1),(j,j))$ or $((j,j),(j+1,j-1))$, hence $N_1[\ldneq \theta]$ is spanned by the height-$j$ invariant(s).
Further, by Proposition \ref{p connected component of degree-preserving moves} (b) and the fact that the action of the global Kashiwara operators  $\crystal{F}_V, \crystal{F}_W$ on NST preserve degree, there holds $\nsbr{Y}_\beta \tsr N_1[\ldneq \theta] \tsr \nsbr{Y}_\delta \subseteq (\nsbr{Y}_{\alpha})_{h+1}$.
\end{proof}

Now we can show that combinatorics of the previous subsection is actually relevant to the algebraic objects $\nsbr{X}_\nu$ and $\gr(\nsbr{X}_\nu)$.
\begin{proposition}
\label{p main canonical basis}
Let  $T,T'$ be SNST of shape $\nu'$ for some $\nu \vdash_\dx r$, and let $\mathbf{T}, \mathbf{T'}$ be the NSTC containing  $T$ and $T'$, respectively.
\begin{list}{\emph{(\roman{ctr})}} {\usecounter{ctr} \setlength{\itemsep}{1pt} \setlength{\topsep}{2pt}}
\item If  $T \equiv T'$ (equivalently, $\mathbf{T} = \mathbf{T'}$), then $\varpi_\nu(T) = \varpi_\nu(T')$ and we identify this element of $ \nsbr{X}_\nu$ with the NSTC $\mathbf{T}$.
\item If  $T \to 0$ is an integral move, then  $\mathbf{T} = 0$ as an element of $\nsbr{X}_\nu$.
\item If $T$ is dishonest, then $\grin_h(\mathbf{T}) = 0$ in  $(\nsbr{X}_\nu)_h / (\nsbr{X}_\nu)_{h+1}$, where $h = \deg(T)$.
\item The set $\grin_h(\pNSTC(\nu)_{h})$ spans $\gr(\nsbr{X}_\nu)_h$.  In particular, $\grin(\pNSTC(\nu))$ spans $\gr(\nsbr{X}_\nu)$.
\item The set $\pNSTC(\nu)_{\geq h}$ spans $(\nsbr{X}_\nu)_h$. In particular, $\pNSTC(\nu)$ spans $\nsbr{X}_\nu$.
\end{list}
\end{proposition}
Note that part (i) of the proposition implies that if  $T \equiv T'$ ($T,T'\in \text{SNST}(\nu')$) in the notation of the previous subsection, then  $T \equiv T'$ mod  $\nsbr{Y}_{\gdneq \nu'}$.
\begin{proof}
By the top case of \eqref{e two cases ABC-ABC} of Theorem \ref{t heartsQ good for submodules}, if $T \to T'$ is an integral move, then $\varpi_\nu(T) = \varpi_\nu(T')$. Since an NSTC is just a connected component in the undirected graph consisting of degree-preserving moves (Proposition \ref{p connected component of degree-preserving moves}(c)) and degree-preserving  moves are integral, (i) follows.  This implies (ii) as well.

Next suppose $T \to T'$ is a graded move and $\deg(T) = h$ (and, since the move is graded, $\deg(T') > h$). Then $cT - c'T'$ is of the form  $B \heartns \tilde{C} \heartns D$  in the notation of Theorem \ref{t heartsQ good for submodules} and Definition \ref{d tgraph definition}, so $B \heartns \tilde{C} \heartns D \in \nsbr{Y}_{\gdneq \nu'} + (\nsbr{Y}_{\nu'})_{h+1}$ by \eqref{e two cases ABC-ABC}, hence $T \in (\nsbr{Y}_{\gdneq \nu'})_h + (\nsbr{Y}_{\nu'})_{h+1}$, which proves (iii).

By (i), $\pNSTC(\nu)$ is a well-defined subset of $\nsbr{X}_\nu$ and, moreover, by Proposition \ref{p connected component of degree-preserving moves}, $\pNSTC(\nu)_{\geq h}$ is a well-defined subset of $(\nsbr{X}_\nu)_h$. Then (iii) and the fact that $\text{NST}(\nu')_{\geq h}$ is a basis of  $(\nsbr{Y}_{\nu'})_h$ prove (iv).
Statement (v) follows easily from (iv).
\end{proof}

\begin{example}
\label{ex associated graded needed}
The theorem above does not hold without the $\gr$: suppose  $\alpha = (1,1,1)$, $\beta = (1)$,  $\gamma = (1,1)$,  $\delta=()$,  $B = \ctableausmalla{3}$, $\tilde{C} = \ctableausmall{2}{1}$, and consider the NST
 $B \heartns \tilde{C} = \ctableausmallc{3}{2}{1}$.
Since  $\ctableausmall{2}{1} \in \nsbr{Y}_{\gdneq \gamma}$, we would like to conclude that  $\varpi_{\alpha'}(\ctableausmallc{3}{2}{1}) = 0$ in  $\nsbr{X}_{\alpha'}$, however this is only true in  $\gr(\nsbr{X}_{\alpha'})$.  This can be seen by expressing  $\ctableausmalla{3} \heartg \ctableausmall{2}{1} \in \text{NST}(\beta, \gdneq\gamma)$ in terms of  $\text{NST}(\alpha)$ as
\[ \ctableaua{3} \heartg \ctableau{2}{1} = \ctableauc{3}{2}{1} + \frac{1}{[2]} \ctableaua{1} \tsr \ctableau{4}{1} = \ctableauc{3}{2}{1} + \frac{1}{[2]} \ctableauc{1}{4}{1} . \]
The next subsection will give a nice way of doing such computations, but for now we can verify it by checking that the right-hand side is $\br{\cdot}$-invariant, which is clear, and that
\[
\begin{array}{cl}
\ctableauc{3}{2}{1} + \frac{1}{[2]} \ctableaua{1} \tsr \ctableau{4}{1} &= \cvw{211}{121} + \frac{1}{[2]} \cvw{121}{121} \\[.1in]
&=\cvw{2}{1} \tsr \cvw{11}{21} - \ui \cvw{1}{1} \tsr \cvw{21}{21} - \u^{-2}\cvw{1}{1} \tsr \cvw{12}{21} + \frac{1}{[2]} \cvw{1}{1} \tsr \cvw{21}{21} \vspace{.1in} \\
&= \ctableaua{3}\tsr \ctableau{2}{1}-\u^{-2}\ctableaua{1} \tsr (\ctableau{2}{3} + \frac{1}{[2]} \ctableau{4}{1}),
\end{array}\]
which lies in  $\nsbr{Y}_{(\beta,\gdneq \gamma)}$ and is congruent to $\ctableausmalla{3}\tsr \ctableausmall{2}{1}$ mod $\ui \field_\infty \text{NST}(\beta,\gdneq \gamma)$.

Therefore $\ctableausmallc{3}{2}{1} =  \ctableausmalla{3} \heartg \ctableausmall{2}{1} - \frac{1}{[2]} \ctableausmalla{1} \tsr \ctableausmall{4}{1}$ does not belong to $\nsbr{Y}_{(\beta,\gdneq \gamma)}$, but it does belong to $\nsbr{Y}_{(\beta,\gdneq \gamma)} + (\nsbr{Y}_{\alpha})_1$.  Hence its image under  $\nsbr{Y}_{\alpha} \xrightarrow{\grin_0} \gr(\nsbr{Y}_{\alpha}) \twoheadrightarrow \gr(\nsbr{X}_{\alpha'})$ is 0.
\end{example}

\begin{remark}
\label{r pNSTC alpha generalization}
We can also construct a basis for $\nsbr{Y}_{\gdneq \gamma}$ in the case $\ell(\gamma) = 2$ and  $\gamma_1 < \gamma_2$.  This has a similar form to the basis  $ \text{NST}( \gdneq (\gamma_2,\gamma_1))$.  Theorem \ref{t heartsQ good for submodules} above can be extended to this case and many of the results of this section, including Theorem \ref{t main canonical basis} (i) and (ii), can be extended to the case $\nu'$ is not necessarily a partition (after introducing some more invariant  $ \gdneq \text{NST}$).  We believe that all of Theorem \ref{t main canonical basis} can be extended to this case, but we have not worked out the necessary combinatorics.
This is a more precise version of a special case of Conjecture \ref{cj canonical basis X^r}, as explained in the discussion following the conjecture.
\end{remark}

\subsection{Explicit formulae for nonintegral $\text{NST}(\beta, \gdneq \gamma,\delta)$}
\label{ss explicit formulae for nonintegral}
Using Corollary \ref{c projecting to heartproj basis},
we now make explicit how to express $\text{NST}(\beta, \gdneq \gamma,\delta)$ in terms of $\text{NST}(\alpha)$ for those elements of $\text{NST}(\beta, \gdneq \gamma,\delta)$ corresponding to nonintegral $\text{NST}(\gdneq\gamma)$.

Denote some of the invariants as follows
\[ I_1^V = I_1^W = \ctableau{4}{1}, \ I_2^V = \ctableau{3 \\ 4}{1 \\ 2},\ I_2^W= \ctableau{2 \\ 4}{1 \\3},\  I_3^V= I_3^W = \ctableau{2 \\ 3 \\4}{1 \\ 2 \\3}. \]
For $\gamma = (j,j)$,  $j \in [3]$, define
\[\text{NST}_V(\gamma) := \{T \in \text{NST}(\gamma): \varphi_V(T)+\varepsilon_V(T)=2, \varphi_W(T)+\varepsilon_W(T)=0\}  \sqcup \{I_j^V\}. \]
Note that $\text{NST}_V(\gamma)$ spans an upper based $U_q(\sl(V))$-submodule of $(\nsbr{Y}_\gamma, \text{NST}(\gamma))$ that is isomorphic to $(\Res_{U_q(\sl(V))}V^{\tsr 2},B^2_V)$ (where $(V^{\tsr 2}, B^2_V)$ is as in \textsection\ref{ss upper canonical basis of bT}).
For each  $C \in \text{NST}_V(\gamma)$, let $\tilde{C} := \pi^{\nsbr{Y}_{\gamma}}_{((j+1,j-1),(j,j))}(C)$ be the corresponding element of  $\text{NST}(\gdneq \gamma)$ (see the discussion before Corollary \ref{c projecting to heartproj basis}), which is a nonintegral  $\gdneq \text{NST}$.   Denote the set of such $\tilde{C}$ by  $\text{NST}_V(\gdneq \gamma)$.  This is just the horizontal $\g_V$-string in Figure \ref{f straightening11}, \ref{f straightening22a}, or \ref{f straightening33} for $j=1,2,3$, respectively.
Define $\text{NST}_W(\gamma)$ and $\text{NST}_W(\gdneq \gamma)$ similarly, with the roles of $W$-word and  $V$-word interchanged and  $I_j^W$ in place of  $I_j^V$.

Since $\{T \in \text{NST}(\alpha): T|_{[i,i+1]} \in \text{NST}_V(\gamma)) \}$ spans an upper based $U_q(\sl(V))$-submodule of  $(\nsbr{Y}_\alpha, \text{NST}(\alpha))$, and this is a direct sum of upper based $U_q(\sl(V))$-modules of the form in Corollary \ref{c projecting to heartproj basis} (restrict the  $U_q(\g_V)$-modules of the corollary to  $U_q(\sl(V))$), with $B_t$ of the corollary equal to  $\text{NST}_V(\gamma)$, we obtain
\begin{corollary}
\label{c projecting to heartproj basis Kronecker}
Maintain the notation above, with $\gamma = (j,j)$,  $j \in [3]$, and let $B \in \text{NST}(\beta), C \in \text{NST}_V(\gamma), D \in \text{NST}(\delta)$. Let  $T =B \heartns C \heartns D$ and  $\mathbf{k'}$ be the unpaired  $V$-word of  $C$.
Then
\[ B \heartg \tilde{C} \heartg D =  \begin{cases}
B \heartns C \heartns D + \frac{1}{[2]} B' \heartns I_j^V \heartns D & \text{if \, $\mathbf{k'} = 11$},\\
B \heartns C \heartns D + \frac{1}{[2]} B \heartns I_j^V \heartns D & \text{if \, $\mathbf{k'} = 12$},\\
B \heartns C \heartns D + \frac{1}{[2]} B \heartns I_j^V \heartns D' & \text{if \, $\mathbf{k'} = 22$},
\end{cases}
\]
where  $B' \in \text{NST}(\beta)\sqcup \{0\},  D' \in \text{NST}(\delta) \sqcup \{0\}$ are determined by the graphical calculus as in Corollary \ref{c projecting to heartproj basis}, and depend on which (if any) $V$-arcs of  $T$ are paired with  $k'_1, k'_2$ in the  $V$-diagram of  $T$.

A similar statement holds for $C \in \text{NST}_W(\gamma)$ by considering  $W$-words instead of  $V$-words.
\end{corollary}
\begin{example} 
The following example corresponds to Corollary \ref{c projecting to heartproj basis Kronecker} for $\beta = (), \gamma = (2,2), \delta = (1,1)$, and to the case $\mathbf{k'}= 22$:
\[  \myvcenter{\ensuremath{\column{3 \\ 4}\column{3 \\ 2} \raisebox{-10pt}{\heartg} \column{1}\column{1}}} = \ctableaud{3 \\ 4}{3 \\ 2}{1}{1} + \frac{1}{[2]} \ctableaud{3 \\ 4}{1 \\ 2}{3}{1} \ (\text{in } \nsbr{Y}_{(2,2,1,1)}).\]
Hence
\[ \ctableaud{3 \\ 4}{3 \\ 2}{1}{1} = - \frac{1}{[2]} \ctableaud{3 \\ 4}{1 \\ 2}{3}{1} =  0 \text{ in } \nsbr{X}_{(2,2,1,1)'}.\]
Compare this to
\[  \myvcenter{\ensuremath{\column{3 \\ 4}\column{3 \\ 2} \raisebox{-10pt}{\heartg} \column{1}\column{2}}} = \ctableaud{3 \\ 4}{3 \\ 2}{1}{2} + \frac{1}{[2]} \ctableaud{3 \\ 4}{1 \\ 2}{3}{2} \ (\text{in } \nsbr{Y}_{(2,2,1,1)}).\]
Combining this with
\[\myvcenter{\ensuremath{\column{3 \\ 4}\column{1 \\ 2} \raisebox{-10pt}{\heartg} \raisebox{-10pt}{$($}\column{3}\column{2} \raisebox{-10pt}{$+ \frac{1}{[2]}$}\column{4}\column{1}\raisebox{-10pt}{$)$}}} = \ctableaud{3 \\ 4}{1 \\ 2}{3}{2} + \frac{1}{[2]}\ctableaud{3 \\ 4}{1 \\ 2}{4}{1} \ (\text{in } \nsbr{Y}_{(2,2,1,1)})
\]
yields
\[ \ctableaud{3 \\ 4}{3 \\ 2}{1}{2} = \frac{1}{[2]^2} \ctableaud{3 \\ 4}{1 \\ 2}{4}{1} \text{ in } \nsbr{X}_{(2,2,1,1)'}.\]
\end{example}

\begin{remark}
Corollary \ref{c projecting to heartproj basis Kronecker} is a more precise result than the bottom case of \eqref{e two cases ABC-ABC} of Theorem \ref{t heartsQ good for submodules}, however we believe the method of proof of the theorem to be valuable for its potential use outside the two-row case.
\end{remark}

We now have the tools to easily prove Proposition \ref{p invariant moves}.  Recall that this states that the invariant  $ \text{NST}( \gdneq\gamma)$ actually belong to $\nsbr{Y}_{\gdneq \gamma}$.
\begin{proof}[Proof of Proposition \ref{p invariant moves}]
The statement for Figure \ref{f straightening111} follows from
\be
\label{e 412 241}
-\frac{1}{[2]} \ctableauc{4}{1}{2} \equiv \ctableauc{2}{3}{2} \equiv -\frac{1}{[2]}\ctableauc{2}{4}{1} \, ,
\ee
where the equivalences are mod $\nsbr{Y}_{\gdneq (1,1,1)}$ and are by Corollary \ref{c projecting to heartproj basis Kronecker}. Next, $\Uqt$ applied to $\ctableausmallc{4}{1}{2} - \ctableausmallc{2}{4}{1} \in \nsbr{Y}_{\gdneq (1,1,1)}$ yields the $\Uqt$-module in Figure \ref{f straightening111}, hence this is a $\Uqt$-submodule of $\nsbr{Y}_{\gdneq (1,1,1)}$.

The statement for Figure \ref{f straightening333} has a similar proof to that for Figure \ref{f straightening111} because \newline $\Res_{U_q(\sl(V)\oplus\sl(W))} \nswedge{3 }{X} \cong \Res_{U_q(\sl(V)\oplus\sl(W))} \nswedge{1 }{X}$.

Let us show the statement for Figure \ref{f straightening3221} for  $t=1$, the general case being similar.  The  $t=1$ case follows from
\be
-\frac{1}{[2]}\ctableaud{2\\3\\4}{2\\4}{1\\3}{1} \equiv \ctableaud{2\\3\\4}{2\\3}{2\\3}{1} \equiv -\ctableaud{2\\3\\4}{3\\2}{3\\2}{1} \equiv \frac{1}{[2]} \ctableaud{2\\3\\4}{3\\4}{1\\2}{1} \equiv \frac{1}{[2]}\ctableaud{2\\3\\4}{2\\4}{1\\3}{1},
\ee
where the equivalences are mod $\nsbr{Y}_{\gdneq (3,2,2,1)}$, the first and third equivalence are by Corollary \ref{c projecting to heartproj basis Kronecker}, the second is by
moves defined by Figures \ref{f straightening21} and \ref{f straightening32}, and the fourth by a move defined by Figure \ref{f straightening22c}.  Here we are implicitly using Proposition \ref{p main canonical basis} (i).
\end{proof}

\subsection{A basis for the two-row Kronecker problem}
After two preliminary lemmas, we state the technical version of our main theorem about the two-row Kronecker problem. We give the main body of the proof in this  subsection, although it depends on combinatorics and a detailed case-by-case analysis given in future sections.
The reader may want to postpone a careful reading of the lemmas until seeing how they are applied in the proof.

\begin{lemma}
\label{l basis to local basis for NSTC}
Let $B$ be a basis of a module $N$ in $\Oint{\gl_2}$
and $G$ a digraph with vertex set $B$ that is a disjoint union of directed paths.
Let $\tilde{F}_*$, (resp. $\tilde{E}_*$) be the function from  $B$ to  $B \sqcup \{0\}$ that takes  $b \in B$ to the vertex obtained by following the edge leaving (resp. going to) $b$ if it exists and to 0 otherwise.  Define functions  $\varphi, \varepsilon: B \to \ZZ_{\geq 0}$ by
\[
\varphi(b) := \max\{i: \tilde{F}_*^i(b) \neq 0\}, \qquad \varepsilon(b) := \max\{i: \tilde{E}_*^i(b) \neq 0\}.
\]
Suppose that
\be \label{e FE action}
\begin{array}{ccl}
 F(b) &=& \displaystyle [\varphi(b)] \tilde{F}_*(b) +  \sum_{\stackrel{b' \in B,}{\varphi(b') + \varepsilon(b') < \varphi(b) + \varepsilon(b)}} a^-_{b' b} b', \quad a^-_{b' b} \in \QQA, \ \br{a^-_{b' b}} = a^-_{b' b}, \ \deg(a^-_{b' b}) < \varphi(b),  \vspace{3mm}\\
E(b) &=& \displaystyle [\varepsilon(b)] \tilde{E}_*(b) +  \sum_{\stackrel{b' \in B,}{\varphi(b') + \varepsilon(b') < \varphi(b) + \varepsilon(b)}} a^+_{b' b} b', \quad a^+_{b' b} \in \QQA,  \ \br{a^+_{b' b}} = a^+_{b' b}, \ \deg(a^+_{b' b}) < \varepsilon(b),
\end{array}
\ee
for any  $b \in B$.  Then the pair $(N,B)$ satisfies (c) and (d) of Definition \ref{d upper based}.
\end{lemma}
\begin{proof}
The $\br{\cdot}$-invariance of $a^-_{b' b}, a^+_{b' b}$ easily implies (c).

Next, note that it follows from the form of \eqref{e FE action} that $N$ is filtered by the submodules $N_{\leq k} := \field \{b \in B: \varphi(b)+\varepsilon(b) \leq k\}$ and the subquotient  $N_{\leq k}/N_{\leq k-1}$ is isomorphic to $\bigoplus_{\lambda_1 - \lambda_2 = k} N[\lambda]$.  Set  $\pi^N_k = \sum_{\lambda_1-\lambda_2 = k} \pi^N_\lambda$.  Recall that  $N[\lambda]$ is the  $V_\lambda$-isotypic component of  $N$ and  $\pi^N_\lambda : N \to N$ is the projector with image $N[\lambda]$.

To prove that  $(N,B)$ satisfies (d), let $\L(N) = \field_\infty B$ as in Definition \ref{d upper based}. To show that $\L(N)$ is an upper crystal lattice at $\u = \infty$, we first show that for any $b \in B$, $\crystalusual{F}(\pi^{N}_k(b)) \in \L(N)$ with $k = \varphi(b)+\varepsilon(b)$. Using the filtration of $N$ just mentioned, we see that
\be\label{e pi b equiv b}
\pi^{N}_k(b) = \frac{[k-j]!F^j}{[k]!} \frac{E^j}{[j]!} \, b \equiv b,
\ee
where $j := \varepsilon(b)$, and the equivalence is mod $\ui\L(N)$ and is proved using \eqref{e FE action} to evaluate the middle expression.
Hence
\be\label{e crystal F on pi b}
\crystalusual{F}(\pi^{N}_k(b)) = \pi^{N}_k(\tilde{F}_*(b)) \equiv \tilde{F}_*(b) \in \L(N),
\ee
where the equivalence is mod $\ui \L(N)$. To show that $\crystalusual{F}(\L(N)) \subseteq \L(N)$ we must also show that $\crystalusual{F}(b - \pi^N_k(b)) \in \L(N)$. But this follows from what was just shown applied inductively, as $b - \pi^N_k(b) \in \ui\L(N) \cap N_{\leq k-1}$.
The proof that $\crystalusual{E}(\L(N)) \subseteq \L(N)$ is similar. Finally, it follows from \eqref{e pi b equiv b} and \eqref{e crystal F on pi b} and similar statements for $E$ in place of $F$ that $(\L(N), \B)$ is an upper crystal basis at $\u = \infty$ (where $\B$ is the image of $B$ in $\L(N) / \ui\L(N)$).
\end{proof}

For the statement of Theorem \ref{t main canonical basis}, the integral forms needed to define specializations at $\u=1$ must be chosen carefully (recall that if  $N$ is a  $\field$-module and $N_\mathbf{A}$ is an  $\mathbf{A}$-submodule of $N$ that is understood from context, then $N|_{\u=1}$ is defined to be $\QQ \tsr_{\mathbf{A}} N_\mathbf{A}$, the map $\mathbf{A} \to \QQ$ given by $\u \mapsto 1$).
Define the following integral forms and basis of  $\nsbr{Y}_{\gdneq^i \nu'}$:
\be\label{e integral forms}
\begin{array}{cll}
\nsbr{Y}_{\nu'}^{S \mathbf{A}} &:=& \mathbf{A} \text{SNST}(\nu'), \vspace{.03in} \\
\text{SNST}(\gdneq^i \nu') & :=& \parbox[t]{10.7cm}{
$\mathbf{A}\Big\{({\textstyle -\frac{1}{[2]}})^{\deg(\tilde{T})} \tilde{T}: \tilde{T} \in \text{NST}(\beta, \gdneq\gamma,\delta) \Big\},$
where $\nu' =(\beta, \gamma,\delta)$ as in  \textsection\ref{ss agrees in associated graded} and $\gamma = (\nu'_i, \nu'_{i+1})$,  }
\vspace{.06in} \\
\nsbr{Y}_{\gdneq^i \nu'}^{S \mathbf{A}} &:=& \mathbf{A} \text{SNST}(\gdneq^i \nu'), \\[1mm]
\nsbr{Y}_{\gdneq \nu'}^{S \mathbf{A}} & := & \sum_{i =1}^{l-1} \nsbr{Y}_{\gdneq^i \nu'}^{S \mathbf{A}}, \vspace{.03in} \\
\nsbr{X}_\nu^{'\mathbf{A}} & := & \nsbr{Y}_{\nu'}^{S \mathbf{A}}/\nsbr{Y}_{\gdneq \nu'}^{S \mathbf{A}}, \vspace{.03in} \\
\nsbr{X}_\nu^\mathbf{A} &:=& \nsbr{Y}_{\nu'}^{S \mathbf{A}}/(\nsbr{Y}_{\gdneq \nu'} \cap \nsbr{Y}_{\nu'}^{S \mathbf{A}}) = \varpi(\nsbr{Y}_{\nu'}^{S \mathbf{A}}) = \mathbf{A} \text{NSTC}(\nu) = \mathbf{A} \pNSTC(\nu).
\end{array}
\ee
For the second line, note that $\deg(x)$ is defined for any  $x \in \nsbr{Y}_{\nu'}$ in \textsection\ref{ss invariants}.
On the last line, the second equality follows from  Proposition \ref{p main canonical basis} (i); the third equality holds because those  $\text{NSTC}(\nu)$ that are dishonest because of an integral move are 0 in  $\nsbr{X}_\nu$ by Proposition \ref{p main canonical basis} (ii), and those that are dishonest because of a nonintegral move lie in $\mathbf{A} \pNSTC(\nu)$ by Corollary \ref{c projecting to heartproj basis Kronecker} and Proposition \ref{p connected component of degree-preserving moves} (b).

\begin{remark} \label{r integral form trickiness}
It is true that $\nsbr{X}_\nu^{'\mathbf{A}} = \nsbr{X}_\nu^\mathbf{A}$, however this is not yet clear, and in fact, this can fail outside the two-row case.  We have that  $\nsbr{Y}_{\gdneq \nu'}^{S \mathbf{A}} \subseteq (\nsbr{Y}_{\gdneq \nu'} \cap \nsbr{Y}_{\nu'}^{S \mathbf{A}})$.  Hence there is a canonical surjection $s_X : \nsbr{X}_\nu^{'\mathbf{A}} \twoheadrightarrow \nsbr{X}_\nu^\mathbf{A}$.   This is all we can say in general, without the  $\dv = \dw =2$  assumption.


It will be shown in Theorem \ref{t main canonical basis} that  $\nsbr{X}_\nu^\mathbf{A}$ has  $\mathbf{A}$-basis  $\pNSTC(\nu)$ and that this basis has the correct size $|\text{SSYT}_{\dx}(\nu)|$.  Thus there are no relations amongst  $\pSNST(\nu')$ not already accounted for by Proposition \ref{p main canonical basis}.  Also, from the proof of Proposition \ref{p invariant moves}, we see that all the identifications of  $\pSNST(\nu')$ made to get  $\pNSTC(\nu)$ actually belong to $\nsbr{Y}_{\gdneq \nu'}^{S \mathbf{A}}$.  Thus the surjection  $s_X$ is actually an equality.  However, we do not need this since once we know  $\pNSTC(\nu)$ is an  $\mathbf{A}$-basis of  $\nsbr{X}_\nu^\mathbf{A}$, we can just use this integral form of  $\nsbr{X}_\nu$ and forget about $\nsbr{X}_\nu^{'\mathbf{A}}$.
\end{remark}


The next lemma roughly says that the quotient of a vector space with basis by relations that only involve two basis elements is easily understood in terms of the components of a graph.

\begin{lemma}
\label{l u=1 okay}
Let  $Y$ be a  $\field$-vector space with basis $B = \{b_1,\ldots,b_s\}$ and set  $Y_\mathbf{A} = \mathbf{A} B$. Let $r_i = c_i b_{j_i} + c_i' b_{j_i'}$,  $i \in [t]$,  $j_i, j_i' \in [s]$ be elements of  $Y_\mathbf{A}$ with $c_i, c_i' \in \mathbf{A}$ and $c_i \neq 0$; let $M$ (resp. $M_\mathbf{A}$) be the $\field$-submodule (resp. $\mathbf{A}$-submodule) of  $Y$ spanned by $r_1, \ldots, r_t$. Let  $X$ (resp. $X_\mathbf{A}$) be the quotient  $Y/M$ (resp. $Y_\mathbf{A}/M_\mathbf{A}$). Suppose the $c_i, c_i'$ have no poles or zeros at $\u = 1$ except for those $c_i'$ that are 0, and, for any well-defined product
\[p = \prod_{j \in S} \frac{c_j}{c_j'} \prod_{j \in S'} \frac{c_j'}{c_j}, \qquad S \sqcup S' \subseteq [t],\]
$p = 1$ if and only if $p|_{\u=1} = 1$. Then  $\dim_\field X = \dim_\QQ X|_{\u=1}$ and the exact sequence
\[ 0 \to M_\mathbf{A} \to Y_\mathbf{A} \to X_\mathbf{A} \to 0 \]
remains exact after tensoring with $\QQ$, the map  $\mathbf{A} \to \QQ$ given by $\u \mapsto 1$.
\end{lemma}
\begin{proof}
Let $G$ be the weighted digraph with vertex set $B \sqcup \{0\}$ and, for each  $i \in [t]$,
\[
\begin{cases}
\text{a directed edge $b_{j_i} \to b_{j_i'}$ with weight  $\frac{c_i'}{c_i}$ and} &  \vspace{-.03in}\\
\text{a directed edge $b_{j_i'} \to b_{j_i}$ with weight  $\frac{c_i}{c_i'}$} &  \text{if } c_i' \neq 0, \vspace{.04in}\\
\text{a directed edge $b_{j_i} \to 0$ with weight $c_i$} & \text{if } c_i' = 0.
\end{cases}
\]
Let  $G|_{\u=1}$ be the same digraph with edge weights evaluated at $\u=1$ (by the hypotheses of the lemma, these evaluations are well-defined elements of  $\QQ \setminus \{0\}$).

We say that a component $C$ of the underlying undirected graph of  $G$ (or of  $G|_{\u=1}$) is \emph{honest} if  $0 \notin V(C)$, and, for every directed cycle in $G$ with vertex contained in $V(C)$, the product of its edge weights is 1.  Then  $\dim_\field X$ (resp.  $\dim_\QQ X|_{\u=1}$) is the number of honest components of $G$ (resp.  $G|_{\u=1}$).  The hypotheses of the lemma certainly imply that  $G$ and  $G|_{\u=1}$ have the same number of honest components.

To prove the statement about exactness, first observe that applying the functor $\QQ \tsr_\mathbf{A} \cdot$ to the exact sequence above is the same as first applying  $\QQ \tsr_\ZZ \cdot$ and  then applying  $\QQ \tsr_{\QQA} \cdot$.  Applying  $\QQ \tsr_\ZZ \cdot$ yields the sequence
\[ 0 \to \QQ \tsr_\ZZ M_\mathbf{A} \to \QQ \tsr_\ZZ Y_\mathbf{A} \to \QQ \tsr_\ZZ X_\mathbf{A} \to 0, \]
which is exact because localizations are flat.  Since torsion-free  $\QQA$-modules are free, $\QQ \tsr_\ZZ M_\mathbf{A}$ is a free  $\QQA$-module, hence $\dim_\QQ M|_{\u=1} = \rank_\QQA \QQ \tsr_\ZZ M_\mathbf{A} = \dim_\field M$.  We also have $\dim_\QQ Y|_{\u=1} = \dim_\field Y$ and  we just proved $\dim_\field X = \dim_\QQ X|_{\u=1}$.  Since
\[ 0 \to \QQ \tsr_\mathbf{A} M_\mathbf{A} \to \QQ \tsr_\mathbf{A} Y_\mathbf{A} \to \QQ \tsr_\mathbf{A} X_\mathbf{A} \to 0 \]
is right exact, it is exact by the dimension count
\[\dim_\QQ M|_{\u=1} = \dim_\field M = \dim_\field Y - \dim_\field X = \dim_\QQ Y|_{\u=1} - \dim_\QQ X|_{\u=1}.\]
\end{proof}

\begin{definition}
\label{d crystal components pNSTC}
A  $\Uqvw$-crystal component (resp. -cell) $\G$ of $(\nsbr{X}_\nu, \pNSTC(\nu))$ is a \emph{$(\lambda, \mu)$-crystal component (resp. cell) of $\pNSTC(\nu)$} if its highest weight $\pNSTC$ has weight $(\lambda, \mu)$.
This means that if $\G$ is drawn as in Figure \ref{f crystal component 321}, then it occupies a grid of width $\lambda_1 - \lambda_2 +1$ and height  $\mu_1-\mu_2+1$.

A  $\Uqt$-cell $\G$ of $(\nsbr{X}_\nu, \pNSTC(\nu))$ is a \emph{$\{\lambda, \mu\}$-cell (resp.  $\varepsilon \lambda$-cell) of $\pNSTC(\nu)$} if the corresponding cellular subquotient is isomorphic to $\X_{\{\lambda,\mu\}}$ (resp. $\X_{\varepsilon \lambda}$) (see \textsection\ref{ss representation theory Uqt Oqt} for notation).

A crystal component or cell of  $\pNSTC(\nu)$ is \emph{invariant-free} if every (equivalently, any)  $\pNSTC$ it contains is invariant-free.
\end{definition}

We now state our main result on the two-row Kronecker problem.  This is a stronger and more technical version of Theorem \ref{t intro main theorem advertisement}.
 We give most of the proof now, although it depends on several results that we prove later.
One is Corollary \ref{c basis correct size} (see \textsection \ref{ss semistandard bijection}), in which we use a counting argument to show that $|\pNSTC(\nu)| = |\text{SSYT}_{\dx}(\nu)|$.  This, together with what we have proved so far, is enough to obtain a combinatorial formula for two-row Kronecker coefficients.
The stronger statements (v) and (vi) require a more detailed combinatorial understanding of the relations satisfied by $\varpi_\nu(\text{NST}(\nu'))$ and the rather involved case-by-case analysis of the $\F_{(j)V}(T)$ given in \textsection\ref{ss action of Chevalley generators}.

\begin{theorem}
\label{t main canonical basis}
The pair $(\pNSTC(\nu), \nsbr{X}_\nu)$ yields a ``crystal basis-theoretic'' solution to the two-row Kronecker problem.
Precisely, we prove the following.
\begin{list}{\emph{(\roman{ctr})}} {\usecounter{ctr} \setlength{\itemsep}{1pt} \setlength{\topsep}{2pt}}
\item $\dim_\field \nsbr{X}_\nu = \dim_\QQ \QQ \tsr_\mathbf{A} \nsbr{X}_\nu^{'\mathbf{A}}$.
\item $\QQ \tsr_\mathbf{A} \nsbr{X}_\nu^{'\mathbf{A}} \cong \Res_{U^\tau} (X_\nu|_{\u = 1})$ as $U^\tau$-modules (where $U^\tau := U(\gl_2) \wr \S_2$).
\item The set $\pNSTC(\nu)_{\geq h}$ is an  $\mathbf{A}$-basis of $(\nsbr{X}_\nu^\mathbf{A})_h$. In particular, $\pNSTC(\nu)$ is an  $\mathbf{A}$-basis of $\nsbr{X}_\nu^\mathbf{A}$.
\item The pair $(\gr(\nsbr{X}_\nu), \grin(\pNSTC(\nu)))$ is an upper based $\Uqvw$-module.
\item The pair $(\nsbr{X}_\nu, \pNSTC(\nu))$ is an upper based $\Uqvw$-module.
\item The upper based  $\Uqvw$-modules $(\gr(\nsbr{X}_\nu), \grin(\pNSTC(\nu)))$ and $(\nsbr{X}_\nu, \pNSTC(\nu))$ are upper based $\Uqt$-modules.
\item The Kronecker coefficient $g_{\lambda\mu \nu}$ is the number of highest weight $\pNSTC$ of shape $\nu$ and weight $(\lambda,\mu)$, or equivalently, the number of $(\lambda, \mu)$-crystal components of  $\pNSTC(\nu)$.
\item The symmetric or exterior Kronecker coefficient $g_{\varepsilon \, \lambda\nu}$ is the number of highest weight $\pNSTC$  $\mathbf{T}$ of shape $\nu$ and weight $(\lambda,\lambda)$ such that  $\tau \mathbf{T} = \varepsilon \mathbf{T}$, or equivalently, the number of $\varepsilon \lambda$-cells
    of  $\pNSTC(\nu)$.  Moreover, the condition $\tau \mathbf{T} = \varepsilon \mathbf{T}$ is equivalent to $(-1)^{\nu_3+\arc{3-2}(\mathbf{T})} = \varepsilon$ (the notation $\arc{3-2}$ is introduced after Definition \ref{d 2-1 arc}).
\end{list}
\end{theorem}

\begin{remark}\label{r main canbas}
\
\begin{list}{(\arabic{ctr})} {\usecounter{ctr} \setlength{\itemsep}{1pt} \setlength{\topsep}{2pt}}
\item It follows from (iii) that the degree of a $\pNSTC$ as an element of the filtered module $\nsbr{X}_\nu$ is the same as its degree defined after Proposition \ref{p connected component of degree-preserving moves} (the degree of a dishonest NSTC as an element of  the filtered $\nsbr{X}_\nu$ is more difficult to determine).
Thus  $\bigsqcup_{h \geq 0} \grin_h(\pNSTC(\nu)_{h})$ and  $\grin(\pNSTC(\nu))$ are identical, so we can safely use the latter as a shorthand for the former.
\item The  $\Uqvw$-cells of $(\gr(\nsbr{X}_\nu), \grin(\pNSTC(\nu)))$ and $(\nsbr{X}_\nu, \pNSTC(\nu))$ coincide under the bijection $\grin(\pNSTC(\nu)) \cong \pNSTC(\nu)$.  The same goes for $\Uqt$-cells.  The $\Uqvw$-cells of $(\gr(\nsbr{X}_\nu), \grin(\pNSTC(\nu)))$ and $(\nsbr{X}_\nu, \pNSTC(\nu))$ also coincide with their crystal components (this is always true for upper based $\Uqvw$-modules---see  \textsection\ref{ss global canonical bases}). We give a fairly explicit description of these cells in Corollaries \ref{c NSTC cells} and \ref{c NSTC tau cells} after we have a better combinatorial understanding of $\pNSTC$.
\item It follows from \textsection\ref{ss nonstandard two-row case} that the $\nsbr{Y}_{\gdneq \gamma}$ for $\ell(\gamma) =2$ are $\O(GL_q(\nsbr{X}))$-comodules, hence $\nsbr{X}_\nu$ is a $\O(GL_q(\nsbr{X}))$-comodule. It also follows from (vi) and \textsection\ref{ss nonstandard two-row case} that  the $\Uqt$-cells and  $\field \nsSchur{r}$-cells of  $(\nsbr{X}_\nu, \pNSTC(\nu))$ are the same except that $(\nsbr{X}_{(r)},\pNSTC((r)))$ is a $\field \nsSchur{r}$-cell and the union of  $\lfloor \frac{r}{2} \rfloor +1$  $\Uqt$-cells (recall that $\field \nsSchur{r}$ is the algebra dual to the coalgebra $\O(M_q(\nsbr{X}))_r$). See the comment after Corollary \ref{c NSTC tau cells} for more details.
\item The pairs $(\gr(\nsbr{X}_\nu), \grin(\pNSTC(\nu)))$ and  $(\nsbr{X}_\nu, \pNSTC(\nu))$ are not isomorphic as upper based $\Uqvw$-modules.
\item The proof of (iv) and (v) uses Lemma \ref{l basis to local basis for NSTC}, which uses information about a basis to deduce the existence of a (local) crystal basis. This may seem backward since one usually first proves the existence of a (local) crystal basis to deduce the existence of a global crystal basis. However, because we have explicit formulae for the action of the Chevalley generators on NST in the two-row case (Proposition \ref{p F action on heartsQ basis}),
    our proof is essentially the same as, and perhaps somewhat more convenient than, one that first proves the existence of a (local) crystal basis.
\item The proof of (v) might be easier if we could construct a basis for all of $\nsbr{Y}_{\gdneq \nu'}$ in which $(\nsbr{X}_\nu, \pNSTC(\nu))$ occurs as a  cellular quotient (see  Conjecture \ref{cj canonical basis X^r}).
\end{list}
%
%
\end{remark}

\begin{proof}[Most of the proof of Theorem  \ref{t main canonical basis}]
\ \newline
\vspace{-1.2mm}
Statement (i): by the definitions \eqref{e integral forms}, we have the exact sequence of $\mathbf{A}$-modules
\be \label{e exact integral form}
0 \to \nsbr{Y}_{\gdneq \nu'}^{S \mathbf{A}} \to \nsbr{Y}_{\nu'}^{S \mathbf{A}} \to  \nsbr{X}^{'\mathbf{A}}_\nu  \to 0.
\ee
Since localizations are flat, tensoring with $\field$ yields the exact sequence
\[ 0 \to \nsbr{Y}_{\gdneq \nu'} \to \nsbr{Y}_{\nu'} \to  \nsbr{X}_{\nu} \to 0.\]
Then Lemma \ref{l u=1 okay} applied with  $Y = \nsbr{Y}_{\nu'}$, $B = \text{SNST}(\nu')$, $M = \nsbr{Y}_{\gdneq \nu'}$, and $\{r_i : i \in [t]\} = \bigsqcup_{i \in [l-1]} \text{SNST}(\gdneq^i \nu')$ yields the desired $\dim_\field \nsbr{X}_{\nu} = \dim_\QQ \QQ \tsr_\mathbf{A} \nsbr{X}_\nu^{'\mathbf{A}}$. Here we have used the top case of \eqref{e two cases ABC-ABC} of Theorem \ref{t heartsQ good for submodules} and Corollary \ref{c projecting to heartproj basis Kronecker} to show that any product $p$ of Lemma \ref{l u=1 okay} is $\pm [2]^k$, $k \in \ZZ$, so the hypotheses of the lemma are satisfied.

Statement (ii): the above application of Lemma \ref{l u=1 okay} also implies that \eqref{e exact integral form} remains exact after tensoring with $\QQ$ (the map $\mathbf{A} \to \QQ$ given by $\u \mapsto 1$), i.e., $\QQ \tsr_\mathbf{A} \nsbr{X}_\nu^{'\mathbf{A}} \cong \nsbr{Y}_{\nu'}|_{\u=1} / \nsbr{Y}_{\gdneq \nu'}|_{\u=1}$.
This quotient is isomorphic to $\Res_{U^\tau} (X_\nu|_{\u=1})$: this is true in the $\ell(\nu') \leq 2$ case because the
$\nsbr{Y}_{\gdneq \nu'}$ were defined to make this true; also note that the decomposition of $\nsbr{Y}_{\gdneq \nu'}|_{\u = 1}$ into $U^\tau$-modules is multiplicity free for $\ell(\nu') = 2$; the case  $\ell(\nu') > 2$ then follows because the fact $\nsbr{Y}_{\gdneq \nu'}|_{q=1} = \sum_{i=1}^{l-1} \nsbr{Y}_{\gdneq^i \nu'}|_{q=1}$ implies that $\nsbr{Y}_{\nu'}|_{\u=1} / \nsbr{Y}_{\gdneq \nu'}|_{\u=1}$ matches the definition of the Schur functor $L_{\nu'} X$ given in \cite[Chapter 2]{Weyman} (see \textsection\ref{ss the approach of Adsul, Sohoni, and Subrahmanyam}).

Statement (iii):  Proposition \ref{p main canonical basis} (v) implies the left-hand inequality, (i) and (ii) imply the middle equality, and Proposition \ref{p NSTC semistandard bijection} implies the right-hand inequality of
\[|\pNSTC(\nu)| \geq \dim_\field \nsbr{X}_\nu = |\text{SSYT}_{\dx}(\nu)| \geq |\pNSTC(\nu)|,\]
hence equality must hold throughout.  Thus $\pNSTC(\nu)$ is a basis of $\nsbr{X}_\nu$, and, since any nontrivial relation with  coefficients in $\mathbf{A}$  satisfied by the $\pNSTC(\nu)$ is also a relation over  $\field$, we conclude that $\pNSTC(\nu)$ is an $\mathbf{A}$-basis of  $\nsbr{X}_\nu^\mathbf{A}$.
Now this further implies by Proposition \ref{p main canonical basis} (iv) and (v) that $\pNSTC(\nu)_{\geq h}$ is an  $\mathbf{A}$-basis for $(\nsbr{X}_\nu^\mathbf{A})_h$ (where $(\nsbr{X}_\nu^\mathbf{A})_h  := \nsbr{X}_\nu^\mathbf{A} \cap (\nsbr{X}_\nu)_h $) and $\grin_h(\pNSTC(\nu)_{h})$ is an $\mathbf{A}$-basis for $\gr(\nsbr{X}_\nu^\mathbf{A})_h$.

Statement (iv): we need to check conditions (a)--(d) of Definition \ref{d upper based} for each pair $P_h := (\gr(\nsbr{X}_\nu)_h, \grin_h(\pNSTC(\nu)_{h}))$. First of all, we have seen in Proposition-Definition \ref{pd gr Y alpha} that $(\gr(\nsbr{Y}_{\nu'})_h, \grin_h(\text{NST}(\nu')_{h}))$ is a weak upper based $\Uqt$-module. Scaling this basis by the factor $(-\frac{1}{[2]})^h$ yields the weak upper based $\Uqt$-module $(\gr(\nsbr{Y}_{\nu'})_h, \grin_h(\pSNST(\nu')_{h}))$.  By Proposition \ref{p main canonical basis} (i) and (iii), the pair $(\gr(\nsbr{X}_\nu)_h, \grin_h(\pNSTC(\nu)_{h}))$ is then obtained from this one by quotienting by $\gr(\nsbr{Y}_{\gdneq \nu'})_h$, which amounts to identifying  $\pSNST(\nu')$ in the same strong component of  $\tgraph(\nu)$ and getting rid of those $\pSNST(\nu')$ that become 0, which are exactly the dishonest ones.  Given this, condition (a) is clear and condition (b) for $P_h$ is easy to check from condition (b) for $(\gr(\nsbr{Y}_{\nu'})_h, \grin_h(\text{NST}(\nu')_{h}))$.

To prove that (c) and (d) hold, we apply Lemma \ref{l basis to local basis for NSTC} (we apply the lemma to show that (d) holds for  $P_h$ as a  $U_q(\g_V)$-module with basis; that (d) holds for  $P_h$ as a  $U_q(\g_W)$-module with basis is similar):
first note that
\refstepcounter{equation}
\begin{enumerate}[label={(\theequation)}]
\label{e V word same}
\item if $T \to T'$ is a degree-preserving move at  $[i,i+t-1]$, then the unpaired $V$-diagram of $T$ is the same as that of $T'$.  Moreover, the  $V$-diagram of  $T$ and  $T'$ are identical outside  columns  $i$ through  $i+t-1$.
\end{enumerate}
Then, for $\mathbf{T} \in \pNSTC(\nu)_{h}$, define $\tilde{F}_*(\mathbf{T})$ to be the  $\pNSTC$ containing $\F_{(\varphi_V(T))V}(T)$ for any $T \in \mathbf{T}$.
To show that this is well-defined we must show that if  $T \to T'$ is a degree-preserving move at  $[i,i+t-1]$, then $\F_{(\varphi_V(T))V}(T) \to \F_{(\varphi_V(T))V}(T')$ is a degree-preserving move.  Given \eqref{e V word same}, this is evident from Figures \ref{f straightening21}, \ref{f straightening22b}, \ref{f straightening22c}, \ref{f straightening32}, \ref{f straightening111}, and \ref{f straightening333}
if $T$ and  $\F_{(\varphi_V(T))V}(T)$ differ in the  $j$-th column and  $j \in [i,i+t-1]$,  and is clear if  $j \notin [i,i+t-1]$.
The graph  $G$ and $\tilde{E}_*$ of the lemma are then determined uniquely so that the conditions of the lemma are satisfied.  The conditions in \eqref{e FE action} on the coefficients $a^-_{b' b}$ and $a^+_{b' b}$ hold by Proposition
\ref{p F action on heartsQ basis} and Proposition \ref{p main canonical basis} (i), (iii).

Statement (v): the proof is similar to that of (iv). The following modifications will suffice:
to show condition (b) for the pair $(\nsbr{X}_\nu, \pNSTC(\nu))$, we need check that the structure coefficients for the action of $F_V^{(m)}, F_W^{(m)}, E_V^{(m)}, E_W^{(m)}$ in the basis $\pNSTC(\nu)$ lie in $\mathbf{A}$.  This follows from condition (b) for  $(\nsbr{Y}_{\nu'},\text{NST}(\nu'))$, Corollary \ref{c projecting to heartproj basis Kronecker} together with Proposition \ref{p connected component of degree-preserving moves} (b), and Proposition \ref{p main canonical basis} (i), (ii). This is easy because scaling the elements of $\text{NST}(\nu')_{h}$ by the factor $(-\frac{1}{[2]})^h$ only makes it easier to have the structure coefficients lie in $\mathbf{A}$.
The main difficulty is showing that the degree bounds  on $a^-_{b' b}$ and $a^+_{b' b}$ given in \eqref{e FE action} hold.  These follow from Theorem \ref{t positivity} (see  \textsection\ref{ss action of Chevalley generators}).

Statement (vi): the given upper based $\Uqvw$-modules are weak upper based $\Uqt$-modules since  $(\nsbr{Y}_{\nu'}, \text{NST}(\nu'))$ is a weak upper based $\Uqt$-module.  That they are in fact upper based  $\Uqt$-modules is the contents of Proposition \ref{p NSTC is upper based tau module}.

Statement (vii): this follows from (ii), (v), and the general facts about upper based  $\Uqvw$-modules in  \eqref{e based module filtration}.

Statement (viii): this follow from (ii), (vi), and the general facts about upper based  $\Uqt$-modules in  \textsection\ref{ss upper based Uqt}.
\end{proof}

\section{Straightened NST and semistandard tableaux}
\label{s Straightened NST and semistandard tableaux}
In this section we study the equivalence classes  $\pNSTC$ of  $\pSNST$ combinatorially.  We define a lexicographic order on  $\pSNST$ and give explicit necessary conditions on the columns of a $\pSNST$ for it to be the smallest $\pSNST$ in its class---we call a $\pSNST$ and its corresponding NST satisfying these conditions \emph{straightened}.  We then exhibit a bijection between straightened NST of shape  $\nu'$ and SSYT$_{\dx}(\nu)$, completing the proof of Theorem \ref{t main canonical basis} (iii) and (iv) and showing that these necessary conditions are also sufficient.

\subsection{Lexicographic order on NST}
We define a total order $<$ on nonstandard columns, which is very similar to the order that columns in a semistandard tableau must satisfy:
\begin{eqnarray}
\label{e ordering}
\ctableaua{1 \\ 2 \\ 3 \\ 4} < \ctableaua{1 \\ 2 \\ 3 }< \ctableaua{1 \\ 2 \\ 4 } < \ctableaua{1 \\ 3 \\ 4 } < \ctableaua{2 \\ 3 \\ 4 }<  \ctableaua{1 \\ 2} < \ctableaua{1 \\ 3} < \ctableaua{2 \\ 3} < \ctableaua{3 \\ 2} <  \ctableaua{2 \\ 4}< \ctableaua{3 \\ 4} < \ctableaua{1 }<  \ctableaua{2 } < \ctableaua{3 } < \ctableaua{4 }.
\end{eqnarray}
We consider the lexicographic order on $\pSNST(\nu')$ induced from this order: for  $T, T' \in \pSNST(\nu')$, $T < T'$ if  $T|_j < T'|_j$ and $T|_{[j-1]} = T'|_{[j -1]}$, $j \in [l]$.  We also consider this order on $\text{NST}(\nu')$.

A nonstandard column of height $2$ is of type  $V$ (resp.  $W$) if it does not contain an internal $V$-arc (resp. $W$-arc):
\[
\begin{array}{ccccccc}
\column{1 \\ 2} & \column{3 \\ 2} & \column{3 \\ 4} & \quad &\column{1 \\ 3} & \column{2 \\ 3} &  \column{2 \\ 4} \vspace{.07in} \\
\multicolumn{3}{c}{\text{type  $V$}} & \quad & \multicolumn{3}{c}{\text{type $W$}}
\end{array}.
\]

We will occasionally write height-3 columns in a more compact form as
\[\ctableaua{\dual{1}} := \ctableaua{1 \\ 2 \\ 3}, \ \ctableaua{\dual{2}} := \ctableaua{1 \\ 2 \\ 4 } , \ \ctableaua{\dual{3}} := \ctableaua{1 \\ 3 \\ 4 } , \ \ctableaua{\dual{4}} := \ctableaua{2 \\ 3 \\ 4 }. \]
For an NST or SNST  $T$, we sometimes use $m$ with various subscripts to denote the number of columns of each type comprising $T$, i.e.  $m_1$ is the number of  columns of  $T$  equal to $\ctableausmalla{1}$, $m_{\substack{2 \\ 3}}$ is the number of  columns of  $T$  equal to $\ctableausmalla{2 \\ 3}$, etc.

We begin with a basic result about the order and number of columns in an honest SNST.
\begin{proposition}
Let $T$ be an honest SNST and let $m_1, m_2$, etc. denote the number of columns of $T$ of each type, as above.
\label{p easy straightenings}
\begin{list}{\emph{(\roman{ctr})}} {\usecounter{ctr} \setlength{\itemsep}{1pt} \setlength{\topsep}{2pt}}
\item If two honest SNST differ by moving some type $V$ columns past type $W$ columns, then they are equivalent.
\item If two honest SNST differ by moving a height-2 invariant past some columns of height $2$, then they are equivalent.
\item The type $V$ columns of the invariant-free part of $T$ are weakly increasing; $m_{\substack{3 \\ 2}} \leq 1$.
\item The type $W$ columns of the invariant-free part of $T$ are weakly increasing; $m_{\substack{2 \\ 3}} \leq 1$.
\item The height-1 columns of the invariant-free part of $T$ are weakly increasing; at least one of $m_2$ and $m_3$ is 0.
\item The height-3 columns of the invariant-free part of $T$ are weakly increasing; at least one of $m_{\dual{2}}$ and $m_{\dual{3}}$ is 0.
\end{list}
\end{proposition}
\begin{proof}
Statement (i) follows from the moves defined by Figure \ref{f straightening22b}.
Statement (ii) follows from (i) and the moves defined by Figure \ref{f straightening22c} (which simply correspond to replacing a contiguous subtabloid equal to a height-2 invariant with the other height-2 invariant).

Given (i) and (ii),  $T$ is equivalent to a scaled nonstandard tabloid $T'$ satisfying:
(1) the type $V$ columns of  $T'$ that remain in its invariant-free part form a contiguous subtabloid, (2) all of its height-2 invariant column pairs are contiguous and occur at the end of its subtabloid of height-2 columns, and (3) the type $V$ columns of its invariant-free part are in the same order as the type $V$ columns of the invariant-free part of $T$. Since  $T$ is honest, so is  $T'$ and then (iii) is immediate from the moves defined by Figure \ref{f straightening22a}.  The proof of (iv) is similar.

By the moves defined by Figure \ref{f straightening11}, $T$ is honest implies $T$ does not contain a contiguous subtabloid $\ctableausmall{i}{j}$ with $i > j$ unless $\ctableausmall{i}{j} = \ctableausmall{4}{1}$. By the invariant moves defined by Figure \ref{f straightening111}, $T$ is equivalent to a $\pSNST$ whose height-1 columns are of the form \[\myvcenter{\ensuremath{\column{1} \downdots \column{1}\column{2}\downdots\column{2}\column{3}\downdots\column{3}\column{4}\downdots\column{4}
\column{4}\column{1}\downdots\column{4}\column{1}}}.\]
Also, by the moves defined by Figure \ref{f straightening11}, $T$ does not contain $\ctableausmall{2}{3}$ or $\ctableausmall{3}{2}$ as a contiguous subtabloid, so at least one of $m_2, m_3$ is zero.  This proves (v).  The proof of (vi) is similar using the moves defined by Figure \ref{f straightening33}.
\end{proof}

\begin{proposition}
\label{p difficult straightenings}
Let   $\mathbf{T}$ be an invariant-free honest NSTC of shape  $\nu' \vdash r$ and  $T$ the lexicographically minimum SNST in this class. Then
\begin{list}{} {\usecounter{ctr} \setlength{\itemsep}{1pt} \setlength{\topsep}{2pt}}
\item[\emph{(0.1)}] the height-$j$ columns of  $T$ are weakly increasing for all $j \in [\dx]$,

\item[\emph{(1.1)}] at least one of $m_2$ and $m_3$ is zero,
\item[\emph{(1.2)}] $m_1 > 0$ implies $m_{\substack{3 \\ 4}} = m_{\substack{2 \\ 4}} = 0$,
\item[\emph{(1.3)}] $m_2 > 0$ implies $m_{\substack{3 \\ 4}} = 0$,
\item[\emph{(1.4)}] $m_3 > 1$ implies $m_{\substack{2 \\ 4}} = 0$,
\item[\emph{(1.5)}] $m_1 > 0$ implies $m_{\substack{3 \\ 2}} = 0$,

\item[\emph{(2.1)}] $m_{\substack{2 \\ 3}} \leq 1$ and $m_{\substack{3 \\ 2}} \leq 1$,

\item[\emph{(3.1)}] at least one of $m_{\dual{2}}$ and $m_{\dual{3}}$ is zero,
\item[\emph{(3.2)}] $m_{\dual{4}} > 0$ implies $m_{\substack{1 \\ 2}} = m_{\substack{1 \\ 3}} = 0$,
\item[\emph{(3.3)}] $m_{\dual{3}} > 0$ implies $m_{\substack{1 \\ 2}} = 0$,
\item[\emph{(3.4)}] $m_{\dual{2}} > 1$ implies $m_{\substack{1 \\ 3}} = 0$,
\item[\emph{(3.5)}] $m_{\dual{4}} > 0$ implies  $m_{\substack{3 \\ 2}} =0$,

\item[\emph{(4.1)}] at least one of $m_{\dual{4}}$ and $m_1$ is zero.
\end{list}
\end{proposition}
\begin{proof}
Statements (0.1), (1.1), and (3.1) follow from Proposition \ref{p easy straightenings} and the invariant-free assumption.
Statement (2.1) is a restatement of parts of Proposition \ref{p easy straightenings} (iii) and (iv).
Given (0.1), statements (3.2) and (3.3) are by the moves defined by Figure \ref{f straightening32}.

To prove (3.4), suppose the contrary, that $m_{\dual{2}} > 1$ and  $m_{\substack{1 \\ 3}} > 0 $.  Given (0.1), (3.1), and (3.2), we can apply the move
\[ \myvcenter{\ensuremath{\downdots\column{1\\2\\4}\column{1 \\ 2 \\ 4}\column{1 \\ 3}\downdots}}  \longleftrightarrow -\myvcenter{\ensuremath{\downdots\column{1\\2\\4}\column{1 \\ 3 \\ 4}\column{1 \\ 2}\downdots}} \]
to  $T$, which, by Proposition \ref{p easy straightenings} (vi),  contradicts that  $T$ is honest.

If $m_{\dual{4}} > 0$ and $m_{\substack{2 \\ 3}} > 0$ and $m_{\substack{3 \\ 2}} > 0$, then by (3.2) and (0.1) we can apply the sequence of moves
\[ \myvcenter{\ensuremath{\downdots\column{2\\3\\4}\column{2 \\ 3}\column{3 \\ 2}\downdots}}  \longleftrightarrow -\myvcenter{\ensuremath{\downdots\column{2\\3\\4}\column{3 \\ 2}\column{3 \\ 2}\downdots}} \longrightarrow 0, \]
which contradicts that $\mathbf{T}$ is honest. This proves (3.5).

Statements (1.1)--(1.5) have proofs similar to (3.1)--(3.5), using Figure \ref{f straightening21} instead of Figure \ref{f straightening32} and Proposition \ref{p easy straightenings} (v) instead of (vi).

Finally, for statement (4.1), assume for a contradiction that $m_{\dual{4}}$ and $m_1$ are positive. By (0.1), (1.2), (1.5), (3.2), and (3.5) $m_{\substack{1 \\ 2}} = m_{\substack{1 \\ 3}} = m_{\substack{3 \\ 2}} = m_{\substack{2 \\ 4}} = m_{\substack{3 \\ 4}} = 0$. If $m_{\substack{2 \\ 3}} = 0$ \vspace{1mm} then $T$ is dishonest by the moves defined by Figure \ref{f straightening31}. If $m_{\substack{2 \\ 3}} = 1$ then the moves
\[ \myvcenter{\ensuremath{\downdots\column{2\\3\\4}\column{2 \\ 3}\column{1}\downdots}}  \longleftrightarrow -\myvcenter{\ensuremath{\downdots\column{2\\3\\4}\column{3 \\ 2}\column{1}\downdots}} \longleftrightarrow
-\myvcenter{\ensuremath{\downdots\column{2\\3\\4}\column{2 \\ 3}\column{1}\downdots}}, \]
show that $\mathbf{T}$ is nonorientable, contradicting the assumption that it is honest.
\end{proof}

The invariant moves and Proposition \ref{p easy straightenings} (ii) reduce the analysis of  $\tgraph(\nu)$ to the invariant-free case, as we now verify.

\begin{proposition}
\label{p straightening non invariant-free}
Let   $\mathbf{\check{T}}$ be an honest NSTC of shape $\nu$ and  $\check{T}$ a  SNST in this class that is lexicographically minimum. Then the invariant-free part  $T$ of $\check{T}$ satisfies the conditions of Proposition \ref{p difficult straightenings}.
\end{proposition}
\begin{proof}
By Proposition \ref{p easy straightenings} (ii) and invariant moves, if $T \to T'$ is any move, then there is a path in  $\tgraph(\nu)$ from $\check{T}$ to the
SNST$(\nu')$ obtained from $\check{T}$ by keeping its invariant column pairs fixed and replacing its invariant-free part with $T'$.
Note that if  $T \to T' = 0$ is a move defined by Figure \ref{f straightening31}, then a move defined by Figure \ref{f straightening3221} is needed to conclude that  $\check{T}$ has a path to $0$. It follows that $T$ satisfies the conditions of Proposition \ref{p difficult straightenings}.
\end{proof}

An NST or SNST $\check{T}$ is \emph{straightened} if its invariant-free part  $T$ satisfies the conditions of Proposition \ref{p difficult straightenings} and its invariant column pairs are positioned to make  $\check{T}$ as small as possible in lexicographic order, i.e., height-3 invariant column pairs are contiguous and lie between the last $\ctableaua{\dual{3}}$ column of $T$ and the first $\ctableaua{\dual{4}}$ column of $T$, height-2 invariant column pairs are contiguous, are all of the form  $\ctableausmall{2 \\ 4}{1 \\ 3}$, and lie between the last $\ctableausmalla{3 \\ 2}$ and the first $\ctableausmalla{2 \\ 4}$ of  $T$, and height-1 invariant column pairs are contiguous and lie between the last $\ctableausmalla{3}$ and the first $\ctableausmalla{4}$ of $T$.  \emph{Straightening} an NST  $T$ is the process of following moves from  $T$ to the straightened representative of its class, or following moves that show it to be dishonest.

We have shown in this subsection that, as sets,
\[
\parbox[t]{13.5cm}{\pNSTC($\nu$) $\cong$ lexicographically minimum representatives of $\pNSTC(\nu)$ $\subseteq$ straightened elements of $\pSNST(\nu')$.}
\]
In the next subsection, we will show that this containment is actually an equality.

\subsection{Bijection with semistandard tableaux}
In this subsection we prove the following combinatorial result, completing the proof of Theorem \ref{t main canonical basis} (iii) and (iv).

\label{ss semistandard bijection}
\begin{proposition}
\label{p NSTC semistandard bijection}
There is a bijection between straightened NST of shape $\nu'$ and SSYT$_{\dx}(\nu)$.
\end{proposition}
\begin{proof}
We define a map $f$ from straightened NST of shape $\nu'$ to SSYT$_{\dx}(\nu)$ and check that it has an inverse (we omit a fully explicit description of the inverse and check this somewhat informally).
To define  $f$, let $\check{T}$ be an arbitrary straightened NST of shape  $\nu'$ with invariant record $(i_4, i_3, i_2, i_1)$ and invariant-free part  $T$.  As above, let $m_{\dual{1}}, m_{\dual{2}}, \dots$ denote the number of columns of each type occurring in $T$.  The semistandard tableau  $f(\check{T})$ will be defined through pictures and by specifying the number of columns $m'_{\dual{1}}, m'_{\dual{2}}, \dots$ of each type it is to contain.  Regardless of the values of $(i_4, i_3, i_2, i_1)$ and $m_{\dual{1}}, m_{\dual{2}}, \dots$, we always set $m'_{\dual{1}} = m_{\dual{1}}$, $m'_{\dual{4}} = m_{\dual{4}}$, $m'_1 = m_1$,  $m'_4 = m_4$, and there is no choice for the height-4 columns of $f(\check{T})$.  The remainder of $f(\check{T})$ is defined as follows: a right, middle, and left subtableaux of this remainder (as shown in the pictures below, to the right of the  $\mapsto$) are defined from  $i_1$ and a right subtabloid of  $T$,  $i_2$ and a middle subtabloid of  $T$, and  $i_3$ and a left subtabloid of $T$ (as shown in the pictures below, to the left of the $\mapsto$), respectively.
We freely use Proposition \ref{p difficult straightenings} to break up the definition of $f$ into cases---the parenthetical comments in the cases are consequences of this proposition.

If  $m_1 > 0$, then the right subtableau of $f(\check{T})$ is defined from  $i_1$ and the right subtabloid of  $T$  as follows:
\[ \raisebox{-3pt}{\myvcenter{\ensuremath{\myunderbrace{\column{1} \downdots\downdots \column{1}}_{m_1} \myunderbrace{\column{2} \downdots \column{2}}_{m_2}\myunderbrace{\column{3} \downdots \column{3}}_{m_3} }}}
\ \mapsto \
\raisebox{-3pt}{\myvcenter{\ensuremath{\myunderbrace{\column{1} \downdots\downdots \column{1}}_{m_1} \myunderbrace{\column{2} \downdots \column{2}}_{i_1+m_2}\myunderbrace{\column{3} \downdots \column{3}}_{i_1+m_3} }}}.\]
Here, and elsewhere in the paper, the two sets of dots is an added visual aid to indicate that this type of column appears at least once.

If  $m_1 = 0$, then the right subtableau of  $f(\check{T})$ is defined from $i_1$ and the right subtabloid of  $T$ as follows:
\[ \raisebox{-5pt}{\myvcenter{\ensuremath{\myunderbrace{\column{2 \\ 4} \downdots \column{2 \\ 4}}_{m_{\substack{2 \\ 4}}} \myunderbrace{\column{3 \\ 4} \downdots \column{3 \\ 4}}_{m_{\substack{3 \\ 4}}} \myunderbrace{\column{2} \downdots \column{2}}_{m_2}\myunderbrace{\column{3} \downdots \column{3}}_{m_3}}}} \ \mapsto \]
\[
\begin{cases}%
\myvcenter{\ensuremath{\myunderbrace{\column{2 \\ 4} \downdots \downdots \column{2 \\ 4}}_{m_{\substack{2 \\ 4}}} \myunderbrace{\column{3 \\ 4} \downdots \column{3 \\ 4}}_{m_{\substack{3 \\ 4}}} \myunderbrace{\column{3} \downdots \column{3}}_{2i_1 + m_3}}} & \text{if  $m_{\substack{2 \\ 4}} > 0$ and  $m_2 = 0$ (and  $m_3 \leq 1$),} \\
\myvcenter{\ensuremath{\myunderbrace{\column{2 \\ 4} \downdots\downdots \column{2 \\ 4}}_{m_{\substack{2 \\ 4}}} \myunderbrace{\column{2} \downdots \downdots \column{2}}_{i_1 + m_2}\myunderbrace{\column{3} \downdots \column{3}}_{i_1}}} & \text{if $m_{\substack{2 \\ 4}} > 0$ and  $m_2 > 0$ (and  $m_{\substack{3 \\ 4}} = m_3 = 0$),} \\
\myvcenter{\ensuremath{\myunderbrace{\column{2 \\ 4} \downdots \column{2 \\ 4}}_{m_{\substack{3 \\ 4}}} \myunderbrace{\column{2} \downdots \downdots \column{2}}_{i_1 + m_2}\myunderbrace{\column{3} \downdots \downdots \column{3}}_{i_1 + m_3}}} & \text{if  $m_{\substack{2 \\ 4}} = 0$ and  $i_1  >0$,}\\
\myvcenter{\ensuremath{\myunderbrace{\column{3 \\ 4} \downdots \column{3 \\ 4}}_{m_{\substack{3 \\ 4}}} \myunderbrace{\column{2} \downdots \column{2}}_{m_2}\myunderbrace{\column{3} \downdots \column{3}}_{m_3}}} & \text{if $m_{\substack{2 \\ 4}} = 0$ and  $i_1 = 0$.}
\end{cases}
\]
Note that the last case actually is a semistandard tableau because  $m_{\substack{3 \\ 4}} > 0$ implies  $m_2 = 0$.

Let us check that  $i_1$ and the right subtabloid of  $T$ can be recovered from the right subtableau of $f(\check{T})$.  First observe that we can determine from $f(\check{T})$ which case applies: the first case applies if $m'_{\substack{2 \\ 4}}> 0$ and $m'_2 = 0$,  the second if $m'_{\substack{2 \\ 4}} > 0 $ and  $m'_2 > m'_3$, the third if ($m'_{\substack{2 \\ 4}} > 0$ and $0 < m'_2 \leq m'_3$) or ($m'_{\substack{2 \\ 4}} = 0$ and  $\min(m'_2,m'_3) > 0$), and the fourth if  $m'_{\substack{2 \\ 4}} = 0$ and $\min(m'_2,m'_3) = 0$.  Next, we can recover  $i_1$ by $i_1 = \lfloor \frac{m'_3}{2} \rfloor$ if the first case applies, and
$i_1 = \min(m'_2,m'_3)$ otherwise.  Also,  in the first case $m_3 \leq 1$, so  $m_3$ is determined by $m_3 \equiv m'_3 \mod 2$.  In the remaining cases, the right subtabloid of  $T$ is easily recovered from the right subtableau of $f(\check{T})$.

Next, the middle subtableau of $f(\check{T})$ is defined from $i_2$ and  the middle subtabloid of  $T$ as follows:
\[
\raisebox{-6pt}{ \myvcenter{\ensuremath{ \myunderbrace{\column{2\\ 3} }_{m_{\substack{2 \\ 3}}} \myunderbrace{\column{3\\ 2} }_{m_{\substack{3 \\ 2}}} }}}
 \ \mapsto \
\begin{cases}%
 \myvcenter{\ensuremath{  \myunderbrace{\column{1\\4} \downdots \column{1\\4}}_{2i_2+m_{\substack{2 \\ 3}}} }} & \text{if  $m_{\dual{4}} = 0$ and  $m_1 > 0$ (and  $m_{\substack{3 \\ 2}} = m_{\substack{2 \\ 4}} = m_{\substack{3 \\ 4}}= 0$),}  \\
 \myvcenter{\ensuremath{ \myunderbrace{\column{2\\3} \downdots \column{2\\3}}_{2i_2+m_{\substack{2 \\ 3}}} }}
 & \text{if  $m_{\dual{4}} > 0$ and $m_1 = 0$ (and  $m_{\substack{1 \\ 2}} = m_{\substack{1 \\ 3}}= m_{\substack{3 \\ 2}} = 0$),} \\
  \myvcenter{\ensuremath{
   \myunderbrace{\column{1\\4} \downdots \column{1\\4}}_{2i_2+m_{\substack{3 \\ 2}}} }}
 & \text{if  $m_{\dual{4}} = 0, m_1 = 0$, and $m_{\substack{2 \\ 3}} = 0$,} \\
  \myvcenter{\ensuremath{
   \myunderbrace{\column{2\\3} \downdots\downdots \column{2\\3}}_{2i_2+m_{\substack{3 \\ 2}}+m_{\substack{2 \\ 3}}}
   }}
 & \text{if  $m_{\dual{4}} = 0, m_1 = 0$, and $m_{\substack{2 \\ 3}} > 0$}.
 \end{cases}
\]
Note that by Proposition \ref{p difficult straightenings}, exactly one of these conditions is satisfied.
The integer $i_2$ and the middle subtabloid of  $T$ can be recovered from the middle subtableau of $f(\check{T})$ as follows: since $m_{{\dual{4}}} = m'_{{\dual{4}}}$,  $m_1 = m'_1$,  which of $m'_1, m'_{\dual{4}}, m'_{\substack{2 \\ 3}}$ are $0$ determines which case applies. Then $i_2 = \lfloor \frac{1}{2}(m'_{\substack{1 \\ 4}}+m'_{\substack{2 \\ 3}}) \rfloor$ in the first three cases  and $i_2 = \lfloor \frac{1}{2}(m'_{\substack{2 \\ 3}}-1) \rfloor$ in the last case.  Also,  $m_{\substack{2 \\ 3}} \leq 1$ and $m_{\substack{3 \\ 2}} \leq 1$, so  $m_{\substack{2 \\ 3}}$ and  $m_{\substack{3 \\ 2}}$ are determined by $m'_{\substack{1 \\ 4}} \equiv m_{\substack{2 \\ 3}} \mod 2$ in the first case, $m'_{\substack{2 \\ 3}} \equiv m_{\substack{2 \\ 3}} \mod 2$ in the second, $m'_{\substack{1 \\ 4}} \equiv m_{\substack{3 \\ 2}} \mod 2$ in the third, and $m'_{\substack{2 \\ 3}}-1 \equiv m_{\substack{3 \\ 2}} \mod 2$ and  $m_{\substack{2 \\ 3}} = 1$ in the fourth.

The columns of height-3 are handled similarly to the height-1 columns.  If  $m_{\dual{4}} > 0$, then the left subtableau of  $f(\check{T})$ is defined from  $i_3$ and the left subtabloid of  $T$ as follows:
\[\raisebox{-5pt}{\myvcenter{\ensuremath{\myunderbrace{\column{1 \\ 2 \\4} \downdots \column{1 \\ 2 \\4}}_{m_{\dual{2}}} \myunderbrace{\column{1 \\ 3\\4} \downdots \column{1\\3\\4}}_{m_{\dual{3}}}\myunderbrace{\column{2 \\ 3 \\4} \downdots\downdots \column{2 \\ 3 \\4}}_{m_{\dual{4}}}}}}
\ \mapsto \
\raisebox{-5pt}{\myvcenter{\ensuremath{\myunderbrace{\column{1 \\ 2 \\4} \downdots \column{1 \\ 2 \\4}}_{i_3 +m_{\dual{2}}} \myunderbrace{\column{1 \\ 3\\4} \downdots \column{1\\3\\4}}_{i_3 +m_{\dual{3}}}\myunderbrace{\column{2 \\ 3 \\4} \downdots\downdots \column{2 \\ 3 \\4}}_{m_{\dual{4}}}}}}.
\]
If  $m_{\dual{4}} = 0$, then the left subtableau of  $f(\check{T})$ is defined from  $i_3$ and the left subtabloid of  $T$ as follows:
\[ \raisebox{-5pt}{\myvcenter{\ensuremath{ \myunderbrace{\column{1 \\ 2 \\4} \downdots \column{1 \\ 2 \\4}}_{m_{\dual{2}}}\myunderbrace{\column{1 \\ 3 \\4} \downdots \column{1 \\ 3 \\4}}_{m_{\dual{3}}} \myunderbrace{\column{1\\2} \downdots \column{1\\2}}_{m_{\substack{1 \\ 2}}} \myunderbrace{\column{1\\3} \downdots \column{1 \\ 3}}_{m_{\substack{1 \\ 3}}} }}} \  \mapsto \]
\[
\begin{cases}%
\myvcenter{\ensuremath{ \myunderbrace{\column{1 \\ 2 \\4} \downdots \column{1 \\ 2 \\4}}_{2i_3+m_{\dual{2}}} \myunderbrace{\column{1\\2} \downdots \column{1\\2}}_{m_{\substack{1 \\ 2}}} \myunderbrace{\column{1\\3} \downdots\downdots \column{1 \\ 3}}_{m_{\substack{1 \\ 3}}} }} & \text{if  $m_{\substack{1 \\ 3}} > 0$ and  $m_{\dual{3}} = 0$ (and  $m_{\dual{2}} \leq 1$),} \\
\myvcenter{\ensuremath{ \myunderbrace{\column{1 \\ 2 \\4} \downdots \column{1 \\ 2 \\4}}_{i_3}\myunderbrace{\column{1 \\ 3 \\4} \downdots \downdots \column{1 \\ 3 \\4}}_{i_3+m_{\dual{3}}} \myunderbrace{\column{1\\3} \downdots\downdots \column{1 \\ 3}}_{m_{\substack{1 \\ 3}}} }} & \text{if $m_{\substack{1 \\ 3}} > 0$ and  $m_{\dual{3}}  > 0$ (and  $m_{\substack{1 \\ 2}} = m_{\dual{2}}  = 0$),} \\
\myvcenter{\ensuremath{ \myunderbrace{\column{1 \\ 2 \\4} \downdots\downdots \column{1 \\ 2 \\4}}_{i_3 +m_{\dual{2}}}\myunderbrace{\column{1 \\ 3 \\4} \downdots\downdots \column{1 \\ 3 \\4}}_{i_3+m_{\dual{3}}} \myunderbrace{\column{1\\3} \downdots \column{1 \\ 3}}_{m_{\substack{1 \\ 2}}} }}  & \text{if  $m_{\substack{1 \\ 3}} = 0$ and  $i_3  >0$,}\\
\myvcenter{\ensuremath{ \myunderbrace{\column{1 \\ 2 \\4} \downdots \column{1 \\ 2 \\4}}_{m_{\dual{2}}}\myunderbrace{\column{1 \\ 3 \\4} \downdots \column{1 \\ 3 \\4}}_{m_{\dual{3}}} \myunderbrace{\column{1\\2} \downdots \column{1 \\ 2}}_{m_{\substack{1 \\ 2}}} }}  & \text{if $m_{\substack{1 \\ 3}} = 0$ and  $i_3 = 0$.}
\end{cases}
\]
The integer $i_3$ and the left subtabloid of  $T$ can be recovered from the left subtableau of $f(\check{T})$ as follows:
the first case applies if $m'_{\substack{1 \\ 3}} > 0$ and $m'_{\dual{3}} = 0$, the second if $m'_{\substack{1 \\ 3}} > 0 $ and  $m'_{\dual{3}} > m'_{\dual{2}}$, the third if ($m'_{\substack{1 \\ 3}} > 0$ and $0 < m'_{\dual{3}} \leq m'_{\dual{2}}$) or ($m'_{\substack{1 \\ 3}} = 0$ and  $\min(m'_{\dual{3}},m'_{\dual{2}}) > 0$), and the fourth if  $m'_{\substack{1 \\ 3}} = 0$ and $\min(m'_{\dual{3}},m'_{\dual{2}}) = 0$.  In the first case,
$i_3 = \lfloor \frac{m'_{\dual{2}}}{2} \rfloor$, and otherwise $i_3 = \min(m'_{\dual{3}},m'_{\dual{2}})$.  In the first case, $m_{\dual{2}}$ is the element of $\{0,1\}$ with the same parity as $m'_{\dual{2}}$.  Otherwise, $m_{\dual{2}}, m_{\dual{3}},m_{\substack{1 \\ 2}}$, and $m_{\substack{1 \\ 3}}$ are easily recovered from $m'_{\dual{2}}, m'_{\dual{3}},m'_{\substack{1 \\ 2}}$, and $m'_{\substack{1 \\ 3}}$ once the correct case is known.
%
\end{proof}

The proof of Theorem \ref{t main canonical basis} (iii) and Propositions \ref{p straightening non invariant-free} and \ref{p NSTC semistandard bijection} establish the following.
\begin{corollary}\label{c basis correct size}
As sets,
\[
\parbox[t]{13.5cm}{
$\pNSTC(\nu) \cong$ lexicographically minimum representatives of $\pNSTC(\nu)$ $=$ straightened elements of $\pSNST(\nu') \cong$ SSYT$_{\dx}(\nu)$.
}
\]
\end{corollary}

\subsection{Invariant-free straightened highest weight NST}
It is not difficult to analyze the straightening conditions of Proposition \ref{p difficult straightenings} to determine exactly the form of the invariant-free straightened highest weight NST.  This will be used to obtain explicit formulae for Kronecker coefficients in  \textsection\ref{s explicit formulae for Kronecker coefficients}.
\begin{proposition}
\label{p highest weight straightened NST}
Let $T$ be an invariant-free NST. 
Then $T$ is straightened and highest weight if and only if
\[T = \raisebox{-5pt}{\myvcenter{\ensuremath{\myunderbrace{\column{1 \\ 2 \\ 3} \downdots \column{1 \\ 2 \\ 3}}_{m_{{\dual{1}}}} \myunderbrace{\column{1 \\ 2 \\ 4} \downdots \column{1 \\ 2 \\ 4}}_{m_{{\dual{2}}}} \myunderbrace{\column{1 \\ 3 \\ 4} \downdots \column{1 \\ 3 \\ 4}}_{m_{{\dual{3}}}} \myunderbrace{\column{1 \\ 2} \downdots \column{1 \\ 2}}_{m_{\substack{1 \\ 2}}} \myunderbrace{\column{1 \\ 3} \downdots \column{1 \\ 3}}_{m_{\substack{1 \\ 3}}} \myunderbrace{\column{2 \\ 3}}_{m_{\substack{2 \\ 3}}} \myunderbrace{\column{1} \downdots \column{1}}_{m_1}}}}, \] where
\begin{list}{\emph{(\alph{ctr})}} {\usecounter{ctr} \setlength{\itemsep}{1pt} \setlength{\topsep}{2pt}}
\item  at least one of $m_{{\dual{2}}}, m_{{\dual{3}}}$ is zero,
\item  $m_{\substack{1 \\ 2}} > 0$ implies $m_{{\dual{3}}} = 0$,
\item  $m_{\substack{1 \\ 3}} > 0$ implies $m_{{\dual{2}}} \leq 1$,
\item  $m_{\substack{2 \\ 3}} \leq 1$,
\item  $m_{\substack{2 \\ 3}} = 1$  implies $m_1 > 0$,
\item $m_{{\dual{2}}} \leq m_1 + m_{\substack{1 \\ 3}}$,
\item $m_{{\dual{3}}} \leq m_1 + m_{\substack{1 \\ 2}}$.
\end{list}
\end{proposition}
\begin{proof}
If  $T$ is straightened, then it satisfies (a)--(d). If it is also highest weight, then  $m_4 = m_3 = m_2 = 0$ as shown in the picture of  $T$ above, and (e)--(g) hold. Next,  $m_{\substack{3 \\ 2}} = m_{\substack{2 \\ 4}} = m_{\substack{3 \\ 4}} = 0$ because if  $m_1 > 0$, this is by the straightening conditions (1.2) and (1.5) and if  $m_1 =0$, this is by the highest weight assumption.  Finally, $m_{ \dual{4}}= 0$ by (4.1) if  $m_1 > 0$, by (3.2) if $m_{\substack{1 \\ 2}} > 0$ or  $m_{\substack{1 \\ 3}} > 0$, and by the highest weight assumption otherwise.
This proves the ``only if'' direction.  For the ``if'' direction, one checks directly and easily that if  $T$ has the form above, then $T$ is straightened and highest weight.
\end{proof}

\section{A Kronecker graphical calculus and applications}
\label{s A Kronecker graphical calculus}
In \textsection\ref{ss Kronecker graphical calculus} we give a description of the  $\Uqvw$-crystal components of $(\gr(\nsbr{X}_\nu)$, $\grin(\pNSTC(\nu)))$ in terms of arcs.
We hope this to be the beginnings of a Kronecker version of the $U_q(\sl_2)$ graphical calculus of \cite{FK}. This graphical description of crystal components will make it easier to obtain explicit formulae for Kronecker coefficients in  \textsection\ref{s explicit formulae for Kronecker coefficients}.  We also use it to write down the action of the Chevalley generators on  $\pNSTC$ (\textsection\ref{ss action of Chevalley generators}) and the action of  $\tau$ on  $\pNSTC$ (\textsection\ref{ss The action of tau on pNSTC}).

\subsection{Kronecker graphical calculus}
\label{ss Kronecker graphical calculus}
Though it is not strictly necessary for the results in the next two sections, we believe it to be useful to give a description of crystal components that is independent of any $\pNSTC$ in the component and any representative $\pSNST$ in its class. We hope this to be the beginnings of a graphical calculus that describes $\O(M_q(\nsbr{X}))$-comodules and morphisms between them in terms of their $\pNSTC$ bases or some generalization thereof.  However, to more fully develop such a theory it seems that we need a canonical basis for all of $\nsbr{X}^{\tsr r}$ as detailed in Conjecture \ref{cj canonical basis X^r}.

\begin{definition}\label{d 2-1 arc}
A \emph{$k$-$l$  $V$-arc} of an NST or SNST is an external $V$-arc between a height-$k$ column and a height-$l$ column. We define \emph{$k$-$l$  $W$-arcs} similarly, and a \emph{$k$-$l$  arc} is either a  $k$-$l$ $V$-arc or a  $k$-$l$ $W$-arc.
We also define a 3-2-1 arc to be a $k$-$l$ arc and a $k$-$l'$  arc that share a  height-$k$ column and such that  $\{k,l,l'\} =  \{1,2,3\}$. A 3-2-1 arc is a 3-2-1  $V$-arc (resp.  $W$-arc) if both of its arcs are  $V$-arcs (resp.  $W$-arcs) or if the longer of its two arcs is a  $V$-arc (resp.  $W$-arc) (see Example \ref{ex 3-2-1 arc}).


A \emph{pure} $k$-$l$  $V$-arc is a $k$-$l$  $V$-arc that is not part of a 3-2-1 arc.  Pure  $k$-$l$ $W$-arcs and pure  $k$-$l$ arcs are defined similarly.
\end{definition}
We write
\[\arc{$k$-$l$ $V$}(T), \arc{$k$-$l$ $W$}(T), \arc{$k$-$l$}(T), \arc{3-2-1 $V$}(T), \arc{3-2-1 $W$}(T), \arc{3-2-1}(T) \]
for the number of pure $k$-$l$  $V$-arcs, pure $k$-$l$  $W$-arcs, pure $k$-$l$ arcs, 3-2-1  $V$-arcs, 3-2-1  $W$-arcs, and 3-2-1 arcs of  $T$, respectively.
Also let \[\arc{ext}(T)\]
denote the total number of external arcs of $T$.  Finally, let
\[ \exfreev(T), \exfreew(T) \]
be the number of type $V$ columns of  $T$ not at the end of an external arc, type $W$ columns of  $T$ not at the end of an external arc, respectively.

\begin{example}
\label{ex 3-2-1 arc}
Here are three equivalent SNST with their external $V$- and $W$-arcs drawn.
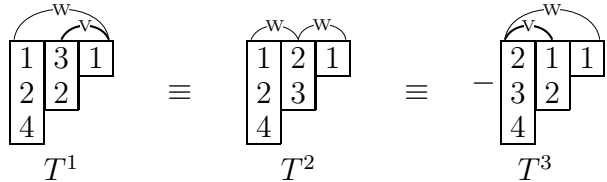
\setlength{\cellsizeCol}{2.5ex}
\begin{center}
\begin{tikzpicture}[xscale=.9]
\tikzstyle{column} = [inner sep = -4pt]
\tikzstyle{edgeV} = [draw,-,black,thick]
\tikzstyle{edgeW} = [draw,-,black,very thin]
\tikzstyle{edge} = [draw,very thin,postaction={draw,dashed,dash pattern= on 2pt off 2pt,thick}]
\tikzstyle{LabelStyleH} = [text=black, fill =white, inner sep = -.8pt]

\begin{scope}
\foreach \y/\z in {.7/-4*.35} {
    \draw[edgeV, bend left=65] (\z+2*.7,\y) to node[LabelStyleH]{{\TinyV}} (\z+3*.7,\y);
    \draw[edgeW, bend left=82] (\z+1*.7,\y) to node[LabelStyleH]{{\TinyW}} (\z+3*.7,\y);
    \node[column] (theNode) at (0,0) {$\myvcenter{\ensuremath{\columnL{1 \\ 2\\4} \columnL{3 \\2} \columnL{1}}}$};
    \node[column] (label) at (0,-1) {$T^1$};
}
\end{scope}

\begin{scope}[xshift= 1.75cm]
    \node[column] (theNode) at (0,0) {$\equiv$};
\end{scope}

\begin{scope}[xshift= 3.5cm]
\foreach \y/\z in {.7/-4*.35} {
    \draw[edgeW, bend left=65] (\z+1*.7,\y) to node[LabelStyleH]{{\TinyW}} (\z+2*.7,\y);
    \draw[edgeW, bend left=82] (\z+2*.7,\y) to node[LabelStyleH]{{\TinyW}} (\z+3*.7,\y);
    \node[column] (theNode) at (0,0) {$\myvcenter{\ensuremath{\columnL{1 \\ 2\\4} \columnL{2 \\3} \columnL{1}}}$};
    \node[column] (label) at (0,-1) {$T^2$};
}
\end{scope}

\begin{scope}[xshift= 5.25cm]
    \node[column] (theNode) at (0,0) {$\equiv$};
\end{scope}

\begin{scope}[xshift= 7cm]
\foreach \y/\z in {.7/-5*.35 -.1} {
    \draw[edgeV, bend left=65] (\z+2*.7,\y) to node[LabelStyleH]{{\TinyV}} (\z+3*.7,\y);
    \draw[edgeW, bend left=82] (\z+2*.7,\y) to node[LabelStyleH]{{\TinyW}} (\z+4*.7,\y);
    \node[column] (theNode) at (0,0) {$\pad{-}\myvcenter{\ensuremath{\columnL{2 \\ 3\\4} \columnL{1 \\2} \columnL{1}}}$};
    \node[column] (label) at (0,-1) {$T^3$};
}
\end{scope}

\end{tikzpicture}
\setlength{\cellsizeCol}{2.1ex}
\end{center}

There is one 3-2-1  $W$-arc in each $T^i$, and there are no pure $k$-$l$ $V$- or $W$-arcs in any of the $T^i$.
\end{example}

\begin{proposition}\label{p diagram data}
From the diagram of a $\pSNST(\nu')$ $T$ we extract the following information:
\begin{list}{\emph{(\Alph{ctr})}} {\usecounter{ctr} \setlength{\itemsep}{1pt} \setlength{\topsep}{2pt}}
\item (the unpaired $V$-diagram, the unpaired $W$-diagram)
\item the invariant record $(i_4, i_3, i_2, i_1)$ of $T$,
\item $(\arc{2-1}(T),\arc{3-2}(T))$,
\item $(\arc{3-1}(T), \arc{3-2-1}(T),|\exfreev(T) - \exfreew(T)|)$,
\item $(\arc{3-1  $V$}(T),  \arc{3-1  $W$}(T),\arc{3-2-1  $V$}(T), \arc{3-2-1  $W$}(T), \exfreev(T),\exfreew(T))$,
\end{list}
The data above is constant on $\pNSTC$, (B)--(E) are constant on $\Uqvw$-crystal components of $(\gr(\nsbr{X}_\nu),\grin(\pNSTC(\nu)))$, and (B)--(D) are constant on $\Uqt$-cells of \newline $(\gr(\nsbr{X}_\nu),\grin(\pNSTC(\nu)))$.
\end{proposition}
Note that the crystal components and cells of $(\gr(\nsbr{X}_\nu),\grin(\pNSTC(\nu)))$ are the same and coincide with those of  $(\nsbr{X}_\nu,\pNSTC(\nu))$, but we do not yet know that the latter is an upper based  $\Uqvw$- or  $\Uqt$-module.
\begin{proof}
Proposition \ref{p connected component of degree-preserving moves} (a) says exactly that (B) is constant on NSTC.
Next, one checks that degree-preserving moves preserve the number of internal and external  $V$- and  $W$-arcs.  Thus (A) is constant on NSTC.  Also, observe that an honest NST does not contain any  $k$-$k$ arcs,  $k \in [\dx]$, except for those that are part of an invariant column pair.
With these facts in mind, it is easy to check that for each degree-preserving move, (C)--(E) are constant on  $\pNSTC$.  For instance, for a degree-preserving move of the form below, we have drawn the $V$- and $W$-arcs (the arcs that have one end on the dots may not exist--call them potential arcs).
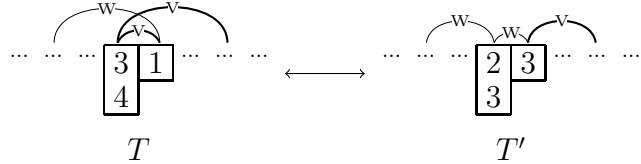
\setlength{\cellsizeCol}{2.5ex}
\begin{center}
\begin{tikzpicture}[xscale=.9]
\tikzstyle{column} = [inner sep = -4pt]
\tikzstyle{edgeV} = [draw,-,black,thick]
\tikzstyle{edgeW} = [draw,-,black,very thin]
\tikzstyle{edge} = [draw,very thin,postaction={draw,dashed,dash pattern= on 2pt off 2pt,thick}]
\tikzstyle{aedge} = [draw, <->, black]
\tikzstyle{LabelStyleH} = [text=black, fill =white, inner sep = -.8pt]

\begin{scope}
\foreach \y/\z in {.41/-9*.25} {
    \draw[edgeV, bend left=65] (\z+4*.5-.06,\y) to node[LabelStyleH]{{\TinyV}} (\z+5*.5+.06,\y);
    \draw[edgeW, bend right=90] (\z+5*.5+.06,\y) to node[LabelStyleH]{{\TinyW}} (\z+2*.5,\y);
    \draw[edgeV, bend left=90] (\z+4*.5-.06,\y) to node[LabelStyleH]{{\TinyV}} (\z+7*.5+.06,\y);
    \node[column] (theNode) at (0,0) {$\myvcenter{\ensuremath{\downdots\downdots\downdots\columnL{3 \\ 4} \columnL{1}\downdots\downdots\downdots}}$};
    \node[column] (label) at (0,-1) {$T$};
}
\end{scope}

\begin{scope}
\node (T) at (2,0){};
\node (Tp) at (3.5,0){};
\draw[aedge] (T) to (Tp);
\end{scope}

\begin{scope}[xshift=5.5cm]
\foreach \y/\z in {.41/-9*.25} {
    \draw[edgeW, bend left=65] (\z+4*.5,\y) to node[LabelStyleH]{{\TinyW}} (\z+5*.5,\y);
    \draw[edgeW, bend right=82] (\z+4*.5,\y) to node[LabelStyleH]{{\TinyW}} (\z+2*.5,\y);
    \draw[edgeV, bend left=82] (\z+5*.5,\y) to node[LabelStyleH]{{\TinyV}} (\z+7*.5,\y);
    \node[column] (theNode) at (0,0) {$\myvcenter{\ensuremath{\downdots\downdots\downdots\columnL{2 \\ 3} \columnL{3}\downdots\downdots\downdots}}$};
    \node[column] (label) at (0,-1) {$T'$};
}
\end{scope}
\end{tikzpicture}
\end{center}
This turns the shown 2-1  $V$-arc of $T$ into the shown 2-1 $W$-arc of $T'$, but (C)--(E) remain constant: the potential $V$-arc does not exist because, as just mentioned, $T'$ honest implies that it does not contain a 1-1  $V$-arc not part of an invariant column pair; the potential $W$-arc cannot have its undetermined end on a height-2 column because $T'$ honest implies that it does not contain a 2-2  $W$-arc not part of an invariant column pair;  if the potential $W$-arc has its undetermined end on a height-3 column, then it is part of a 3-2-1  $W$-arc in $T$ and $T'$.


Since the action of $\crystal{F}_V$ and $\crystal{F}_W$ on NST does not modify external $V$- and $W$-arcs, (B)--(E) are constant on $\Uqvw$-crystal components. This, together with the fact that the action of $\tau$ on $\pSNST$ interchanges $V$- and  $W$-arcs (and sometimes multiplies  by  $-1$), proves that (B)--(D) are constant on $\Uqt$-cells.
\end{proof}

\begin{figure}
\setlength{\cellsizeCol}{2.5ex}
\begin{tikzpicture}[xscale=.9]
\tikzstyle{column} = [inner sep = -4pt]
\tikzstyle{edgeV} = [draw,-,black,thick]
\tikzstyle{edgeW} = [draw,-,black,very thin]
\tikzstyle{edge} = [draw,very thin,postaction={draw,dashed,dash pattern= on 2pt off 2pt,thick}]
\tikzstyle{LabelStyleH} = [text=black, fill =white, inner sep = -.8pt]

\begin{scope}[xshift=-4.32cm, yshift = 2cm]
\foreach \y/\z in {.68/-15*.253} {
    \draw[edgeV, bend left=70] (\z+4*.506,\y) to node[LabelStyleH]{{\tiny $V$}} (\z+13*.506,\y);
    \draw[edgeV, bend left=70] (\z+6*.506,\y) to node[LabelStyleH]{{\tiny $V$}} (\z+12*.506,\y);
    \draw[edgeV, bend left=70] (\z+7*.506,\y) to node[LabelStyleH]{{\tiny $V$}} (\z+11*.506,\y);
    \node[column] (theNode) at (0,0) {
    $\myvcenter{\ensuremath{\columnL{1 \\ 2\\3}\downdots\columnL{1\\2\\3}\columnL{1\\3\\4} \downdots\downdots \columnL{1 \\3 \\4}
    \columnL{1\\3} \downdots \columnL{1\\3}\columnL{1} \downdots \downdots \columnL{1}}}$
    };
}
\end{scope}

\begin{scope}[xshift=4.35cm, yshift = 2cm]
\foreach \y/\z in {.68/-15*.253} {
    \draw[edgeW, bend left=70] (\z+4*.506,\y) to node[LabelStyleH]{{\tiny $W$}} (\z+13*.506,\y);
    \draw[edgeW, bend left=70] (\z+6*.506,\y) to node[LabelStyleH]{{\tiny $W$}} (\z+12*.506,\y);
    \draw[edgeW, bend left=70] (\z+7*.506,\y) to node[LabelStyleH]{{\tiny $W$}} (\z+11*.506,\y);
    \node[column] (theNode) at (0,0) {
    $\myvcenter{\ensuremath{\columnL{1 \\ 2\\3}\downdots\columnL{1\\2\\3}\columnL{1\\2\\4} \downdots\downdots \columnL{1 \\2 \\4}
    \columnL{1\\2} \downdots \columnL{1\\2}\columnL{1} \downdots \downdots \columnL{1}}}$
    };
}
\end{scope}

\begin{scope}[yshift=0cm]
\foreach \y/\z in {.68/-16*.253} {
      \node[column] (theNode) at (0,0) {
    $\myvcenter{\ensuremath{\columnL{1 \\ 2\\3}\downdots\columnL{1\\2\\3}
    \columnL{1\\2} \downdots \columnL{1\\2}\columnL{1\\3} \downdots \columnL{1\\3}\columnL{1} \downdots \columnL{1}}}$
    };
}
\end{scope}

\begin{scope}
    \node[column] (label) at (0,-1.2) {{\small $\hat{g}^0_{\lambda\mu\nu}$}};
\end{scope}

\begin{scope}[xshift=-4.25cm, yshift=-3.15cm]
\foreach \y/\z in {.68/-15*.253} {
    \draw[edge, bend left=70] (\z+10*.506,\y) to (\z+11*.506,\y);
    \node[column] (theNode) at (0,0) {
    $\myvcenter{\ensuremath{\columnL{1 \\ 2 \\3} \downdots \columnL{1 \\2 \\3}
    \columnL{1\\2} \downdots \columnL{1\\2} \columnL{1\\3} \downdots \columnL{1\\3} \columnL{2\\3} \columnL{1} \downdots\downdots \columnL{1}}}$
    };
    \node[column] (label) at (0,-1) {{\small $\hat{g}^{\text{2-1}}_{\lambda\mu\nu}$}};
}
\end{scope}

\begin{scope}[xshift=4.25cm, yshift=-3.15cm]
\foreach \y/\z in {.68/-15*.253} {
    \draw[edge, bend left=70] (\z+4*.506,\y) to (\z+8*.506,\y);
    \node[column] (theNode) at (0,0) {
    $\myvcenter{\ensuremath{\columnL{1 \\ 2 \\3} \downdots \columnL{1 \\2 \\3}
    \columnL{1\\2\\4} \columnL{1\\2} \downdots \columnL{1\\2} \columnL{1\\3} \downdots\downdots \columnL{1\\3} \columnL{1} \downdots \columnL{1}}}$
    };
    \node[column] (label) at (0,-1) {{\small $\hat{g}^{\text{3-2}}_{\lambda\mu\nu}$}};
}
\end{scope}

\begin{scope}[yshift=-6.8cm, xshift=-4.35cm]
\foreach \y/\z in {.68/-16*.253} {
    \draw[edgeV, bend left=70] (\z+4*.506,\y) to node[LabelStyleH]{{\tiny $V$}} (\z+14*.506,\y);
    \draw[edgeV, bend left=70] (\z+6*.506,\y) to node[LabelStyleH]{{\tiny $V$}} (\z+13*.506,\y);
    \draw[edgeV, bend left=70] (\z+7*.506,\y) to node[LabelStyleH]{{\tiny $V$}} (\z+12*.506,\y);
    \draw[edgeW, bend left=70] (\z+11*.506,\y) to node[LabelStyleH]{{\tiny $W$}} (\z+12*.506,\y);
    \node[column] (theNode) at (0,0) {
    $\myvcenter{\ensuremath{\columnL{1 \\ 2\\3} \downdots \columnL{1 \\2 \\3}
    \columnL{1\\3\\4} \downdots\downdots \columnL{1\\3\\4} \columnL{1\\3} \downdots \columnL{1\\3} \columnL{2\\3}\columnL{1} \downdots\downdots \columnL{1}}}$
    };
}
\end{scope}

\begin{scope}[yshift=-6.8cm]
    \node[column] (label) at (0,-1.25) {{\small $\hat{g}^{\text{3-2-1}}_{\lambda\mu\nu}$}};
\end{scope}
\begin{scope}[yshift=-6.8cm, xshift=4.35cm]
\foreach \y/\z in {.68/-16*.253} {
    \draw[edgeW, bend left=70] (\z+4*.506,\y) to node[LabelStyleH]{{\tiny $W$}} (\z+14*.506,\y);
    \draw[edgeW, bend left=70] (\z+6*.506,\y) to node[LabelStyleH]{{\tiny $W$}} (\z+13*.506,\y);
    \draw[edgeW, bend left=70] (\z+7*.506,\y) to node[LabelStyleH]{{\tiny $W$}} (\z+11*.506,\y);
    \draw[edgeW, bend left=70] (\z+11*.506,\y) to node[LabelStyleH]{{\tiny $W$}} (\z+12*.506,\y);

    \node[column] (theNode) at (0,0) {
    $\myvcenter{\ensuremath{\columnL{1 \\ 2\\3}\downdots\columnL{1\\2\\3}\columnL{1\\2\\4} \downdots\downdots \columnL{1 \\2 \\4}
    \columnL{1\\2} \downdots \columnL{1\\2} \columnL{2\\3} \columnL{1} \downdots\downdots \columnL{1}}}$
    };
}
\end{scope}

\begin{scope}[yshift=-10.1cm]
\foreach \y/\z in {.68/-17*.253} {
    \draw[edge, bend left=70] (\z+4*.506,\y) to (\z+8*.506,\y);
    \draw[edge, bend left=70] (\z+12*.506,\y) to (\z+13*.506,\y);
    \node[column] (theNode) at (0,0) {
    $\myvcenter{\ensuremath{\columnL{1 \\ 2 \\3} \downdots \columnL{1 \\2 \\3}
    \columnL{1\\2\\4} \columnL{1\\2} \downdots \columnL{1\\2} \columnL{1\\3} \downdots\downdots \columnL{1\\3} \columnL{2\\3} \columnL{1} \downdots\downdots \columnL{1}}}$
    };
    \node[column] (label) at (0,-1) {{\small $\hat{g}^{\text{3-2, 2-1}}_{\lambda\mu\nu}$}};
}
\end{scope}

\end{tikzpicture}
\caption{The Kronecker graphical calculus for straightened highest weight invariant-free NST, drawn with the conventions of \eqref{e dot conventions exactly once}. The eight cases are grouped into five cases corresponding to the different types of invariant-free Kronecker coefficients, as described in  \textsection\ref{s explicit formulae for Kronecker coefficients}.}
\label{f kronecker graphical calculus}
\setlength{\cellsizeCol}{2.1ex}
\end{figure}
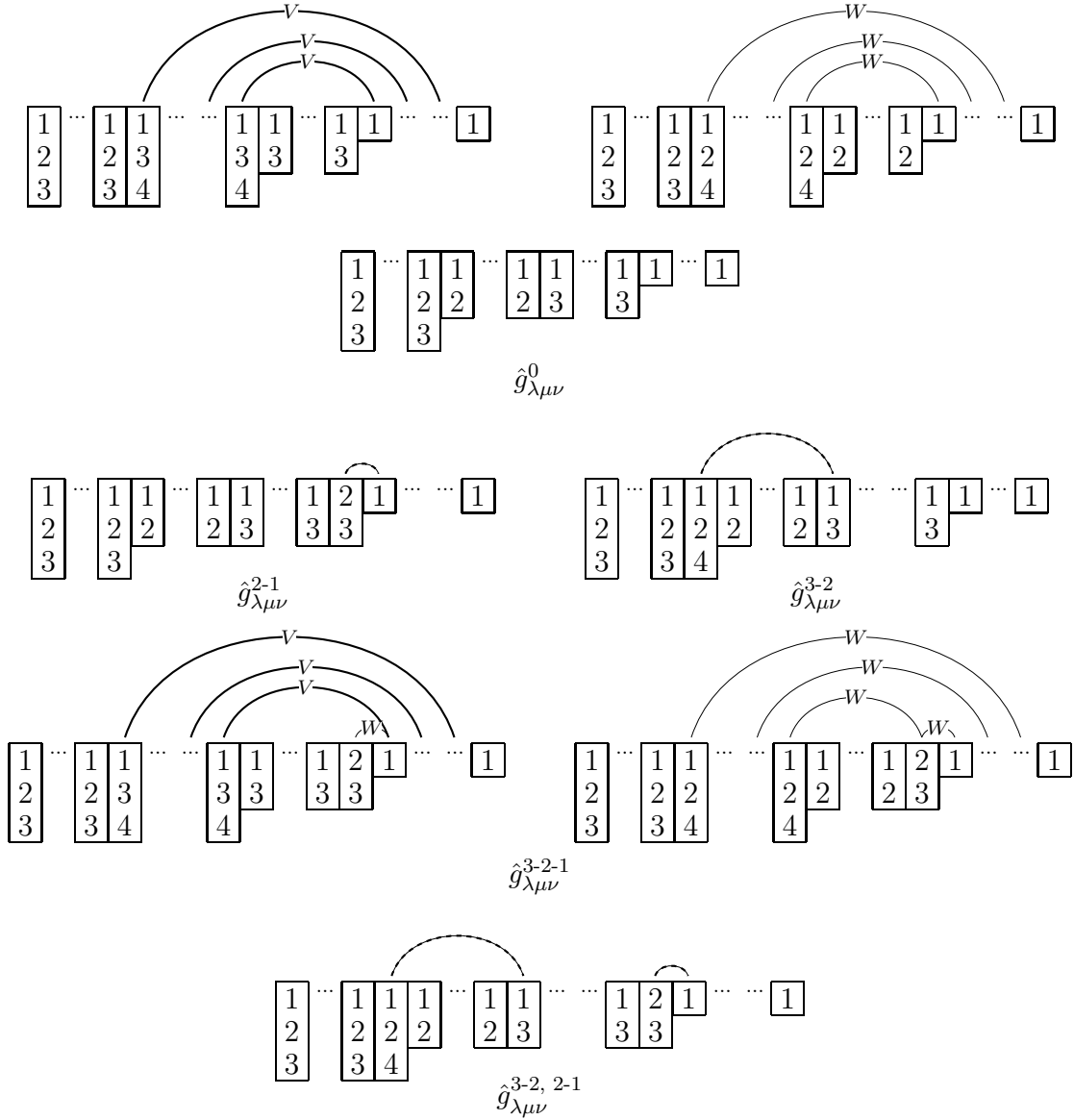

Proposition \ref{p highest weight straightened NST} implies that any invariant-free straightened highest weight NST belongs to exactly one of the eight cases of Figure \ref{f kronecker graphical calculus}, which are drawn with the following conventions:
\be \label{e dot conventions exactly once}
\parbox{13.6cm}{two sets of dots indicates that that type of column appears at least once; one set of dots indicates that that type of column can appear any number of times; a column appearing once (and with no dots) indicates that that type of column appears exactly once.  All external arcs are drawn
(pure 2-1 and 3-2 arcs are not labeled as $V$-arcs or $W$-arcs, as indicated by the funny dashed style, because these are not constant on $\pNSTC$).}
\ee
These eight cases are grouped into five cases corresponding to the different types of invariant-free Kronecker coefficients $\hat{g}_{\lambda \mu \nu}^*$, as described in the next section (these coefficients give a convenient way of decomposing Kronecker coefficients into a sum of smaller nonnegative coefficients).

Propositions \ref{p NSTC semistandard bijection}, \ref{p highest weight straightened NST}, and \ref{p diagram data} have the following corollary, which partially realizes our original goal of obtaining a bijection \eqref{e bijection goal} that is some kind of ``Kronecker analog'' of the RSK correspondence.
\begin{corollary}
\label{c diagram kronecker}
The map from $\pSNST(\nu')$ to (A),(B),(C),(E) of Proposition \ref{p diagram data} has fibers given by $\pNSTC(\nu)$. Pre-composing this map with the bijection of Proposition \ref{p NSTC semistandard bijection} yields a bijection
\be \label{e obtained bijection}
\begin{array}{ccccl}
\hspace{-1mm} SSYT_{\dx}(\nu) &\cong& \pNSTC(\nu) &\xrightarrow{\cong}& \bigsqcup_{\lambda,\mu} SSYT_{\dv}(\lambda) \times SSYT_{\dw}(\mu) \times \mathbf{g}_{\lambda \mu \nu} \\
\hspace{-1mm} f(\mathbf{T}) &\longleftrightarrow &\mathbf{T}& \mapsto & \qquad\quad\quad P(\mathbf{k}), \qquad P(\mathbf{l}), \quad \ \ \text{\small(B),(C),(E)},
\end{array}
\ee
where $\mathbf{k}, \mathbf{l}$ are the $V$- and $W$-word of any $T \in \mathbf{T}$ and $\mathbf{g}_{\lambda \mu \nu}$ is a subset of $\ZZ_{\geq 0}^{12}$ that depends on  $\lambda, \mu, \nu$, but not on $\mathbf{T}$, and has cardinality $g_{\lambda \mu \nu}$ (here, (B),(C), and (E) are encoded as a 12-tuple of nonnegative integers).
%
\end{corollary}
Moreover, the set $\mathbf{g}_{\lambda \mu \nu}$ is not hard to read off from Figure \ref{f kronecker graphical calculus}, and we make this even more explicit in the next section.
\begin{remark}
We have only partially realized the goal of obtaining a bijection as in \eqref{e bijection goal} because we really want $\text{SYT}(\nu)$-many bijections that are all slightly different, but similar to \eqref{e obtained bijection}.  We expect these bijections to be realized algebraically by finding a basis for $ \nsbr{X}^{\tsr r}$ whose  $\Uqt$-cells can be partitioned into  $\text{SYT}(\nu)$-many cellular subquotients, called fat cells in Conjecture \ref{cj canonical basis X^r}, each similar to (but not necessarily isomorphic as a based module to) $(\nsbr{X}_\nu, \pNSTC(\nu))$. See Conjecture \ref{cj canonical basis X^r} for more details.
\end{remark}

By projecting the right-hand side of \eqref{e obtained bijection} onto  $\bigsqcup_{\lambda, \mu} \mathbf{g}_{\lambda \mu \nu}$, we also obtain
a nice description of $\Uqvw$-crystal components in terms of the Kronecker graphical calculus.  Let us assume Theorem \ref{t main canonical basis} for this corollary so that we can state it in terms of the basis $\pNSTC(\nu)$ rather than $\grin(\pNSTC(\nu))$.
\begin{corollary}\label{c NSTC cells}
The $\Uqvw$-module with basis $(\nsbr{X}_\nu, \pNSTC(\nu))$ decomposes into $\Uqvw$-crystal components (or $\Uqvw$-cells) as
\[ \pNSTC(\nu) = \bigsqcup_{\lambda, \mu \vdash_2 r, \ \zeta \in \mathbf{g}_{\lambda \mu \nu}} \nsbr{\Lambda}_{\nu,\zeta},\]
where $\nsbr{\Lambda}_{\nu, \zeta}$ consists of those $\mathbf{T} \in \pNSTC(\nu)$ such that the 12-tuple of $\mathbf{T}$ is equal to $\zeta$.
\end{corollary}
Note that the sets  $\{\mathbf{g}_{\lambda \mu \nu}\}_{\lambda, \mu}$ for fixed  $\nu$ are actually disjoint as subsets of $\ZZ_{\geq 0}^{12}$ (this will be seen in \eqref{e l1 + m1 formulae}), and this was used implicitly in the definition of  $\nsbr{\Lambda}_{\nu,\zeta} $.

In light of this corollary and Proposition \ref{p diagram data}, it makes sense to write $\arc{2-1}(\G)=\arc{2-1}(\mathbf{T})=\arc{2-1}(T), \arc{3-2}(\G)=\arc{3-2}(\mathbf{T})=\arc{3-2}(T)$, etc. for  $\mathbf{T}$ the $\pNSTC$ containing an  $\pSNST$  $T$ and $\G$ the $\Uqvw$-crystal component containing  $\mathbf{T}$.

\subsection{Action of the Chevalley generators on $\pNSTC$}
\label{ss action of Chevalley generators}

\newcommand{\DRAW}[2]{
\raisebox{#1}{\ensuremath{
\vcenter{\hbox{
\begin{tikzpicture}[scale=.4]
\tikzstyle{column} = [inner sep = -4pt]
\tikzstyle{edgeV} = [draw,-,black,thick]
\tikzstyle{edgeW} = [draw,-,black,very thin]
\tikzstyle{edge} = [draw,very thin,postaction={draw,dashed,dash pattern= on 2pt off 2pt,thick}]
\tikzstyle{LabelStyleH} = [text=black, fill =white, inner sep = -.8pt]
#2
\end{tikzpicture}
}}}}}

\newcommand{\DRAWccbdbbaa}{
\DRAW{1.6mm}{
\foreach \y/\z in {1.14/-9*.37} {
    \draw[edgeW, bend left=79] (\z+1*.74,\y) to (\z+8*.74,\y);
    \draw[edgeW, bend left=79] (\z+2*.74,\y) to (\z+6*.74,\y);
    \draw[edgeW, bend left=79] (\z+6*.74,\y) to (\z+7*.74,\y);
    \node[column] (theNode) at (0,0) {$\myvcenter{\ensuremath{\columnL{ \\ \\ } \columnL{ \\ \\ } \columnL{ \TinyV \\ } \downdots \columnL{ \TinyV \\ } \columnL{ \TinyW \\ } \columnL{}\columnL{}}}$};
}}
}

\newcommand{\DRAWcbba}{
\DRAW{1.6mm}{
\foreach \y/\z in {1.14/-5*.37} {
    \draw[edgeW, bend left=79] (\z+1*.74,\y) to (\z+3*.74,\y);
    \draw[edgeW, bend left=79] (\z+3*.74,\y) to (\z+4*.74,\y);
    \node[column] (theNode) at (0,0) {$\myvcenter{\ensuremath{\columnL{ \\ \\ } \columnL{ \TinyV \\ } \columnL{ \TinyW \\ } \columnL{}}}$};
}}
}

\newcommand{\DRAWTWOcbba}{
\DRAW{1.6mm}{
\foreach \y/\z in {1.14/-5*.37} {
    \draw[edgeW, bend left=79] (\z+1*.74,\y) to (\z+2*.74,\y);
    \draw[edgeW, bend left=79] (\z+3*.74,\y) to (\z+4*.74,\y);
    \node[column] (theNode) at (0,0) {$\myvcenter{\ensuremath{\columnL{ \\ \\ } \columnL{\TinyW \\ } \columnL{\TinyW \\ } \columnL{}}}$};
}}
}

\newcommand{\DRAWccba}{
\DRAW{1.6mm}{
\foreach \y/\z in {1.14/-5*.37} {
    \draw[edgeW, bend left=70] (\z+1*.74,\y) to (\z+2*.74,\y);
    \draw[edgeV, bend left=79] (\z+1*.74-.14,\y) to (\z+2*.74+.14,\y);
    \draw[edge, bend left=79] (\z+3*.74,\y) to (\z+4*.74,\y);
    \node[column] (theNode) at (0,0) {$\myvcenter{\ensuremath{\columnL{\\ \\} \columnL{\\ \\}\columnL{ \\ }\columnL{}}}$};
}}
}

\newcommand{\DRAWccbaa}{
\DRAW{1.6mm}{
\foreach \y/\z in {1.14/-6*.37} {
    \draw[edgeW, bend left=79] (\z+1*.74,\y) to (\z+5*.74,\y);
    \draw[edgeW, bend left=79] (\z+2*.74-.14,\y) to (\z+3*.74,\y);
    \draw[edgeW, bend left=79] (\z+3*.74,\y) to (\z+4*.74+ .14,\y);
    \node[column] (theNode) at (0,0) {$\myvcenter{\ensuremath{\columnL{ \\ \\ } \columnL{ \\ \\ } \columnL{ \TinyW \\ } \columnL{}\columnL{}}}$};
}}
}

\newcommand{\DRAWFOURccaa}{
\DRAW{1.6mm}{
\foreach \y/\z in {1.14/-5*.37} {
    \draw[edgeW, bend left=70] (\z+1*.74,\y) to (\z+2*.74,\y);
    \draw[edgeV, bend left=79] (\z+1*.74-.14,\y) to (\z+2*.74+.14,\y);
    \draw[edgeW, bend left=70] (\z+3*.74,\y) to (\z+4*.74,\y);
    \draw[edgeV, bend left=79] (\z+3*.74-.14,\y) to (\z+4*.74+.14,\y);
    \node[column] (theNode) at (0,0) {$\myvcenter{\ensuremath{\columnL{ \\ \\ } \columnL{ \\ \\ } \columnL{}\columnL{}}}$};
}}
}

\newcommand{\DRAWFOURccbaa}{
\DRAW{1.6mm}{
\foreach \y/\z in {1.14/-6*.37} {
    \draw[edgeW, bend left=70] (\z+1*.74,\y) to (\z+2*.74,\y);
    \draw[edgeV, bend left=79] (\z+1*.74-.14,\y) to (\z+2*.74+.14,\y);
    \draw[edgeW, bend left=70] (\z+4*.74,\y) to (\z+5*.74,\y);
    \draw[edgeV, bend left=79] (\z+4*.74-.14,\y) to (\z+5*.74+.14,\y);
    \node[column] (theNode) at (0,0) {$\myvcenter{\ensuremath{\columnL{ \\ \\ } \columnL{ \\ \\ } \columnL{ \TinyV \\ } \columnL{}\columnL{}}}$};
}}
}

\newcommand{\DRAWTHREEcbba}{
\DRAW{1.6mm}{
\foreach \y/\z in {1.14/-5*.37} {
    \draw[edgeW, bend left=79] (\z+1*.74,\y) to (\z+4*.74,\y);
    \draw[edge, bend left=70] (\z+2*.74,\y) to (\z+3*.74,\y);
    \draw[edge, bend left=79] (\z+2*.74-.14,\y) to (\z+3*.74+.14,\y);
    \node[column] (theNode) at (0,0) {$\myvcenter{\ensuremath{\columnL{ \\ \\ } \columnL{  \\ } \columnL{ \\ } \columnL{}}}$};
}}
}

\newcommand{\DRAWTHREEVcbba}{
\DRAW{1.6mm}{
\foreach \y/\z in {1.14/-5*.37} {
    \draw[edgeV, bend left=79] (\z+1*.74,\y) to (\z+4*.74,\y);
    \draw[edge, bend left=70] (\z+2*.74,\y) to (\z+3*.74,\y);
    \draw[edge, bend left=79] (\z+2*.74-.14,\y) to (\z+3*.74+.14,\y);
    \node[column] (theNode) at (0,0) {$\myvcenter{\ensuremath{\columnL{ \\ \\ } \columnL{  \\ } \columnL{ \\ } \columnL{}}}$};
}}
}

\newcommand{\DRAWcbdbaa}{
\DRAW{1.6mm}{
\foreach \y/\z in {1.14/-7*.37} {
    \draw[edgeW, bend left=79] (\z+1*.74,\y) to (\z+5*.74,\y);
    \node[column] (theNode) at (0,0) {$\myvcenter{\ensuremath{\columnL{ \\ \\ } \columnL{ \TinyV \\ } \downdots \columnL{ \TinyV \\ } \columnL{}\columnL{}}}$};
}}
}

\newcommand{\DRAWccbdba}{
\DRAW{1.6mm}{
\foreach \y/\z in {1.14/-7*.37} {
    \draw[edgeW, bend left=79] (\z+2*.74,\y) to (\z+6*.74,\y);
    \node[column] (theNode) at (0,0) {$\myvcenter{\ensuremath{\columnL{ \\ \\ } \columnL{ \\ \\ } \columnL{ \TinyV \\ } \downdots \columnL{ \TinyV \\ } \columnL{}}}$};
}}}

\newcommand{\DRAWccbdbaa}{
\DRAW{1.6mm}{
\foreach \y/\z in {1.14/-8*.37} {
    \draw[edgeW, bend left=79] (\z+1*.74,\y) to (\z+7*.74,\y);
    \draw[edgeW, bend left=79] (\z+2*.74,\y) to (\z+6*.74,\y);
    \node[column] (theNode) at (0,0) {$\myvcenter{\ensuremath{\columnL{ \\ \\ } \columnL{ \\ \\ } \columnL{ \TinyV \\ } \downdots \columnL{ \TinyV \\ } \columnL{}\columnL{}}}$};
}}
}

\newcommand{\DRAWccaa}{
\DRAW{1.6mm}{
\foreach \y/\z in {1.14/-5*.37} {
    \draw[edgeW, bend left=79] (\z+1*.74,\y) to (\z+4*.74,\y);
    \draw[edgeW, bend left=79] (\z+2*.74,\y) to (\z+3*.74,\y);
    \node[column] (theNode) at (0,0) {$\myvcenter{\ensuremath{\columnL{ \\ \\ } \columnL{ \\ \\ }\columnL{}\columnL{}}}$};
}}
}

\newcommand{\DRAWTWOcbdba}{
\DRAW{1.6mm}{
\foreach \y/\z in {1.14/-6*.37} {
    \draw[edgeW, bend left=79] (\z+1*.74,\y) to (\z+4*.74,\y);
    \draw[edgeW, bend left=79] (\z+4*.74,\y) to (\z+5*.74,\y);
    \node[column] (theNode) at (0,0) {$\myvcenter{\ensuremath{\columnL{ \\ \\ } \columnL{ \TinyV \\ } \downdots \columnL{ \TinyV \\ } \columnL{}}}$};
}}
}

\newcommand{\DRAWcbdba}{
\DRAW{1.6mm}{
\foreach \y/\z in {1.14/-6*.37} {
    \draw[edgeW, bend left=79] (\z+1*.74,\y) to (\z+5*.74,\y);
    \node[column] (theNode) at (0,0) {$\myvcenter{\ensuremath{\columnL{ \\ \\ } \columnL{ \TinyV \\ } \downdots \columnL{ \TinyV \\ } \columnL{}}}$};
}}
}

\newcommand{\DRAWcbaa}{
\DRAW{1.6mm}{
\foreach \y/\z in {1.14/-5*.37} {
    \draw[edgeW, bend left=70] (\z+3*.74,\y) to (\z+4*.74,\y);
    \draw[edgeV, bend left=79] (\z+3*.74-.14,\y) to (\z+4*.74+.14,\y);
    \draw[edgeW, bend left=79] (\z+1*.74,\y) to (\z+2*.74,\y);
    \node[column] (theNode) at (0,0) {$\myvcenter{\ensuremath{\columnL{ \\ \\ }\columnL{ \TinyW \\ }\columnL{} \columnL{}}}$};
}}
}

\newcommand{\DRAWcbastuff}[3]{
\DRAW{1.6mm}{
\foreach \y/\z in {#1/-#2*.37} {
    \draw[edgeW, bend left=79] (\z+2*.74,\y) to (\z+3*.74+.14,\y);
    \draw[edgeW, bend left=79] (\z+1*.74-.14,\y) to (\z+2*.74,\y);
    \node[column] (theNode) at (0,0) {$\myvcenter{\ensuremath{\columnL{ \\ \\ } \columnL{ \TinyW \\ }\columnL{}#3}}$};
}}
}

\newcommand{\DRAWstuffcba}[3]{
\DRAW{1.6mm}{
\foreach \y/\z in {#1/-#2*.37} {
    \draw[edgeW, bend left=79] (\z+#2*.74-2*.74,\y) to (\z+#2*.74-1*.74+.14,\y);
    \draw[edgeW, bend left=79] (\z+#2*.74-3*.74 -.14,\y) to (\z+#2*.74-2*.74,\y);
    \node[column] (theNode) at (0,0) {$\myvcenter{\ensuremath{#3\columnL{ \\ \\ } \columnL{ \TinyW \\ }\columnL{}}}$};
}}
}

\newcommand{\DRAWstuffaa}[3]{
\DRAW{1.6mm}{
\foreach \y/\z in {#1/-#2*.37} {
    \draw[edgeW, bend left=70] (\z+#2*.74-2*.74,\y) to (\z+#2*.74-1*.74,\y);
    \draw[edgeV, bend left=79] (\z+#2*.74-2*.74-.14,\y) to (\z+#2*.74-1*.74+.14,\y);
    \node[column] (theNode) at (0,0) {$\myvcenter{\ensuremath{#3\columnL{} \columnL{}}}$};
}}
}

\newcommand{\DRAWstuffbb}[3]{
\DRAW{1.6mm}{
\foreach \y/\z in {#1/-#2*.37} {
    \draw[edge, bend left=70] (\z+#2*.74-2*.74,\y) to (\z+#2*.74-1*.74,\y);
    \draw[edge, bend left=79] (\z+#2*.74-2*.74-.14,\y) to (\z+#2*.74-1*.74+.14,\y);
    \node[column] (theNode) at (0,0) {$\myvcenter{\ensuremath{#3\columnL{ \\ } \columnL{ \\ }}}$};
}}
}

\newcommand{\DRAWccstuff}[3]{
\DRAW{1.6mm}{
\foreach \y/\z in {#1/-#2*.37} {
    \draw[edgeW, bend left=70] (\z+1*.74,\y) to (\z+2*.74,\y);
    \draw[edgeV, bend left=79] (\z+1*.74-.14,\y) to (\z+2*.74+.14,\y);
    \node[column] (theNode) at (0,0) {$\myvcenter{\ensuremath{\columnL{ \\ \\ } \columnL{ \\ \\ }#3}}$};
}}
}

\newcommand{\DRAWbbstuff}[3]{
\DRAW{1.6mm}{
\foreach \y/\z in {#1/-#2*.37} {
    \draw[edge, bend left=70] (\z+1*.74,\y) to (\z+2*.74,\y);
    \draw[edge, bend left=79] (\z+1*.74-.14,\y) to (\z+2*.74+.14,\y);
    \node[column] (theNode) at (0,0) {$\myvcenter{\ensuremath{\columnL{ \\ } \columnL{ \\ }#3}}$};
}}
}

\newcommand{\DRAWbastuff}[3]{
\DRAW{1.6mm}{
\foreach \y/\z in {#1/-#2*.37} {
    \draw[edge, bend left=70] (\z+1*.74,\y) to (\z+2*.74,\y);
    \node[column] (theNode) at (0,0) {$\myvcenter{\ensuremath{\columnL{ \\ } \columnL{}#3}}$};
}}
}

\newcommand{\DRAWbaWstuff}[3]{
\DRAW{1.6mm}{
\foreach \y/\z in {#1/-#2*.37} {
    \draw[edgeW, bend left=70] (\z+1*.74,\y) to (\z+2*.74,\y);
    \node[column] (theNode) at (0,0) {$\myvcenter{\ensuremath{\columnL{\TinyW \\ } \columnL{}#3}}$};
}}
}

\newcommand{\DRAWcbstuff}[3]{
\DRAW{1.6mm}{
\foreach \y/\z in {#1/-#2*.37} {
    \draw[edgeW, bend left=79] (\z+1*.74,\y) to (\z+2*.74,\y);
    \node[column] (theNode) at (0,0) {$\myvcenter{\ensuremath{\columnL{\\ \\} \columnL{ \TinyW \\ }#3}}$};
}}
}

\newcommand{\DRAWcastuff}[3]{
\DRAW{1.6mm}{
\foreach \y/\z in {#1/-#2*.37} {
    \draw[edgeW, bend left=79] (\z+1*.74,\y) to (\z+2*.74,\y);
    \node[column] (theNode) at (0,0) {$\myvcenter{\ensuremath{\columnL{\\ \\} \columnL{}#3}}$};
}}
}

\newcommand{\DRAWstuffcb}[3]{
\DRAW{1.6mm}{
\foreach \y/\z in {#1/-#2*.37} {
    \draw[edgeW, bend left=79] (\z+#2*.74-2*.74,\y) to (\z+#2*.74-1*.74,\y);
    \node[column] (theNode) at (0,0) {$\myvcenter{\ensuremath{#3\columnL{\\ \\} \columnL{ \TinyW \\ }}}$};
}}
}

\newcommand{\DRAWstuffca}[3]{
\DRAW{1.6mm}{
\foreach \y/\z in {#1/-#2*.37} {
    \draw[edgeW, bend left=79] (\z+#2*.74-2*.74,\y) to (\z+#2*.74-1*.74,\y);
    \node[column] (theNode) at (0,0) {$\myvcenter{\ensuremath{#3\columnL{\\ \\} \columnL{}}}$};
}}
}

\newcommand{\DRAWstuffba}[3]{
\DRAW{1.6mm}{
\foreach \y/\z in {#1/-#2*.37} {
    \draw[edge, bend left=79] (\z+#2*.74-2*.74,\y) to (\z+#2*.74-1*.74,\y);
    \node[column] (theNode) at (0,0) {$\myvcenter{\ensuremath{#3\columnL{\\} \columnL{}}}$};
}}
}

\newcommand{\DRAWstuffbaW}[3]{
\DRAW{1.6mm}{
\foreach \y/\z in {#1/-#2*.37} {
    \draw[edgeW, bend left=79] (\z+#2*.74-2*.74,\y) to (\z+#2*.74-1*.74,\y);
    \node[column] (theNode) at (0,0) {$\myvcenter{\ensuremath{#3\columnL{\TinyW\\} \columnL{}}}$};
}}
}

\newcommand{\DRAWaa}{
\DRAW{1.6mm}{
\foreach \y/\z in {.42/-3*.37} {
    \draw[edgeW, bend left=70] (\z+1*.74,\y) to (\z+2*.74,\y);
    \draw[edgeV, bend left=79] (\z+1*.74-.14,\y) to (\z+2*.74+.14,\y);
    \node[column] (theNode) at (0,0) {$\myvcenter{\ensuremath{\columnL{} \columnL{}}}$};
}}
}

\newcommand{\DRAWbb}{
\DRAW{1.6mm}{
\foreach \y/\z in {.79/-3*.37} {
    \draw[edge, bend left=70] (\z+1*.74,\y) to (\z+2*.74,\y);
    \draw[edge, bend left=79] (\z+1*.74-.14,\y) to (\z+2*.74+.14,\y);
    \node[column] (theNode) at (0,0) {$\myvcenter{\ensuremath{\columnL{\\ } \columnL{ \\}}}$};
}}
}
\newcommand{\DRAWcc}{
\DRAW{1.6mm}{
\foreach \y/\z in {1.14/-3*.37} {
    \draw[edgeW, bend left=70] (\z+1*.74,\y) to (\z+2*.74,\y);
    \draw[edgeV, bend left=79] (\z+1*.74-.14,\y) to (\z+2*.74+.14,\y);
    \node[column] (theNode) at (0,0) {$\myvcenter{\ensuremath{\columnL{\\ \\} \columnL{\\ \\}}}$};
}}
}

\newcommand{\DRAWcaV}{
\DRAW{1.6mm}{
\foreach \y/\z in {1.14/-3*.37} {
    \draw[edgeV, bend left=79] (\z+1*.74-.14,\y) to (\z+2*.74+.14,\y);
    \node[column] (theNode) at (0,0) {$\myvcenter{\ensuremath{\columnL{\\ \\} \columnL{}}}$};
}}
}

\newcommand{\DRAWcba}{
\DRAW{1.6mm}{
\foreach \y/\z in {1.14/-4*.37} {
    \draw[edgeW, bend left=79] (\z+2*.74,\y) to (\z+3*.74+.14,\y);
    \draw[edgeW, bend left=79] (\z+1*.74-.14,\y) to (\z+2*.74,\y);
    \node[column] (theNode) at (0,0) {$\myvcenter{\ensuremath{\columnL{\\ \\} \columnL{\TinyW\\}\columnL{}}}$};
}}
}

\newcommand{\DRAWcbaV}{
\DRAW{1.6mm}{
\foreach \y/\z in {1.14/-4*.37} {
    \draw[edgeW, bend left=79] (\z+2*.74,\y) to (\z+3*.74+.14,\y);
    \draw[edgeV, bend left=79] (\z+1*.74-.14,\y) to (\z+3*.74+.14,\y);
    \node[column] (theNode) at (0,0) {$\myvcenter{\ensuremath{\columnL{\\ \\} \columnL{\TinyW\\}\columnL{}}}$};
}}
}

\newcommand{\DRAWONEcba}{
\DRAW{1.6mm}{
\foreach \y/\z in {1.14/-4*.37} {
    \draw[edgeW, bend left=79] (\z+1*.74,\y) to (\z+3*.74+.14,\y);
    \node[column] (theNode) at (0,0) {$\myvcenter{\ensuremath{\columnL{\\ \\} \columnL{\TinyV\\}\columnL{}}}$};
}}
}

\newcommand{\DRAWba}{
\DRAW{1.6mm}{
\foreach \y/\z in {.79/-3*.37} {
    \draw[edge, bend left=79] (\z+1*.74,\y) to (\z+2*.74,\y);
    \node[column] (theNode) at (0,0) {$\myvcenter{\ensuremath{\columnL{\\ } \columnL{}}}$};
}}
}

\newcommand{\DRAWcb}{
\DRAW{1.6mm}{
\foreach \y/\z in {1.14/-3*.37} {
    \draw[edgeW, bend left=79] (\z+1*.74,\y) to (\z+2*.74,\y);
    \node[column] (theNode) at (0,0) {$\myvcenter{\ensuremath{\columnL{\\ \\} \columnL{ \TinyW \\ }}}$};
}}
}
\setlength{\cellsizeCol}{1.6ex}

For $g \in \Uqt$ and $\mathbf{T} \in \pNSTC(\nu)$, define the structure coefficients $a^g_{\mathbf{T}'\mathbf{T}}$ by $g \mathbf{T} = \sum_{\mathbf{T}' \in \pNSTC(\nu)} a^g_{\mathbf{T}'\mathbf{T}}\mathbf{T}'$.
We now determine the structure coefficients $a^g_{\mathbf{T}'\mathbf{T}}$ when $g$ is one of the Chevalley generators $F_V, F_W, E_V, E_W$. This requires a somewhat involved case-by-case analysis.

Let $\mathbf{T}$ be a $\pNSTC$. By Proposition \ref{p F action on heartsQ basis}, $F_V \mathbf{T} = \sum_{j=1}^{\varphi_V(\mathbf{T})} [j]\F_{(j)V}(\mathbf{T})$. Here,  $\F_{(j)V}(\mathbf{T})$ is defined to be the  NSTC containing $\F_{(j)V}(T)$ for any  $T \in \mathbf{T}$ (this definition is sound by the same proof given for the case  $j = \varphi_V(\mathbf{T})$ in the proof of Theorem \ref{t main canonical basis} (iv)). The drawings in cases (A)--(G) below describe $\F_{(j)V}(\mathbf{T})$, for $j < \varphi_V(\mathbf{T})$, in terms of  $\pNSTC$.  Cases (A)--(G) correspond to the eight cases in Figure \ref{f kronecker graphical calculus}, except that case (B) combines the left case for $\hat{g}^{0}_{\lambda\mu\nu}$ and the left case for $\hat{g}^{\text{3-2-1}}_{\lambda\mu\nu}$.

In the drawings below, the left-hand side represents part of some  $T \in \mathbf{T}$ and the right-hand side represents part of $\F_{(j)V}(\mathbf{T})$, expressed in terms of $\pNSTC$ (it turns out that $\F_{(j)V}(\mathbf{T})$ is always equal to an honest NSTC or to 0, though a priori we only know it to be some  $\mathbf{A}$-linear combination of  $\pNSTC$).
Note that $\F_{(j)V}(T)$, $j < \varphi_V(\mathbf{T})$, has one more external $V$-arc than $T$, for any  $T \in \mathbf{T}$.  Internal to case (*), subcases are labeled in the format (*.$k$-$l$) to indicate that the external $V$-arc created in the change from  $T$ to $\F_{(j)V}(T)$ is a  $k$-$l$  $V$-arc, and  (*.$k$-$l$.1), (*.$k$-$l$.2), etc. is used if there is more than one way to add an external  $k$-$l$ $V$-arc. In general,  $k$ and  $l$ depend on the choice of $T \in \mathbf{T}$, but we specify as little information about  $T$ as possible.
One way this is done is that only the columns involved in the change from $\mathbf{T}$ to  $\F_{(j)V}(\mathbf{T})$ are drawn; so, for instance, unless specified otherwise, there may be some type $V$ or type $W$ columns that are not shown that lie between height-3 and height-1 columns that are shown.
Other conventions for the drawings are that  $V$-arcs are thick,  $W$-arcs are thin, and arcs that we do not want to specify have the funny dashed style.

For example, the drawing $\DRAWONEcba \tto \DRAWcba  \vspace{2mm}$ in subcase (C.2-1) indicates that
\setlength{\cellsizeCol}{2.1ex}
if $T = \ctableausmallc{1 \\ 2 \\ 4}{1 \\ 2}{1}$, then $\F_{(3)V}(T) = \ctableausmallc{1 \\ 2 \\ 4}{3 \\ 2}{1} \equiv \ctableausmallc{1 \\ 2 \\ 4}{2 \\ 3}{1}$; another specific example covered by this subcase is that
$T = \ctableausmallh{1\\2 \\ 3}{1 \\ 2 \\ 4}{1 \\ 2 \\ 4}{1 \\ 2 \\ 4}{1 \\ 2}{1 \\ 2}{1}{3}$ implies
$\F_{(8)V}(T) = \ctableausmallh{1\\2\\3}{1 \\ 2 \\4}{1 \\ 2 \\ 4}{1 \\ 2 \\ 4}{1 \\ 2}{3 \\ 2}{1}{3} \equiv
\ctableausmallh{1 \\ 2 \\ 3}{1\\2\\4}{1 \\ 2 \\ 4}{1 \\ 2 \\ 4}{1 \\ 2}{2 \\ 3}{1}{3}$.   \vspace{3mm}
\setlength{\cellsizeCol}{1.6ex}
As another example of how to interpret the subcases, the drawing $\myvcenter{\ensuremath{\columnL{ \\ \\ }\columnL{\TinyV \\ }}} \tto -\DRAWcb$ appears in (A.3-2) and (D.3-2); in (A.3-2) it means that if  $\mathbf{T}$ has no external arcs and shape  $[n_4, n_3, n_2, n_1]$ (see \textsection\ref{ss type A combinatorics preliminaries} for notation), then $\F_{(n_3)V}(\mathbf{T})$ has a pure 3-2 arc and no 2-1 arcs, and in (D.3-2) it means that if  $\mathbf{T}$ has a pure 2-1 arc and no 3-2 arcs and shape  $[n_4, n_3, n_2, n_1]$, then $\F_{(n_3)V}(\mathbf{T})$ has a pure 3-2 arc and a pure 2-1 arc.  See the examples interspersed between the cases for more about how to interpret these drawings.

The subcases are fairly redundant, but we include them all for completeness.  The subcases that are genuinely different from all previous ones are marked by bold labels.

\bigskip
Case (A): $\mathbf{T}$ has no external arcs:

\setlength{\extrarowheight}{4pt}
\begin{longtable}[l]{@{\hspace{1.5cm}}ll}
\textbf{(A.1-1)} &  $\myvcenter{\ensuremath{\columnL{}\columnL{}}} \tto -\frac{1}{[2]}\DRAWaa$ or 0,\\[2mm]
\textbf{(A.2-1)} &  $\myvcenter{\ensuremath{\columnL{\TinyV \\ }\columnL{}}} \tto \DRAWba$,\\
\textbf{(A.2-2)} &  $\myvcenter{\ensuremath{\columnL{\TinyV \\ }\columnL{\TinyV \\ }}} \tto -\frac{1}{[2]}\DRAWbb$ or 0,\\[2mm]
\textbf{(A.3-1)} &  $\myvcenter{\ensuremath{\columnL{ \\ \\ }\columnL{}}} \tto \DRAWcaV$ (if $\exfreev(\mathbf{T}) = 0$),\\[2mm]
\textbf{(A.3-2)} &  $\myvcenter{\ensuremath{\columnL{ \\ \\ }\columnL{\TinyV \\ }}} \tto -\DRAWcb$,\\[1mm]
\textbf{(A.3-3)} &  $\myvcenter{\ensuremath{\columnL{ \\ \\ }\columnL{ \\ \\ }}} \tto -\frac{1}{[2]}\DRAWcc$ or 0.
\end{longtable}

\bigskip
Case (B): $\mathbf{T}$ has a pure 3-1 $V$-arc or a 3-2-1  $V$-arc (combines two cases from Figure \ref{f kronecker graphical calculus}):

\begin{longtable}[l]{@{\hspace{1.5cm}}ll}
(B.1-1) &  $\myvcenter{\ensuremath{\columnL{}\columnL{}}} \tto -\frac{1}{[2]}\DRAWaa$ or 0,\\[2mm]
(B.3-1) &  $\myvcenter{\ensuremath{\columnL{ \\ \\ }\columnL{}}} \tto \DRAWcaV$,\\[2mm]
(B.3-3) &  $\myvcenter{\ensuremath{\columnL{ \\ \\ }\columnL{ \\ \\ }}} \tto -\frac{1}{[2]}\DRAWcc$ or 0.
\end{longtable}

\bigskip
Case (C): $\mathbf{T}$ has at least one pure 3-1 $W$-arc and no 3-2 or 2-1 arcs:

\begin{longtable}[l]{@{\hspace{1.5cm}}ll}
(C.1-1.1) &  $\myvcenter{\ensuremath{\columnL{}\columnL{}}} \tto -\frac{1}{[2]}\DRAWaa$ or 0,\\
\textbf{(C.1-1.2)} &  $\DRAWcastuff{1.14}{4}{\columnL{}} \tto -\frac{1}{[2]}\DRAWstuffaa{1.14}{4}{\columnL{ \\ \\ }}$,\\[3mm]
\textbf{(C.1-1.3)} &  $\DRAWccaa \tto 0$,\\[4mm]
\textbf{(C.2-1)} &  $\DRAWONEcba \tto \DRAWcba$,\\[3mm]
(C.2-2) &  $\myvcenter{\ensuremath{\columnL{\TinyV \\ }\columnL{\TinyV \\ }}} \tto -\frac{1}{[2]}\DRAWbb$ or 0,\\[2mm]
\textbf{(C.3-1)} &  $\DRAWcastuff{1.14}{3}{} \tto 0$ (if $\exfreev(\mathbf{T}) = 0$),\\[3mm]
\textbf{(C.3-2)} &  $\DRAWONEcba \tto -\DRAWcba$,\\[3mm]
\textbf{(C.3-3.1)} &  $\DRAWccaa \tto \frac{1}{[2]^2}\DRAWFOURccaa$ or 0,\\[4mm]
\textbf{(C.3-3.2)} &  $\DRAWstuffca{1.14}{4}{\columnL{\\ \\}} \tto -\frac{1}{[2]}\DRAWccstuff{1.14}{4}{\columnL{}}$,\\[3mm]
(C.3-3.3) &  $\myvcenter{\ensuremath{\columnL{ \\ \\ }\columnL{ \\ \\ }}} \tto -\frac{1}{[2]}\DRAWcc$ or 0.
\end{longtable}

\bigskip
\setlength{\cellsizeCol}{2.1ex}
For example, if $T = \ctableausmallh{1 \\ 2 \\ 4}{1 \\ 2 \\ 4}{1 \\ 2 \\ 4}{1 \\ 2}{1 \\ 2}{1}{1}{1}$, then subcase (C.2-1) indicates that
\[\F_{(7)V}(T) = \ctableauh{1 \\ 2 \\ 4}{1 \\ 2 \\ 4}{1 \\ 2 \\ 4}{1 \\ 2}{3 \\ 2}{1}{1}{1}
\equiv \ctableauh{1 \\ 2 \\ 4}{1 \\ 2 \\ 4}{1 \\ 2 \\ 4}{1 \\ 2}{2 \\ 3}{1}{1}{1}, \]
and subcase (C.3-2) indicates that
\[\F_{(3)V}(T) = \ctableauh{1 \\ 2 \\ 4}{1 \\ 2 \\ 4}{2 \\ 3 \\ 4}{1 \\ 2}{1 \\ 2}{1}{1}{1}
\equiv -\ctableauh{1 \\ 2 \\ 4}{1 \\ 2 \\ 4}{1 \\ 2 \\ 4}{1 \\ 2}{2 \\ 3}{1}{1}{1}. \]
Set $T' = \ctableausmallh{1 \\ 2 \\ 4}{1 \\ 2 \\ 4}{1 \\ 2 \\ 4}{1 \\ 2}{2 \\ 3}{1}{1}{1}$ and let $\mathbf{T}'$ (resp. $\mathbf{T}$) be the $\pNSTC$ containing $T'$ (resp. $T$). Examining all of case (C), we observe that these are the only contributions to the structure coefficient
$a^{F_V}_{\mathbf{T}'\mathbf{T}}$, hence $a^{F_V}_{\mathbf{T}'\mathbf{T}} = [7] - [3]$.
\setlength{\cellsizeCol}{1.6ex}

\bigskip
Case (D): $\mathbf{T}$ has a pure 2-1 arc and no 3-2 arcs:

\begin{longtable}[l]{@{\hspace{1.5cm}}ll}
(D.1-1.1) &  $\myvcenter{\ensuremath{\columnL{}\columnL{}}} \tto -\frac{1}{[2]}\DRAWaa$ or 0,\\
\textbf{(D.1-1.2)} &  $\DRAWbaWstuff{.79}{4}{\columnL{}} \tto -\frac{1}{[2]}\DRAWstuffaa{.79}{4}{\columnL{\TinyW \\ }}$,\\[1mm]
\textbf{(D.2-1)} &  $\DRAWstuffbaW{.79}{4}{\columnL{\TinyV \\ }} \tto -\frac{1}{[2]}\DRAWbbstuff{.79}{4}{\columnL{}}$,\\[2mm]
(D.2-2) &  $\myvcenter{\ensuremath{\columnL{\TinyV \\ }\columnL{\TinyV \\ }}} \tto -\frac{1}{[2]}\DRAWbb$ or 0,\vspace{1mm}\\
\textbf{(D.3-1)} &  $\DRAWstuffbaW{1.14}{4}{\columnL{\\ \\}} \tto \DRAWcbaV$ (if $\exfreev(\mathbf{T}) = 0$),\vspace{2mm}\\[2mm]
(D.3-2) &  $\myvcenter{\ensuremath{\columnL{ \\ \\ }\columnL{\TinyV \\ }}} \tto -\DRAWcb$,\\[1mm]
(D.3-3) &  $\myvcenter{\ensuremath{\columnL{ \\ \\ }\columnL{ \\ \\ }}} \tto -\frac{1}{[2]}\DRAWcc$ or 0.
\end{longtable}

\bigskip
Case (E): $\mathbf{T}$ has a pure 3-2 arc and no 2-1 arcs:

\begin{longtable}[l]{@{\hspace{1.5cm}}ll}
(E.1-1) &  $\myvcenter{\ensuremath{\columnL{}\columnL{}}} \tto -\frac{1}{[2]}\DRAWaa$ or 0,\\[1mm]
(E.2-1) &  $\myvcenter{\ensuremath{\columnL{\TinyV \\ }\columnL{}}} \tto \DRAWba$,\\[1mm]
(E.2-2) &  $\myvcenter{\ensuremath{\columnL{\TinyV \\ }\columnL{\TinyV \\ }}} \tto -\frac{1}{[2]}\DRAWbb$ or 0,\\[2mm]
\textbf{(E.3-1)} &  $\DRAWcbstuff{1.14}{4}{\columnL{}} \tto -\DRAWcbaV$ (if $\exfreev(\mathbf{T}) = 0$),\\[3mm]
\textbf{(E.3-2)} &  $\DRAWcbstuff{1.14}{4}{\columnL{\TinyV \\ }} \tto \frac{1}{[2]}\DRAWstuffbb{1.14}{4}{\columnL{ \\ \\ }}$,\\[3mm]
\textbf{(E.3-3.1)} &  $\DRAWstuffcb{1.14}{4}{\columnL{ \\ \\ }} \tto -\frac{1}{[2]}\DRAWccstuff{1.14}{4}{\columnL{ \TinyW \\ }}$,\\[3mm]
(E.3-3.2) &  $\myvcenter{\ensuremath{\columnL{ \\ \\ }\columnL{ \\ \\ }}} \tto -\frac{1}{[2]}\DRAWcc$ or 0.
\end{longtable}

\bigskip
Case (F): $\mathbf{T}$ has a 3-2-1  $W$-arc:

\begin{longtable}[l]{@{\hspace{1.5cm}}ll}
 (F.1-1.1) &  $\myvcenter{\ensuremath{\columnL{}\columnL{}}} \tto -\frac{1}{[2]}\DRAWaa$ or 0,\\
 (F.1-1.2) &  $\DRAWcastuff{1.14}{4}{\columnL{}} \tto -\frac{1}{[2]}\DRAWstuffaa{1.14}{4}{\columnL{ \\ \\ }}$,\\[3mm]
 (F.1-1.3) &  $\DRAWccaa \tto 0$,\\[4mm]
 \textbf{(F.1-1.4)} &  $\DRAWccbaa \tto  -\frac{1}{[2]^2}\DRAWFOURccbaa$,\\[5mm]
 \textbf{(F.1-1.5)} &  $\DRAWcbastuff{1.14}{5}{\columnL{}} \tto -\frac{1}{[2]} \DRAWcbaa$ (if $\arc{3-1 $W$}(\mathbf{T}) = 0$),\\[2mm]
 \textbf{(F.2-1)} &  $\DRAWcbba \tto -\frac{1}{[2]}\DRAWTHREEcbba$,\\[3mm]
 (F.2-2) &  $\myvcenter{\ensuremath{\columnL{\TinyV \\ }\columnL{\TinyV \\ }}} \tto -\frac{1}{[2]}\DRAWbb$ or 0,\\[2mm]
 \textbf{(F.3-1)} &  $\DRAWcba \tto 0$ (if $\exfreev(\mathbf{T}) = 0$),\\[2mm]
 \textbf{(F.3-2)} &  $\DRAWcbba \tto \frac{1}{[2]}\DRAWTHREEcbba$, \\[3mm]
 \textbf{(F.3-3.1)} &  $\DRAWstuffcba{1.14}{5}{\columnL{\\ \\}} \tto -\frac{1}{[2]}\DRAWccba$ (if $\arc{3-1 $W$}(\mathbf{T}) = 0$),\\[2mm]
 \textbf{(F.3-3.2)} &  $\DRAWccbaa \tto \frac{1}{[2]^2}\DRAWFOURccbaa$,\\[4mm]
 (F.3-3.3) &  $\DRAWccaa \tto \frac{1}{[2]^2}\DRAWFOURccaa$ or 0,\\[4mm]
 (F.3-3.4) &  $\DRAWstuffca{1.14}{4}{\columnL{\\ \\}} \tto -\frac{1}{[2]}\DRAWccstuff{1.14}{4}{\columnL{}}$,\\[4mm]
 (F.3-3.5) &  $\myvcenter{\ensuremath{\columnL{ \\ \\ }\columnL{ \\ \\ }}} \tto -\frac{1}{[2]}\DRAWcc$ or 0.
\end{longtable}

\bigskip
\setlength{\cellsizeCol}{2.1ex}
For example, if $T = \ctableausmallf{1 \\ 2 \\ 3}{1 \\ 2 \\ 4}{1 \\ 2 \\ 4}{2 \\ 3}{1}{1}$, then subcase (F.1-1.4) indicates that
\[\F_{(4)V}(T) = \ctableauf{1 \\ 2 \\ 3}{1 \\ 2 \\ 4}{1 \\ 2 \\ 4}{2 \\ 3}{3}{1}
\equiv -\frac{1}{[2]}\ctableauf{1 \\ 2 \\ 3}{1 \\ 2 \\ 4}{1 \\ 2 \\ 4}{1 \\ 3}{4}{1}\equiv \frac{1}{[2]}\ctableauf{1 \\ 2 \\ 3}{1 \\ 2 \\ 4}{1 \\ 3 \\ 4}{1 \\ 2}{4}{1}
\equiv -\frac{1}{[2]^2}\ctableauf{1 \\ 2 \\ 3}{2 \\ 3 \\ 4}{1 \\ 2 \\ 3}{1 \\ 2}{4}{1}, \]
and subcase (F.3-3.2) indicates that
\[\F_{(2)V}(T) = \ctableauf{1 \\ 2 \\ 3}{2 \\ 3 \\ 4}{1 \\ 2 \\ 4}{2 \\ 3}{1}{1}
\equiv -\frac{1}{[2]}\ctableauf{1 \\ 2 \\ 3}{2 \\ 3 \\ 4}{1 \\ 2 \\ 3}{2 \\ 4}{1}{1} \equiv
-\frac{1}{[2]} \ctableauf{1 \\ 2 \\ 3}{2 \\ 3 \\ 4}{1 \\ 2 \\ 3}{3 \\ 2}{2}{1} \equiv
\frac{1}{[2]^2}\ctableauf{1 \\ 2 \\ 3}{2 \\ 3 \\ 4}{1 \\ 2 \\ 3}{1 \\ 2}{4}{1}. \]
Set $T' = \frac{1}{[2]^2}\ctableausmallf{1 \\ 2 \\ 3}{2 \\ 3 \\ 4}{1 \\ 2 \\ 3}{1 \\ 2}{4}{1}$ and let $\mathbf{T}'$ (resp. $\mathbf{T}$) be the $\pNSTC$ containing $T'$ (resp. $T$). Examining all of case (F), we see that these are the only cases that contribute to $a^{F_V}_{\mathbf{T}'\mathbf{T}}$, hence $a^{F_V}_{\mathbf{T}'\mathbf{T}} = [2] - [4]$.
\setlength{\cellsizeCol}{1.6ex}

\bigskip
Case (G): $\mathbf{T}$ has a pure 3-2 arc and a pure 2-1 arc:

\begin{longtable}[l]{@{\hspace{1.5cm}}ll}
(G.1-1.1) &  $\myvcenter{\ensuremath{\columnL{}\columnL{}}} \tto -\frac{1}{[2]}\DRAWaa$ or 0,\\
(G.1-1.2) &  $\DRAWbaWstuff{.79}{4}{\columnL{}} \tto -\frac{1}{[2]}\DRAWstuffaa{.79}{4}{\columnL{\TinyW \\ }}$,\\[1mm]
(G.2-1) &  $\DRAWstuffbaW{.79}{4}{\columnL{\TinyV \\ }} \tto -\frac{1}{[2]}\DRAWbbstuff{.79}{4}{\columnL{}}$,\\[1mm]
\textbf{(G.3-1)} &   $\DRAWTWOcbba \tto \frac{1}{[2]} \DRAWTHREEVcbba$ (if $\exfreev(\mathbf{T}) = 0$), \\[4mm]
(G.3-2) &  $\DRAWcbstuff{1.14}{4}{\columnL{\TinyV \\ }} \tto \frac{1}{[2]}\DRAWstuffbb{1.14}{4}{\columnL{ \\ \\ }}$,\\[3mm]
(G.3-3.1) &  $\DRAWstuffcb{1.14}{4}{\columnL{ \\ \\ }} \tto -\frac{1}{[2]}\DRAWccstuff{1.14}{4}{\columnL{ \TinyW \\ }}$, \\[4mm]
(G.3-3.2) &  $\myvcenter{\ensuremath{\columnL{ \\ \\ }\columnL{ \\ \\ }}} \tto -\frac{1}{[2]}\DRAWcc$ or 0.
\end{longtable}

\setlength{\cellsizeCol}{2.1ex}
\setlength{\extrarowheight}{0pt}

The next theorem summarizes the findings of this case-by-case analysis.
We choose to state the theorem in the language of Corollary \ref{c diagram kronecker}; recall that by this corollary,
a $\pNSTC$  $\mathbf{T}$ is determined by its unpaired $V$- and $W$-diagrams and a 12-tuple whose
first four entries are the invariant-record of  $\mathbf{T}$ and next eight entries are
{\small
\[\arc{2-1}(\mathbf{T}),\arc{3-2}(\mathbf{T}),\arc{3-1  $V$}(\mathbf{T}),  \arc{3-1  $W$}(\mathbf{T}),\arc{3-2-1  $V$}(\mathbf{T}), \arc{3-2-1  $W$}(\mathbf{T}), \exfreev(\mathbf{T}),\exfreew(\mathbf{T}).\]
}

\begin{theorem}
\label{t positivity}
For $\pNSTC$ $\mathbf{T}',\mathbf{T}$ of shape $\nu = [n_4,n_3,n_2,n_1]$, the structure coefficient
$a^{F_V}_{\mathbf{T}'\mathbf{T}}$
\[
\left\{ \hspace{-4mm}\parbox{15cm}{
\begin{align}
\small &= [\varphi_V(\mathbf{T})]
&& \text{if \, $\mathbf{T}'=\F_{(\varphi_V(\mathbf{T}))V}(\mathbf{T})$,} \label{e positivity1}\\
&= [n_3+2n_2] - [n_3]
&&
\begin{array}{l@{\ }c}
\mathbf{T} & (i_4,i_3,i_2,i_1 , \ \ \ \ \ \ \ \ \ \ 0, 0 ,  0, \ \  k \ \ \  , 0,0,\ \ n_2\ \ \ ,0) \label{e positivity2}\\
\mathbf{T}' & (i_4,i_3,i_2,i_1 , \ \ \ \ \ \ \ \ \ \ 0, 0 ,  0, k-1, 0,1,n_2-1,0)
\end{array}
\\[2mm]
&= [n_3+2n_2-2]-[n_3] &&
\begin{array}{l@{\ }c}
\mathbf{T}  & (i_4,i_3,\ \ i_2 \ \ \ ,i_1 , \ \ \ \ \  0, 0 ,  0, \ \ k \ \ \  , 0,1,n_2-1,0) \\
\mathbf{T}' & (i_4,i_3,i_2+1,i_1 , \ \ \ \ \ 0, 0 ,  0, k+1, 0,0,n_2-1,0)
\end{array} \label{e positivity3}\\
&= [n_3 - 1] - [n_3+ 2n_2 -1] &&
\begin{array}{l@{\ }c}
\mathbf{T}  & (i_4, \ \  i_3 \ \ \ , i_2, \ \ i_1\  \ \ ,  0, 0 ,  0, \ \ k \ \ \  , 0,1,n_2-1,0) \\
\mathbf{T}' & (i_4,i_3+1,i_2,i_1+1 ,  0, 0 ,  0, k-1, 0,0,\ \ n_2 \ \ \ ,0)
\end{array} \label{e positivity4}\\
&\in \textstyle \bigcup_{j=0}^{\varphi_V(\mathbf{T})-1} \{-[j],[j]\} && \text{otherwise,} \label{e positivity5}
\end{align}
} \notag\right.
\]
where $i_j$ and  $k$ are nonnegative integers, and the conditions for \eqref{e positivity2}--\eqref{e positivity4} are that  $\mathbf{T}$ and  $\mathbf{T}'$ are determined by the 12-tuples shown and the unpaired $V$-diagrams (resp.  $W$-diagrams) of  $\mathbf{T}$ and  $\mathbf{T}'$ have the same number of 2's.

Similarly, the structure coefficient
\[
a^{E_V}_{\mathbf{T}'\mathbf{T}}
\begin{cases}
= [\varepsilon_V(\mathbf{T})]
& \parbox[t]{11cm}{\small if $\mathbf{T}'=\E_{(\varepsilon_V(\mathbf{T}))V}(\mathbf{T})$,} \\
= [n_1] - [n_1+2n_2]
& \text{same 12-tuples as \eqref{e positivity2},} \\
= [n_1]-[n_1+2n_2-2] & \text{same 12-tuples as \eqref{e positivity3},}\\
= [n_1+ 2n_2 -1] - [n_1-1] & \text{same 12-tuples as \eqref{e positivity4},}\\
\in \bigcup_{j=0}^{\varepsilon_V(\mathbf{T})-1} \{-[j],[j]\} & \parbox[t]{11cm}{\small otherwise.}
\end{cases}
\]
For the middle three cases, we also require that the $V$-words (resp.  $W$-words) of  $\mathbf{T}$ and  $\mathbf{T}'$ have the same number of 1's.

Similar statements hold with $W$ in place of  $V$---the coefficients remain the same, though the tuples in \eqref{e positivity2}--\eqref{e positivity4} must have their entries in positions 7 and 8, 9 and 10, and 11 and 12 swapped to accommodate interchanging  $V$ and  $W$.

In particular, for  $g = F_V, F_W, E_V,$ or $E_W$, the structure coefficient $a^g_{\mathbf{T}'\mathbf{T}}$ is a  $\br{\cdot}$-invariant Laurent polynomial in $\u$ with all nonnegative or all nonpositive coefficients.
\end{theorem}
\begin{proof}
The case-by-case analysis shows that $\F_{(j)V}(\mathbf{T})$ is always equal (in $\nsbr{X}_\nu$) to some honest NSTC or to 0.  The theorem then follows by determining those $\F_{(j')V}(\mathbf{T})$ that are proportional to another $\F_{(j)V}(\mathbf{T})$.  The only such $\F_{(j')V}(\mathbf{T})$ are those corresponding to subcases (C.2-1) and (C.3-2), (F.2-1) and (F.3-2), and (F.1-1.4) and (F.3-3.2), which yield \eqref{e positivity2}, \eqref{e positivity3}, and \eqref{e positivity4}, respectively.
\end{proof}

The next example shows that the structure coefficients $a^g_{\mathbf{T}'\mathbf{T}}$, for $g$ a canonical basis element of $\Uqvw$, are not always polynomials in $\u$ with all nonnegative or all nonpositive coefficients.  Note that this also implies that there is no way to readjust the signs of $\pNSTC$ so that the  $a^{F_V}_{\mathbf{T}'\mathbf{T}}$ have only nonnegative coefficients.
\begin{example}\label{ex not positive}
\setlength{\oldSize}{\cellsizeCol}
\cellsizeCol=1.5ex
Let
$T = \myvcenter{\ensuremath{ \tiny \columnL{1 \\ 2\\3}\downdots\columnL{1\\2\\3}\columnL{1\\2\\4} \columnL{1\\2} \downdots\downdots \columnL{1\\2}\columnL{1}\columnL{1}}}$ and
$T'= -\overbtwo\myvcenter{\tiny \ensuremath{\columnL{1 \\ 2\\3}\downdots\columnL{1\\2\\3}\columnL{1\\2\\4} \columnL{2\\3}\columnL{1\\2} \downdots \columnL{1\\2}\columnL{4}\columnL{1}}}$
\setlength{\cellsizeCol}{\oldSize}
be  $\pSNST$ of shape $\nu = [0, n_3, n_2, 2]$, where the  double dots indicate that there is at least one column of this type.  Let $\mathbf{T}$ (resp.  $\mathbf{T}'$) be the $\pNSTC$ containing $T$ (resp.  $T'$). The coefficient $a^{F_V^{(2)}}_{\mathbf{T}'\mathbf{T}}$ is computed as follows. Let
\[\mathbf{T}^1 = \myvcenter{\ensuremath{\column{1 \\ 2\\3}\downdots\column{1\\2\\3}\column{1\\2\\4} \column{2\\3}\column{1\\2} \downdots \column{1\\2}\column{1}\column{1}}},
\quad \mathbf{T}^2 =  -\overbtwo\myvcenter{\ensuremath{\column{1 \\ 2\\3}\downdots\column{1\\2\\3}\column{1\\2\\3} \column{1\\2} \downdots\downdots \column{1\\2}\column{4}\column{1}}}.\]
There holds
\be \label{e aFV computation}
\begin{array}{ccl}
a^{F_V^{(2)}}_{\mathbf{T}'\mathbf{T}} &=& \overbtwo \sum_{\mathbf{T}''} a^{F_V}_{\mathbf{T}'\mathbf{T}''}a^{F_V}_{\mathbf{T}''\mathbf{T}} \\[2mm]
 &=& \overbtwo\big( a^{F_V}_{\mathbf{T}'\mathbf{T}^1}a^{F_V}_{\mathbf{T}^1\mathbf{T}} + a^{F_V}_{\mathbf{T}'\mathbf{T}^2}a^{F_V}_{\mathbf{T}^2\mathbf{T}} \big) \\[2mm]
& =& \overbtwo \Big(\, [n_3+2n_2-1]([n_3+2n_2]-[n_3]) + (-[n_3])[n_3+2n_2+1] \, \Big),
\end{array}
\ee
where the third equality uses \eqref{e positivity2}. To justify the second equality, note that  $a^{F_V}_{\mathbf{D}' \mathbf{D}}$ is nonzero only if $\mathbf{D}'$ has at least as many height-$j$ invariants as  $\mathbf{D}$. Given this, we can read off from case (C) that $\mathbf{T}^1$ and $\mathbf{T}^2$ are the only possibilities for $\mathbf{T}''$ that contribute to the sum.

There are shapes $\nu$ for which the coefficient $a^{F_V^{(2)}}_{\mathbf{T}'\mathbf{T}}$ is not a polynomial in $\u$ with all nonnegative or all nonpositive coefficients. For example, if $n_3 = 3$ and $n_2 = 2$, then $a^{F_V^{(2)}}_{\mathbf{T}'\mathbf{T}} = \u^{10} - \u^4 - \u^{-4} + \u^{-10}$.
\end{example}

\subsection{The action of $\tau$ on $\pNSTC$ }
\label{ss The action of tau on pNSTC}
We have enough information now about straightened highest weight NST to prove Theorem \ref{t main canonical basis} (vi), finally completing the proof of the theorem.
\begin{proposition}
\label{p NSTC is upper based tau module}
Let  $\mathbf{T}$ be a highest weight $\pNSTC$ of shape  $\nu$ and weight $(\lambda, \lambda)$.  Then
\be
\tau(\mathbf{T}) =
\begin{cases}
(-1)^{\nu_3}\mathbf{T} & \text{ if \, $\arc{3-2}(\mathbf{T}) = 0$,} \\
(-1)^{\nu_3-1}\mathbf{T} & \text{ if \, $\arc{3-2}(\mathbf{T}) = 1$.}
\end{cases}
\ee
Hence the weak upper based  $\Uqt$-module  $(\nsbr{X}_\nu, \pNSTC(\nu))$ is an upper based $\Uqt$-module.
\end{proposition}
\begin{proof}
We compute $\tau(\mathbf{T})$ by computing the action of $\tau$ on the straightened representative $T$ of  $\mathbf{T}$ using Proposition \ref{p crystal action on Y} and then straightening the result.  This is a straightforward computation as the invariant-free part  $\hat{T}$ of $T$ is represented by one of the eight cases in Figure \ref{f kronecker graphical calculus}.  Note that $\lambda = \mu$ implies that the left and right cases for $\hat{g}^{0}_{\lambda\mu\nu}$ and the two cases for $\hat{g}^{\text{3-2-1}}_{\lambda\mu\nu}$ cannot occur, and, in the remaining cases, $\exfreev(\hat{T}) = \exfreew(\hat{T})$.
Finally,  the computation of $\tau(\hat{T})$ gives the desired result for $\tau(\mathbf{T})$ by observing that $\tau$ fixes height-$j$ invariants for $j \in [3]$ and takes the height-4 invariant to its negative.
\end{proof}

Define an index set $\mathscr{P}^\tau_{r,2}$ for the $\Uqt$-irreducibles having nonzero multiplicity in $X^{\tsr r}$ (see Proposition \ref{p schur-weyl duality tau}):
\[
\mathscr{P}^\tau_{r,2} :=  \{ \{\lambda,\mu\}: \lambda, \mu \in \mathscr{P}_{r,2}, \, \lambda \neq \mu\} \sqcup\{+\lambda: \lambda \in \mathscr{P}_{r,2}\} \sqcup \{-\lambda: \lambda \in \mathscr{P}'_{r,2}\}.
\]
Let $\pi : \ZZ_{\geq 0}^{12} \twoheadrightarrow \ZZ_{\geq 0}^9$ be the projection from (B), (C), (E) to (B), (C), (D) of Proposition \ref{p diagram data} obtained by projecting (E) onto (D) in the obvious way and leaving (B) and (C) fixed.
For $\alpha \in \mathscr{P}^\tau_{r,2}$ and $\nu \vdash_\dx r$, define subsets $\mathbf{g}^\tau_{\alpha \nu}$ of  $\ZZ_{\geq 0}^9$ by
\be
\mathbf{g}^\tau_{\alpha \nu} :=
\begin{cases}
\pi(\mathbf{g}_{\lambda \mu \nu}) = \pi(\mathbf{g}_{\mu \lambda \nu}) & \text{if $\alpha = \{\lambda,\mu\}$}, \\
\pi(\mathbf{g}_{\lambda \lambda \nu}) \cap \{\zeta \in \ZZ_{\geq 0}^9: \varepsilon = (-1)^{\nu_3+\zeta_{\text{3-2}}}\}& \text{if $\alpha = \varepsilon \lambda$},
\end{cases}
\ee
where, in the second case, $\zeta_{\text{3-2}}$ is the  $\arc{3-2}(\cdot)$ coordinate of  $\zeta$.
Corollary \ref{c NSTC cells} together with the previous proposition then yields
\begin{corollary} \label{c NSTC tau cells}
The $\Uqt$-module with basis $(\nsbr{X}_\nu, \pNSTC(\nu))$ decomposes into $\Uqt$-cells as
\[ \pNSTC(\nu) = \bigsqcup_{\alpha \in \mathscr{P}^\tau_{r,2}, \ \zeta \in \mathbf{g}^\tau_{\alpha \nu}} \nsbr{\Lambda}^\tau_{\nu,\zeta},\]
where $\nsbr{\Lambda}^\tau_{\nu,\zeta}$ is the set of $\mathbf{T} \in \pNSTC(\nu)$ such that the 9-tuple of $\mathbf{T}$ given by (B)--(D) of Proposition \ref{p diagram data} is equal to $\zeta$.
Moreover, similar to the comment after Corollary \ref{c NSTC cells}, \eqref{e l1 + m1 formulae} implies that given $\nsbr{\Lambda}^\tau_{\nu,\zeta}$, $\alpha$ can be determined from $\nu$ and $\zeta$ and the  $\Uqt$-module  $\field \nsbr{\Lambda}^\tau_{\nu,\zeta}$ is isomorphic to $\X_\alpha$.
\end{corollary}
Recall that  $\field\nsSchur{r}$ is the algebra dual to the coalgebra  $ \O(M_q(\nsbr{X}))_r$.
It follows from \textsection\ref{ss nonstandard two-row case} that the $\Uqt$-cells and  $\field \nsSchur{r}$-cells of  $(\nsbr{X}_\nu, \pNSTC(\nu))$ are the same except that $(\nsbr{X}_{(r)},\pNSTC((r)))$ is a $\field \nsSchur{r}$-cell and the union
\[ \bigsqcup_{\zeta} (\field \nsbr{\Lambda}^\tau_{(r),\zeta}, \nsbr{\Lambda}^\tau_{(r),\zeta}) = \bigsqcup_{\lambda \vdash_2 r} (\X_{+\lambda}, B_V(\lambda) \tsrvw B_W(\lambda)) \]
of  $\lfloor \frac{r}{2} \rfloor +1$  $\Uqt$-cells.

\section{Explicit formulae for Kronecker coefficients}
\label{s explicit formulae for Kronecker coefficients}
Here we deduce explicit formulae for Kronecker coefficients by counting the number of  $(\lambda, \mu)$-cells of $(\nsbr{X}_\nu, \pNSTC)$.  This count can be simplified by writing it as a sum of what we call invariant-free Kronecker coefficients, which correspond to counting invariant-free  $\Uqvw$-cells.
We then use the Kronecker graphical calculus to organize the invariant-free Kronecker coefficients into nonnegative sums of smaller coefficients, and we use this to give a fairly simple, explicit formula for two-row Kronecker coefficients.
This is the first obviously positive formula for these coefficients.
We also give an elegant formula for symmetric and exterior Kronecker coefficients
(\textsection\ref{ss The symmetric and exterior Kronecker coefficients}).
In  \textsection\ref{ss comparisons with the formulae}, we compare our formulae to ones in \cite{BWZ} and use them to determine exactly when two-row Kronecker coefficients vanish, reproducing a result of \cite{BOR}.

\subsection{Invariant-free Kronecker coefficients and explicit formulae}
\label{ss invariant free Kronecker coefficients}
We can now give a fairly simple and explicit description of two-row Kronecker coefficients.

\begin{definition}
The \emph{invariant-free Kronecker coefficient} $\hat{g}_{\lambda\mu\nu}$ is the number of invariant-free $(\lambda,\mu)$-cells
of $\pNSTC(\nu)$.
We write $\hat{g}_{\lambda\mu\nu}$ as the sum of five terms according to the Kronecker graphical calculus described above (see Figure \ref{f kronecker graphical calculus}).
\[
\begin{array}{ccl}
  \text{Contribution to } \hat{g}_{\lambda\mu\nu}  & & \text{The number of invariant-free $(\lambda,\mu)$-cells} \\
 & & \text{of $\pNSTC(\nu)$ containing}  \\
  \noalign{\hrule height \arrayrulewidth}
  \hat{g}^0_{\lambda\mu\nu} & &\text{no 3-2 or 2-1 arcs.} \\
  \hat{g}^{\text{2-1}}_{\lambda\mu\nu} & &\text{a pure 2-1 arc but no 3-2 arcs.} \\
  \hat{g}^{\text{3-2}}_{\lambda\mu\nu} & &\text{a pure 3-2 arc but no 2-1 arcs.} \\
  \hat{g}^{\text{3-2-1}}_{\lambda\mu\nu} && \text{a 3-2-1 arc.} \\
  \hat{g}^{\text{3-2, 2-1}}_{\lambda\mu\nu}& &\text{a pure 3-2 arc and a pure 2-1 arc.}
\end{array}
\]
We use $\hat{g}_{\lambda\mu\nu}^*$ to refer to any of these five types of invariant-free Kronecker coefficients.
\end{definition}
Thus, letting $\lambda,\mu,\nu = [l_2,l_1], [m_2,m_1], [n_4, n_3,n_2,n_1]$, there holds
\be
\label{e g = invariant-free g}
g_{\lambda\mu\nu} = \sum_{i_1, i_2, i_3} \hat{g}_{\hat{\lambda}\hat{\mu}\hat{\nu}}  =
\sum_{i_1, i_2, i_3} \hat{g}_{\hat{\lambda}\hat{\mu}\hat{\nu}}^0 + \hat{g}_{\hat{\lambda}\hat{\mu}\hat{\nu}}^{\text{2-1}}+ \hat{g}_{\hat{\lambda}\hat{\mu}\hat{\nu}}^{\text{3-2}}+\hat{g}_{\hat{\lambda}\hat{\mu}\hat{\nu}}^{\text{3-2-1}} + \hat{g}_{\hat{\lambda}\hat{\mu}\hat{\nu}}^{\text{3-2, 2-1}},
\ee
where
\be \label{e hat lambda definition}
\begin{array}{ccl}
\hat{\lambda} &=& [l_2 - i_1 - 2i_2 - 3i_3 - 2n_4, l_1], \\
\hat{\mu} &=& [m_2 - i_1 - 2i_2 - 3i_3 - 2n_4, m_1], \\
\hat{\nu} &=& [0, n_3 - 2i_3, n_2 - 2i_2, n_1 - 2i_1],
\end{array}
\ee
and the sum is over all  $i_1, i_2, i_3$ such that $\hat{\lambda}, \hat{\mu}, \hat{\nu}$ are partitions.

Remarkably, the invariant-free Kronecker coefficients are at most  $2$. Moreover, the coefficients $\hat{g}_{\lambda \mu \nu}^*$ are at most  $1$. We now determine exactly when each of these is 0 or 1.

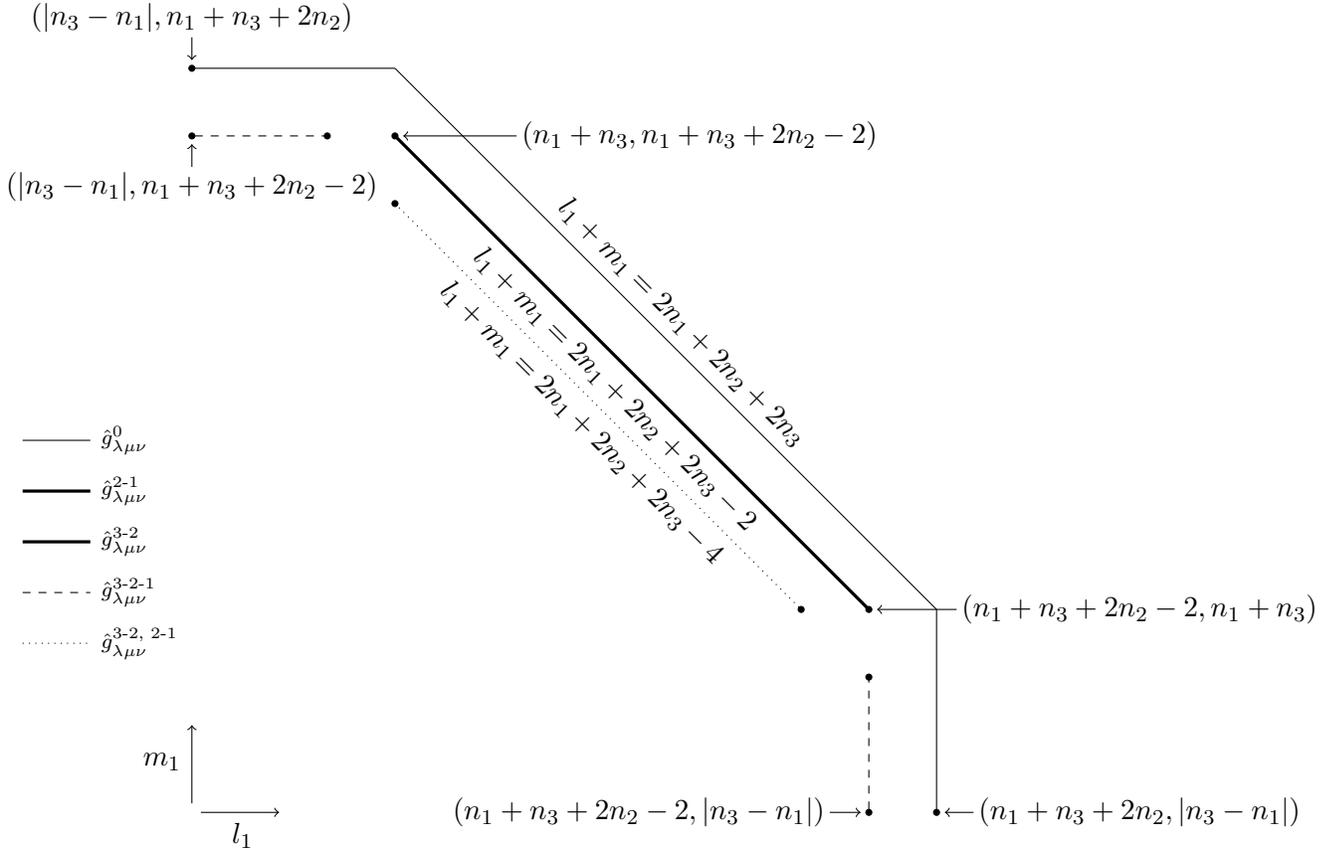
\begin{figure}
\begin{tikzpicture}[scale=.45]
\tikzstyle{vertexShow}=[shape=circle, draw, inner sep=.8pt, outer sep=0pt, fill = black]
\tikzstyle{vertexA}=[shape=coordinate, inner sep=-1pt, outer sep=0pt]
\tikzstyle{vertexHidden}=[shape=circle, inner sep=-1pt, outer sep=2pt]
\tikzstyle{vertexB}=[shape=circle, draw, inner sep=2pt, outer sep=3pt]
\tikzstyle{vertexBound}=[inner sep=0pt, outer sep=2pt, fill=white]
\tikzstyle{edge} = [draw,->,black, shorten >= 2pt]
\tikzstyle{edgeA} = [draw,-,black]
\tikzstyle{edgeT} = [draw,-,black,very thick]
\tikzstyle{edgeD} = [draw,-,black,dotted]
\tikzstyle{edgeW} = [draw,-,black,dashed]
\tikzstyle{LabelStyleH} = [text=black, anchor=north]
\tikzstyle{LabelStyleV} = [text=black, anchor=east]
\tikzstyle{LabelStyleDLeft} = [text=black, anchor=north, sloped]
\tikzstyle{LabelStyleDRight} = [text=black, anchor=south, sloped]

\node[vertexBound] (b264) at (32,4) {{\small $(n_1 + n_3 + 2n_2, |n_3 - n_1|)$}};
\node[vertexBound] (b244) at (17.2,4) {{\small $(n_1 + n_3 + 2n_2 - 2, |n_3 - n_1|)$}};
\node[vertexShow] (264) at (26, 4) {};
\node[vertexA] (266) at (26, 6) {};
\node[vertexA] (268) at (26, 8) {};
\node[vertexA] (2610) at (26, 10) {};
\node[vertexShow] (1024) at (10, 24) {};

\node[vertexA] (2412) at (24, 12) {};
\node[vertexA] (1224) at (12, 24) {};
\node[vertexShow] (2410) at (24, 10) {};
\node[vertexA] (1222) at (12, 22) {};

\node[vertexShow] (244) at (24, 4) {};
\node[vertexShow] (248) at (24, 8) {};
\node[vertexShow] (824) at (8, 24) {};
\node[vertexShow] (424) at (4, 24) {};

\node[vertexShow] (2210) at (22, 10) {};
\node[vertexShow] (1022) at (10, 22) {};

\foreach \x / \y in {24 / 12, 22 / 14, 20 / 16, 18 / 18, 16 / 20, 14 / 22, 12 / 24} {
        \node[vertexA] (\x\y) at (\x, \y) {};
        \node[vertexA] (a\x\y) at (\x, \y -2) {};
}

\node[vertexBound] (b426) at (4,27.5) {{\small $(|n_3 - n_1|, n_1 + n_3 + 2n_2)$}};
\node[vertexBound] (b424) at (4,22.5) {{\small $(|n_3 - n_1|, n_1 + n_3 + 2n_2 - 2)$}};
\node[vertexShow] (426) at (4, 26) {};
\node[vertexA] (626) at (6, 26) {};
\node[vertexA] (826) at (8, 26) {};
\node[vertexA] (1026) at (10, 26) {};

\draw[edge] (b264) to (264);
\draw[edge] (b426) to (426);
\draw[edge] (b244) to (244);
\draw[edge] (b424) to (424);

\draw[edgeD] (2210) to node[LabelStyleDLeft]{{\small $l_1+m_1 = 2n_1+2n_2+2n_3-4$}} (1022);
\draw[edgeW] (244) to (248);
\draw[edgeW] (424) to (824);

\draw[edgeA] (264) to (2610);
\draw[edgeA] (426) to (1026);
\draw[edgeA] (1026) to node[LabelStyleDRight]{{\small $l_1 + m_1 = 2n_1+2n_2+2n_3$}} (2610);
\draw[edgeT] (1024) to node [LabelStyleDLeft]{{\small $l_1 + m_1 = 2n_1+2n_2+2n_3 - 2$}} (2410);

\node[vertexBound] (b1024) at (19,24) {{\small $(n_1+n_3, n_1 + n_3 + 2n_2 -2)$}};
\node[vertexBound] (b2410) at (32,10) {{\small $(n_1 + n_3 + 2n_2 -2, n_1+n_3)$}};

\draw[edge] (b1024) to (1024);
\draw[edge] (b2410) to (2410);

\node[vertexA, label=right:\tiny{$\hat{g}^0_{\lambda\mu\nu}$}] (l1) at (1,15) {};
\node[vertexA, label=right:\tiny{$\hat{g}^{\text{2-1}}_{\lambda\mu\nu}$}] (l4) at (1,13.5) {};
\node[vertexA, label=right:\tiny{$\hat{g}^{\text{3-2}}_{\lambda\mu\nu}$}] (l3) at (1,12) {};
\node[vertexA, label=right:\tiny{$\hat{g}^{\text{3-2-1}}_{\lambda\mu\nu}$}] (l2) at (1,10.5) {};
\node[vertexA, label=right:\tiny{$\hat{g}^{\text{3-2, 2-1}}_{\lambda\mu\nu}$}] (l5) at (1,9) {};

\node[vertexA] (bl1) at (-1,15) {};
\node[vertexA] (bl4) at (-1,13.5) {};
\node[vertexA] (bl3) at (-1,12) {};
\node[vertexA] (bl2) at (-1,10.5) {};
\node[vertexA] (bl5) at (-1,9) {};

\draw[edgeA] (bl1) to (l1);
\draw[edgeW] (bl2) to (l2);
\draw[edgeT] (bl3) to (l3);
\draw[edgeT] (bl4) to (l4);
\draw[edgeD] (bl5) to (l5);

\node[vertexHidden] (legO) at (4,4) {};
\node[vertexHidden] (legN) at (4,7) {};
\node[vertexHidden] (legE) at (7,4) {};

\draw[edge] (legO) to node[LabelStyleV]{{\small $m_1$}} (legN);
\draw[edge] (legO) to node[LabelStyleH]{{\small $l_1$}} (legE);

\end{tikzpicture}
\caption{Polytopes for the five types of invariant-free Kronecker coefficients.}
\label{f invariant-free g polytope}
\end{figure}

\begin{theorem}\label{t consequence for Kronecker}
Maintain the notation above. The two-row Kronecker coefficients are given by \eqref{e g = invariant-free g} and each type of invariant-free Kronecker coefficient is 0 or 1; it is 1 if and only if  $(l_1,m_1)$ lies in the corresponding (one-dimensional) polytope shown in Figure \ref{f invariant-free g polytope}, $l_1 \equiv m_1 \equiv r \mod 2$, and
\be \label{e invariant-free active}
\begin{array}{cllll}
\text{ [no extra condition]} & \text{for type } \hat{g}^0_{\lambda\mu\nu},\\
\text{$n_1 \geq 1$ and  $n_2 \geq 1$} & \text{for type } \hat{g}^{\text{2-1}}_{\lambda\mu\nu}, \\
\text{$n_2 \geq 1$ and  $n_3 \geq 1$} & \text{for type } \hat{g}^{\text{3-2}}_{\lambda\mu\nu},\\
\text{$n_1 \geq 1$, $n_2 \geq 1$, and  $n_3 \geq 1$} & \text{for type } \hat{g}^{\text{3-2-1}}_{\lambda\mu\nu}, \\
\text{$n_1 \geq 1$, $n_2 \geq 2 $, and  $n_3 \geq 1$} & \text{for type } \hat{g}^{\text{3-2, 2-1}}_{\lambda\mu\nu}.
\end{array}
\ee
\end{theorem}
\begin{proof}
Let $\G$ be an invariant-free $(\lambda,\mu)$-cell of $\pNSTC(\nu)$.  The arcs of  $\G$ are given by one of the eight cases in Figure \ref{f kronecker graphical calculus}.  The partitions  $\lambda$ and  $\mu$ are can be expressed in terms of  $\nu$ and the arcs of  $\G$:
\[
\begin{array}{ccl}
\frac{l_1 + m_1}{2} &= & n_3 + n_2 + n_1 - \arc{ext}(\G), \vspace{1pt}\\
\frac{l_1 - m_1}{2} &= & \exfreev(\G) - \arc{3-1 $V$}(\G) - \arc{3-2-1  $V$}(\G) - (\exfreew(\G) - \arc{3-1 $W$}(\G) - \arc{3-2-1  $W$}(\G)) .
\end{array}\]
Then for each case of Figure \ref{f kronecker graphical calculus} we have
\be
\label{e l1 + m1 formulae}
\begin{array}{lrlrl}
\hat{g}^0_{\lambda\mu\nu} \text{\small (middle)}& \frac{l_1 + m_1}{2} = & n_3 + n_2+n_1 & \frac{l_1 - m_1}{2} = & \exfreev(\G) - \exfreew(\G), \\
\hat{g}^0_{\lambda\mu\nu} \text{\small (left)} & m_1 \ \ = & n_3 +2 n_2+n_1 & l_1\ \ = & n_1 + n_3 - 2 \arc{3-1}(\G), \\
\hat{g}^0_{\lambda\mu\nu} \text{\small (right)}& l_1 \ \ = & n_3 +2 n_2+n_1 & m_1 \ \ = & n_1 + n_3 - 2 \arc{3-1}(\G), \\
\hat{g}^{\text{2-1}}_{\lambda\mu\nu} & \frac{l_1 + m_1}{2} = & n_3 +n_2+n_1-1 & \frac{l_1 - m_1}{2} = & \exfreev(\G) - \exfreew(\G), \\
\hat{g}^{\text{3-2}}_{\lambda\mu\nu} & \frac{l_1 + m_1}{2} = & n_3 +n_2+n_1-1 & \frac{l_1 - m_1}{2} = & \exfreev(\G) - \exfreew(\G), \\
\hat{g}^{\text{3-2-1}}_{\lambda\mu\nu} \text{\small (left)}& m_1\ \ = & n_3 + 2 n_2+n_1-2 & l_1 \ \ = &  n_1+n_3-2- 2 \arc{3-1}(\G), \\
\hat{g}^{\text{3-2-1}}_{\lambda\mu\nu} \text{\small (right)}& l_1 \ \ = & n_3 + 2 n_2+n_1-2 & m_1 \ \ = &  n_1+n_3-2- 2 \arc{3-1}(\G), \\
\hat{g}^{\text{3-2, 2-1}}_{\lambda\mu\nu}& \frac{l_1 + m_1}{2} = & n_3 +n_2+n_1-2 & \frac{l_1 - m_1}{2}  = & \exfreev(\G) - \exfreew(\G).
\end{array}
\ee
The theorem follows by simply recording the contribution to  $\hat{g}^*_{\lambda \mu \nu}$ for each of the eight cases.  There is only one free parameter in each case: $\arc{3-1}(\G)$ in the $\hat{g}^{\text{3-2-1}}_{\lambda\mu\nu}$ cases and the left and right cases of $\hat{g}^{0}_{\lambda\mu\nu}$ and $\exfreev(\G) - \exfreew(\G)$ in the other cases.  The eight line segments of Figure \ref{f invariant-free g polytope} are thus obtained from the eight cases.

The endpoints of each line segment are read off easily from \eqref{e l1 + m1 formulae} and the constraints
\[
\begin{array}{ccl}
\arc{3-1}(\G)+\arc{3-2-1}(\G) &\leq &\min(n_3, n_1), \\
|\exfreev(\G) - \exfreew(\G)|& \leq & n_2 - \arc{2-1}(\G) - \arc{3-2}(\G) - \arc{3-2-1}(\G).
\end{array}
\]
The additional conditions in \eqref{e invariant-free active} are clearly necessary and it is not hard to see that they are the only additional conditions needed.
\end{proof}

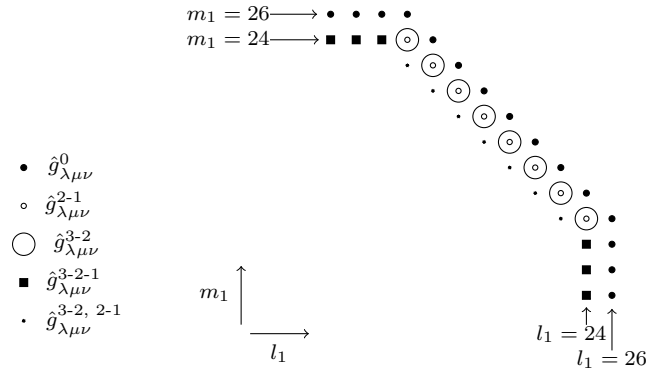
\begin{figure}
\begin{tikzpicture}[scale=.17]
\tikzstyle{vertexA}=[shape=circle, draw, inner sep=.8pt, outer sep=4pt, fill = black]
\tikzstyle{vertexB}=[shape=rectangle, draw, inner sep=1.5pt, outer sep=3pt, fill=black]
\tikzstyle{vertexC}=[shape=circle, draw, inner sep=3pt, outer sep=3pt]
\tikzstyle{vertexD}=[shape=star, star points=7, draw, inner sep=.1pt, outer sep=3pt, fill=black]
\tikzstyle{vertexE}=[shape=circle, draw, inner sep=.7pt, outer sep=3pt]
\tikzstyle{vertexBound}=[inner sep=0pt, outer sep=0pt, fill=white]
\tikzstyle{vertexHidden}=[shape=circle, inner sep=-1pt, outer sep=2pt]
\tikzstyle{edge} = [draw,->,black]
\tikzstyle{LabelStyleH} = [text=black, anchor=north]
\tikzstyle{LabelStyleV} = [text=black, anchor=east]

\node[vertexA, label=right:\tiny{$\hat{g}^0_{\lambda\mu\nu}$}] (l1) at (-20,14) {};
\node[vertexE, label=right:\tiny{$\hat{g}^{\text{2-1}}_{\lambda\mu\nu}$}] (l4) at (-20,11) {};
\node[vertexC, label=right:\tiny{$\hat{g}^{\text{3-2}}_{\lambda\mu\nu}$}] (l3) at (-20,8) {};
\node[vertexB, label=right:\tiny{$\hat{g}^{\text{3-2-1}}_{\lambda\mu\nu}$}] (l2) at (-20,5) {};
\node[vertexD, label=right:\tiny{$\hat{g}^{\text{3-2, 2-1}}_{\lambda\mu\nu}$}] (l5) at (-20,2) {};


\node[vertexA] (264) at (26, 4) {};
\node[vertexA] (266) at (26, 6) {};
\node[vertexA] (268) at (26, 8) {};
\node[vertexA] (2610) at (26, 10) {};
\node[vertexB] (244) at (24, 4) {};
\node[vertexB] (246) at (24, 6) {};
\node[vertexB] (248) at (24, 8) {};

\node[vertexE] (1024) at (10, 24) {};
\node[vertexC] (1024) at (10, 24) {};

\foreach \x / \y in {24 / 12, 22 / 14, 20 / 16, 18 / 18, 16 / 20, 14 / 22, 12 / 24} {
        \node[vertexA] (\x\y) at (\x, \y) {};
        \node[vertexE] (a\x\y) at (\x, \y -2) {};
        \node[vertexC] (c\x\y) at (\x, \y -2) {};
        \node[vertexD] (d\x\y) at (\x-2, \y -2) {};
}

\node[vertexA] (426) at (4, 26) {};
\node[vertexA] (626) at (6, 26) {};
\node[vertexA] (826) at (8, 26) {};
\node[vertexA] (1026) at (10, 26) {};
\node[vertexB] (424) at (4, 24) {};
\node[vertexB] (624) at (6, 24) {};
\node[vertexB] (824) at (8, 24) {};

\node[vertexBound] (m26) at (-4, 26) {{\tiny $m_1 = 26$}};
\node[vertexBound] (m24) at (-4, 24) {{\tiny$m_1 = 24$}};
\node[vertexBound] (l26) at (26, -1) {{\tiny$l_1 = 26$}};
\node[vertexBound] (l24) at (24, 1) {{\tiny$l_1 = 24$ \hphantom{22}}};

\draw[edge] (m26) to (426);
\draw[edge] (m24) to (424);
\draw[edge] (l26) to (264);
\draw[edge] (l24) to (244);

\node[vertexHidden] (legO) at (-3,1) {};
\node[vertexHidden] (legN) at (-3,7) {};
\node[vertexHidden] (legE) at (3,1) {};

\draw[edge] (legO) to node[LabelStyleV]{{\tiny $m_1$}} (legN);
\draw[edge] (legO) to node[LabelStyleH]{{\tiny $l_1$}} (legE);

\end{tikzpicture}
\caption{The contributions to the invariant-free Kronecker coefficient $\hat{g}_{\lambda\mu\nu}$ for $\nu = [0,7,8,3]$. The vertex styles distinguish the five types of invariant-free Kronecker coefficients.}
\label{f invariant-free polytope example}
\end{figure}

\subsection{The symmetric and exterior Kronecker coefficients}
\label{ss The symmetric and exterior Kronecker coefficients}
Maintain the notation  from  \textsection\ref{ss Schur-Weyl duality Uqt S2Hr} and Definition \ref{d crystal components pNSTC}.
We now obtain an elegant formula for symmetric and exterior Kronecker coefficients.
Let $\hat{g}_{\varepsilon \lambda \nu}^*$ be the number of invariant-free $\varepsilon \lambda$-cells of $\pNSTC(\nu)$ containing arcs as specified by  $*$.

\begin{corollary}
\label{c consequence for Kronecker tau}
Maintain the notation above and that of \eqref{e g = invariant-free g} and \eqref{e hat lambda definition}. Then
\be \label{e g = invariant-free g sym ext}
g_{\varepsilon\lambda \nu} = \sum_{i_1, i_2, i_3} \hat{g}_{\varepsilon'\hat{\lambda}\hat{\nu}}^0 + \hat{g}_{\varepsilon'\hat{\lambda}\hat{\nu}}^{\text{2-1}}+ \hat{g}_{\varepsilon'\hat{\lambda}\hat{\nu}}^{\text{3-2}}+\hat{g}_{\varepsilon'\hat{\lambda}\hat{\nu}}^{\text{3-2, 2-1}},
\ee
where $\varepsilon' = (-1)^{n_4}\varepsilon$ (we have identified the symbol  $+$ with  $+1$ and  $-$ with  $-1$).
Moreover, the coefficients $\hat{g}_{\varepsilon\lambda\nu}^*$ are 0 or 1, and they are 1 if and only if the following conditions are met:
\be \label{e constraints g* symmetric exterior}
\begin{array}{cl}
\hat{g}^0_{\varepsilon\lambda\nu}  & \text{$n_2$ is even, $\varepsilon = (-1)^{n_3}$, and} \vspace{-2pt}\\
 & l_1 = n_1 + n_2 + n_3, \vspace{2pt}\\
\hat{g}^{\text{2-1}}_{\varepsilon\lambda\nu} & \text{$n_2$ is odd, $\varepsilon = (-1)^{n_3}$, and} \vspace{-2pt}\\
 & l_1 = n_1 + n_2 + n_3 - 1, \quad n_1 \geq 1, n_2 \geq 1,\vspace{2pt}\\
\hat{g}^{\text{3-2}}_{\varepsilon\lambda\nu} & \text{$n_2$ is odd, $\varepsilon = (-1)^{n_3 + 1}$, and} \vspace{-2pt}\\
 & l_1 = n_1 + n_2 + n_3 - 1, \quad n_2 \geq 1, n_3 \geq 1,\vspace{2pt}\\
\hat{g}^{\text{3-2, 2-1}}_{\varepsilon\lambda\nu}  & \text{$n_2$ is even, $\varepsilon = (-1)^{n_3 + 1}$, and} \vspace{-2pt}\\
 &  l_1 = n_1 + n_2 + n_3 - 2, \quad n_1 \geq 1, n_2 \geq 2, n_3 \geq 1.
\end{array}
\ee
\end{corollary}
\begin{proof}
This follows from the discussion above, Theorem \ref{t consequence for Kronecker}, and Proposition \ref{p NSTC is upper based tau module}.  The parity conditions on $n_2$ come from the condition $l_1 \equiv m_1 \equiv r \mod 2$ from Theorem \ref{t consequence for Kronecker}, the fact that $n_1 + n_3 \equiv r \mod 2$, and the constraint on $l_1$ in each case.
\end{proof}

We can now obtain particularly nice formulae for symmetric and exterior Kronecker coefficients by assembling them into generating functions and explicitly evaluating \eqref{e g = invariant-free g sym ext}.  Define the \emph{symmetric} (resp. \emph{exterior}) \emph{Kronecker generating function}
\be
g_{\varepsilon\nu}(x) := \sum_{\lambda \vdash_2 r} g_{\varepsilon\lambda\nu}x^{l_1}, \quad \varepsilon  = + \text{ (resp. }\varepsilon = -).
\ee

For $k,l \in \ZZ,$  $k \leq l$, define $\llbracket k,l\rrbracket  = x^{l} + x^{l-2} + \dots + x^{k'}$, where $k'$ is $k$ if $k' \equiv l \mod 2$ and $k' + 1$ otherwise. Also set $\llbracket^\varepsilon k,l\rrbracket$ to be $\llbracket k,l\rrbracket $ if $(-1)^{k-l} = \varepsilon$ and $0$ otherwise. Lastly, set $\llbracket l\rrbracket  := \llbracket 0,l\rrbracket $ and $\llbracket ^\varepsilon l\rrbracket  := \llbracket ^\varepsilon 0, l\rrbracket $.
\begin{corollary}\label{c kronecker generating function sym ext}
The symmetric and exterior Kronecker generating functions are given by
\[
g_{\varepsilon\nu}(x) = {\small \begin{cases}
 \llbracket n_1\rrbracket  \llbracket  n_2\rrbracket  \llbracket  n_3\rrbracket  & \text{if } \hphantom{-\ }(-1)^{n_2} = \hphantom{- }(-1)^{n_3 + n_4} \varepsilon = 1, \\
 \llbracket n_1 - 1\rrbracket  \llbracket  n_2-1\rrbracket  \llbracket  n_3\rrbracket x & \text{if } -(-1)^{n_2} = \hphantom{- }(-1)^{n_3 + n_4} \varepsilon = 1, \\
  \llbracket n_1\rrbracket  \llbracket  n_2-1\rrbracket  \llbracket  n_3-1\rrbracket x & \text{if } -(-1)^{n_2} = -(-1)^{n_3 + n_4} \varepsilon = 1, \\
   \llbracket n_1-1\rrbracket  \llbracket  n_2-2\rrbracket  \llbracket  n_3 -1\rrbracket x^2 & \text{if } \hphantom{-\ }(-1)^{n_2} = -(-1)^{n_3 + n_4} \varepsilon = 1.
\end{cases}}
\]
\end{corollary}
\begin{proof}
We first prove that
\be
\label{e symmetric and exterior Kronecker generating function big}
\begin{array}{ccl}
g_{\varepsilon\nu}(x)& = &\llbracket n_1\rrbracket  \llbracket ^+n_2\rrbracket  \llbracket ^{\varepsilon'} n_3\rrbracket  + \llbracket n_1-1\rrbracket  \llbracket ^+n_2-1\rrbracket  \llbracket ^{\varepsilon'} n_3\rrbracket  x \ + \vspace{.1in}\\
&& \llbracket n_1\rrbracket  \llbracket ^+n_2-1\rrbracket  \llbracket ^{\varepsilon'} n_3-1\rrbracket  x + \llbracket n_1-1\rrbracket  \llbracket ^{+} n_2-2\rrbracket  \llbracket ^{\varepsilon'} n_3-1\rrbracket  x^2,
\end{array}
\ee
where $\varepsilon' = (-1)^{n_4}\varepsilon$.
This is straightforward from Corollary \ref{c consequence for Kronecker tau}.  Each term of \eqref{e symmetric and exterior Kronecker generating function big} corresponds to one of \eqref{e g = invariant-free g sym ext}. For example, for the  $\hat{g}_{\varepsilon' \hat{\lambda}\hat{\nu}}^{\text{2-1}}$ contribution we have
\begin{align*}
\sum_{\lambda \vdash_2 r} \ \sum_{i_1, i_2, i_3} \hat{g}_{\varepsilon' \hat{\lambda}\hat{\nu}}^{\text{2-1}} x^{l_1}  &=
\sum_{i_1, i_2, i_3} x^{\hat{n}_1+\hat{n}_2+\hat{n}_3-1}  \sum_{\lambda \vdash_2 r} \hat{g}_{\varepsilon' \hat{\lambda}\hat{\nu}}^{\text{2-1}}  =
x^{-1} \prod_{j=1}^3 \sum_{i_j} x^{\hat{n}_j} \Big( \sum_{\lambda \vdash_2 r} \hat{g}_{\varepsilon' \hat{\lambda}\hat{\nu}}^{\text{2-1}} \Big) \\
&= x^{-1} \llbracket 1,n_1\rrbracket  \llbracket ^+ 1,n_2 \rrbracket  \llbracket ^{\varepsilon'} n_3\rrbracket  = \llbracket n_1-1\rrbracket  \llbracket ^+n_2-1\rrbracket  \llbracket ^{\varepsilon'} n_3\rrbracket  x,
\end{align*}
which accounts for the second term of \eqref{e symmetric and exterior Kronecker generating function big}.  Here we have set $\hat{\lambda} = [\hat{l}_2,\hat{l}_1]$, $\hat{\nu} = [0, \hat{n}_3,\hat{n}_2, \hat{n}_1]$, and the first equality uses $l_1 = \hat{l}_1$ and the second that  $\sum_{\lambda \vdash_2 r} \hat{g}_{\varepsilon' \hat{\lambda}\hat{\nu}}^{\text{2-1}}$ is independent of  $\hat{\nu}$ and is either 0 or 1, hence equal to its cube. (The sum $\sum_{\lambda \vdash_2 r} \hat{g}_{\varepsilon' \hat{\lambda}\hat{\nu}}^{\text{3-2, 2-1}}$ does depend on $\hat{\nu}$, but this can be dealt with by changing the summation bounds on  $ i_2$.)

The result then follows from \eqref{e symmetric and exterior Kronecker generating function big} by noting that for each of the four possibilities for  $(-1)^{n_2}, (-1)^{n_3 + n_4} \varepsilon$, exactly one term on the right-hand side is nonzero.
\end{proof}

\subsection{Comparisons with other formulae}
\label{ss comparisons with the formulae}
The recent paper \cite{BWZ} gives a very nice explicit formula for the Kronecker coefficients $g_{[d,0] \mu \nu}$, where  $r$ is even and $d = r/2$.
Maintain the notation of \eqref{e g = invariant-free g} and \eqref{e hat lambda definition} and define the generating function
\be
g_{[d, 0]  \nu}(x) := \sum_{\mu \vdash_2 r} g_{[d,0] \mu \nu}x^{m_1}.
\ee
Then the main result of \cite{BWZ}, rephrased in our notation, is
\be \label{e BWZ main result}
\begin{array}{ccl}
g_{[d, 0]  \nu}(x^{1/2}) &=& \displaystyle\sum_{k = 0}^d \Big( \sum_{j=0}^k \chi\{n_3 - j, n_2 - (k-j), n_1 - j \text{ are even and nonnegative}\} \  + \vspace{.07in}\\
& & \displaystyle \sum_{j=1}^k \chi\{n_3 - j, n_2 - (k+1-j), n_1 - j \text{ are even and nonnegative}\} \Big) x^k,
\end{array}
\ee
where $\chi\{P\}$ is equal to $1$ if $P$ is true and 0 otherwise.

This result can be reproduced from Theorem \ref{t consequence for Kronecker} as it follows from the theorem that
\be \label{e BWZ my version}
\begin{array}{c}
g_{[d, 0]  \nu}(x^{1/2}) = \displaystyle\sum_{\stackrel{m_1 = 0,}{m_1 \text{ even}}}^r \sum_{i_1, i_2, i_3}  \hat{g}_{\hat{\lambda} \hat{\mu} \hat{\nu}} (x^{1/2})^{m_1}
=  \displaystyle  \sum_{k:= \frac{m_1}{2} = 0}^d \Big(\sum_{i_1, i_2, i_3}  \hat{g}_{\hat{\lambda} \hat{\mu} \hat{\nu}}^0 +  \hat{g}_{\hat{\lambda} \hat{\mu} \hat{\nu}}^{\text{3-2-1}} \Big) x^k.
\end{array}
\ee
Here we have set  $\lambda = [d,0]$ and have used that only the reduced Kronecker coefficients  $\hat{g}_{\hat{\lambda} \hat{\mu} \hat{\nu}}^0$ and
$\hat{g}_{\hat{\lambda} \hat{\mu} \hat{\nu}}^{\text{3-2-1}}$
contribute to $g_{\lambda \mu \nu}$.
The right-hand sides of \eqref{e BWZ main result}  and \eqref{e BWZ my version} are readily seen to be equal by noting that
\be \label{e BWZ condition}
\begin{array}{ccl}
\hat{g}_{\hat{\lambda} \hat{\mu} \hat{\nu}}^0 &=& \chi\{\hat{n}_1 = \hat{n}_3\} \chi\{m_1 = 2\hat{n}_1 + 2\hat{n}_2\}, \\
\hat{g}_{\hat{\lambda} \hat{\mu} \hat{\nu}}^{\text{3-2-1}} &=& \chi\{\hat{n}_1 \geq 1\}\chi\{\hat{n}_2 \geq 1\}\chi\{\hat{n}_1 = \hat{n}_3\} \chi\{m_1 = 2\hat{n}_1 + 2\hat{n}_2 - 2\}
\end{array}
\ee
(where $\hat{\lambda}, \hat{\mu}, \hat{\nu}$ are as in the right-hand side of \eqref{e BWZ my version}) and identifying  $j$ with  $ \hat{n}_1 = \hat{n}_3$ and  $\hat{n}_2$ with $k-j$ (resp.  $k+1-j$) for the top (resp. bottom) line of \eqref{e BWZ main result}.  Interestingly, not only do our formulae coincide in this case, but they also decompose Kronecker coefficients into a sum of smaller nonnegative quantities in a similar way.

Using similar arguments to those for Proposition \ref{c kronecker generating function sym ext} and with the notation of the proposition, \eqref{e BWZ my version} can be converted into the following compact expression.
\begin{proposition}
\label{p Kronecker consequence l1 = 0}
The Kronecker coefficients for $\lambda = [d,0]$ are given by
\[
g_{[d, 0]  \nu}(x^{1/2}) =
\llbracket \min(n_1,n_3)\rrbracket \llbracket n_2\rrbracket + \llbracket \min(n_1,n_3)-1 \rrbracket \llbracket n_2-1\rrbracket x.
\]
\end{proposition}
\begin{proof}
From \eqref{e BWZ my version} and \eqref{e BWZ condition}, we obtain
\[
\begin{array}{ccl}
g_{[d, 0]  \nu}(x^{1/2}) &=&   \displaystyle \sum_{\stackrel{i_1, i_2, i_3,}{\hat{n}_3 = \hat{n}_1}}  \hat{g}_{\hat{\lambda} \hat{\mu} \hat{\nu}}^0 x^{\hat{n}_1+\hat{n}_2} +  \hat{g}_{\hat{\lambda} \hat{\mu} \hat{\nu}}^{\text{3-2-1}} x^{\hat{n}_1+\hat{n}_2-1}  \\
&=& \llbracket \min(n_1,n_3)\rrbracket \llbracket n_2\rrbracket + \llbracket  1, \min(n_1,n_3) \rrbracket \llbracket 1, n_2\rrbracket x^{-1}.
\end{array}
\]
\end{proof}

In \cite{BOR}, the authors give an explicit description of which two-row Kronecker coefficients are zero.
We can use the explicit formulae for Kronecker coefficients established in this section to reproduce this result.  (Note that the conditions in (7) of  \cite{BOR} involve a description of a cone $\Delta$ from \cite{Bravyi}, which is roughly the cone generated by positive Kronecker coefficients; also, there is a mistake in the 2008 arXiv version of \cite{BOR} that will be corrected in a later version.)
\begin{proposition}
Let  $\lambda,\mu,\nu = [l_2,l_1], [m_2,m_1], [n_4, n_3,n_2,n_1]$ as above.
The Kronecker coefficient $g_{\lambda \mu \nu}$ is 0 if and only if at least one of the following conditions is satisfied
{\small
\renewcommand{\minalignsep}{2pt}
\begin{align}
\label{e g = 0 1} & \frac{l_1 + m_1}{2} > n_1 + n_2 + n_3; \\
\label{e g = 0 2} & \frac{|l_1-m_1|}{2} > \min(n_1,n_3)+n_2;  \\
\label{e g = 0 3} & n_1 = n_3 = 0 \text{ and } \frac{l_1 + m_1}{2} \not\equiv  n_2 \mod 2;\\
\label{e g = 0 4} & \min(l_1,m_1) = 0, \ \min(n_1,n_2,n_3) = 0, \text{ and } \frac{\max(l_1, m_1)}{2} \not\equiv n_1+n_2 \mod 2;\\
\label{e g = 0 5} & \{l_1,m_1\} = \{0,2\}, \text{ and } n_1, n_2, n_3 \text{ are even};\\
\label{e g = 0 6} & l_1 = m_1 = 0, \text{ and } \text{$n_1$ or $n_2$ is odd}.
\end{align}
}
\end{proposition}
Note that if  $\min(l_1,m_1) = 0$, then  $l_1$ and $m_1$ are even, and  $n_1$ and  $n_3$ have the same parity.
\begin{proof}
By Theorem \ref{t consequence for Kronecker},  $g_{\lambda \mu \nu} = 0$ if \eqref{e g = 0 1} or \eqref{e g = 0 2} holds.  For the remainder of the proof, assume that \eqref{e g = 0 1} and \eqref{e g = 0 2} do not hold.

Assume in addition that $n_1, n_3$ are not both 0, and  $l_1, m_1$ are not 0.
By reducing to the $\lambda = \mu$ case, we will show that $g_{\lambda \mu \nu} > 0$.  If  $\lambda = \mu$ (equivalently, $l_1 = m_1$), then we can see directly from Corollary \ref{c kronecker generating function sym ext} (keeping in mind the parity condition $r \equiv l_1 \equiv m_1 \equiv n_1 + n_3 \mod 2$) that
$g_{\lambda \mu \nu} = g_{+ \lambda \nu} + g_{- \lambda \nu} > 0$.
Moreover, if  $n_2$ is even, then there is a $(\lambda,\lambda)$-cell  $\G$ of $\pNSTC(\nu)$ such that  $\arc{ext}(\G) = 0$.

Now suppose (without loss of generality) that $l_1 >  m_1$.  Set $\hat{\lambda} = [l_2,m_1]$ and
\[ \hat{\nu} = [\hat{n}_4, \hat{n}_3,\hat{n}_2, \hat{n}_1] =
\begin{cases}
[n_4,n_3,n_2 - \frac{(l_1 - m_1)}{2},n_1] & \text{if  $n_2 \geq \frac{l_1 - m_1}{2}$}, \\
[n_4,n_3 - (\frac{l_1 -m_1}{2} - n_2),0,n_1- (\frac{l_1 -m_1}{2} - n_2)] & \text{if  $n_2 < \frac{l_1 - m_1}{2}$}.
\end{cases}
\]
By the previous paragraph, there is a $(\hat{\lambda},\hat{\mu})$-cell  $\hat{\G}$ of $\pNSTC(\hat{\nu})$ such that if  $ \hat{n}_2 = 0$, then $\arc{ext}(\hat{\G}) = 0$.
Then by Corollary \ref{c diagram kronecker} and Figure \ref{f kronecker graphical calculus}, there is a $(\lambda,\mu)$-cell  $\G$ of $\pNSTC(\nu)$ obtained from $ \hat{\G}$ by setting $\exfreev(\G) = \exfreev(\hat{\G}) + \min(n_2,\frac{l_1 - m_1}{2})$, and, if  $n_2 < \frac{l_1 - m_1}{2}$, $\arc{3-1  $W$}(\G) = \arc{3-1  $W$}(\hat{\G}) + (\frac{l_1 -m_1}{2} - n_2)$, and otherwise keeping the data (B),(C),(E) of Proposition \ref{p diagram data} the same for $\G$ and $\hat{\G}$.

Finally, it is a direct check using Theorem \ref{t consequence for Kronecker} that if  $n_1 = n_3 = 0$, then  $g_{\lambda \mu \nu}$ is 0 if \eqref{e g = 0 3} holds and 1 otherwise.  If  $\min(l_1,m_1)= 0$, then we can read off when  $g_{\lambda \mu \nu} = 0$ from Proposition \ref{p Kronecker consequence l1 = 0}, which accounts for \eqref{e g = 0 4}, \eqref{e g = 0 5}, and \eqref{e g = 0 6}.
\end{proof}

\section{Future work}
\label{s future work}

\subsection{A canonical basis for  $\nsbr{X}^{\tsr r}$}
Recall that  $\field\nsSchur{r}$ is the algebra dual to the coalgebra  $ \O(M_q(\nsbr{X}))_r$ (see Theorem \ref{t nonstandard schur-weyl duality})
and the index set $\nsP_{r,2}$ from \eqref{e definition nsp r2} is
\[
\nsP_{r,2} =  \{ \{\lambda,\mu\}: \lambda, \mu \in \mathscr{P}_{r,2}, \, \lambda \neq \mu\} \sqcup\{+\lambda: \lambda \in \mathscr{P}'_{r,2}\} \sqcup \{-\lambda: \lambda \in \mathscr{P}'_{r,2}\} \sqcup \{\nsbr{\epsilon}_+\}.
\]
Recall from the introduction that we seek a basis  $\nsbr{B}^r$ of $\nsbr{X}^{\tsr r} = \nsbr{\bT}$ (assume here that $\dv = \dw =2$) so that the  $(\field \nsSchur{r},\nsH_r)$-bimodule  with basis $(\nsbr{\bT},\nsbr{B}^r)$ is compatible (in the precise sense of \textsection\ref{ss cells}) with the decomposition
\be \label{e future work ns decomp}
\nsbr{X}^{\tsr r} \cong \bigoplus_{\alpha \in \nsP_{r,2}} \nsbr{\X}_\alpha \tsr \nsbr{M}_\alpha,
\ee
and so that the $(U(\g_X),\QQ\S_r)$-bimodule with basis $(\nsbr{\bT}|_{q=1},\nsbr{B}^r|_{q=1})$ is compatible with the decomposition
\be \label{e future work q=1 decomposition}
\nsbr{X}^{\tsr r}|_{q=1}\cong \bigoplus_{\nu \vdash_\dx r} X_\nu|_{q=1} \tsr M_\nu|_{q=1}.
\ee
We now state a more detailed version of this conjecture.

For any $\alpha \vDash_l^\dx r$, let $J_\alpha \subseteq S$ be as in \textsection\ref{ss type A combinatorics preliminaries}.  Define the \emph{nonstandard $J_\alpha$-descent space} to be  $\nsbr{Y}_\alpha \subseteq \nsbr{X}^{\tsr r}$.  For any  $b \in \nsbr{X}^{\tsr r}$, the \emph{nonstandard descent set} of  $b$ is the maximal  $J$ such that  $b$ belongs to the nonstandard $J$-descent space.

For  $\alpha\in\nsP_{r,2}$, we write $(\lambda,\mu) \in \alpha$  if
\[
\begin{array}{l}
\alpha = \nsbr{\epsilon}_+  \text{ and } \lambda = \mu \vdash_\dv r,   \\
\alpha = \pm \nu  \text{ and }  \nu = \lambda = \mu, \text{ or }\\
\alpha =  \{\lambda,\mu\}.
\end{array}
\]
Define the partial order  $\preceq$ on $\nsP_{r,2}$ by
\begin{align*}
\alpha^1 \preceq \alpha^2 & \text{ if $(\lambda^i, \mu^i) \in \alpha^i$,  $i=1,2$ for some  $\lambda^i, \mu^i$ such that $\lambda^1 \ld \lambda^2$ and  $\mu^1 \ld \mu^2$} \\
& \text{with the exceptions that  $+ \lambda \not \preceq -\lambda$, $- \lambda \not \preceq +\lambda$, and $\nsbr{\epsilon}_+ \not \preceq \alpha$ for all $\alpha \neq \nsbr{\epsilon}_+$.}
\end{align*}
Note that this implies $\alpha \preceq \nsbr{\epsilon}_+$ for all $\alpha \in \nsP_{r,2}$.

The multiplicity $m_{\alpha\nu}$ ($\alpha \in \nsP_{r,2}$, $\nu \vdash_\dx r $) was defined in the introduction to be the multiplicity of the $\S_r$-irreducible  $M_\nu|_{q=1}$ in $\nsbr{M}_\alpha|_{\u =1}$.
By \eqref{e ns Kronecker decomposition}, \eqref{e intro n lambda mu multiplicities}, and Theorem \ref{t intro main theorem advertisement}, this is also equal to the multiplicity of $\nsbr{\X}_\alpha$ in  $\nsbr{X}_\nu$ (the $\nsbr{\X}_\alpha$ are defined before Corollary \ref{c irreducible nsOM comodules}).

\begin{conjecture} \label{cj canonical basis X^r}
There is a basis $\nsbr{B}^r$ of $\nsbr{\bT}$ making  $(\nsbr{\bT}, \nsbr{B}^r)$ into an upper based  $ \Uqt$-module satisfying (i)--(vi) below.
To each $b \in \nsbr{B}^r$ there is associated
\[
\begin{array}{l}
\lambda(b) \vdash_\dv r,\ \mu(b) \vdash_\dw r,\ \nu(b) \vdash_\dx r,\ \alpha(b)\in \nsP_{r,2},\   \\
 P_V(b) \in \text{SSYT}_\dv(\lambda(b)),\  P_W(b) \in \text{SSYT}_\dw(\mu(b)),\ \\
 Q^1(b) \in \text{SYT}(\nu(b)), \ \zeta(b) \in \mathbf{g}_{\alpha(b) \nu(b)},
\end{array}
\]
such that  $(\lambda(b), \mu(b)) \in \alpha(b)$.  Here, $\mathbf{g}_{\alpha \nu}$ is a set of cardinality  $m_{\alpha \nu}$.
Define the following subsets of  $\nsbr{B}^r$:
\[\begin{array}{l@{\ :=\ }l}
\nsbr{\Gamma}_{\alpha, P_V,P_W} & \{b: \alpha(b) = \alpha,\ P_V(b) = P_V,\ P_W(b) = P_W\},\\
\nsbr{\Gamma}^1_{\alpha,P_V,P_W,\nu,\zeta} & \{b: \alpha(b) = \alpha,\ P_V(b) = P_V, \ P_W(b) = P_W,\ \nu(b) = \nu,\ \zeta(b) = \zeta\},\\
\nsbr{\Lambda}_{\alpha,Q,\zeta} & \{b: \alpha(b) = \alpha,\ Q^1(b)=Q,\ \zeta(b) = \zeta\}, \\
\nsbr{\Lambda}^1_Q & \{b: Q^1(b) = Q\}.
\end{array}
\]
\begin{list}{\emph{(\roman{ctr})}} {\usecounter{ctr} \setlength{\itemsep}{1pt} \setlength{\topsep}{2pt}}
\item The nonstandard  $J$-descent spaces of $\nsbr{\bT}$ are spanned by subsets of $\nsbr{B}^r$, i.e. for each $\alpha \vDash_l^\dx r$,  $\nsbr{Y}_\alpha$ is a  $\field \nsSchur{r}$-cellular submodule of $(\nsbr{\bT},\nsbr{B}^r)$.
\item The decomposition of $\nsbr{B}^r$ into  $\nsH_r$-cells is
\[\nsbr{B}^r = \bigsqcup_\stack{\alpha \in \nsP_{r,2}}{ (\sh(P_V), \sh(P_W)) \in \alpha} \nsbr{\Gamma}_{\alpha,P_V,P_W},\]
and $\mathbf{A}\nsbr{\Gamma}_{\alpha,P_V,P_W} \cong \nsbr{M}^\mathbf{A}_\alpha$.
The partial order on cells is refined by $\prec$, i.e.
$\nsbr{\Gamma}_{\alpha,P_V,P_W} \kloneq{\nsbr{B}^r} \nsbr{\Gamma}_{\alpha',P_V',P_W'}$ implies  $\alpha \prec \alpha'$ and  $P_V$ (resp.  $P_W$) has the same content as  $P_V'$ (resp. $P_W$).
\item We say that a $\QQ \S_r$-cell of $(\nsbr{\bT}|_{\u=1},\nsbr{B}^r|_{\u=1})$ is a \emph{quasi-cell}.  The decomposition into quasi-cells is
\[\nsbr{B}^r|_{q=1} = \bigsqcup_\stack{\alpha \in \nsP_{r,2},\ (\sh(P_V), \sh(P_W)) \in \alpha}{\nu\vdash_\dx r,\ \zeta \in \mathbf{g}_{\alpha \nu}} \nsbr{\Gamma}^1_{\alpha,P_V,P_W,\nu,\zeta},\]
and $\QQ \nsbr{\Gamma}^1_{\alpha,P_V,P_W,\nu,\zeta}|_{q=1} \cong \QQ M_{\nu}|_{q=1}$. The partial order on quasi-cells is refined by dominance order in  $\nu$.
\item The decomposition of $\nsbr{B}^r$ into  $\field \nsSchur{r}$-cells is
\[\nsbr{B}^r = \bigsqcup_\stack{\alpha \in \nsP_{r,2},\ \nu \vdash_\dx r}{Q \in \text{SYT}(\nu),\ \zeta \in \mathbf{g}_{\alpha \nu}} \nsbr{\Lambda}_{\alpha,Q,\zeta},\]
and  $\field \nsbr{\Lambda}_{\alpha,Q,\zeta} \cong \nsbr{\X}_\alpha$.
The partial order on  $ \field \nsSchur{r}$-cells is refined by  $\prec$.
\item The set of $\field \nsSchur{r}$-cells $\{\nsbr{\Lambda}_{\alpha,Q,\zeta}\}$ of $(\nsbr{\bT},\nsbr{B}^r)$  can be partitioned into  $\field \nsSchur{r}$-cellular subquotients $\nsbr{\Lambda}^1_Q = \bigsqcup_{\alpha, \zeta} \nsbr{\Lambda}_{\alpha,Q,\zeta}$ called \emph{fat cells}.  The decomposition into fat cells is given by
\[\nsbr{B}^r=\bigsqcup_{\nu \vdash_\dx r,\ Q \in \text{SYT}(\nu)} \nsbr{\Lambda}^1_Q,\]
and $\QQ \nsbr{\Lambda}^1_Q|_{q=1} \cong \Res_{U^\tau} (X_\nu|_{\u=1})$.
Moreover, each $\nsbr{\Lambda}^1_Q$ is a subset of the nonstandard  $R(Q)$-descent space, where
\[R(Q) = \{ s_i : i + 1 \text{ is strictly to the south of $i$ in $Q$}\}\]
(this is the same as the descent set in \eqref{e tableau descent set}).
The partial order on fat cells at  $q=1$ is refined by dominance order in $\sh(Q)$, i.e.
$\nsbr{\Lambda}^1_Q|_{q=1} \kloneq{\nsbr{B}^r|_{q=1}} \nsbr{\Lambda}^1_{Q'}|_{q=1}$ implies  $\sh(Q) \ldneq \sh(Q')$.
\item In the case $Q = \transpose{(Z_{\nu'}^*)}$, the fat cell $(\field \nsbr{\Lambda}^1_{Q},  \nsbr{\Lambda}^1_{Q})$ is equal to
$(\nsbr{X}_{\nu}, \pNSTC(\nu))$ (after perhaps modifying the sign convention \eqref{e sign convention pNSTC} for $\pNSTC(\nu)$).
\end{list}
\end{conjecture}
\setlength{\cellsize}{1.5ex}
Several remarks are now in order.
One of the difficulties in constructing  $\nsbr{B}^r$ is that the integral form $\nsbr{\bT}_{\mathbf{A}} := \mathbf{A}\nsbr{B}^r$ will not be equal to  $X^{\tsr r}_\mathbf{A}$, and the lattice $\nsbr{\L} := \field_\infty \nsbr{B}^r$ will not be equal to  $\L_V \tsrvw_{\field_\infty} \L_W$ (in the notation of  \textsection\ref{s notation for GLV GLW}).
We expect that $\nsbr{\bT}_\mathbf{A}$ and $ \nsbr{\L}$ will be close to the $\mathbf{A}$ and  $\field_\infty$-span of $\text{SNST}((1^r))$, or at least something of the same flavor; see Example \ref{ex ns Schur-Weyl duality canonical basis r4} for what happens in the $r=4$ case.
Note that it follows from the general theory of crystal bases that once  $\nsbr{\bT}_{\mathbf{A}}$ and
$\nsbr{\L}$ are specified, it only suffices to specify the  image of the highest weight elements of $\nsbr{B}^r$ in  $\nsbr{\L}/\ui \nsbr{\L}$.

The assumption that $(\nsbr{\bT}, \nsbr{B}^r)$ is an upper based $\Uqt$-module implies that each $\Uqt$-cell of  $(\nsbr{\bT},\nsbr{B}^r)$ is isomorphic to one of the irreducible upper based $\Uqt$-modules from \eqref{e Uqt canonical bases Z lambda mu}.
It then follows from  \textsection\ref{ss nonstandard two-row case} that the  $\Uqt$-cells and  $\field \nsSchur{r}$-cells of  $(\nsbr{\bT}, \nsbr{B}^r)$ are the same except that $\nsbr{\Lambda}_{\nsbr{\epsilon}_+,Z^*_{(r)},\zeta}$ is a union of  $\lfloor \frac{r}{2} \rfloor +1$  $\Uqt$-cells, where $Z_{(r)}^*$ is the SYT of shape $(r)$.
The reason that $\Uqt$ is mentioned in the conjecture is that we have a theory of based modules for $\Uqt$, but not for  $\field \nsSchur{r}$.  If such a theory is developed for  $\field \nsSchur{r}$ or for the hypothetical nonstandard enveloping algebra, then this conjecture should be strengthened to accommodate it.

The set $\mathbf{g}_{\alpha \nu}$ above could be taken to be $\mathbf{g}^\tau_{\alpha \nu}$ from Corollary \ref{c consequence for Kronecker tau} in the case $\nu \neq (r)$.  We may want to allow this set to depend on the SYT  $Q$.

Regarding (ii), it can be shown that $(\nsbr{\bT}, \nsbr{B}^r)$ an upper based  $ \Uqt$-module implies $\nsbr{\Gamma}_{\alpha,P_V,P_W} \cong \nsbr{\Gamma}_{\alpha,P'_V,P'_W}$ for all $\alpha \in \nsP_{r,2}$ and SSYT $P_V,P_V',P_W,P_W'$ such that $(\sh(P_V), \sh(P_W)) \in \alpha$,  $(\sh(P_V'), \sh(P_W')) \in \alpha$.

Regarding (iv), the statement about partial order follows from  $(\nsbr{\bT}, \nsbr{B}^r)$ being an upper based $ \Uqt$-module, except in the case one of cells is  $\nsbr{\Lambda}_{\nsbr{\epsilon}_+,Z_{(r)}^*,\zeta}$,   However, requirement (i) implies that this cell is a maximal element for the partial order on  $\field \nsSchur{r}$-cells.

Requirement (vi) can be strengthened to say that if  $\alpha \vDash_l^\dx r$ and $\nu'$ is the partition obtained from $\alpha$ by sorting its parts and $Q$ is the unique  SYT$(\nu)$ with  $R(C_Q) = J_\alpha$, then there is a straightforward generalization $\pNSTC^\prime(\alpha)$  of  $\pNSTC(\nu)$, as in Remark \ref{r pNSTC alpha generalization}, such that
$(\field \nsbr{\Lambda}^1_{Q},  \nsbr{\Lambda}^1_{Q})$ is equal to $(\nsbr{Y}_{\alpha}/\nsbr{Y}_{\gdneq \alpha}, \pNSTC^\prime(\alpha))$.

Regarding (v), Example \ref{ex ns Schur-Weyl duality canonical basis r4} shows that we should not demand that all the fat cells of the same shape (the shape being the shape of Q) are isomorphic as  $\field \nsSchur{r}$-modules with basis.  This example also has two quasi-cells of the same shape that are not isomorphic as  $\QQ \S_4$-modules with basis.

Although we have not done many computations outside of the two-row case, we are hopeful that
Conjecture \ref{cj canonical basis X^r} holds for general $d = \dv = \dw$.  For this generalization, $\nsP_{r,2}$ would need to be replaced by an index set $\nsP_{r,d}$ parameterizing irreducible representations of $\field \nsH_{r,d}$.

\begin{example}
\label{ex ns Schur-Weyl duality canonical basis r3}
Recall nonstandard Schur-Weyl duality in the two-row case for  $r=3$:
\[
\nsbr{X}^{\tsr 3}\cong \nssym{3}{X} \tsr \nsbr{\epsilon}_+ \ \oplus \ \nsbr{\X}_{\{(3), (2,1)\}} \tsr \nsbr{M}_{\{(3), (2,1)\}} \ \oplus \ \nsbr{\X}_{+(2,1)} \tsr S' \nsbr{M}_{(2,1)} \ \oplus \ \nswedge{3}{X} \tsr \nsbr{\epsilon}_-.
\]
Define the basis $\nsbr{B}^3$ of  $\nsbr{X}^{\tsr 3}$ to be the union of the following  $\field \nsSchur{r}$-cells.
\begin{longtable}[l]{@{\hspace{3.5cm}}lll}
$\nsbr{\Gamma}_{-(2,1),\tiny \tableau{1 \\ 2 \\3}} $ & $:=$ & $ \text{\pSNST}((3)), $\\
$\nsbr{\Gamma}_{+(2,1),\tiny \tableau{1 & 3 \\ 2}} $ & & $  \text{is defined below},$\\
$\nsbr{\Gamma}_{+(2,1),\tiny \tableau{1 &2\\3}} $ & & $ \text{is defined below}, $\\
$\nsbr{\Gamma}_{\{(3),(2,1)\},\tiny \tableau{1 & 3 \\2}}  $ & $:=$ & $\text{ext-free}\big(\text{\pSNST}((2,1))\big), $\\
$\nsbr{\Gamma}_{\{(3),(2,1)\},\tiny \tableau{1 & 2 \\3}}  $ & $:=$ & $ \text{ext-free}\big(\text{\pSNST}((1,2))\big), $\\
$\nsbr{\Gamma}_{\nsbr{\epsilon}_+,\tiny \tableau{1 & 2 &3}} $ & $:=$ & $ \text{Lex}(\text{\pSNST}((1,1,1))),$
\end{longtable}
where  $\text{Lex}(\pSNST(\nu'))$ denotes the set of straightened $\pSNST$ of shape $\nu'$ and ext-free denotes those  $\pSNST$ with no external arcs.  Here we have suppressed $\zeta$ from the notation  $\nsbr{\Gamma}_{\alpha,Q,\zeta}$ because all the sets  $\mathbf{g}_{\alpha \nu}$ have size 0 or 1.

This basis satisfies the requirements of Conjecture \ref{cj canonical basis X^r}.  There are four fat cells---the two fat cells of shape  $(2,1)$ are the union of two $\field \nsSchur{r}$-cells.
In this case, the quasi-cells are the same as the  $\nsH_3$-cells.
Each  $\nsH_3$-cell of  $\nsbr{X}^{\tsr 3}$ is isomorphic to a right $\nsH_3$-cell of the cellular basis $\nsbr{\Cbasis}^3$ of $\field\nsH_3$ defined in \eqref{e canonical basis definition m odd}.

\begin{center}
\begin{tikzpicture}[xscale = 2, yscale= 1.6]
\tikzstyle{vertex}=[inner sep=0pt, outer sep=3pt]
\tikzstyle{edge} = [draw, thick, ->,black]
\tikzstyle{aedge} = [draw, thick, <->,black]
\tikzstyle{LabelStyleH} = [text=black, anchor=south]
\tikzstyle{LabelStyleV} = [text=black, anchor=east]
\tikzstyle{LabelStyleVw} = [text=black, anchor=west]

\begin{scope}[xshift=0cm]
\node[vertex] (v1) at (1,1){$\frac{1}{2}\left(\myvcenter{\ensuremath{\column{2 \\ 3}\column{1}}} + \myvcenter{\ensuremath{\column{3 \\ 2}\column{1}}}\right)$};
\node[vertex] (v2) at (-1,1){$\frac{1}{2}\left(\myvcenter{\ensuremath{\column{2 \\ 3}\column{3}}} + \myvcenter{\ensuremath{\column{3 \\ 4}\column{1}}}\right)$};
\node[vertex] (v3) at (-1,-1){$\frac{1}{2}\left(\myvcenter{\ensuremath{\column{2 \\ 4}\column{3}}} + \myvcenter{\ensuremath{\column{3 \\ 4}\column{2}}}\right)$};
\node[vertex] (v4) at (1,-1){$\frac{1}{2}\left(\myvcenter{\ensuremath{\column{2 \\ 4}\column{1}}} + \myvcenter{\ensuremath{\column{3 \\ 2}\column{2}}}\right)$};
\draw[edge] (v1) to node[LabelStyleV]{$1$} (v4);
\draw[edge] (v1) to node[LabelStyleH]{$1$} (v2);
\draw[edge] (v2) to node[LabelStyleV]{$1$} (v3);
\draw[edge] (v4) to node[LabelStyleH]{$1$} (v3);
\node[vertex] (label) at (0,-1.6) {$\nsbr{\Gamma}_{+(2,1),\tiny \tableau{1 & 3 \\ 2}}$};
\end{scope}
\begin{scope}[xshift=4cm]
\node[vertex] (v1) at (1,1){$\frac{1}{2}\left(\myvcenter{\ensuremath{\column{3}\column{1 \\ 2}}} + \myvcenter{\ensuremath{\column{2}\column{3 \\ 1}}}\right)$};
\node[vertex] (v2) at (-1,1){$\frac{1}{2}\left(\myvcenter{\ensuremath{\column{3}\column{3 \\ 2}}} + \myvcenter{\ensuremath{\column{4}\column{1 \\ 3}}}\right)$};
\node[vertex] (v3) at (-1,-1){$\frac{1}{2}\left(\myvcenter{\ensuremath{\column{4}\column{3\\2}}} + \myvcenter{\ensuremath{\column{4}\column{2 \\ 3}}}\right)$};
\node[vertex] (v4) at (1,-1){$\frac{1}{2}\left(\myvcenter{\ensuremath{\column{4}\column{1\\2}}} + \myvcenter{\ensuremath{\column{2}\column{2 \\3}}}\right)$};
\draw[edge] (v1) to node[LabelStyleV]{$1$} (v4);
\draw[edge] (v1) to node[LabelStyleH]{$1$} (v2);
\draw[edge] (v2) to node[LabelStyleV]{$1$} (v3);
\draw[edge] (v4) to node[LabelStyleH]{$1$} (v3);
\node[vertex] (label) at (0,-1.6) {$\nsbr{\Gamma}_{+(2,1),\tiny \tableau{1 & 2 \\ 3}}$};
\end{scope}
\end{tikzpicture}
\end{center}
\end{example}

\begin{example}
\label{ex ns Schur-Weyl duality canonical basis r4}
\setlength{\cellsizeCol}{1.5ex}
We describe a basis  $\nsbr{B}^4$ of  $\nsbr{X}^{\tsr 4}$ satisfying the requirements of Conjecture \ref{cj canonical basis X^r}.  As remarked above,
it is enough to specify the integral form  $ \nsbr{\bT}_\mathbf{A}$, the lattice  $ \nsbr{\L}$, and the highest weight elements of $\nsbr{B}^4$.
Define the integral form $\nsbr{\bT}_\mathbf{A}$ and the lattice $\nsbr{\L}$ to be the $\mathbf{A}$ and $\field_\infty$-span of $\text{SNST}'((1^4))$, respectively, where
\[\text{SNST}'((1^4)) := \bigsqcup_{T \in \text{NST}((1^4))}\Big\{({\textstyle -\frac{1}{[2]}})^{\deg'(T)}T, -({\textstyle -\frac{1}{[2]}})^{\deg'(T)}T\Big\}, \]
just as in Definition \ref{ss NSTC} and  $\deg'$ is a slightly different definition of degree:
it agrees with the earlier definition except that $\deg'({ \ctableausmalld{4}{2}{3}{1}})=\deg'({\ctableausmalld{4}{3}{2}{1}}) := 1$ (with the earlier definition, they have degree 0).

The highest weight elements of $ \nsbr{B}^4$ are partitioned into the following $\nsH_4$-cells:
\begin{longtable}[l]{@{\hspace{1.5cm}}l@{\ = \ }l}
$\nsbr{\Gamma}_{-(2,2), {\tiny \tableau{1 & 1 \\ 2 & 2},\tableau{1 & 1 \\ 2 & 2}}} $&$ \left\{-\overbtwo \ctableausmalla{1\\ 2\\3 \\ 4} \right\}, $\\[1.3mm]
$\nsbr{\Gamma}_{+(2,2), {\tiny \tableau{1 & 1 \\ 2 & 2},\tableau{1 & 1 \\ 2 & 2}}} $&$ \left\{-\overbtwo \ctableausmall{3 \\ 4}{1 \\ 2}, -\overbtwo \ctableausmall{2 \\ 4}{1 \\ 3} \right\}, $\\[1.3mm]
$\nsbr{\Gamma}_{-(3,1), \tiny \tableau{1 & 1 & 1\\ 2},\tableau{1 & 1 & 1\\ 2}} $&$ \left\{\ctableausmall{1 \\ 2\\3}{1}, \ctableausmall{1\\3}{1\\2}-\ctableausmall{1\\2}{1\\3}, \ctableausmall{1}{1\\2\\3}\right\}, $\\[1.3mm]
$\nsbr{\Gamma}_{\{(3,1),(2,2)\},\tiny \tableau{1 & 1 & 1\\ 2},\tableau{1 & 1 \\ 2 & 2}} $&$ \text{the 6 elements in Figure \ref{f nonstandardX4 3122}}, $\\[1.3mm]
$\nsbr{\Gamma}_{\{(3,1),(2,2)\},\tiny \tableau{1 & 1 \\ 2 & 2}, \tableau{1 & 1 & 1\\ 2}} $&$ \left\{\tau(b): b \text{ is an element in Figure \ref{f nonstandardX4 3122}}\right\}, $\\[1.7mm]
$\nsbr{\Gamma}_{+(3,1), \tiny \tableau{1 & 1 &1\\ 2},\tiny \tableau{1 & 1 &1\\ 2}}$&$ \text{the 5 elements in Figure \ref{f nonstandardX4 3131}}, $\\[1.7mm]
$\nsbr{\Gamma}_{\{(4),(2,2)\},\tiny \tableau{1 & 1 & 1&1},\tableau{1 & 1 \\ 2 & 2}} $&$ \left\{\ctableausmall{1 \\ 2}{1\\2}, \ctableausmallc{2}{1\\2}{1}\right\}, $\\[1.7mm]
$\nsbr{\Gamma}_{\{(4),(2,2)\},\tiny \tableau{1 & 1 \\ 2 & 2}, \tableau{1 & 1 & 1&1}} $&$ \left\{\ctableausmall{1 \\ 3}{1\\3}, \ctableausmallc{3}{1\\3}{1}\right\}, $\\[1.7mm]
$\nsbr{\Gamma}_{\{(4),(3,1)\},\tiny \tableau{1 & 1 & 1&1},\tableau{1 & 1 &1\\ 2}} $&$ \left\{\ctableausmallc{1\\2}{1}{1},\ctableausmallc{1}{1\\2}{1},\ctableausmallc{1}{1}{1\\2} \right\}, $\\[1.7mm]
$\nsbr{\Gamma}_{\{(4),(3,1)\},\tiny \tableau{1 & 1 &1\\ 2},\tableau{1 & 1 & 1&1}} $&$ \left\{\ctableausmallc{1\\3}{1}{1},\ctableausmallc{1}{1\\3}{1},\ctableausmallc{1}{1}{1\\3} \right\}, $\\[1.7mm]
$\nsbr{\Gamma}_{\nsbr{\epsilon}_+, \tiny \tableau{1 & 1 & 1&1},\tableau{1 & 1 & 1&1}}$&$ \left\{ \ctableausmalld{1}{1}{1}{1}\right\}, $\\[1.7mm]
$\nsbr{\Gamma}_{\nsbr{\epsilon}_+, \tiny \tableau{1 & 1 &1\\ 2},\tableau{1 & 1 &1\\ 2}}$&$ \left\{ -\overbtwo \ctableausmalld{1}{1}{4}{1}\right\}, $\\[1.7mm]
$\nsbr{\Gamma}_{\nsbr{\epsilon}_+, \tiny \tableau{1 & 1 \\ 2 & 2},\tableau{1 & 1 \\ 2 & 2}}$&$ \left\{ \frac{1}{[2]^2}\ctableausmalld{4}{1}{4}{1}  \right\}.$
\end{longtable}

As evidence that this basis is nice, the coefficients that show up for the action of $F_V,F_W,E_V$, and $E_W$ lie in
\[
\frac{1}{2}\ZZ\{[2],[3],[2]^2,[4],a_2\},
\]
and the coefficients for the action of $\sQ_i$ lie in
\[
\frac{1}{2}\ZZ\{[2],[2]^2,a_2,c_0\},
\]
where $a_2=[2]^2-2$ and $c_0=[2]^2-4$.

Figures \ref{f nonstandardX4 3122} and \ref{f nonstandardX4 3131} are examples of $\nsH_4$-cells that are not quasi-cells.
The top row of Figure \ref{f nonstandardX4 3122} is the quasi-cell $\nsbr{\Gamma}^1_{\{(3,1),(2,2)\},Z_{(3,1)},Z_{(2,2)},(2,1,1)}$, which spans a  $\QQ \S_4$-cellular submodule of $\QQ\nsbr{\Gamma}_{\{(3,1),(2,2)\},Z_{(3,1)},Z_{(2,2)}}|_{q=1}$ isomorphic to $M_{(2,1,1)}|_{q=1}$ (we are omitting $\zeta \in \mathbf{g}_{\alpha \nu}$ from the notation because all Kronecker coefficients for  $r=4$ are 1 or 0).
The bottom row is a quasi-cell, which spans a $\QQ \S_4$-cellular quotient of $\QQ\nsbr{\Gamma}_{\{(3,1),(2,2)\},Z_{(3,1)},Z_{(2,2)}}|_{q=1}$ isomorphic to $M_{(3,1)}|_{q=1}$.
The top row of Figure \ref{f nonstandardX4 3131} is the quasi-cell $\nsbr{\Gamma}^1_{+(3,1),Z_{(3,1)},Z_{(3,1)},(2,2)}$, and the bottom row is the quasi-cell $\nsbr{\Gamma}^1_{+(3,1),Z_{(3,1)},Z_{(3,1)},(3,1)}$.

The following fat cells of  $(\nsbr{X}^{\tsr 4}, \nsbr{B}^4)$ are not isomorphic (as  $\field \nsSchur{r}$-modules with basis)
\[ \pNSTC((2,1,1)) = \nsbr{\Lambda}^1_{\tiny \tableau{1 & 4 \\ 2 \\ 3}} \neq \nsbr{\Lambda}^1_{\tiny \tableau{1 & 3 \\ 2 \\ 4}}.  \]
The fat cell $\nsbr{\Lambda}^1_{\tiny \tableau{1 & 4 \\ 2 \\ 3}}$ is the union of the  $\field \nsSchur{r}$-cell  $\nsbr{\Lambda}_{-(3,1),\tiny \tableau{1 & 4 \\ 2 \\ 3}}$ (which has highest weight
$\ctableausmall{1\\2\\3}{1}$) and the  $\field \nsSchur{r}$-cell $\nsbr{\Lambda}_{\{(3,1),(2,2)\},\tiny \tableau{1 & 4 \\ 2 \\ 3}}$ (which has highest weights  $-\ctableausmall{1\\2\\4}{1}, \ctableausmall{1\\3\\4}{1}$);
the fat cell $\nsbr{\Lambda}^1_{\tiny \tableau{1 & 3 \\ 2 \\ 4}}$ is the union of the $\field \nsSchur{r}$-cell $\nsbr{\Lambda}_{-(3,1),\tiny \tableau{1 & 3 \\ 2 \\ 4}}$ (which has highest weight $\ctableausmall{1\\3}{1\\2}-\ctableausmall{1\\2}{1\\3}$) and the $\field \nsSchur{r}$-cell $\nsbr{\Lambda}_{\{(3,1),(2,2)\},\tiny \tableau{1 & 3 \\ 2 \\ 4}}$ (which has highest weights $\ctableausmall{3\\2}{1\\2}, \ctableausmall{2\\3}{1\\3}$).

\setlength{\cellsize}{2.5ex}
\setlength{\cellsizeCol}{2.1ex}
\begin{figure}[h]
\begin{tikzpicture}[xscale = 5, yscale= 3.1]
\tikzstyle{vertex}=[inner sep=0pt, outer sep=3pt]
\tikzstyle{edge} = [draw, thick, ->,black]
\tikzstyle{aedge} = [draw, thick, <->,black]
\tikzstyle{LabelStyleH} = [text=black, anchor=south, near start]
\tikzstyle{LabelStyleV} = [text=black, anchor=east, near start]
\tikzstyle{LabelStyleVw} = [text=black, anchor=west, near start]

\node[vertex] (v12) at (-1,1){$  {-\myvcenter{\ensuremath{\column{1 \\ 2 \\ 4}\column{1}}}}\atop{\vspace{5mm}\{s_1,s_2\}}$};
\node[vertex] (v13) at (0,1){$  {\myvcenter{\ensuremath{\column{3 \\ 2}\column{1 \\ 2}}}}\atop{\vspace{5mm}\{s_1,s_3\}}$};
\node[vertex] (v23) at (1,1){$  {\myvcenter{\ensuremath{\column{2}\column{1 \\ 2 \\ 3}}}}\atop{\vspace{5mm}\{s_2,s_3\}}$};
\node[vertex] (v3)  at (-1,-1){$  {-\frac{1}{[2]}\myvcenter{\ensuremath{\column{4}\column{1}\column{1 \\ 2}}}}\atop{\vspace{5mm}\{s_3\}}$};
\node[vertex] (v2)  at (0,-1){$  {\myvcenter{\ensuremath{\column{2}\column{2 \\ 3}\column{1}}}}\atop{\vspace{5mm}\{s_2\}}$};
\node[vertex] (v1)  at (1,-1){$  {-\frac{1}{[2]}\myvcenter{\ensuremath{\column{1 \\ 2}\column{4}\column{1}}}}\atop{\vspace{5mm}\{s_1\}}$};

\draw[edge, bend left=10] (v12) to node[auto]{\tiny ${s_3,[2]^2}$} (v13);
\draw[edge, bend left=10] (v13) to node[auto]{\tiny ${s_2,1}$} (v12);
\draw[edge, bend left=10] (v13) to node[auto]{\tiny ${s_2,1}$} (v23);
\draw[edge, bend left=10] (v23) to node[auto]{\tiny ${s_1,[2]^2}$} (v13);
\draw[edge] (v3) to node[auto]{\tiny ${s_2,-1}$} (v12);
\draw[edge, bend left=5] (v2) to node[auto]{\tiny ${s_1,s_3,a_2}$} (v13);
\draw[edge, bend left=5] (v13) to node[auto]{\tiny ${s_2,c_0}$} (v2);
\draw[edge] (v1) to node[auto]{\tiny ${s_2,-1}$} (v23);
\draw[edge] (v2) to node[auto]{\tiny ${s_1,1}$} (v12);
\draw[edge] (v2) to node[below right]{\tiny ${s_3,1}$} (v23);
\draw[edge, bend left=10] (v3) to node[auto]{\tiny ${s_2,2}$} (v2);
\draw[edge, bend left=10] (v2) to node[auto]{\tiny ${s_3,a_2}$} (v3);
\draw[edge, bend left=10] (v2) to node[auto]{\tiny ${s_1,a_2}$} (v1);
\draw[edge, bend left=10] (v1) to node[auto]{\tiny ${s_2,2}$} (v2);
\end{tikzpicture}
\caption{
The $\nsH_4$-cell $\nsbr{\Gamma}_{\{(3,1),(2,2)\},Z_{(3,1)},Z_{(2,2)}}$ of $\nsbr{B}^4$.
Below each basis element is its nonstandard descent set; if $s_i$ is in the nonstandard descent set of $b$, then $b\sQ_i=[2]^2 b$. The action of $\sQ_i$ on the basis elements without a nonstandard descent at $i$ is given by the edges and their labels, where $a_2=[2]^2-2,\ c_0=[2]^2-4$.
}
\label{f nonstandardX4 3122}
\end{figure}
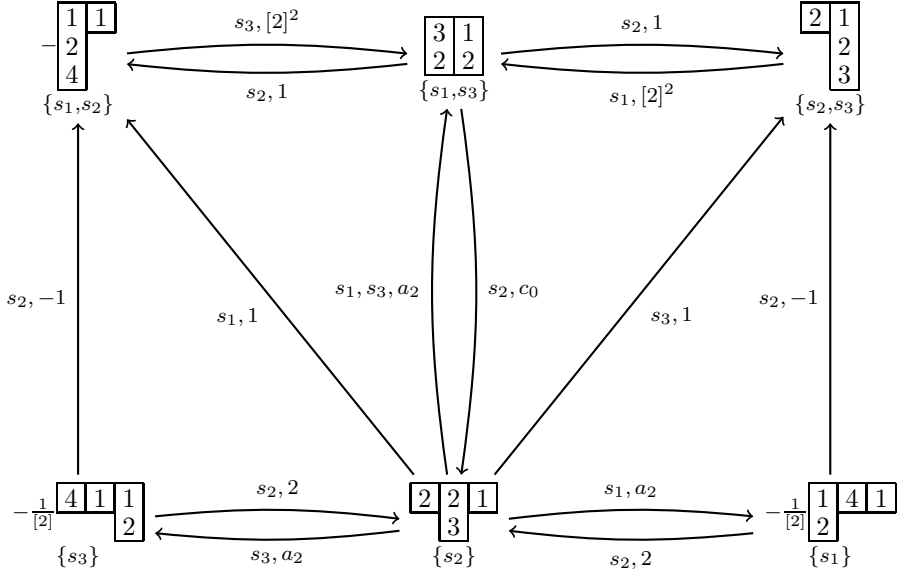

\begin{figure}[h]
\begin{tikzpicture}[xscale = 5, yscale= 3.1]
\tikzstyle{vertex}=[inner sep=0pt, outer sep=3pt]
\tikzstyle{edge} = [draw, thick, ->,black]
\tikzstyle{aedge} = [draw, thick, <->,black]
\tikzstyle{LabelStyleH} = [text=black, anchor=south]
\tikzstyle{LabelStyleV} = [text=black, anchor=east]
\tikzstyle{LabelStyleVw} = [text=black, anchor=west]

\node[vertex] (v2) at (-.75,1){$  {\frac{(1+\tau)}{2}\big(\myvcenter{\ensuremath{\column{1}\column{2 \\3}\column{1}}} \ +\ \myvcenter{\ensuremath{\column{3}\column{1\\2}\column{1}}}\big)} \atop{\vspace{5mm}\{s_2\}}$};
\node[vertex] (v13) at (.75,1){$  {\frac{(1+\tau)}{2}\myvcenter{\ensuremath{\column{1 \\ 3}\column{1 \\ 2}}}}\atop{\vspace{5mm}\{s_1,s_3\}}$};
\node[vertex] (v1)  at (-1,-1){$  {\frac{(1+\tau)}{2}\myvcenter{\ensuremath{\column{2\\3}\column{1}\column{1}}}}\atop{\vspace{5mm}\{s_1\}}$};
\node[vertex] (v22)  at (0,-1){$  {\frac{(1+\tau)}{2}\myvcenter{\ensuremath{\column{3}\column{1 \\ 2}\column{1}}}}\atop{\vspace{5mm}\{s_2\}}$};
\node[vertex] (v3)  at (1,-1){$  {\frac{(1+\tau)}{2}\myvcenter{\ensuremath{\column{1}\column{3}\column{1\\2}}}}\atop{\vspace{5mm}\{s_3\}}$};

\draw[edge, bend left=10] (v2) to node[above]{\tiny $s_1,s_3,-[2]$} (v13);
\draw[edge, bend left=10] (v13) to node[below]{\tiny $ s_2,-[2] $} (v2);
\draw[edge] (v2) to node[left]{\tiny $ s_1,c_0 $} (v1);
\draw[edge, bend left=10] (v1) to node[auto]{\tiny $ s_2,a_2 $} (v22);
\draw[edge, bend left=10] (v22) to node[auto]{\tiny $ s_1,a_2 $} (v1);
\draw[edge] (v1) to node[auto, near start]{\tiny $ s_3,-[2] $} (v13);

\draw[edge] (v22) to node[auto, very near start]{\tiny $ s_3,-[2] $} (v13);
\draw[edge, bend left=10] (v22) to node[auto]{\tiny $ s_3,-2 $} (v3);
\draw[edge, bend left=10] (v3) to node[auto]{\tiny $ s_2,-a_2 $} (v22);
\draw[edge, bend left=6] (v2) to node[auto, near end]{\tiny $ s_3,c_0 $} (v3);
\draw[edge, bend left=6] (v3) to node[auto, near end]{\tiny $ s_2,a_2 $} (v2);
\draw[edge] (v3) to node[right]{\tiny $ s_1,-[2] $} (v13);
\end{tikzpicture}
\caption{The $\nsH_4$-cell
$\nsbr{\Gamma}_{+(3,1),Z_{(3,1)},Z_{(3,1)}}$ of $\nsbr{B}^4$.
Conventions are the same as for Figure \ref{f nonstandardX4 3122}.}
\label{f nonstandardX4 3131}
\end{figure}
\end{example}

We have also been able to construct some $\nsH_r$-cells of a basis satisfying the requirements of Conjecture \ref{cj canonical basis X^r} for  $r = 5$ and  $r=6$.  In addition, we can construct some quasi-cells for  $r=7$ and  $r=8$.

We are also hopeful that the nonstandard Temperley-Lieb quotient $R\nsH_{r,d}$ of  $R\nsH_r$ (see  \textsection\ref{ss nonstandard two-row case}) for suitable  $R$ has a canonical basis $\nsbr{\Cbasis}^{r,d}$ that is a cellular basis in the sense of \cite{GrahamLehrer} and is compatible with nonstandard descent spaces, as described in  \textsection\ref{scanonical}.
We would also like this basis to have right cells isomorphic to the  $\nsH_r$-cells of  $\nsbr{X}^{\tsr r}$ from Conjecture \ref{cj canonical basis X^r}.
We have only been able to construct such a basis for $r=3$ (see \eqref{e canonical basis definition m odd}), and the next easiest case $r=4,\ d=2$ seems to be quite difficult.
We would hope to obtain $\nsbr{\Cbasis}^{r,d}$ by a globalization procedure similar to that used for the Hecke algebra in \cite{KL}, or quantum group representations in  \cite{Kas1,Kas2}, but this may require having a presentation and monomial basis for $\nsH_{r,d}$, which we know to be difficult from \textsection\ref{sb4}.
We would also like a relatively simple description of the lattice  $\field_\infty \nsbr{\Cbasis}^{r,d}$, but even in the $r=3$ case, the only description we have is as the $\field_ \infty$-span of the canonical basis $ \nsbr{\Cbasis}^3$, rather than as the $\field_\infty$-span of certain monomials in the $\sQ_i$, say.

\subsection{Defining $\nsbr{X}_\nu$ outside the two-row case}
\label{ss Defining X nu outside the two-row case}
Let $\dv,\dw$ be arbitrary and let $\nu \vdash_\dx r$, $\ell(\nu') = 2$. It is possible to define a version of $\nsbr{X}_{\nu}$ in this case---this corresponds to the special case of the Kronecker problem in which one of the partitions has two columns and the other two are arbitrary.
As before, for $\alpha \vDash_l^\dx r$, set
\[
\nsbr{Y}_\alpha = \nswedge{{\alpha_1}}{X} \tsr \nswedge{{\alpha_2}}{X} \tsr \dots \tsr \nswedge{{\alpha_l}}{X}.
\]
Following the construction of Schur modules from \cite{ABWeyman,Weyman} mentioned in the introduction, one can define
injective $\O(GL_q(\nsbr{X}))$-comodule homomorphisms $\iota^{L}_{\nu'} : \nsbr{Y}_{\nu'_1+1, \nu'_2-1} \hookrightarrow \nsbr{Y}_{\nu'_1, \nu'_2}$, $\iota^R_{\nu'} : \nsbr{Y}_{\nu'_2-1, \nu'_1+1} \hookrightarrow \nsbr{Y}_{\nu'_1, \nu'_2}$. Then define
\[
\begin{array}{llll}
\nsbr{Y}^L_{\gdneq \nu'} &= \im(\iota^L_{\nu'}),\ & \nsbr{X}^L_{\nu} &= \nsbr{Y}_{\nu'}/ \nsbr{Y}^L_{\gdneq \nu'},\\[1.4mm]
\nsbr{Y}^R_{\gdneq \nu'} &= \im(\iota^R_{\nu'}),\ & \nsbr{X}^R_{\nu} &= \nsbr{Y}_{\nu'}/ \nsbr{Y}^R_{\gdneq \nu'}.
\end{array}
\]
This yields two $\Uqvw$-modules $\nsbr{X}^L_{\nu},\nsbr{X}^R_{\nu}$ that specialize to $\Res_{U(\g_V\oplus\g_W)} X_{\nu}|_{q=1}$ at $q=1$. The modules $\nsbr{Y}_{\nu'}, \nsbr{Y}^L_{\gdneq \nu'},\nsbr{Y}^R_{\gdneq \nu'}$ all come with canonical bases, but we do not yet understand the maps $\iota^{L}_{\nu'}$ and $\iota^{R}_{\nu'}$ combinatorially. This is not an easy task, but it would yield a combinatorial formula for Kronecker coefficients in the case that one of the partitions has two columns and the other two are arbitrary (actually, it would yield two different formulae, one from $\nsbr{X}^L_{\nu}$ and one from $\nsbr{X}^R_{\nu}$).

Extending this approach to  $\ell(\nu') > 2$  meets with serious difficulties.
The $\field$-vector subspaces $\nsbr{Y}^L_{\gdneq \nu'}$ and $\nsbr{Y}^R_{\gdneq \nu'}$ of $\nsbr{Y}_{\nu'}$ are not in general equal, even though they both have integral forms that specialize to the same thing at $q=1$. As a consequence, if we define $\nsbr{X}_\nu$ for $\ell(\nu')>2$ as we have before (as in \eqref{e gdneq alpha def}), using either $\nsbr{Y}^L_{\gdneq \nu'}$ or $\nsbr{Y}^R_{\gdneq \nu'}$ to define $\nsbr{Y}_{\gdneq^i\nu'}$, then the resulting $\nsbr{X}_\nu$ is in general too small, i.e. its $\field$-dimension is less than the $\QQ$-dimension of $X_\nu|_{q=1}$. We hope that understanding the quasi-cells of Conjecture \ref{cj canonical basis X^r} will help us better understand this difficulty and
suggest a way around it.

\section*{Acknowledgments}
The authors thank K.V. Subrahmanyam for instructive comments on an early version of this manuscript and fruitful discussions.
J. Blasiak is extremely grateful to John Stembridge for his generous advice and many detailed discussions.  He also thanks Anton Geraschenko for help clarifying several technical details,  Thomas Lam and Matt Satriano for helpful conversations, and Michael Bennett and Garrett Lyon for help typing and typesetting figures.

\appendix
\section{Reduction system for $\O(M_q(\nsbr{X}))$}
\label{s reduction system}
In this section we reformulate the relations \eqref{eqdefnewquan} for
 $\O(M_q(\nsbr{X}))$  in the form of a reduction system, as mentioned in
\textsection\ref{sdiamondvia}, and show that this does not satisfy the diamond property.

To define the reduction system, define the following total order $\succeq $
on the variables  $\zz{a}{a'}\in\nsbr{Z},\ a',a\in[\dx]$: $\zz{a}{a'}\succeq\zz{b}{b'}$ if $a>b$ or ($a=b$ and $a'\geq b'$), i.e. $(a,a')\geq(b,b')$ lexicographically. We say that a monomial $\zz{a_1}{a_1'}\cdots\zz{a_r}{a_r'}\in\O(M_q(\nsbr{X}))_r$ is {\em descending} if $\zz{a_1}{a_1'}\succeq\cdots\succeq\zz{a_r}{a_r'}$ and {\em nondescending} otherwise.
We shall now see that the set of degree $2$ descending monomials
\be
 B^{\nsbr{Z}} := \{\zz{a}{a'} \zz{b}{b'} : (a,a') \succeq (b,b')\} \subseteq \O(M_q(\nsbr{X}))_2,
\ee
is a basis of $\O(M_q(\nsbr{X}))_2$.

Set  $\Ap = \mathbf{A}[\frac{1}{[2]}]$.  To define specializations at $q =1$, define the following integral forms:
\be \label{e integral form nsOM stuff}
\begin{array}{ccl}
\ssym{2}{V}_\Ap & :=& \Ap \tilde{B}^V_+, \\[1.3mm]
\swedge{2}{V}_\Ap & :=& \Ap \tilde{B}^V_-,\\[1.3mm]
V^{\tsr r}_\Ap & := & \Ap \{\bv_\mathbf{k} : \mathbf{k} \in [\dv]^r\}, \\[1.3mm]
\nssym{2}{X}_\Ap&:=& \ssym{2}{V}_\Ap \tsrvw \ssym{2}{W}_\Ap \oplus \swedge{2}{V}_\Ap\tsrvw \swedge{2}{W}_\Ap, \\[1.3mm]
\nswedge{2}{X}_\Ap&:=& \ssym{2}{V}_\Ap \tsrvw \swedge{2}{W}_\Ap \oplus \swedge{2}{V}_\Ap\tsrvw \ssym{2}{W}_\Ap, \\[1.3mm]
(\nsbr{\mathcal{I}}_2)_\Ap & := & \nssym{2}{X}_\Ap \tsrdual \nswedge{2}{X}^*_\Ap  \oplus \nswedge{2}{X}_\Ap \tsrdual \nssym{2}{X}^*_\Ap, \\[1.3mm]
(\nsbr{\mathcal{I}}_2^\perp)_\Ap & := & \nssym{2}{X}_\Ap \tsrdual \nssym{2}{X}^*_\Ap  \oplus \nswedge{2}{X}_\Ap \tsrdual \nswedge{2}{X}^*_\Ap, \\[1.3mm]
\nsbr{Z}^{\tsr r}_\Ap & := & \Ap \{\zz{a_1}{a_1'} \tsr \cdots \tsr \zz{a_r}{a_r'}  : \mathbf{a}, \mathbf{a'} \in [\dx]^r\},
\end{array}
\ee
where $\tilde{B}^V_+$, $\tilde{B}^V_-$ are as in  \textsection\ref{ss definitions ns symmetric exterior}, and
$\ssym{2}{W}_\Ap$, $\swedge{2}{W}_\Ap$ are defined similarly to $\ssym{2}{V}_\Ap, \swedge{2}{V}_\Ap$ and
$\nssym{2}{X}^*_\Ap, \nswedge{2}{X}^*_\Ap$ are defined similarly to $\nssym{2}{X}_\Ap$, $\nswedge{2}{X}_\Ap$.
These are integral forms of the corresponding  $\field$ vector spaces,
meaning that $\field \tsr_\Ap \ssym{2}{V}_\Ap \cong \ssym{2}{V}$, $\field \tsr_\Ap \nssym{2}{X}_\Ap \cong \nssym{2}{X}$, etc.
($\nssym{2}{X}$ is defined in \eqref{e symwedge2X VW}).

One  checks that  $V^{\tsr 2}_\Ap = \ssym{2}{V}_\Ap \oplus \swedge{2}{V}_\Ap$.  It follows that
\[
\nsbr{Z}^{\tsr 2}_\Ap = (\nsbr{\mathcal{I}}_2)_\Ap \oplus (\nsbr{\mathcal{I}}_2^\perp)_\Ap,
\]
and therefore
\be \label{e nsOM integral form definition}
(\O(M_q(\nsbr{X}))_2)_\Ap := \nsbr{Z}^{\tsr 2}_\Ap/(\nsbr{\mathcal{I}}_2)_\Ap\cong (\nsbr{\mathcal{I}}_2^\perp)_\Ap
\ee
is an integral form of $\O(M_q(\nsbr{X}))_2$.
Additionally, one checks from the explicit formulae for  $\tilde{B}^V_+$ and  $\tilde{B}^V_-$ in \eqref{eqA11} that $\ssym{2}{V}|_{q=1} = \QQ \{\bv_{ij} + \bv_{ji}: 1 \leq i \leq j \leq \dv \}$ and  $\swedge{2}{V}|_{q=1} = \QQ \{\bv_{ij} - \bv_{ji}: 1 \leq i < j \leq \dv\}$ (the formulae in \eqref{eqA11} are only for the  $\dv = \dw = 2$ case, but the general case is just as easy).  Hence,
\begin{align}
\nsbr{\mathcal{I}}_2^\perp|_{q=1} &=  \QQ \{\zz{a}{a'}\zz{b}{b'} + \zz{b}{b'}\zz{a}{a'}:(a,a') \succeq (b,b')\},  \label{e Z2 integral form1}\\
\nsbr{\mathcal{I}}_2|_{q=1} &=  \QQ \{\zz{a}{a'}\zz{b}{b'} - \zz{b}{b'}\zz{a}{a'}:(a,a') \succ (b,b')\}.  \label{e Z2 integral form2}
\end{align}

Now since the degree 2 descending monomials lie in  $\nsbr{Z}^{\tsr 2}_\Ap$, their images in  $ \O(M_q(\nsbr{X}))_2$, i.e.  $B^\nsbr{Z}$, lie in  $(\O(M_q(\nsbr{X}))_2)_\Ap$.
Set $N_{\Ap} := \Ap B^{\nsbr{Z}} \subseteq (\O(M_q(\nsbr{X}))_2)_\Ap$.
 By \eqref{e Z2 integral form1} and \eqref{e Z2 integral form2}, $B^{\nsbr{Z}}|_{q=1}$ is a basis of $\O(M_q(\nsbr{X}))_2|_{q=1}$.  Since any relation satisfied by $B^{\nsbr{Z}}|_{q=1} \subseteq N|_{q=1}$ would yield one of $B^{\nsbr{Z}}|_{q=1} \subseteq \O(M_q(\nsbr{X}))_2|_{q=1}$, $B^{\nsbr{Z}}|_{q=1}$  is also a basis of $N|_{q=1}$. Since a torsion-free $\QQ \tsr_\ZZ \Ap$-module is free,  $\QQ \tsr_\ZZ N_{\Ap}$ is a free $\QQ \tsr_\ZZ \Ap$-module.   Thus $|B^{\nsbr{Z}}| = \dim_\QQ N|_{q=1} = \rank_\QQA \QQ \tsr_\ZZ N_{\Ap}$.   It follows that $\QQ \tsr_\ZZ N_{\Ap}$ is a free  $\QQA$-module with  $\QQA$-basis  $B^{\nsbr{Z}}$.
We also have
\[ |B^\nsbr{Z}| = \binom{\dx^2+1}{2} = \rank_\Ap (\O(M_q(\nsbr{X}))_2)_\Ap = \dim_\field \O(M_q(\nsbr{X}))_2,\]
where the  second equality follows from \eqref{e integral form nsOM stuff}.
Hence  $N_{\Ap}$ is also an integral form of $ \O(M_q(\nsbr{X}))_2$ and, in particular,
the set of degree 2 descending monomials $B^{\nsbr{Z}}$ is a basis of  $\O(M_q(\nsbr{X}))_2$.

This yields a reduction system for $\nsbr{Z}^{\tsr 2}/ \nsbr{\mathcal{I}}_2$
wherein we have
\[ \zz{a}{a'} \zz{b}{b'} = \sum_{i} \alpha_i \zz{a_i}{a_i'} \zz{b_i}{b_i'}, \]
where all $\zz{a_i}{a_i'} \zz{b_i}{b_i'}$ are descending.

By this reduction rule, there is a method of expanding any monomial
$\zz{a}{a'} \zz{b}{b'} \zz{c}{c'}$ as
\[ \zz{a}{a'} \zz{b}{b'} \zz{c}{c'} =\sum_i \alpha_i \zz{a_i}{a_i'} \zz{b_i}{b_i'} \zz{c_i}{c_i'} \]
wherein $(a_i, a_i') \succeq (b_i,b_i') \succeq (c_i,c_i')$. Thus every nondescending monomial
may be expanded into a linear combination of descending monomials.
Unfortunately, this reduction system does not obey the {\em diamond
lemma}. In other words, there exist monomials $\zz{a}{a'} \zz{b}{b'} \zz{c}{c'} $
wherein two different simplifications using the above reduction rules
yield two different expansions into descending monomials.

Consider the monomial
$m=\zz{1}{1} \zz{1}{2} \zz{2}{3}$.
For any monomial $\zz{a}{a'} \zz{b}{b'} \zz{c}{c'} $, if $(a,a')\not \succeq (b,b')$, then we may apply
the reduction system above for the first two monomials and this is denoted
as $(\zz{a}{a'} \zz{b}{b'} )\zz{c}{c'} $. We say that $R_1 $ applies and display the
result. Similar, we say that $R_2 $ applies if $(b,b') \not \succeq (c,c') $
and denote this application by $\zz{a}{a'} (\zz{b}{b'} \zz{c}{c'})$. The monomial
above has two expansions, viz., $R_1 R_2 R_1$ and $R_2 R_1 R_2$,
reading both strings from {\em left to right}.

The first expansion yields
\begin{align*}
 l_{1}&:=(m)R_1 =q \cdot \zz{1}{2}\zz{1}{1}\zz{2}{3},\\
 l_{12}&:=(m)R_1 R_2 =(-1+{q}^{2}) \cdot \zz{1}{2}\zz{2}{1}\zz{1}{3}+q \cdot \zz{1}{2}\zz{2}{3}\zz{1}{1}.
\end{align*}
The expression $l_{12}$ has two monomials, $m_1 $ and $m_2 $.
We have
\[ (m_1 )R_1 =  \zz{2}{1}\zz{1}{2}\zz{1}{3} =
\zz{2}{1} \zz{1}{3} \zz{1}{2}; \]
the last equality follows since $\zz{1}{3}$ and $\zz{1}{2}$ commute, as is easy to show.
\begin{align*}
 (m_2)R_1={\frac {{q}-q^{-1}}{[2]}} \cdot \zz{2}{1}\zz{1}{4}\zz{1}{1}+{\frac {{1}-{q}^{-2}}{[2]}} \cdot \zz{2}{2}\zz{1}{3}\zz{1}{1}+
 {\frac {2}{[2]}} \cdot \zz{2}{3}\zz{1}{2}\zz{1}{1}+{\frac {q^{-1}-{q}}{[2]}} \cdot \zz{2}{4}\zz{1}{1}\zz{1}{1},
\end{align*}
Combining all this, we have $l_{121}=(m)R_1 R_2 R_1 $:
\begin{align*}
 l_{121}&=(-1+{q}^{2}) \cdot \zz{2}{1}\zz{1}{3}\zz{1}{2}+{\frac {{q}^{2}-1}{[2]}} \cdot \zz{2}{1}\zz{1}{4}\zz{1}{1}+{\frac {{q}-{q}^{-1}}{[2]}} \cdot \zz{2}{2}\zz{1}{3}\zz{1}{1}\\&+
 {\frac {2{q}}{[2]}} \cdot \zz{2}{3}\zz{1}{2}\zz{1}{1}+{\frac {1-{q}^{2}}{[2]}} \cdot \zz{2}{4}\zz{1}{1}\zz{1}{1}
\end{align*}
This is a linear combination of descending monomials so no further reductions are needed.

The second expansion is $(m)R_2 R_1 R_2 $.
First, we have
\begin{align*}
l_2 := (m)R_2=\frac{1}{[2]}\left(({q}-q^{-1}) \cdot \zz{1}{1}\zz{2}{1}\zz{1}{4}+{({1}-q^{-2})} \cdot \zz{1}{1}\zz{2}{2}\zz{1}{3}+{2} \cdot \zz{1}{1}\zz{2}{3}\zz{1}{2}+{(q^{-1}-{q})} \cdot \zz{1}{1}\zz{2}{4}\zz{1}{1}\right)
\end{align*}
This has 4 monomials, $m'_1,\dots, m'_4$, and applying $R_1 $ to each monomial yields:
\begin{align*}
 m'_1&=q \cdot \zz{2}{1}\zz{1}{1}\zz{1}{4},\\
 m'_2&=({q}-q^{-1}) \cdot \zz{2}{1}\zz{1}{2}\zz{1}{3}+ \zz{2}{2}\zz{1}{1}\zz{1}{3},\\
 m'_3&=({q}-q^{-1}) \cdot \zz{2}{1}\zz{1}{3}\zz{1}{2}+ \cdot \zz{2}{3}\zz{1}{1}\zz{1}{2},\\
 m'_4&=\frac{1}{[2]}\left({({q}^{2}-1)} \cdot \zz{2}{1}\zz{1}{4}\zz{1}{1}+{({q}-q^{-1})} \cdot \zz{2}{2}\zz{1}{3}\zz{1}{1}
 +{({q}-q^{-1})} \cdot \zz{2}{3}\zz{1}{2}\zz{1}{1}+{2} \cdot \zz{2}{4}\zz{1}{1}\zz{1}{1}\right).
\end{align*}
Whence, $l_{21}=(m)R_2 R_1 $ equals:
\[
\begin{array}{cl@{\ \ +\ \ }l@{\ \ +\ \ }l}
 l_{21}&=\frac{1}{[2]}\big({\left ({q}^{2}-1\right )} \cdot \zz{2}{1}\zz{1}{1}\zz{1}{4}&\frac{[3]-3}{q} \cdot \zz{2}{1}\zz{1}{2}\zz{1}{3}&(2{q}-2q^{-1}) \cdot \zz{2}{1}\zz{1}{3}\zz{1}{2}\\[2mm]
 &+\ \frac{q(3-[3])}{[2]}\cdot \zz{2}{1}\zz{1}{4}\zz{1}{1}&{(1-{q}^{-2})} \cdot \zz{2}{2}\zz{1}{1}\zz{1}{3}&\frac{3-[3]}{[2]} \cdot \zz{2}{2}\zz{1}{3}\zz{1}{1}\\[2mm]
 &+\ {2} \cdot \zz{2}{3}\zz{1}{1}\zz{1}{2}&\frac{3-[3]}{[2]} \cdot \zz{2}{3}\zz{1}{2}\zz{1}{1}&{\left (2{q}^{-1}-2{q}\right )} \cdot \zz{2}{4}\zz{1}{1}\zz{1}{1}\big).
\end{array}
\]
This has 9 monomials, $m''_1,\dots,m''_9$, which on applying $R_2 $ yields:
\begin{align*}
 m''_1&=({q}-q^{-1}) \cdot \zz{2}{1}\zz{1}{3}\zz{1}{2}+ \zz{2}{1}\zz{1}{4}\zz{1}{1},\\
 m''_2&=  \zz{2}{1}\zz{1}{3}\zz{1}{2},\\
 m''_3&=  \zz{2}{1}\zz{1}{3}\zz{1}{2},\\
 m''_4&=  \zz{2}{1}\zz{1}{4}\zz{1}{1},\\
 m''_5&=q \cdot \zz{2}{2}\zz{1}{3}\zz{1}{1},\\
 m''_6&=  \zz{2}{2}\zz{1}{3}\zz{1}{1},\\
 m''_7&=q \cdot \zz{2}{3}\zz{1}{2}\zz{1}{1},\\
 m''_8&=  \zz{2}{3}\zz{1}{2}\zz{1}{1},\\
 m''_9&=  \zz{2}{4}\zz{1}{1}\zz{1}{1}.
\end{align*}
Finally, collating this, we get $l_{212}=(m)R_2 R_1 R_2 $ as follows:
\begin{align*}
 l_{212}&={\frac {q^3 +q -3{q}^{-1}+{q}^{-3}}{[2]}}
\cdot \zz{2}{1}\zz{1}{3}\zz{1}{2}+
{\frac {2{q}-2q^{-1}}{[2]^{2}}} \cdot \zz{2}{1}\zz{1}{4}\zz{1}{1}+ {\frac { 2-2{q}^{-2}}{[2]^{2}}} \cdot \zz{2}{2} \zz{1}{3} \zz{1}{1}\\
&+{\frac {{q}^{2}+4-{q}^{-2}}{[2]^{2}}} \cdot \zz{2}{3}\zz{1}{2}\zz{1}{1} -  {\frac { 2q-2{q}^{-1}}{[2]^{2}}}\cdot \zz{2}{4}\zz{1}{1}\zz{1}{1},
\end{align*}
which is a linear combination of descending monomials. Observe that the expansions $(m)R_1 R_2 R_1$ and $(m)R_2 R_1 R_2 $ do not coincide.

\section{The Hopf algebra $\Oqt$}
\label{s appendix Oqt}
Here we give the details of the construction of $\Oqt$, a quantized coordinate algebra that is Hopf dual to  $\Uqt$.
Let $\cH$ be a bialgebra and $\X$ a left $\cH$-comodule coalgebra via $\beta:\X\to \cH\tsr \X$.  As explained in \cite[\textsection10.2]{KS}, the \emph{left crossed coproduct coalgebra $\X \rtimes \cH$ of $\cH$ and $\X$} is the coalgebra, equal to $\X \tsr \cH$ as a vector space, with comultiplication and counit given by
\begin{align}
\Delta(x\sharp h)&=\sum x_{(1)} \sharp (x_{(2)})^{(-1)} h_{(1)} \tsr (x_{(2)})^{(0)} \sharp h_{(2)}, \label{e coproduct semidirect}\\
\epsilon(x\sharp h)&=\epsilon_\X(x)\epsilon_\cH(h). \notag
\end{align}
Here $\Delta_\X, \Delta_\cH,$ and $\epsilon_\X,\epsilon_\cH$ are the coproducts and counits of  $\X$ and  $\cH$, and we have used the Sweedler notation $\Delta_\cH(h)=\sum h_{(1)}\tsr h_{(2)}$, $\Delta_\X(x)=\sum x_{(1)} \tsr x_{(2)}$, $\beta(x)=\sum x^{(-1)} \tsr x^{(0)}$.  The coalgebra  $\X \rtimes \cH$ is also called the smash coproduct and is denoted $\X \sharp \cH$; we have chosen to use the  $\sharp$ symbol to denote elements of  $\X \rtimes \cH$.  With suitable assumptions, $\X \rtimes \cH$ can be given the structure of a bialgebra. There is a general theorem due to Radford along these lines (see, e.g., \cite[Theorem 10.6.5]{Montgomery}); for our purposes, the following easy result will suffice.

\begin{proposition} \label{p semidirect coproduct}
Maintain the notation above and further assume that  $\cH$ is commutative, $\X$ is a bialgebra, and  $\beta$ makes  $\X$ into a left  $\cH$-comodule algebra.  Then,
giving $\X \rtimes \cH$ the algebra structure of $\X \tsr \cH$ makes it into a bialgebra.
\end{proposition}
\begin{proof}
The assumption that  $\beta$ makes $\X$ into a left $\cH$-comodule algebra means that
\be \label{e Sweedler}
 \sum x^{(-1)} x'^{(-1)} \tsr x^{(0)} x'^{(0)} = \sum (xx')^{(-1)} \tsr (xx')^{(0)}, \ x, x' \in \X.
\ee
We need to check that the coproduct $\Delta$ given in \eqref{e coproduct semidirect} is an algebra homomorphism.
For  $h, h' \in \cH, \ x, x' \in \X$,
\[
\begin{array}{rll}
 \hspace{-.8cm}\Delta(x\sharp h)\Delta(x'\sharp h')
& = &\sum  x_{(1)} x'_{(1)} \sharp (x_{(2)})^{(-1)} h_{(1)} (x'_{(2)})^{(-1)} h'_{(1)} \tsr (x_{(2)})^{(0)} (x'_{(2)})^{(0)} \sharp h_{(2)} h'_{(2)} \\[1.7mm]
&& \text{$\cH$ is commutative} \\[1.7mm]
& = &\sum  x_{(1)} x'_{(1)} \sharp (x_{(2)})^{(-1)} (x'_{(2)})^{(-1)} h_{(1)} h'_{(1)} \tsr (x_{(2)})^{(0)} (x'_{(2)})^{(0)} \sharp h_{(2)} h'_{(2)} \\[1.7mm]
&& \text{$\Delta_\cH$ is an algebra homomorphism} \\[1.7mm]
& = &\sum  x_{(1)} x'_{(1)} \sharp (x_{(2)})^{(-1)} (x'_{(2)})^{(-1)} (hh')_{(1)} \tsr (x_{(2)})^{(0)} (x'_{(2)})^{(0)} \sharp (hh')_{(2)}\\[1.7mm]
&& \text{by \eqref{e Sweedler}} \\[1.7mm]
& = &\sum  x_{(1)} x'_{(1)} \sharp (x_{(2)} x'_{(2)})^{(-1)} (hh')_{(1)} \tsr (x_{(2)} x'_{(2)})^{(0)} \sharp (hh')_{(2)} \\[1.7mm]
&& \text{$\Delta_\X$ is an algebra homomorphism} \\[1.7mm]
& = &\sum  (xx')_{(1)} \sharp ((xx')_{(2)})^{(-1)} (hh')_{(1)} \tsr ((xx')_{(2)})^{(0)} \sharp (hh')_{(2)} \\[1.7mm]
& = &\Delta(hh'\sharp xx').
\end{array}
\]
Additionally, observe that $\X$ a left $\cH$-comodule algebra implies  $\Delta(1 \sharp 1) = 1\sharp 1 \tsr 1 \sharp 1.$
%
\end{proof}

In what follows we let $\tau : \O(GL_q(V))\tsrvw\O(GL_q(W)) \to \O(GL_q(V))\tsrvw\O(GL_q(W))$ denote the involution given by  $\tau(f \tsrvw g) = g \tsrvw f$.
The next proposition follows easily from Proposition \ref{p semidirect coproduct} and the fact that  $\tau$ is a Hopf algebra involution.
\begin{proposition}
\label{p Oqt Hopf algebra technical}
Maintain the notation of  Proposition \ref{p semidirect coproduct}.  Set $\cH = \F(\S_2)$, the Hopf algebra of functions on the 2 element group with values in  $\field$; let  $e, \tau$ be the elements of $\S_2$ ($e$ the identity element) and let  $\dual{e}, \dual{\tau}$ be the dual basis in  $\F(\S_2)$.  Set  $\X = \O(GL_q(V))\tsrvw\O(GL_q(W))$, suppose  $\dv = \dw$, and define  $\beta$ by
\[\beta(x) = \dual{e} \tsr x + \dual{\tau} \tsr \tau(x), \]
for $x \in \X$.
Then  $\cH, \X$, and $\beta$ satisfy the hypotheses of Proposition \ref{p semidirect coproduct}, hence
\[\Oqt := \O(GL_q(V))\tsrvw\O(GL_q(W)) \rtimes \F(\S_2) \]
is a bialgebra with multiplication equal to that of  $\O(GL_q(V))\tsrvw\O(GL_q(W)) \rtimes \F(\S_2)$ and coproduct given by
\begin{align*}
\Delta(x \sharp \dual{e}) &=
\sum x_{(1)} \sharp \dual{e}  \tsr x_{(2)} \sharp \dual{e} +
x_{(1)} \sharp \dual{\tau} \tsr \tau(x_{(2)}) \sharp \dual{\tau}, \\
\Delta(x \sharp \dual{\tau}) &=
\sum x_{(1)} \sharp \dual{e}  \tsr x_{(2)} \sharp \dual{\tau} +
x_{(1)} \sharp \dual{\tau} \tsr  \tau(x_{(2)}) \sharp \dual{e}.
\end{align*}
Here we have used the Sweedler notation $\Delta_\X(x)= \sum x_{(1)} \tsr x_{(2)}$.
Moreover,  $\Oqt$ is a Hopf algebra with antipode $S$ given by
\[S(x \sharp \dual{e}) = S(x) \sharp \dual{e}, \ \  S(x \sharp \dual{\tau}) = \tau(S(x)) \sharp \dual{\tau}.\]
\end{proposition}

%

Since there are nondegenerate Hopf pairings between  $U_q(\g_V)$ and  $\O(GL_q(V))$ \cite[Corollary 54, Chapter 11]{KS} and between
$\field \S_2$ and  $\F(\S_2)$, it is straightforward to check that
\begin{corollary} \label{c Oqt Uqt pairing}
There is a nondegenerate pairing of Hopf algebras $\langle \cdot, \cdot \rangle:  \Uqt \times \Oqt \to \field$
given by  $\langle y a, x \sharp h \rangle = \langle y, x \rangle h(a)$, where $y \in U_q(\g_V)\tsrvw U_q(\g_W)$, $a \in \field \S_2$, $x \in \O(GL_q(V))\tsrvw \O(GL_q(W))$, $h \in \F(\S_2)$.
\end{corollary}

Let $\pi : \Oqt \hookrightarrow \O(GL_q(V))\tsrvw \O(GL_q(W))$,  $x \sharp h \mapsto x\epsilon_\cH(h)$ be the canonical surjection (in the notation of Proposition \ref{p semidirect coproduct}).  Recall the homomorphism $\tilde{\psi}: \O(GL_q(\nsbr{X})) \to \O(GL_q(V))\tsrvw \O(GL_q(W)), \nsbr{\mb{z}} \mapsto {\mathbf u^V} \tsrvw
{\mathbf u^W}$ from Proposition \ref{phomomorph GL}.
We now prove the following stronger result:
\begin{proposition} \label{phomomorph tau}
Let  $\O(GL_q(\nsbr{X}))^{1,1}$ be the sub-Hopf algebra $\bigoplus_{r \in \ZZ} \O(GL_q(V))_r \tsrvw \O(GL_q(W))_r$ of  $\O(GL_q(V))\tsrvw \O(GL_q(W))$ and let
$\tau_1$ (resp.  $\tau_2$) be the algebra involution of  $\O(GL_q(\nsbr{X}))^{1,1}$ determined by  $\tau_1(u_i^j \tsrvw u_k^l) = u_k^j \tsrvw u_i^l$
(resp. $\tau_2(u_i^j \tsrvw u_k^l) = u_i^l \tsrvw u_k^j$).
The map
\be
\label{e psi tau definition}
\tilde{\psi}^\tau : \O(GL_q(\nsbr{X})) \to  \Oqt, \ \ z \mapsto  \tilde{\psi}(z) \sharp \dual{e} + \tau_2(\tilde{\psi}(z)) \sharp \dual{\tau}
\ee
is a Hopf algebra homomorphism, and  $\tilde{\psi}$ factors through  $\tilde{\psi}^\tau$ via $\tilde{\psi} = \pi \circ \tilde{\psi}^\tau$.
\end{proposition}
\begin{proof}
In the proof of Proposition \ref{phomomorph}, it shown that
\[
P_-^\nsbr{X} ({\mathbf u^V} \tsrvw  {\mathbf u^W} \tsr {\mathbf u^V} \tsrvw  {\mathbf u^W})=  ({\mathbf u^V} \tsrvw  {\mathbf u^W} \tsr {\mathbf u^V} \tsrvw  {\mathbf u^W}) P_-^\nsbr{X}.
\]
Since  $P_\pm^V$ is sent to  $P_\pm^W$ under the isomorphism  $\End(V^{\tsr 2}) \cong  \End(W^{\tsr 2})$ induced by  $V \cong W$,
 $P_-^\nsbr{X} = P_-^V\tsrvw  P_+^W + P_+^V\tsrvw  P_-^W = P_-^W\tsrvw  P_+^V + P_+^W\tsrvw  P_-^V$.  Hence, there also holds
\[
P_-^\nsbr{X} \tau_2({\mathbf u^V} \tsrvw  {\mathbf u^W} \tsr {\mathbf u^V} \tsrvw  {\mathbf u^W})=  \tau_2({\mathbf u^V} \tsrvw  {\mathbf u^W} \tsr {\mathbf u^V} \tsrvw  {\mathbf u^W}) P_-^\nsbr{X}.
\]
It follows that  $\tilde{\psi}^\tau$ is a well-defined map.  That it is an algebra homomorphism follows directly from the definition \eqref{e psi tau definition} and the fact that $\dual{e}$ and  $\dual{\tau}$ are orthogonal idempotents.

To show that  $\tilde{\psi}^\tau$ is a coalgebra homomorphism, one first checks directly from the definitions that
\be \label{e tau1 tau2}
1 \tsr \tau_2 \circ \Delta  = \tau_2 \tsr \tau_2 \circ \Delta,\ \
\Delta \circ \tau_2 = 1 \tsr \tau_2 \circ \Delta, \ \
1 \tsr \tau \circ \Delta \circ \tau_2 = \tau_2 \tsr 1 \circ \Delta.
\ee
Here  $\Delta$ is the coproduct of $ \O(GL_q(\nsbr{X}))^{1,1}$.  Then, for any $z \in \O(GL_q(\nsbr{X}))$, there holds
\[
\begin{array}{rl}
\Delta \circ \tilde{\psi}^\tau(z)
= & \Delta \big( \tilde{\psi}(z) \sharp \dual{e} + \tau_2(\tilde{\psi}(z)) \sharp \dual{\tau} \big) \\[1.7mm]
= & \sum   \tilde{\psi}(z)_{(1)} \sharp \dual{e} \tsr \tilde{\psi}(z)_{(2)} \sharp \dual{e}  +
 \tilde{\psi}(z)_{(1)} \sharp \dual{\tau} \tsr \tau(\tilde{\psi}(z)_{(2)}) \sharp \dual{\tau}    \\[1.7mm]
& + \ \tau_2(\tilde{\psi}(z))_{(1)} \sharp \dual{e} \tsr \tau_2(\tilde{\psi}(z))_{(2)} \sharp \dual{\tau} +
\tau_2(\tilde{\psi}(z))_{(1)} \sharp \dual{\tau} \tsr \tau(\tau_2(\tilde{\psi}(z))_{(2)}) \sharp \dual{e} \\[1.7mm]
& \text{by \eqref{e tau1 tau2}}\\[1.7mm]
= & \sum   \tilde{\psi}(z)_{(1)} \sharp \dual{e} \tsr \tilde{\psi}(z)_{(2)} \sharp \dual{e}  +
 \tau_2(\tilde{\psi}(z)_{(1)}) \sharp \dual{\tau} \tsr \tau_2(\tilde{\psi}(z)_{(2)}) \sharp \dual{\tau}    \\[1.7mm]
& + \ \tilde{\psi}(z)_{(1)} \sharp \dual{e} \tsr \tau_2(\tilde{\psi}(z)_{(2)}) \sharp \dual{\tau} +
\tau_2(\tilde{\psi}(z)_{(1)}) \sharp \dual{\tau} \tsr \tilde{\psi}(z)_{(2)} \sharp \dual{e}  \\[1.7mm]
& \text{$\tilde{\psi}$ is a coalgebra homomorphism} \\[1.7mm]
= & \sum   \tilde{\psi}(z_{(1)}) \sharp \dual{e} \tsr \tilde{\psi}(z_{(2)}) \sharp \dual{e}  +
 \tau_2(\tilde{\psi}(z_{(1)})) \sharp \dual{\tau} \tsr \tau_2(\tilde{\psi}(z_{(2)})) \sharp \dual{\tau}    \\[1.7mm]
& + \ \tilde{\psi}(z_{(1)}) \sharp \dual{e} \tsr \tau_2(\tilde{\psi}(z_{(2)})) \sharp \dual{\tau} +
\tau_2(\tilde{\psi}(z_{(1)})) \sharp \dual{\tau} \tsr \tilde{\psi}(z_{(2)}) \sharp \dual{e}  \\[1.7mm]
= & \sum \big( \tilde{\psi}(z_{(1)}) \sharp \dual{e} + \tau_2(\tilde{\psi}(z_{(1)})) \sharp \dual{\tau} \big) \otimes \big( \tilde{\psi}(z_{(2)}) \sharp \dual{e} + \tau_2(\tilde{\psi}(z_{(2)})) \sharp \dual{\tau} \big) \\[1.7mm]
= & \sum \tilde{\psi}^\tau z_{(1)} \otimes \tilde{\psi}^\tau z_{(2)} \\[1.7mm]
= & (\tilde{\psi}^\tau \otimes \tilde{\psi}^\tau) \circ \Delta (z).
\end{array}
\]
This proves that $\tilde{\psi}^\tau$ is a coalgebra homomorphism.
The compatibility of  $\tilde{\psi}^\tau$ with the counits is clear.
Thus $\tilde{\psi}^\tau$ is a bialgebra homomorphism.  Since a bialgebra homomorphism of Hopf algebras is always a Hopf algebra homomorphism \cite[\textsection1.2.4]{KS}, the result follows.
\end{proof}

\bibliographystyle{plain}
\bibliography{mycitations}

\end{document}